\documentclass[12pt]{article}
\usepackage{amsmath,amsthm}
\usepackage{eufrak}

\newcommand{\ar}{\renewcommand{\arraystretch}{1}} 
\DeclareMathAlphabet{\bb}{U}{msb}{m}{n}
\gdef\C{\bb C}

\gdef\dS{\bb S}
\gdef\R{\bb R}

\DeclareMathOperator{\spin}{{\bf Spin}}

\DeclareMathOperator{\Sym}{Sym}

\newcommand{\scr}{\scriptstyle}

\newcommand{\re}{\mbox{\rm Re}\,}
\newcommand{\im}{\mbox{\rm Im}\,}

\newcommand{\cP}{{\cal P}}
\newcommand{\cL}{\mathcal{L}}
\newcommand{\cM}{{\cal M}}

\newcommand{\sA}{{\sf A}}
\newcommand{\sB}{{\sf B}}
\newcommand{\sI}{{\sf I}}

\newcommand{\sX}{{\sf X}}
\newcommand{\sY}{{\sf Y}}

\newcommand{\sL}{\Lambda}

\newcommand{\bx}{{\bf x}}
\newcommand{\by}{{\bf y}}
\newcommand{\bz}{{\bf z}}
\newcommand{\bB}{{\bf B}}

\newcommand{\bE}{{\bf E}}

\newcommand{\fM}{\mathfrak{M}}
\newcommand{\fG}{\mathfrak{G}}

\newcommand{\fL}{\mathfrak{L}}

\newcommand{\fZ}{\mathfrak{Z}}

\newcommand{\fg}{\mathfrak{g}}

\newcommand{\balpha}{\boldsymbol{\alpha}}
\newcommand{\cl}{C\kern -0.2em \ell}

\newcommand{\ld}{\left[}
\newcommand{\rd}{\right]}

\newcommand{\tg}{\tan}
\newcommand{\ch}{\cosh}
\newcommand{\sh}{\sinh}
\newcommand{\tnh}{\tanh}

\newcommand{\ctg}{\cot}
\begin{document}
\title{General Solutions of Relativistic Wave Equations II: Arbitrary
Spin Chains}
\author{V. V. Varlamov\\
{\small\it Department of Mathematics, Siberia State
University of Industry,}\\
{\small\it Kirova 42, Novokuznetsk 654007, Russia}}
\date{}
\maketitle
\begin{abstract}
A construction of relativistic wave equations on the homogeneous spaces
of the Poincar\'{e} group is given for arbitrary spin chains.
Parametrizations of the field functions and harmonic analysis
on the homogeneous spaces are studied. It is shown that a direct product
of Minkowski spacetime and two-dimensional complex sphere is the most
suitable homogeneous space for the physical applications.
The Lagrangian formalism and field equations on the Poincar\'{e}
and Lorentz groups
are considered. A boundary value problem for the relativistically
invariant system is defined. General solutions of this problem are
expressed via an expansion in hyperspherical functions defined on the complex
two-sphere.
\end{abstract}

\section{Introduction}
The theory of relativistic wave equations (RWE) is one of the oldest topics
in theoretical physics. As usual, the theory of RWE closely relates with
higher spin formalisms. However, at present there is no a fully
adequate formalism for description of higher--spin fields (all widely
accepted higher--spin formalisms such as Rarita--Schwinger approach \cite{RS41},
Bargmann--Wigner \cite{BW48} and Gel'fand--Yaglom \cite{GY48} multispinor
theories, and also Joos--Weinberg $2(2j+1)$--component formalism
\cite{Joo62,Wein} have many intrinsic contradictions and difficulties).
On the other hand, it is known that the study of RWE leads naturally to the
fields which depend on both Minkowski space coordinates and some continuous
variables corresponding to spin degrees of freedom
\cite{GT47,BW48,Yuk50,Shi51}. Wave functions of this type can be treated
as the fields on homogeneous spaces of the Poincar\'{e} group. These
fields was first studied by Finkelstein \cite{Fin55}, he gave a classification
and explicit constructions of homogeneous spaces of the Poincar\'{e}
group.
The general form of these fields closely
relates with the structure of the Lorentz and Poincar\'{e} group
representations
\cite{GMS,Nai58,BBTD88,GS01}
and admits the following
factorization $f(x,\bz)=\phi^n(\bz)\psi_n(x)$, where $x\in T_4$ and
$\phi^n(\bz)$ form a basis in the representation space of the Lorentz
group, $T_4$ is the group of four-dimensional translations. In addition to
a general theory of relativistic wave equations, a search of solutions
for RWE has a great importance (see, for example, \cite{BG89}).

In the present work we study solutions of RWE in terms of the fields
defined on the two-dimensional complex sphere. This sphere is a
homogeneous space of the Lorentz group (in this case the field functions
$f(x,\bz)$ reduce to $\phi^n(\bz)$).
In the previous work \cite{Var03},
solutions of RWE were obtained in the form of expansions in
hyperspherical functions for the representations of the type
$(l,0)\oplus(0,\dot{l})$. Solutions (wavefunctions on the group manifold
$\cM_{10}$) of the simplest RWE (Dirac $(1/2,0)\oplus(0,1/2)$ and
Maxwell $(1,0)\oplus(0,1)$ fields) have been given in \cite{Var03c,Var03d}.
In this work we study more general representations
$(l_1,\dot{l}_2)\oplus(l_2,\dot{l}_1)$ (tensor representations of the
proper orthocronous Lorentz group $\fG_+$) which correspond to
arbitrary spin chains. In turn, arbitrary spin chains contain interlocking
representations of the group $\fG_+$ as a particular case.
It is known that the higher spin fields and correspondingly composite
elementary particles can be formulated in terms of interlocking
representations of $\fG_+$ (Bhabha-Gel'fand-Yaglom chains). The stable
composite particle corresponds to an indecomposable equation defined
within interlocking representation, and vice versa, unstable particles
are described by decomposable equations.

The present paper is organized as follows.
In the section 2 we consider some basic facts concerning the group
$SL(2,\C)$. The main object of this section is a local isomorphism
$SL(2,\C)\simeq SU(2)\otimes SU(2)$. We introduce here hyperspherical
functions for the tensor representations and further we study
recurrence relations between these functions. It is shown that matrix
elements of the representations $(l_1,\dot{l}_2)\oplus(l_2,\dot{l}_1)$
are expressed via the hyperspherical functions. In the section 3 we give
a brief introduction to the fields on the Poincar\'{e} group $\cP$.
We consider fields on the group manifold $\cM_{10}$ and on the
eight-dimensional homogeneous space $\cM_8=\R^{1,3}\times\dS^2$ of $\cP$,
where $\R^{1,3}$ is the Minkowski spacetime and $\dS^2$ is the
two-dimensional complex sphere. In parallel, we consider basic facts concerning
harmonic analysis on the homogeneous spaces $\cM_{10}$, $\cM_8$, and also
on the group $SL(2,\C)$ and on the sphere $\dS^2$.
The following logical step consists in definition of the Lagrangian
formalism and field equations on the homogeneous spaces of $\cP$.
The field equations for arbitrary spin are derived in the section 4
by the standard variation
procedure from a selected Lagrangian. The explicit construction of RWE
for the arbitrary spin chains of $\fG_+$ is given in the section 5.
In the section 6 we separate variables in a relativistically invariant
system via the recurrence relations between hyperspherical functions.
First of all, we set up a boundary value problem for RWE, which can be
treated as a Dirichlet problem for a complex ball.
It is shown that solutions of this
problem are expressed via Fourier type series on the two-dimensional complex
sphere.

\section{The group $SL(2,\C)$}\label{SL2C}
As is known, the group $SL(2,\C)$ is an universal covering of the proper
orthochronous Lorentz group $\fG_+$.
The group $SL(2,\C)$ of all complex matrices
\[\ar
\fg=
\begin{pmatrix}
\alpha & \beta\\
\gamma & \delta
\end{pmatrix}
\]
of 2-nd order with the determinant $\alpha\delta-\gamma\beta=1$, is
a {\it complexification} of the group $SU(2)$. The group $SU(2)$ is one of
the real forms of $SL(2,\C)$. The transition from $SU(2)$ to $SL(2,\C)$
is realized via the complexification of three real parameters
$\varphi,\,\theta,\,\psi$ (Euler angles). Let $\theta^c=\theta-i\tau$,
$\varphi^c=\varphi-i\epsilon$, $\psi^c=\psi-i\varepsilon$ be complex
Euler angles,\index{Euler angles!complex} where
\begin{equation}\label{CEA}
{\renewcommand{\arraystretch}{1.05}
\begin{array}{ccccc}
0 &\leq&\re\theta^c=\theta& \leq& \pi,\\
0 &\leq&\re\varphi^c=\varphi& <&2\pi,\\
-2\pi&\leq&\re\psi^c=\psi&<&2\pi,
\end{array}\quad\quad
\begin{array}{ccccc}
-\infty &<&\im\theta^c=\tau&<&+\infty,\\
-\infty&<&\im\varphi^c=\epsilon&<&+\infty,\\
-\infty&<&\im\psi^c=\varepsilon&<&+\infty.
\end{array}}
\end{equation}
The group $SL(2,\C)$ has six one-parameter subgroups
\begin{gather}
a_1(t)=\begin{pmatrix}
\cos\frac{t}{2} & i\sin\frac{t}{2}\\
i\sin\frac{t}{2} & \cos\frac{t}{2}
\end{pmatrix},\quad a_2(t)=\begin{pmatrix}
\cos\frac{t}{2} & -\sin\frac{t}{2}\\
\sin\frac{t}{2} & \cos\frac{t}{2}
\end{pmatrix},\quad a_3(t)=\begin{pmatrix}
e^{\frac{it}{2}} & 0\\
0 & e^{-\frac{it}{2}}
\end{pmatrix},\nonumber\\
b_1(t)=\begin{pmatrix}
\ch\frac{t}{2} & \sh\frac{t}{2}\\
\sh\frac{t}{2} & \ch\frac{t}{2}
\end{pmatrix},\quad b_2(t)=\begin{pmatrix}
\ch\frac{t}{2} & i\sh\frac{t}{2}\\
-i\sh\frac{t}{2} & \ch\frac{t}{2}
\end{pmatrix},\quad b_3(t)=\begin{pmatrix}
e^{\frac{t}{2}} & 0\\
0 & e^{-\frac{t}{2}}
\end{pmatrix}.\nonumber
\end{gather}
The tangent matrices $A_i$ and $B_i$ of these subgroups are defined
as follows
\begin{eqnarray}
A_1&=&\left.\frac{d a_1(t)}{dt}\right|_{t=0}=\frac{i}{2}\begin{pmatrix}
0 & 1\\
1 & 0
\end{pmatrix},\nonumber\\
A_2&=&\left.\frac{d a_2(t)}{dt}\right|_{t=0}=\frac{1}{2}\begin{pmatrix}
0 & -1\\
1 & 0
\end{pmatrix},\nonumber\\
A_3&=&\left.\frac{d a_3(t)}{dt}\right|_{t=0}=\frac{i}{2}\begin{pmatrix}
1 & 0\\
0 & -1
\end{pmatrix},\nonumber\\
B_1&=&\left.\frac{d b_1(t)}{dt}\right|_{t=0}=\frac{1}{2}\begin{pmatrix}
0 & 1\\
1 & 0
\end{pmatrix},\nonumber\\
B_2&=&\left.\frac{d b_2(t)}{dt}\right|_{t=0}=\frac{i}{2}\begin{pmatrix}
0 & 1\\
-1 & 0
\end{pmatrix},\nonumber\\
B_3&=&\left.\frac{d b_3(t)}{dt}\right|_{t=0}=\frac{1}{2}\begin{pmatrix}
1 & 0\\
0 &-1
\end{pmatrix}.\nonumber
\end{eqnarray}
The elements $\sA_i$ and $\sB_i$ form a basis of Lie algebra
$\mathfrak{sl}(2,\C)$ and satisfy the relations
\begin{equation}\label{Com1}
\left.\begin{array}{lll}
\ld\sA_1,\sA_2\rd=\sA_3, & \ld\sA_2,\sA_3\rd=\sA_1, &
\ld\sA_3,\sA_1\rd=\sA_2,\\[0.1cm]
\ld\sB_1,\sB_2\rd=-\sA_3, & \ld\sB_2,\sB_3\rd=-\sA_1, &
\ld\sB_3,\sB_1\rd=-\sA_2,\\[0.1cm]
\ld\sA_1,\sB_1\rd=0, & \ld\sA_2,\sB_2\rd=0, &
\ld\sA_3,\sB_3\rd=0,\\[0.1cm]
\ld\sA_1,\sB_2\rd=\sB_3, & \ld\sA_1,\sB_3\rd=-\sB_2, & \\[0.1cm]
\ld\sA_2,\sB_3\rd=\sB_1, & \ld\sA_2,\sB_1\rd=-\sB_3, & \\[0.1cm]
\ld\sA_3,\sB_1\rd=\sB_2, & \ld\sA_3,\sB_2\rd=-\sB_1. &
\end{array}\right\}
\end{equation}
Let us consider the operators
\begin{gather}
\sX_l=\frac{1}{2}i(\sA_l+i\sB_l),\quad\sY_l=\frac{1}{2}i(\sA_l-i\sB_l),
\label{SL25}\\
(l=1,2,3).\nonumber
\end{gather}
Using the relations (\ref{Com1}), we find that
\begin{equation}\label{Com2}
\ld\sX_k,\sX_l\rd=i\varepsilon_{klm}\sX_m,\quad
\ld\sY_l,\sY_m\rd=i\varepsilon_{lmn}\sY_n,\quad
\ld\sX_l,\sY_m\rd=0.
\end{equation}
Further, introducing generators of the form
\begin{equation}\label{SL26}
\left.\begin{array}{cc}
\sX_+=\sX_1+i\sX_2, & \sX_-=\sX_1-i\sX_2,\\[0.1cm]
\sY_+=\sY_1+i\sY_2, & \sY_-=\sY_1-i\sY_2,
\end{array}\right\}
\end{equation}
we see that in virtue of commutativity of the relations (\ref{Com2}) a
space of an irreducible finite--dimensional representation of the group
$SL(2,\C)$ can be spanned on the totality of $(2l+1)(2\dot{l}+1)$ basis
vectors $\mid l,m;\dot{l},\dot{m}\rangle$, where $l,m,\dot{l},\dot{m}$
are integer or half--integer numbers, $-l\leq m\leq l$,
$-\dot{l}\leq \dot{m}\leq \dot{l}$. Therefore,
\begin{eqnarray}
&&\sX_-\mid l,m;\dot{l},\dot{m}\rangle=
\sqrt{(l+m)(l-m+1)}\mid l,m-1,\dot{l},\dot{m}\rangle
\;\;(m>-l),\nonumber\\
&&\sX_+\mid l,m;\dot{l},\dot{m}\rangle=
\sqrt{(l-m)(l+m+1)}\mid l,m+1;\dot{l},\dot{m}\rangle
\;\;(m<l),\nonumber\\
&&\sX_3\mid l,m;\dot{l},\dot{m}\rangle=
m\mid l,m;\dot{l},\dot{m}\rangle,\nonumber\\
&&\sY_-\mid l,m;\dot{l},\dot{m}\rangle=
\sqrt{(\dot{l}+\dot{m})(\dot{l}-\dot{m}+1)}\mid l,m;\dot{l},\dot{m}-1
\rangle\;\;(\dot{m}>-\dot{l}),\nonumber\\
&&\sY_+\mid l,m;\dot{l},\dot{m}\rangle=
\sqrt{(\dot{l}-\dot{m})(\dot{l}+\dot{m}+1)}\mid l,m;\dot{l},\dot{m}+1
\rangle\;\;(\dot{m}<\dot{l}),\nonumber\\
&&\sY_3\mid l,m;\dot{l},\dot{m}\rangle=
\dot{m}\mid l,m;\dot{l},\dot{m}\rangle.\label{Waerden}
\end{eqnarray}
From the relations (\ref{Com2}) it follows that each of the sets of
infinitesimal operators $\sX$ and $\sY$ generates the group $SU(2)$ and these
two groups commute with each other. Thus, from the relations (\ref{Com2})
and (\ref{Waerden}) it follows that the group $SL(2,\C)$, in essence,
is equivalent locally to the group $SU(2)\otimes SU(2)$.
It should be noted here that the representation basis, defined by the
formulae (\ref{SL25})--(\ref{Waerden}), has an evident physical meaning.
For example, in the case of $(1,0)\oplus(0,1)$--representation space
there is an analogy with the photon spin states. Namely, the operators
$\sX$ and $\sY$ correspond to the right and left polarization states of the
photon. For that reason we will call the canonical basis consisting of the
vectors $\mid lm;\dot{l}\dot{m}\rangle$ as
{\it a helicity basis}.

In the work \cite{Var03} we studied RWE for the fields
$(l,0)\oplus(0,\dot{l})$ which described within the representation
$\boldsymbol{\tau}_{l,0}\oplus\boldsymbol{\tau}_{0,\dot{l}}$ of the
group $\fG_+$. In this work we study more general representations of the
type $\boldsymbol{\tau}_{l,\dot{l}}\oplus\boldsymbol{\tau}_{\dot{l},l}$
(tensor representations),
where $\boldsymbol{\tau}_{l,\dot{l}}=\boldsymbol{\tau}_{l,0}\otimes
\boldsymbol{\tau}_{0,\dot{l}}$ and
$\boldsymbol{\tau}_{\dot{l},l}=\boldsymbol{\tau}_{\dot{l},0}\otimes
\boldsymbol{\tau}_{0,l}$. It is obvious that these representations include
$\boldsymbol{\tau}_{l,0}\oplus\boldsymbol{\tau}_{0,\dot{l}}$ as a
particular case. In general, a tensor structure of the infinitesimal
operators has the form
\begin{eqnarray}
\sA^{l\dot{l}}_i&=&\sA^l_i\otimes\boldsymbol{1}_{2\dot{l}+1}-
\boldsymbol{1}_{2l+1}\otimes\sA^{\dot{l}}_i,\nonumber\\
\sB^{l\dot{l}}_i&=&\sB^l_i\otimes\boldsymbol{1}_{2\dot{l}+1}-
\boldsymbol{1}_{2l+1}\otimes\sB^{\dot{l}}_i,\nonumber\\
\widetilde{\sA}^{l\dot{l}}_i&=&\sA^{\dot{l}}_i\otimes\boldsymbol{1}_{2l+1}-
\boldsymbol{1}_{2\dot{l}+1}\otimes\sA^l_i,\nonumber\\
\widetilde{\sB}^{l\dot{l}}_i&=&\sB^{\dot{l}}_i\otimes\boldsymbol{1}_{2l+1}-
\boldsymbol{1}_{2\dot{l}+1}\otimes\sB^l_i,\label{TenStr}
\end{eqnarray}
where $\sA^l_i$, $\sB^l_i$ and $\sA^{\dot{l}}_i$, $\sB^{\dot{l}}_i$ are
infinitesimal operators of the representations $\boldsymbol{\tau}_{l,0}$
and $\boldsymbol{\tau}_{0,\dot{l}}$, respectively ($i=1,2,3$).
Or, more explicitly,
\begin{multline}
\sA^{l\dot{l}}_1\mid l,m;\dot{l},\dot{m}\rangle=
-\frac{i}{2}\balpha^l_m\mid l,m-1;\dot{l},\dot{m}\rangle
-\frac{i}{2}\balpha^l_{m+1}\mid l,m+1;\dot{l},\dot{m}\rangle+\\
+\frac{i}{2}\balpha^{\dot{l}}_{\dot{m}}\mid l,m;\dot{l},\dot{m}-1\rangle
+\frac{i}{2}\balpha^{\dot{l}}_{\dot{m}+1}\mid l,m;\dot{l},\dot{m}+1\rangle,
\label{Op1}
\end{multline}
\begin{multline}
\sA^{l\dot{l}}_2\mid l,m;\dot{l},\dot{m}\rangle=
\frac{1}{2}\balpha^l_m\mid l,m-1;\dot{l},\dot{m}\rangle
-\frac{1}{2}\balpha^l_{m+1}\mid l,m+1;\dot{l},\dot{m}\rangle-\\
-\frac{1}{2}\balpha^{\dot{l}}_{\dot{m}}\mid l,m;\dot{l},\dot{m}-1\rangle
+\frac{1}{2}\balpha^{\dot{l}}_{\dot{m}+1}\mid l,m;\dot{l},\dot{m}+1\rangle,
\label{Op2}
\end{multline}
\begin{equation}\label{Op3}
\sA^{l\dot{l}}_3\mid l,m;\dot{l},\dot{m}\rangle=
-im\mid l,m;\dot{l},\dot{m}\rangle
+i\dot{m}\mid l,m;\dot{l},\dot{m}\rangle,
\end{equation}
\begin{multline}
\sB^{l\dot{l}}_1\mid l,m;\dot{l},\dot{m}\rangle=
-\frac{1}{2}\balpha^l_m\mid l,m-1;\dot{l},\dot{m}\rangle
-\frac{1}{2}\balpha^l_{m+1}\mid l,m+1;\dot{l},\dot{m}\rangle-\\
-\frac{1}{2}\balpha^{\dot{l}}_{\dot{m}}\mid l,m;\dot{l},\dot{m}-1\rangle
-\frac{1}{2}\balpha^{\dot{l}}_{\dot{m}+1}\mid l,m;\dot{l},\dot{m}+1\rangle,
\label{Op4}
\end{multline}
\begin{multline}
\sB^{l\dot{l}}_2\mid l,m;\dot{l},\dot{m}\rangle=
-\frac{i}{2}\balpha^l_m\mid l,m-1;\dot{l},\dot{m}\rangle
+\frac{i}{2}\balpha^l_{m+1}\mid l,m+1;\dot{l},\dot{m}\rangle-\\
-\frac{i}{2}\balpha^{\dot{l}}_{\dot{m}}\mid l,m;\dot{l},\dot{m}-1\rangle
+\frac{i}{2}\balpha^{\dot{l}}_{\dot{m}+1}\mid l,m;\dot{l},\dot{m}+1\rangle,
\label{Op5}
\end{multline}
\begin{equation}\label{Op6}
\sB^{l\dot{l}}_3\mid l,m;\dot{l},\dot{m}\rangle=
-m\mid l,m;\dot{l},\dot{m}\rangle
-\dot{m}\mid l,m;\dot{l},\dot{m}\rangle,
\end{equation}
\begin{multline}
\widetilde{\sA}^{l\dot{l}}_1\mid l,m;\dot{l},\dot{m}\rangle=
-\frac{i}{2}\balpha^{\dot{l}}_{\dot{m}}\mid l,m;\dot{l},\dot{m}-1\rangle
-\frac{i}{2}\balpha^{\dot{l}}_{\dot{m}+1}\mid l,m;\dot{l},\dot{m}+1\rangle+\\
+\frac{i}{2}\balpha^l_m\mid l,m-1;\dot{l},\dot{m}\rangle
+\frac{i}{2}\balpha^l_{m+1}\mid l,m+1;\dot{l},\dot{m}\rangle,
\label{Op7}
\end{multline}
\begin{multline}
\widetilde{\sA}^{l\dot{l}}_2\mid l,m;\dot{l},\dot{m}\rangle=
\frac{1}{2}\balpha^{\dot{l}}_{\dot{m}}\mid l,m;\dot{l},\dot{m}-1\rangle
-\frac{1}{2}\balpha^{\dot{l}}_{\dot{m}+1}\mid l,m;\dot{l},\dot{m}+1\rangle+\\
+\frac{1}{2}\balpha^l_m\mid l,m-1;\dot{l},\dot{m}\rangle
+\frac{1}{2}\balpha^l_{m+1}\mid l,m+1;\dot{l},\dot{m}\rangle,
\label{Op8}
\end{multline}
\begin{equation}\label{Op9}
\widetilde{\sA}^{l\dot{l}}_3\mid l,m;\dot{l},\dot{m}\rangle=
-i\dot{m}\mid l,m;\dot{l},\dot{m}\rangle
+im\mid l,m;\dot{l},\dot{m}\rangle,
\end{equation}
\begin{multline}
\widetilde{\sB}^{l\dot{l}}_1\mid l,m;\dot{l},\dot{m}\rangle=
\frac{1}{2}\balpha^{\dot{l}}_{\dot{m}}\mid l,m;\dot{l},\dot{m}-1\rangle
+\frac{1}{2}\balpha^{\dot{l}}_{\dot{m}+1}\mid l,m;\dot{l},\dot{m}+1\rangle+\\
+\frac{1}{2}\balpha^l_m\mid l,m-1;\dot{l},\dot{m}\rangle
+\frac{1}{2}\balpha^l_{m+1}\mid l,m+1;\dot{l},\dot{m}\rangle,
\label{Op10}
\end{multline}
\begin{multline}
\widetilde{\sB}^{l\dot{l}}_2\mid l,m;\dot{l},\dot{m}\rangle=
\frac{i}{2}\balpha^{\dot{l}}_{\dot{m}}\mid l,m;\dot{l},\dot{m}-1\rangle
-\frac{i}{2}\balpha^{\dot{l}}_{\dot{m}+1}\mid l,m;\dot{l},\dot{m}+1\rangle+\\
+\frac{i}{2}\balpha^l_m\mid l,m-1;\dot{l},\dot{m}\rangle
-\frac{i}{2}\balpha^l_{m+1}\mid l,m+1;\dot{l},\dot{m}\rangle,
\label{Op11}
\end{multline}
\begin{equation}\label{Op12}
\widetilde{\sB}^{l\dot{l}}_3\mid l,m;\dot{l},\dot{m}\rangle=
\dot{m}\mid l,m;\dot{l},\dot{m}\rangle
+m\mid l,m;\dot{l},\dot{m}\rangle.
\end{equation}
Further, let us define generators $\sX^{l\dot{l}}_-$, $\sX^{l\dot{l}}_+$,
$\sY^{l\dot{l}}_-$, $\sY^{l\dot{l}}_+$ and $\widetilde{\sX}^{l\dot{l}}_-$,
$\widetilde{\sX}^{l\dot{l}}_+$, $\widetilde{\sY}^{l\dot{l}}_-$,
$\widetilde{\sY}^{l\dot{l}}_+$ following the rule (\ref{SL26}).
It is easy to verify that these generators satisfy the relations of the
type (\ref{Waerden}).

On the group $SL(2,\C)$ there exist the following Laplace-Beltrami operators:
\begin{eqnarray}
\sX^2&=&\sX^2_1+\sX^2_2+\sX^2_3=\frac{1}{4}(\sA^2-\sB^2+2i\sA\sB),\nonumber\\
\sY^2&=&\sY^2_1+\sY^2_2+\sY^2_3=
\frac{1}{4}(\widetilde{\sA}^2-\widetilde{\sB}^2-
2i\widetilde{\sA}\widetilde{\sB}).\label{KO}
\end{eqnarray}
At this point, we see that operators (\ref{KO}) contain the well known
Casimir operators $\sA^2-\sB^2$, $\sA\sB$ of the Lorentz group.
Using expressions (\ref{CEA}), we obtain
an Euler parametrization of the Laplace-Beltrami operators,
\begin{eqnarray}
\sX^2&=&\frac{\partial^2}{\partial\theta^c{}^2}+
\ctg\theta^c\frac{\partial}{\partial\theta^c}+\frac{1}{\sin^2\theta^c}\left[
\frac{\partial^2}{\partial\varphi^c{}^2}-
2\cos\theta^c\frac{\partial}{\partial\varphi^c}
\frac{\partial}{\partial\psi^c}+
\frac{\partial^2}{\partial\psi^c{}^2}\right],\nonumber\\
\sY^2&=&\frac{\partial^2}{\partial\dot{\theta}^c{}^2}+
\ctg\dot{\theta}^c\frac{\partial}{\partial\dot{\theta}^c}+
\frac{1}{\sin^2\dot{\theta}^c}\left[
\frac{\partial^2}{\partial\dot{\varphi}^c{}^2}-
2\cos\dot{\theta}^c\frac{\partial}{\partial\dot{\varphi}^c}
\frac{\partial}{\partial\dot{\psi}^c}+
\frac{\partial^2}{\partial\dot{\psi}^c{}^2}\right].\label{KO2}
\end{eqnarray}
Matrix elements $t^{l\dot{l}}_{mn;\dot{m}\dot{n}}(\fg)=
\fM^{l\dot{l}}_{mn;\dot{m}\dot{n}}(\varphi^c,\theta^c,\psi^c)$ of
irreducible representations of $SL(2,\C)$ are
eigenfunctions of the operators (\ref{KO2}),
\begin{eqnarray}
\left[\sX^2+l(l+1)\right]\fM^{l\dot{l}}_{mn;\dot{m}\dot{n}}(\varphi^c,\theta^c,\psi^c)&=&0,\nonumber\\
\left[\sY^2+\dot{l}(\dot{l}+1)\right]\fM^{l\dot{l}}_{mn;\dot{m}\dot{n}}
(\dot{\varphi}^c,\dot{\theta}^c,\dot{\psi}^c)&=&0,\label{EQ}
\end{eqnarray}
where
\begin{equation}\label{HF3'}
\fM^{l\dot{l}}_{mn;\dot{m}\dot{n}}(\fg)=
e^{-i(m\varphi^c+n\psi^c)}\fZ^{l\dot{l}}_{mn;\dot{m}\dot{n}}
(\cos\theta^c,\cos\dot{\theta}^c)e^{i(\dot{m}\dot{\varphi}^c+
\dot{n}\dot{\psi}^c)}.
\end{equation}
Substituting the functions (\ref{HF3'}) into (\ref{EQ}) and
taking into account the operators (\ref{KO2}) and
substitutions $z=\cos\theta^c$,
$\overset{\ast}{z}=\cos\dot{\theta}^c$, we come to the following
differential equations (a complex analog of the Legendre equations):
\begin{eqnarray}
\left[(1-z^2)\frac{d^2}{dz^2}-2z\frac{d}{dz}-
\frac{m^2+n^2-2mnz}{1-z^2}+l(l+1)\right]
\fZ^{l\dot{l}}_{mn;\dot{m}\dot{n}}(z,\overset{\ast}{z})&=&0,\label{Leg1}\\
\left[(1-\overset{\ast}{z}{}^2)\frac{d^2}{d\overset{\ast}{z}{}^2}-
2\overset{\ast}{z}\frac{d}{d\overset{\ast}{z}}-
\frac{\dot{m}^2+\dot{n}^2-2\dot{m}\dot{n}\overset{\ast}{z}}
{1-\overset{\ast}{z}{}^2}+\dot{l}(\dot{l}+1)\right]
\fZ^{l\dot{l}}_{mn;\dot{m}\dot{n}}(z,\overset{\ast}{z})&=&0.\label{Leg2}
\end{eqnarray}
The latter equations have three singular points $-1$, $+1$, $\infty$.
Solutions of (\ref{Leg1})-(\ref{Leg2}) have the form
\begin{equation}
\fZ^{l\dot{l}}_{mn;\dot{m}\dot{n}}(\cos\theta^c,\cos\dot{\theta}^c)=
Z^l_{mn}(\cos\theta^c)Z^{\dot{l}}_{\dot{m}\dot{n}}(\cos\dot{\theta}^c),
\label{HS}
\end{equation}
where
\begin{multline}
Z^l_{mn}(\cos\theta^c)=
\sum^l_{k=-l}i^{m-k}
\sqrt{\Gamma(l-m+1)\Gamma(l+m+1)\Gamma(l-k+1)\Gamma(l+k+1)}\times\\
\cos^{2l}\frac{\theta}{2}\tg^{m-k}\frac{\theta}{2}\times\\[0.2cm]
\sum^{\min(l-m,l+k)}_{j=\max(0,k-m)}
\frac{i^{2j}\tg^{2j}\dfrac{\theta}{2}}
{\Gamma(j+1)\Gamma(l-m-j+1)\Gamma(l+k-j+1)\Gamma(m-k+j+1)}\times\\[0.2cm]
\sqrt{\Gamma(l-n+1)\Gamma(l+n+1)\Gamma(l-k+1)\Gamma(l+k+1)}
\ch^{2l}\frac{\tau}{2}\tnh^{n-k}\frac{\tau}{2}\times\\[0.2cm]
\sum^{\min(l-n,l+k)}_{s=\max(0,k-n)}
\frac{\tnh^{2s}\dfrac{\tau}{2}}
{\Gamma(s+1)\Gamma(l-n-s+1)\Gamma(l+k-s+1)\Gamma(n-k+s+1)}.\label{HS0}
\end{multline}
and $Z^{\dot{l}}_{\dot{m}\dot{n}}$ is a complex conjugate function with
respect to $Z^l_{mn}$.
We will call the functions $\fZ^{l\dot{l}}_{mn;\dot{m}\dot{n}}$ in (\ref{HS}) as
{\it tensor hyperspherical functions}.

Using the formula (\ref{HS}), we find explicit expressions for the
matrices $T_{l\dot{l}}(\theta,\tau)$ of the finite--dimensional representations
$\boldsymbol{\tau}_{\frac{1}{2}\frac{1}{2}}$ and
$\boldsymbol{\tau}_{1\frac{1}{2}}$:
\begin{gather}
T_{\frac{1}{2}\frac{1}{2}}(\theta,\tau)=\ar\begin{pmatrix}
Z^{\frac{1}{2}}_{-\frac{1}{2}-\frac{1}{2}} &
Z^{\frac{1}{2}}_{\frac{1}{2}-\frac{1}{2}} \\
Z^{\frac{1}{2}}_{-\frac{1}{2}\frac{1}{2}} &
Z^{\frac{1}{2}}_{\frac{1}{2}\frac{1}{2}}
\end{pmatrix}\otimes
\begin{pmatrix}
\overset{\ast}{Z}^{\frac{1}{2}}_{-\frac{1}{2}-\frac{1}{2}} &
\overset{\ast}{Z}^{\frac{1}{2}}_{\frac{1}{2}-\frac{1}{2}} \\
\overset{\ast}{Z}^{\frac{1}{2}}_{-\frac{1}{2}\frac{1}{2}} &
\overset{\ast}{Z}^{\frac{1}{2}}_{\frac{1}{2}\frac{1}{2}}
\end{pmatrix}=\nonumber\\
\ar\begin{pmatrix}
\fZ^{\frac{1}{2}\frac{1}{2}}_{-\frac{1}{2}-\frac{1}{2};-\frac{1}{2}-\frac{1}{2}}&
\fZ^{\frac{1}{2}\frac{1}{2}}_{-\frac{1}{2}-\frac{1}{2};\frac{1}{2}-\frac{1}{2}}&
\fZ^{\frac{1}{2}\frac{1}{2}}_{\frac{1}{2}-\frac{1}{2};-\frac{1}{2}-\frac{1}{2}}&
\fZ^{\frac{1}{2}\frac{1}{2}}_{\frac{1}{2}-\frac{1}{2};\frac{1}{2}-\frac{1}{2}}\\
\fZ^{\frac{1}{2}\frac{1}{2}}_{-\frac{1}{2}-\frac{1}{2};-\frac{1}{2}\frac{1}{2}}&
\fZ^{\frac{1}{2}\frac{1}{2}}_{-\frac{1}{2}-\frac{1}{2};\frac{1}{2}\frac{1}{2}}&
\fZ^{\frac{1}{2}\frac{1}{2}}_{\frac{1}{2}-\frac{1}{2};-\frac{1}{2}\frac{1}{2}}&
\fZ^{\frac{1}{2}\frac{1}{2}}_{\frac{1}{2}-\frac{1}{2};\frac{1}{2}\frac{1}{2}}\\
\fZ^{\frac{1}{2}\frac{1}{2}}_{-\frac{1}{2}\frac{1}{2};-\frac{1}{2}-\frac{1}{2}}&
\fZ^{\frac{1}{2}\frac{1}{2}}_{-\frac{1}{2}\frac{1}{2};\frac{1}{2}-\frac{1}{2}}&
\fZ^{\frac{1}{2}\frac{1}{2}}_{\frac{1}{2}\frac{1}{2};-\frac{1}{2}-\frac{1}{2}}&
\fZ^{\frac{1}{2}\frac{1}{2}}_{\frac{1}{2}\frac{1}{2};\frac{1}{2}-\frac{1}{2}}\\
\fZ^{\frac{1}{2}\frac{1}{2}}_{-\frac{1}{2}\frac{1}{2};-\frac{1}{2}\frac{1}{2}}&
\fZ^{\frac{1}{2}\frac{1}{2}}_{-\frac{1}{2}\frac{1}{2};\frac{1}{2}\frac{1}{2}}&
\fZ^{\frac{1}{2}\frac{1}{2}}_{\frac{1}{2}\frac{1}{2};-\frac{1}{2}\frac{1}{2}}&
\fZ^{\frac{1}{2}\frac{1}{2}}_{\frac{1}{2}\frac{1}{2};\frac{1}{2}\frac{1}{2}}
\end{pmatrix},
\label{Tenrep1}
\end{gather}
\[
T_{1\frac{1}{2}}(\theta,\tau)=
\ar\begin{pmatrix}
Z^1_{-1-1} & Z^1_{-10} & Z^1_{-11}\\
Z^1_{0-1} & Z^1_{00} & Z^1_{01}\\
Z^1_{1-1} & Z^1_{10} & Z^1_{11}
\end{pmatrix}\otimes
\begin{pmatrix}
\overset{\ast}{Z}^{\frac{1}{2}}_{-\frac{1}{2}-\frac{1}{2}} &
\overset{\ast}{Z}^{\frac{1}{2}}_{\frac{1}{2}-\frac{1}{2}} \\
\overset{\ast}{Z}^{\frac{1}{2}}_{-\frac{1}{2}\frac{1}{2}} &
\overset{\ast}{Z}^{\frac{1}{2}}_{\frac{1}{2}\frac{1}{2}}
\end{pmatrix}=
\]
\begin{equation}\label{Tenrep2}
\ar\begin{pmatrix}
\fZ^{1\frac{1}{2}}_{-1-1;-\frac{1}{2}-\frac{1}{2}} &
\fZ^{1\frac{1}{2}}_{-1-1;\frac{1}{2}-\frac{1}{2}} &
\fZ^{1\frac{1}{2}}_{-10;-\frac{1}{2}-\frac{1}{2}} &
\fZ^{1\frac{1}{2}}_{-10;\frac{1}{2}-\frac{1}{2}} &
\fZ^{1\frac{1}{2}}_{-11;-\frac{1}{2}-\frac{1}{2}} &
\fZ^{1\frac{1}{2}}_{-11;\frac{1}{2}-\frac{1}{2}} \\
\fZ^{1\frac{1}{2}}_{-1-1;-\frac{1}{2}\frac{1}{2}} &
\fZ^{1\frac{1}{2}}_{-1-1;\frac{1}{2}\frac{1}{2}} &
\fZ^{1\frac{1}{2}}_{-10;-\frac{1}{2}\frac{1}{2}} &
\fZ^{1\frac{1}{2}}_{-10;\frac{1}{2}\frac{1}{2}} &
\fZ^{1\frac{1}{2}}_{-11;-\frac{1}{2}\frac{1}{2}} &
\fZ^{1\frac{1}{2}}_{-11;\frac{1}{2}\frac{1}{2}} \\
\fZ^{1\frac{1}{2}}_{0-1;-\frac{1}{2}-\frac{1}{2}} &
\fZ^{1\frac{1}{2}}_{0-1;\frac{1}{2}-\frac{1}{2}} &
\fZ^{1\frac{1}{2}}_{00;-\frac{1}{2}-\frac{1}{2}} &
\fZ^{1\frac{1}{2}}_{00;\frac{1}{2}-\frac{1}{2}} &
\fZ^{1\frac{1}{2}}_{01;-\frac{1}{2}-\frac{1}{2}} &
\fZ^{1\frac{1}{2}}_{01;\frac{1}{2}-\frac{1}{2}} \\
\fZ^{1\frac{1}{2}}_{0-1;-\frac{1}{2}\frac{1}{2}} &
\fZ^{1\frac{1}{2}}_{0-1;\frac{1}{2}\frac{1}{2}} &
\fZ^{1\frac{1}{2}}_{00;-\frac{1}{2}\frac{1}{2}} &
\fZ^{1\frac{1}{2}}_{00;\frac{1}{2}\frac{1}{2}} &
\fZ^{1\frac{1}{2}}_{01;-\frac{1}{2}\frac{1}{2}} &
\fZ^{1\frac{1}{2}}_{01;\frac{1}{2}\frac{1}{2}} \\
\fZ^{1\frac{1}{2}}_{1-1;-\frac{1}{2}-\frac{1}{2}} &
\fZ^{1\frac{1}{2}}_{1-1;\frac{1}{2}-\frac{1}{2}} &
\fZ^{1\frac{1}{2}}_{10;-\frac{1}{2}-\frac{1}{2}} &
\fZ^{1\frac{1}{2}}_{10;\frac{1}{2}-\frac{1}{2}} &
\fZ^{1\frac{1}{2}}_{11;-\frac{1}{2}-\frac{1}{2}} &
\fZ^{1\frac{1}{2}}_{11;\frac{1}{2}-\frac{1}{2}} \\
\fZ^{1\frac{1}{2}}_{1-1;-\frac{1}{2}\frac{1}{2}} &
\fZ^{1\frac{1}{2}}_{1-1;\frac{1}{2}\frac{1}{2}} &
\fZ^{1\frac{1}{2}}_{10;-\frac{1}{2}\frac{1}{2}} &
\fZ^{1\frac{1}{2}}_{10;\frac{1}{2}\frac{1}{2}} &
\fZ^{1\frac{1}{2}}_{11;-\frac{1}{2}\frac{1}{2}} &
\fZ^{1\frac{1}{2}}_{11;\frac{1}{2}\frac{1}{2}}
\end{pmatrix}.
\end{equation}

%
\subsection{Recurrence relations between the functions
$\fZ^{l\dot{l}}_{mn;\dot{m}\dot{n}}(\cos\theta^c,\cos\dot{\theta}^c)$}
Between the hyperspherical functions $\fZ^{l\dot{l}}_{mn;\dot{m}\dot{n}}$
there exists a wide variety of recurrence relations. The part of them
relates the hyperspherical functions of one and the same order
(with identical $l$ and $\dot{l}$), other part relates the functions of
different orders.

In virtue of the representation (\ref{Waerden}), the recurrence formulae
for the hyperspherical functions of one and the same order follow from
the equalities
\begin{eqnarray}
\sX^{l\dot{l}}_-\fM^{l\dot{l}}_{mn;\dot{m}\dot{n}}=\boldsymbol{\alpha}_n
\fM^{l\dot{l}}_{m,n-1;\dot{m}\dot{n}},&\quad&
\sX^{l\dot{l}}_+\fM^{l\dot{l}}_{mn;\dot{m}\dot{n}}=\boldsymbol{\alpha}_{n+1}
\fM^{l\dot{l}}_{m,n+1;\dot{m}\dot{n}},\label{XG}\\
\sY^{l\dot{l}}_-\fM^{l\dot{l}}_{mn;\dot{m}\dot{n}}=\boldsymbol{\alpha}_{\dot{n}}
\fM^{l\dot{l}}_{mn;\dot{m}\dot{n}-1},&\quad&
\sY^{l\dot{l}}_+\fM^{l\dot{l}}_{mn;\dot{m}\dot{n}}=
\boldsymbol{\alpha}_{\dot{n}+1}
\fM^{l\dot{l}}_{mn;\dot{m}\dot{n}+1}.\label{YG}
\end{eqnarray}
As it is shown in \cite{Var03}, the generators (\ref{SL26})
can be expressed via the complex Euler angles as
\begin{eqnarray}
\sX_+&=&\frac{e^{-i\psi^c}}{2}\left[i\frac{\partial}{\partial\theta}-
\frac{1}{\sin\theta^c}\frac{\partial}{\partial\varphi}
+\ctg\theta^c\frac{\partial}{\partial\psi}-\frac{\partial}{\partial\tau}-
\frac{i}{\sin\theta^c}\frac{\partial}{\partial\epsilon}+
i\ctg\theta^c\frac{\partial}{\partial\varepsilon}\right],\label{SL27}\\
\sX_-&=&\frac{e^{i\psi^c}}{2}\left[i\frac{\partial}{\partial\theta}+
\frac{1}{\sin\theta^c}\frac{\partial}{\partial\varphi}
-\ctg\theta^c\frac{\partial}{\partial\psi}-\frac{\partial}{\partial\tau}+
\frac{i}{\sin\theta^c}\frac{\partial}{\partial\epsilon}-
i\ctg\theta^c\frac{\partial}{\partial\varepsilon}\right],\label{SL28}\\
\sY_+&=&\frac{e^{-i\dot{\psi}^c}}{2}\left[i\frac{\partial}{\partial\theta}-
\frac{1}{\sin\dot{\theta}^c}\frac{\partial}{\partial\varphi}+
\ctg\dot{\theta}^c\frac{\partial}{\partial\psi}+\frac{\partial}{\partial\tau}+
\frac{i}{\sin\dot{\theta}^c}\frac{\partial}{\partial\epsilon}-
i\ctg\dot{\theta}^c\frac{\partial}{\partial\varepsilon}\right],\label{SL29}\\
\sY_-&=&\frac{e^{i\dot{\psi}^c}}{2}\left[i\frac{\partial}{\partial\theta}+
\frac{1}{\sin\dot{\theta}^c}\frac{\partial}{\partial\varphi}
-\ctg\dot{\theta}^c\frac{\partial}{\partial\psi}+\frac{\partial}{\partial\tau}-
\frac{i}{\sin\dot{\theta}^c}\frac{\partial}{\partial\epsilon}+
i\ctg\dot{\theta}^c\frac{\partial}{\partial\varepsilon}\right].\label{SL30}
\end{eqnarray}
Substituting the function
$\fM^{l\dot{l}}_{mn;\dot{m}\dot{n}}=
e^{-m(\epsilon+i\varphi)-n(\varepsilon+i\psi)}
\fZ^{l\dot{l}}_{mn;\dot{m}\dot{n}}(\theta,\tau)
e^{-\dot{m}(\epsilon-i\varpi)-\dot{n}(\varepsilon-i\psi)}$
into the relations (\ref{XG})
and taking into account the operators (\ref{SL27}) and (\ref{SL28}),
we find that
\begin{eqnarray}
i\frac{\partial\fZ^{l\dot{l}}_{mn;\dot{m}\dot{n}}}{\partial\theta}-
\frac{\partial\fZ^{l\dot{l}}_{mn;\dot{m}\dot{n}}}{\partial\tau}-
\frac{2i(m-n\cos\theta^c)}{\sin\theta^c}\fZ^{l\dot{l}}_{mn;\dot{m}\dot{n}}=
2\boldsymbol{\alpha}_n\fZ^{l\dot{l}}_{m,n-1;\dot{m}\dot{n}},\nonumber\\
i\frac{\partial\fZ^{l\dot{l}}_{mn;\dot{m}\dot{n}}}{\partial\theta}-
\frac{\partial\fZ^{l\dot{l}}_{mn;\dot{m}\dot{n}}}{\partial\tau}+
\frac{2i(m-n\cos\theta^c)}{\sin\theta^c}\fZ^{l\dot{l}}_{mn;\dot{m}\dot{n}}=
2\boldsymbol{\alpha}_{n+1}\fZ^{l\dot{l}}_{m,n+1;\dot{m}\dot{n}}.\nonumber
\end{eqnarray}
Since the functions $\fZ^{l\dot{l}}_{mn;\dot{m}\dot{n}}$ are symmetric,
that is, $\fZ^{l\dot{l}}_{mn;\dot{m}\dot{n}}=
\fZ^{l\dot{l}}_{nm;\dot{n}\dot{m}}$, then substituting
$\fZ^{l\dot{l}}_{nm;\dot{n}\dot{m}}$ instead of
$\fZ^{l\dot{l}}_{mn;\dot{m}\dot{n}}$ and replacing $m$ by $n$, and
$n$ by $m$, we obtain
\begin{eqnarray}
i\frac{\partial\fZ^{l\dot{l}}_{mn;\dot{m}\dot{n}}}{\partial\theta}-
\frac{\partial\fZ^{l\dot{l}}_{mn;\dot{m}\dot{n}}}{\partial\tau}-
\frac{2i(n-m\cos\theta^c)}{\sin\theta^c}\fZ^{l\dot{l}}_{mn;\dot{m}\dot{n}}=
2\boldsymbol{\alpha}_m\fZ^{l\dot{l}}_{m-1,n;\dot{m}\dot{n}},\label{RZ1}\\
i\frac{\partial\fZ^{l\dot{l}}_{mn;\dot{m}\dot{n}}}{\partial\theta}-
\frac{\partial\fZ^{l\dot{l}}_{mn;\dot{m}\dot{n}}}{\partial\tau}+
\frac{2i(n-m\cos\theta^c)}{\sin\theta^c}\fZ^{l\dot{l}}_{mn;\dot{m}\dot{n}}=
2\boldsymbol{\alpha}_{m+1}\fZ^{l\dot{l}}_{m+1,n;\dot{m}\dot{n}}.\label{RZ2}
\end{eqnarray}
Analogously, from the relations (\ref{YG}) and generators (\ref{SL29})--
(\ref{SL30}), we have
\begin{eqnarray}
i\frac{\partial\fZ^{l\dot{l}}_{mn;\dot{m}\dot{n}}}{\partial\theta}+
\frac{\partial\fZ^{l\dot{l}}_{mn;\dot{m}\dot{n}}}{\partial\tau}+
\frac{2i(\dot{n}-\dot{m}\cos\dot{\theta}^c)}{\sin\dot{\theta}^c}
\fZ^{l\dot{l}}_{mn;\dot{m}\dot{n}}=
2\boldsymbol{\alpha}_{\dot{m}}\fZ^{l\dot{l}}_{mn;\dot{m}-1,\dot{n}},
\label{RZ3}\\
i\frac{\partial\fZ^{l\dot{l}}_{mn;\dot{m}\dot{n}}}{\partial\theta}+
\frac{\partial\fZ^{l\dot{l}}_{mn;\dot{m}\dot{n}}}{\partial\tau}-
\frac{2i(\dot{n}-\dot{m}\cos\dot{\theta}^c)}{\sin\dot{\theta}^c}
\fZ^{l\dot{l}}_{mn;\dot{m}\dot{n}}=
2\boldsymbol{\alpha}_{\dot{m}+1}\fZ^{l\dot{l}}_{mn;\dot{m}+1,\dot{n}}.
\label{RZ4}
\end{eqnarray}

Further, for the conjugate representations we have
\begin{eqnarray}
\widetilde{\sX}^{l\dot{l}}_-\overset{\ast}{\fM}{}^{l\dot{l}}_{mn;\dot{m}\dot{n}}
=\balpha_n\overset{\ast}{\fM}{}^{l\dot{l}}_{m,n-1;\dot{m}\dot{n}},&\quad&
\widetilde{\sX}^{l\dot{l}}_+\overset{\ast}{\fM}{}^{l\dot{l}}_{mn;\dot{m}\dot{n}}
=\balpha_{n+1}\overset{\ast}{\fM}{}^{l\dot{l}}_{m,n+1;\dot{m}\dot{n}},
\label{CXG}\\
\widetilde{\sY}^{l\dot{l}}_-\overset{\ast}{\fM}{}^{l\dot{l}}_{mn;\dot{m}\dot{n}}
=\balpha_{\dot{n}}\overset{\ast}{\fM}{}^{l\dot{l}}_{m,n-1;\dot{m}\dot{n}},
&\quad&
\widetilde{\sY}^{l\dot{l}}_+\overset{\ast}{\fM}{}^{l\dot{l}}_{mn;\dot{m}\dot{n}}
=\balpha_{\dot{n}+1}\overset{\ast}{\fM}{}^{l\dot{l}}_{m,n+1;\dot{m}\dot{n}},
\label{CYG}
\end{eqnarray}
where
$\widetilde{\sX}^{l\dot{l}}_-\sim\sY^{l\dot{l}}_-$,
$\widetilde{\sX}^{l\dot{l}}_+\sim\sY^{l\dot{l}}_+$,
$\widetilde{\sY}^{l\dot{l}}_-\sim\sX^{l\dot{l}}_-$,
$\widetilde{\sY}^{l\dot{l}}_+\sim\sX^{l\dot{l}}_+$.
Substituting now the function
$\overset{\ast}{\fM}{}^{l\dot{l}}_{mn;\dot{m}\dot{n}}=
e^{-m(\epsilon-i\varphi)-n(\varepsilon-i\psi)}
\overset{\ast}{\fZ}{}^{l\dot{l}}_{mn;\dot{m}\dot{n}}(\theta,\tau)
e^{-\dot{m}(\epsilon+i\varphi)-\dot{n}(\varepsilon+i\psi)}$
into the relations (\ref{CXG})--(\ref{CYG}), we obtain
\begin{eqnarray}
i\frac{\partial\overset{\ast}{\fZ}{}^{l\dot{l}}_{mn;\dot{m}\dot{n}}}
{\partial\theta}+
\frac{\partial\overset{\ast}{\fZ}{}^{l\dot{l}}_{mn;\dot{m}\dot{n}}}
{\partial\tau}+
\frac{2i(n-m\cos\dot{\theta}^c)}{\sin\dot{\theta}^c}
\overset{\ast}{\fZ}{}^{l\dot{l}}_{mn;\dot{m}\dot{n}}=
2\boldsymbol{\alpha}_{m}
\overset{\ast}{\fZ}{}^{l\dot{l}}_{m-1,n;\dot{m}\dot{n}},
\label{RZ5}\\
i\frac{\partial\overset{\ast}{\fZ}{}^{l\dot{l}}_{mn;\dot{m}\dot{n}}}
{\partial\theta}+
\frac{\partial\overset{\ast}{\fZ}{}^{l\dot{l}}_{mn;\dot{m}\dot{n}}}
{\partial\tau}-
\frac{2i(n-m\cos\dot{\theta}^c)}{\sin\dot{\theta}^c}
\overset{\ast}{\fZ}{}^{l\dot{l}}_{mn;\dot{m}\dot{n}}=
2\boldsymbol{\alpha}_{m+1}
\overset{\ast}{\fZ}{}^{l\dot{l}}_{m+1,n;\dot{m}\dot{n}}.
\label{RZ6}
\end{eqnarray}
\begin{eqnarray}
i\frac{\partial\overset{\ast}{\fZ}{}^{l\dot{l}}_{mn;\dot{m}\dot{n}}}
{\partial\theta}-
\frac{\partial\overset{\ast}{\fZ}{}^{l\dot{l}}_{mn;\dot{m}\dot{n}}}
{\partial\tau}-
\frac{2i(\dot{n}-\dot{m}\cos\theta^c)}{\sin\theta^c}
\overset{\ast}{\fZ}{}^{l\dot{l}}_{mn;\dot{m}\dot{n}}=
2\boldsymbol{\alpha}_{\dot{m}}
\overset{\ast}{\fZ}{}^{l\dot{l}}_{mn;\dot{m}-1,\dot{n}},
\label{RZ7}\\
i\frac{\partial\overset{\ast}{\fZ}{}^{l\dot{l}}_{mn;\dot{m}\dot{n}}}
{\partial\theta}-
\frac{\partial\overset{\ast}{\fZ}{}^{l\dot{l}}_{mn;\dot{m}\dot{n}}}
{\partial\tau}+
\frac{2i(\dot{n}-\dot{m}\cos\theta^c)}{\sin\theta^c}
\overset{\ast}{\fZ}{}^{l\dot{l}}_{mn;\dot{m}\dot{n}}=
2\boldsymbol{\alpha}_{\dot{m}+1}
\overset{\ast}{\fZ}{}^{l\dot{l}}_{mn;\dot{m}+1,\dot{n}}.
\label{RZ8}
\end{eqnarray}

\section{Fields on the Poincar\'{e} group}
Fields on the Poincar\'{e} group present itself a natural generalization
of the concept of wave function. These fields (generalized wave functions)
were introduced independently by several authors
\cite{GT47,BW48,Yuk50,Shi51} mainly in connection with constructing
relativistic wave equations. The following logical step was done by
Finkelstein \cite{Fin55}, he suggested to consider the wave function
depending both the coordinates on the Minkowski spacetime and some
continuous variables corresponding to spin degrees of freedom
(internal space). In essence, this generalization consists in replacing
the Minkowski space by a larger space on which the Poincar\'{e} group
acts. If this action is to be transitive, one is lead to consider the
homogeneous spaces of the Poincar\'{e} group. All the homogeneous spaces
of this type were listed by Finkelstein \cite{Fin55} and by Bacry and
Kihlberg \cite{BK69} and the fields on these spaces were considered
in the works \cite{Lur64,BN67,NB67,Kih68,Kih70,Tol96,GS01}.

A homogeneous space $\cM$ of a group $G$ has the following properties:\\
a) It is a topological space on which the group $G$ acts continuously,
that is, let $y$ be a point in $\cM$, then $gy$ is defined and is again
a point in $\cM$ ($g\in G$).\\
b) This action is transitive, that is, for any two points $y_1$ and
$y_2$ in $\cM$ it is always possible to find a group element $g\in G$
such that $y_2=gy_1$.\\
There is a one-to-one correspondence between the homogeneous spaces of
$G$ and the coset spaces of $G$. Let $H_0$ be a maximal subgroup of
$G$ which leaves the point $y_0$ invariant, $gy_0=y_0$, $g\in H_0$, then
$H_0$ is called the stabilizer of $y_0$. Representing now any group
element of $G$ in the form $g=g_cg_0$, where $g_0\in H_0$ and
$g_c\in G/H_0$, we see that, by virtie of the transitivity property,
any point $y\in\cM$ can be given by $y=g_cg_0y_0=g_cy$. Hence it follows
that the elements $g_c$ of the coset space give a parametrization
of $\cM$. The mapping $\cM\leftrightarrow G/H_0$ is continuous since
the group multiplication is continuous and the action on $\cM$ is
continuous by definition. The stabilizers $H$ and $H_0$ of two different
points $y$ and $y_0$ are conjugate, since from $H_0g_0=g_0$,
$y_0=g^{-1}y$, it follows that $gH_0g^{-1}y=y$, that is, $H=gH_0g^{-1}$.

Coming back to the Poincar\'{e} group $\cP$, we see that the enumeration
of the different homogeneous spaces $\cM$ of $\cP$ amounts to an
enumeration of the subgroups of $\cP$ up to a conjugation. Following to
Finkelstein, we require that $\cM$ always contains the Minkowski space
$\R^{1,3}$ which means that four parameters of $\cM$ can be denoted by
$x\,(x^\mu)$. This means that the stabilizer $H$ of a given point in $\cM$
can never contain an element of the translation subgroup of $\cP$.
Thus, the stabilizer must be a subgroup of the proper Lorentz
group $\fG_+$.

In such a way, studying different subgroups of $\fG_+$, we obtain a full
list of homogeneous spaces $\cM=\cP/H$ of the Poincar\'{e} group.
In the present paper we restrict ourselves by a consideration of the
following two homogeneous spaces:
\begin{eqnarray}
\cM_{10}&=&\R^{1,3}\times\fL_6,\quad H=0;\nonumber\\
\cM_8&=&\R^{1,3}\times\dS^2,\quad H=\Omega^c_\psi;\nonumber
\end{eqnarray}
Hence it follows that a group manifold of the Poincar\'{e} group,
$\cM_{10}=\R^{1,3}\times\fL_6$, is a maximal homogeneous space of $\cP$,
$\fL_6$ is a group manifold of the Lorentz group. The fields on the
manifold $\cM_{10}$ were considered by Lur\c{c}at \cite{Lur64}.
These fields depend on all the ten parameters of $\cP$:
\[
\boldsymbol{\psi}(x,\fg)=\psi(x)\psi(\fg)=\psi(x_0,x_1,x_2,x_3)
\psi(\fg_1,\fg_2,\fg_3,\fg_4,\fg_5,\fg_6),
\]
where an explicit form of $\psi(x)$ is given by the exponentials, and the
functions $\psi(\fg)$ are expressed via the generalized hyperspherical
functions $\fM^{l\dot{l}}_{mn;\dot{m}\dot{n}}(\fg)$ in the case of finite
dimensional representations.

The following eight-dimensional homogeneous space
$\cM_8=\R^{1,3}\times\dS^2$ is a direct product of the Minkowski space
$\R^{1,3}$ and the complex two-sphere $\dS^2$. In this case the stabilizer
$H$ consists of the subgroup $\Omega^c_\psi$ of the diagonal matrices
$\begin{pmatrix}
e^{\frac{i\psi^c}{2}} & 0\\
0 & e^{-\frac{i\psi^c}{2}}
\end{pmatrix}$. Bacry and Kihlberg \cite{BK69} claimed that the space
$\cM_8$ is the most suitable for a description of both half-integer and
integer spins. The fields, defined in $\cM_8$, depend on the eight
parameters of $\cP$:
\begin{equation}\label{WF}
\boldsymbol{\psi}(x,\varphi^c,\theta^c)=\psi(x)\psi(\varphi^c,\theta^c)=
\psi(x_0,x_1,x_2,x_3)\psi(\varphi,\epsilon,\theta,\tau),
\end{equation}
where the functions $\psi(\varphi^c,\theta^c)$ are expressed via the
associated hyperspherical functions defined on the
surface of the complex two-sphere $\dS^2$.
\subsection{Harmonic analysis on $SU(2)\otimes SU(2)\odot T_4$}
In this subsection we will consider Fourier series on the Poincar\'{e} group
$\cP$. First of all, the group $\cP$ has the same number of connected
components as with the Lorentz group. Later on we will consider only the
component $\cP^\uparrow_+$ corresponding the connected component
$L^\uparrow_+$ (so called special Lorentz group, see \cite{RF}).
As is known, an universal covering $\overline{\cP^\uparrow_+}$ of the group
$\cP^\uparrow_+$ is defined by a semidirect product
$\overline{\cP^\uparrow_+}=SL(2,\C)\odot T_4\simeq\spin_+(1,3)\odot T_4$,
where $T_4$ is a subgroup of four-dimensional translations.
Since the Poincar\'{e} group is a 10-parameter group, then an invariant
measure on this group has a form
\[
d^{10}\balpha=d^6\fg d^4x,
\]
where $d^6\fg$ is the Haar measure on the Lorentz group. Or, taking
into account (\ref{HA1}), we obtain
\begin{equation}\label{HMP}
d\balpha=\sin\theta^c\sin\dot{\theta}^cd\theta d\varphi d\psi d\tau
d\epsilon d\varepsilon dx_1 dx_2 dx_3 dx_4,
\end{equation}
where $x_i\in T_4$.

Thus, an invariant integration on the group $SL(2,\C)\odot T_4$ is defined
by the formula
\[
\int\limits_{SL(2,\C)\odot T_4}f(\balpha)d^{10}\balpha=
\int\limits_{SL(2,\C)}\int\limits_{T_4}f(x,\fg)d^4xd^6\fg,
\]
where $f(\balpha)$ is a finite continuous function on $SL(2,\C)\odot T_4$.

In the case of finite-dimensional representations we come again to a local
isomorphism $SL(2,\C)\odot T_4\simeq SU(2)\otimes SU(2)\odot T_4$.
In this case basis representation functions of the Poincar\'{e} group
are defined by symmetric polynomials of the form
\begin{gather}
p(x,z,\bar{z})=\sum_{\substack{(\alpha_1,\ldots,\alpha_k)\\
(\dot{\alpha}_1,\ldots,\dot{\alpha}_r)}}\frac{1}{k!\,r!}
a^{\alpha_1\cdots\alpha_k\dot{\alpha}_1\cdots\dot{\alpha}_r}(x)
z_{\alpha_1}\cdots z_{\alpha_k}\bar{z}_{\dot{\alpha}_1}\cdots
\bar{z}_{\dot{\alpha}_r},\label{SF2}\\
(\alpha_i,\dot{\alpha}_i=0,1)\nonumber
\end{gather}
where the coefficients
$a^{\alpha_1\cdots\alpha_k\dot{\alpha}_1\cdots\dot{\alpha}_r}$
depend on the variables $x^\alpha$ ($\alpha=0,1,2,3$)
(the parameters of $T_4$).
The functions (\ref{SF2}) should be considered as {\it the functions on
the Poincar\'{e} group}.\index{function!on the Poincar\'{e} group}
Some applications of these functions contained
in \cite{GS01,Vas96}.
The group $T_4$ is an Abelian group formed by a direct
product of the four one-dimensional translation groups, $T_1$, where
$T_1$ is isomorphic to an additive group of real numbers $\R$ (usual
Fourier analysis is formulated in terms of the group $\R$). Hence it
follows that all irreducible representations of $T_4$ are one-dimensional
and expressed via the exponentials. Thus, the basis functions
(matrix elements) of the finite-dimensional representations of $\cP$
have the form
\begin{equation}\label{BFP}
t^{l\dot{l}}_{mn;\dot{m}\dot{n}}(\balpha)=
e^{-ipx}\fM^{l\dot{l}}_{mn;\dot{m}\dot{n}}(\fg),
\end{equation}
where $x=(x_1,x_2,x_3,x_4)$, and
$\fM^{l\dot{l}}_{mn;\dot{m}\dot{n}}(\fg)$ is the
generalized hyperspherical
function (\ref{HF3'}).

Let us consider now the configuration space $\cM_8=\R^{1,3}\times \dS^2$.
In this case the Fourier series on $\cM_8$ can be defined as follows
\begin{equation}\label{FM8}
f(\balpha)=\sum^{+\infty}_{p=-\infty}e^{ipx}\sum^\infty_{l=0}
\sum^l_{m=-l}\sum^{\dot{l}}_{\dot{m}=-\dot{l}}
\alpha^{l\dot{l}}_{m\dot{m}}\fM^{m\dot{m}}_{l\dot{l}}
(\varphi,\epsilon,\theta,\tau,0,0),
\end{equation}
where
\[
\alpha^{l\dot{l}}_{m\dot{m}}=\frac{(-1)^m(2l+1)(2\dot{l}+1)}{32\pi^4}
\int\limits_{\dS^2}\int\limits_{T_4}f(\balpha)e^{-ipx}
\fM^{m\dot{m}}_{l\dot{l}}(\varphi,\epsilon,\theta,\tau,0,0)d^4xd^4\fg,
\]
and $d^4\fg=\sin\theta^c\sin\dot{\theta}^cd\theta d\varphi d\tau d\epsilon$
is the Haar measure on $\dS^2$, $f(\balpha)$ is the square integrable
function on $\cM_8$, such that
\[
\int\limits_{\dS^2}\int\limits_{T_4}|f(\balpha)|^2d^4xd^4\fg<+\infty.
\]
\subsection{Harmonic analysis on the group $SL(2,\C)$}
First of all, on the group $SL(2,\C)$
there exists an
invariant measure $d\fg$, that is, such a
measure that for any finite continuous function
$f(\fg)$ on $SL(2,\C)$ the following equality
\[
\int f(\fg)d\fg=\int f(\fg_0\fg)d\fg=\int f(\fg\fg_0)d\fg=\int f(\fg^{-1})d\fg
\]
holds.
Now we express
the Haar measure (left or right) in terms of the parameters (\ref{CEA}),
\begin{equation}\label{HA1}
d\fg=\sin\theta^c\sin\dot{\theta}^cd\theta d\varphi d\psi d\tau d\epsilon
d\varepsilon.
\end{equation}
Thus, an invariant integration on the group $SL(2,\C)$
is defined by the formula
\[
\int\limits_{SL(2,\C)}f(g)d\fg=\frac{1}{32\pi^4}
\int\limits^{+\infty}_{-\infty}
\int\limits^{+\infty}_{-\infty}
\int\limits^{+\infty}_{-\infty}
\int\limits^{2\pi}_{-2\pi}
\int\limits^{2\pi}_0
\int\limits^\pi_0
f(\theta,\varphi,\psi,\tau,\epsilon,\varepsilon)
\sin\theta^c\sin\dot{\theta}^cd\theta d\varphi d\psi d\tau d\epsilon
d\varepsilon.
\]
When we consider finite-dimensional (spinor) representations of
$SL(2,\C)$, we come naturally to a local isomorphism
$SU(2)\otimes SU(2)\simeq SL(2,\C)$ considered by many authors
\cite{HS70,Ryd85}.
Since a dimension of the spinor representation
$\boldsymbol{\tau}_{l\dot{l}}$ of
$SU(2)\otimes SU(2)$ is equal to $(2l+1)(2\dot{l}+1)$, then the functions
$\sqrt{(2l+1)(2\dot{l}+1)}t^{l\dot{l}}_{mn;\dot{m}\dot{n}}(\fg)$
form a full orthogonal normalized
system on this group with respect to the invariant measure $d\fg$.
At this point,
the indices $l$ and $\dot{l}$ run
all possible integer or half-integer non-negative values,
and the indices $m$, $n$ and $\dot{m}$, $\dot{n}$ run the values
$-l,-l+1,\ldots,l-1,l$ and $-\dot{l},-\dot{l}+1,\ldots,\dot{l}-1,\dot{l}$.
In virtue of (\ref{HF3'}) the matrix elements
$t^{l\dot{l}}_{mn;\dot{m}\dot{n}}$ are expressed
via the generalized hyperspherical function
$t^{l\dot{l}}_{mn;\dot{m}\dot{n}}(\fg)=
\fM^{l\dot{l}}_{mn;\dot{m}\dot{n}}(\varphi,\epsilon,
\theta,\tau,\psi,\varepsilon)$.
Therefore,
\begin{equation}\label{HA2}
\int\limits_{SU(2)\otimes SU(2)}\fM^{l\dot{l}}_{mn;\dot{m}\dot{n}}(\fg)
\overline{\fM^{l\dot{l}}_{mn;\dot{m}\dot{n}}(\fg)}d\fg=
\frac{32\pi^4}{(2l+1)(2\dot{l}+1)}
\delta(\fg^\prime-\fg),
\end{equation}
where $\delta(\fg^\prime-\fg)$ is a $\delta$-function on the group
$SU(2)\otimes SU(2)$. An explicit form of $\delta$-function is
\begin{multline}
\delta(\fg^\prime-\fg)=\delta(\varphi^\prime-\varphi)
\delta(\epsilon^\prime-\epsilon)\delta(\cos\theta^\prime\ch\tau^\prime-
\cos\theta\ch\tau)\times\\
\times\delta(\sin\theta^\prime\sh\tau^\prime-\sin\theta\sh\tau)
\delta(\psi^\prime-\psi)\delta(\varepsilon^\prime-\varepsilon).\nonumber
\end{multline}
Substituting into (\ref{HA2}) the expression
\[
\fM^{l\dot{l}}_{mn;\dot{m}\dot{n}}(\fg)=e^{-m(\epsilon+i\varphi)-
n(\varepsilon+i\psi)}
\fZ^{l\dot{l}}_{mn;\dot{m}\dot{n}}(\theta,\tau)e^{-\dot{m}(\epsilon-i\varphi)-
\dot{n}(\varepsilon-i\psi)}
\]
and taking into account (\ref{HA1}), we obtain
\begin{multline}
\int\limits_{SU(2)\otimes SU(2)}
\fZ^{l\dot{l}}_{mn;\dot{m}\dot{n}}(\theta,\tau)
\overline{\fZ^{s\dot{s}}_{pq;\dot{p}\dot{q}}(\theta,\tau)}
e^{-(m+p)\epsilon}e^{-i(m-p)\varphi}e^{-(\dot{m}+\dot{p})\epsilon}
e^{i(\dot{m}-\dot{p})\varphi}
\times e^{-(n+q)\varepsilon}\times\\
\times e^{-i(n-q)\psi}e^{-(\dot{n}+\dot{q})\varepsilon}
e^{i(\dot{n}-\dot{q})\psi}
\sin\theta^c\sin\dot{\theta}^cd\theta d\varphi d\psi d\tau d\epsilon
d\varepsilon=\frac{32\pi^4\delta_{ls}\delta_{\dot{l}\dot{s}}
\delta_{mp}\delta_{nq}\delta_{\dot{m}\dot{p}}\delta_{\dot{n}\dot{q}}
\delta(\fg^\prime-\fg)}{(2l+1)(2\dot{l}+1)}.\nonumber
\end{multline}

Thus, any square integrable function $f(\varphi^c,\theta^c,\psi^c)$ on the
group $SU(2)\otimes SU(2)$, such that
\[
\int\limits_{SU(2)\otimes SU(2)}
|f(\varphi^c,\theta^c,\psi^c)|^2
\sin\theta^c\sin\dot{\theta}^cd\theta d\varphi d\psi d\tau d\epsilon
d\varepsilon
<+\infty,
\]
is expanded into a convergent (on an average) Fourier series on
$SU(2)\otimes SU(2)$,
\begin{multline}
f(\varphi^c,\theta^c,\psi^c)=\sum^\infty_{l,\dot{l}=0}
\sum^l_{m=-l}\sum^l_{n=-l}\sum^{\dot{l}}_{\dot{m}=-\dot{l}}
\sum^{\dot{l}}_{\dot{n}=-\dot{l}}
\alpha^{l\dot{l}}_{mn;\dot{m}\dot{n}}\times\\
\times e^{-m(\epsilon+i\varphi)-n(\varepsilon+i\psi)}
\fZ^{l\dot{l}}_{mn;\dot{m}\dot{n}}(\cos\theta^c,\cos\dot{\theta}^c)
e^{-\dot{m}(\epsilon-i\varphi)-\dot{n}(\varepsilon-i\psi)},
\label{CFS}
\end{multline}
where
\begin{multline}
\alpha^{l\dot{l}}_{mn;\dot{m}\dot{n}}=
\frac{(-1)^{m-n}(2l+1)(2\dot{l}+1)}{32\pi^4}\times\\
\int\limits_{SU(2)\otimes SU(2)}
f(\varphi^c,\theta^c,\psi^c)e^{i(m\varphi^c+n\psi^c)}
\fZ^{l\dot{l}}_{mn;\dot{m}\dot{n}}(\cos\theta^c,\cos\dot{\theta}^c)
e^{-i(\dot{m}\dot{\phi}^c+\dot{n}\dot{\psi}^c)}
\sin\theta^c\sin\dot{\theta}^cd\theta d\varphi d\psi d\tau d\epsilon
d\varepsilon.
\nonumber
\end{multline}
The Parseval equality for the case of $SU(2)\otimes SU(2)$ is defined
as follows
\begin{multline}
\sum^\infty_{l,\dot{l}=0}\sum^l_{m=-l}\sum^l_{n=-l}
\sum^{\dot{l}}_{\dot{m}=-\dot{l}}\sum^{\dot{l}}_{\dot{n}=-\dot{l}}
|\alpha^{l\dot{l}}_{mn;\dot{m}\dot{n}}|^2=\\
=\frac{(2l+1)(2\dot{l}+1)}{32\pi^4}
\int\limits_{SU(2)\otimes SU(2)}
|f(\varphi^c,\theta^c,\psi^c)|^2
\sin\theta^c\sin\dot{\theta}^cd\theta d\varphi d\psi d\tau d\epsilon
d\varepsilon.
\nonumber
\end{multline}
About convergence of Fourier series of the type (\ref{CFS}) see \cite{BD70}.

In like manner we can define Fourier series on the two-dimensional
complex sphere via the associated hyperspherical functions.
An expansion of the functions on the surface of the two-dimensional
sphere has an important meaning for the subsequent physical
applications.

So, let $f(\varphi^c,\theta^c)$ be a function on the complex two-sphere
$\dS^2$, such that
\[
\int\limits_{\dS^2}
|f(\varphi^c,\theta^c)|^2
\sin\theta^c\sin\dot{\theta}^cd\theta d\varphi d\tau d\epsilon <+\infty,
\]
then $f(\varphi^c,\theta^c)$ is expanded into a convergent Fourier series
on $\dS^2$,
\[
f(\varphi^c,\theta^c)=\sum^\infty_{l,\dot{l}=0}\sum^l_{m=-l}
\sum^{\dot{l}}_{\dot{m}=-\dot{l}}
\alpha^{l\dot{l}}_{m\dot{m}}e^{-m(\epsilon+i\varphi)}
\fZ^{m\dot{m}}_{l\dot{l}}(\cos\theta^c,\cos\dot{\theta}^c)
e^{-\dot{m}(\epsilon-i\varphi)},
\]
where
\[
\alpha^{l\dot{l}}_{m\dot{m}}=\frac{(-1)^m(2l+1)(2\dot{l}+1)}{32\pi^4}
\int\limits_{\dS^2}
f(\varphi^c,\theta^c)e^{im\varphi^c}
\fZ^{m\dot{m}}_{l\dot{l}}(\cos\theta^c,\cos\dot{\theta}^c)
e^{-i\dot{m}\dot{\varphi}^c}
\sin\theta^c\sin\dot{\theta}^cd\theta d\varphi d\tau d\epsilon,
\]
and $\fZ^{m\dot{m}}_{l\dot{l}}(\cos\theta^c,\cos\dot{\theta}^c)$
is an associated hyperspherical function,
$d\fg=\sin\theta^c\sin\dot{\theta}^cd\theta d\varphi d\tau d\epsilon$ is
the Haar measure on the sphere $\dS^2$. Correspondingly, the Parseval
equality on $\dS^2$ has the form
\[
\sum^\infty_{l,\dot{l}=0}\sum^l_{m=-l}\sum^{\dot{l}}_{\dot{m}=-\dot{l}}
|\alpha^{l\dot{l}}_{m\dot{m}}|^2=
\frac{(2l+1)(2\dot{l}+1)}{32\pi^4}
\int\limits_{\dS^2}
|f(\varphi^c,\theta^c)|^2
\sin\theta^c\sin\dot{\theta}^cd\theta d\varphi d\tau d\epsilon.
\]
\section{Lagrangian formalism and field equations on the\protect\newline
Poincar\'{e} group}
We will start with a more general homogeneous space of the group $\cP$,
$\cM_{10}=\R^{1,3}\times\fL_6$ (group manifold of the Poincar\'{e} group).
Let $\cL(\balpha)$ be a Lagrangian on the group manifold $\cM_{10}$
(in other words, $\cL(\balpha)$ is a 10-dimensional
point function), where $\balpha$ is the parameter set of this group.
Then an integral for $\cL(\balpha)$ on some 10-dimensional volume $\Omega$
of the group manifold we will call {\it an action on the Poincar\'{e}
group}:
\[
A=\int\limits_\Omega d\balpha\cL(\balpha),
\]
where $d\balpha$ is a Haar measure on the group $\cP$ (see (\ref{HMP})).

Let $\boldsymbol{\psi}(\balpha)$ be a function on
the group manifold $\cM_{10}$ (now it is
sufficient to assume that $\boldsymbol{\psi}(\balpha)$
is a square integrable function
on the Poincar\'{e} group) and let
\begin{equation}\label{ELE}
\frac{\partial\cL}{\partial\boldsymbol{\psi}}-\frac{\partial}{\partial\balpha}
\frac{\partial\cL}{\partial\frac{\partial\boldsymbol{\psi}}{\partial\balpha}}=0
\end{equation}
be Euler-Lagrange equations on $\cM_{10}$ (more precisely speaking, the equations
(\ref{ELE}) act on the tangent bundle
$T\cM_{10}=\underset{\balpha\in\cM_{10}}{\cup}T_{\balpha}\cM_{10}$
of the manifold $\cM_{10}$,
see \cite{Arn}). Let us introduce a Lagrangian $\cL(\balpha)$ depending on
the field function $\boldsymbol{\psi}(\balpha)$ as follows
\[
\cL(\balpha)=-\frac{1}{2}\left(\boldsymbol{\psi}^\ast(\balpha)B_\mu
\frac{\partial\boldsymbol{\psi}(\balpha)}{\partial\balpha_\mu}-
\frac{\partial\boldsymbol{\psi}^\ast(\balpha)}
{\partial\balpha_\mu}B_\mu\boldsymbol{\psi}(\balpha)\right)
-\kappa\boldsymbol{\psi}^\ast(\balpha)B_{11}\boldsymbol{\psi}(\balpha),
\]
where $B_\nu$ ($\nu=1,2,\ldots,10$) are square matrices. The number of
rows and columns in these matrices is equal to the number of components
of $\boldsymbol{\psi}(\balpha)$, $\kappa$ is a non-null real constant.

Further, if $B_{11}$ is non-singular, then we can introduce the matrices
\[
\Gamma_\mu=B^{-1}_{11}B_\mu,\quad \mu=1,2,\ldots,10,
\]
and represent the Lagrangian $\cL(\balpha)$ in the form
\begin{equation}\label{Lagrange}
\cL(\balpha)=-\frac{1}{2}\left(\overline{\boldsymbol{\psi}}(\balpha)\Gamma_\mu
\frac{\partial\boldsymbol{\psi}(\balpha)}{\partial\balpha_\mu}-
\frac{\overline{\boldsymbol{\psi}}(\balpha)}{\partial\balpha_\mu}\Gamma_\mu
\boldsymbol{\psi}(\balpha)\right)-
\kappa\overline{\boldsymbol{\psi}}(\balpha)\boldsymbol{\psi}(\balpha),
\end{equation}
where
\[
\overline{\boldsymbol{\psi}}(\balpha)=\boldsymbol{\psi}^\ast(\balpha)B_{11}.
\]

Varying independently $\psi(x)$ and $\overline{\psi}(x)$, we obtain from
(\ref{Lagrange}) in accordance with (\ref{ELE}) the following equations:
\begin{equation}\label{FET}
\begin{array}{ccc}
\Gamma_i\dfrac{\partial\psi(x)}{\partial x_i}+\kappa\psi(x)&=&0,\\
\Gamma^T_i\dfrac{\partial\overline{\psi}(x)}{\partial x_i}-
\kappa\overline{\psi}(x)&=&0.
\end{array}\quad(i=1,\ldots,4)
\end{equation}
Analogously, varying independently $\psi(\fg)$ and $\overline{\psi}(\fg)$
one gets
\begin{equation}\label{FEL}
\begin{array}{ccc}
\Gamma_k\dfrac{\partial\psi(\fg)}{\partial\fg_k}+\kappa\psi(\fg)&=&0,\\
\Gamma^T_k\dfrac{\overline{\psi}(\fg)}{\partial\fg_k}-
\kappa\overline{\psi}(\fg)&=&0,
\end{array}\quad(k=1,\ldots,6)
\end{equation}
where
\[
\psi(\fg)=\begin{pmatrix}
\psi(\fg)\\
\dot{\psi}(\fg)
\end{pmatrix},\quad
\Gamma_k=\begin{pmatrix}
0 & \overset{\ast}{\Lambda}{}^{l\dot{l}}_k\\
\Lambda^{l\dot{l}}_k & 0
\end{pmatrix}.
\]
The doubling of representations, described by a bispinor
$\psi(\fg)=(\psi(\fg),\dot{\psi}(\fg))^T$, is the well known feature of
the Lorentz group representations \cite{GMS,Nai58}.
Since an universal covering $SL(2,\C)$ of the
proper orthochronous Lorentz group is a complexification of the group
$SU(2)$ (see the section \ref{SL2C}), then it is more
convenient to express six parameters $\fg_k$ of the Lorentz group via
three parameters $a_1$, $a_2$, $a_3$ of the group $SU(2)$. It is obvious that
$\fg_1=a_1$, $\fg_2=a_2$, $\fg_3=a_3$, $\fg_4=ia_1$, $\fg_5=ia_2$,
$\fg_6=ia_3$. Then the first equation from (\ref{FEL}) can be written as
\begin{eqnarray}
\sum^3_{j=1}\Lambda^{l\dot{l}}_j\frac{\partial\psi}{\partial a_j}-
i\sum^3_{j=1}\Lambda^{l\dot{l}}_j\frac{\partial\psi}{\partial a^\ast_j}+
\dot{k}^c\dot{\psi}&=&0,\nonumber\\
\sum^3_{j=1}\overset{\ast}{\Lambda}{}^{l\dot{l}}_j\frac{\partial\dot{\psi}}
{\partial\widetilde{a}_j}+i\sum^3_{j=1}
\overset{\ast}{\Lambda}{}^{l\dot{l}}_j\frac{\partial\dot{\psi}}
{\partial\widetilde{a^\ast}_j}+k^c\psi&=&0.\label{CT}
\end{eqnarray}
where $a^\ast_1=-i\fg_4$, $a^\ast_2=-i\fg_5$, $a^\ast_3=-i\fg_6$, and
$\widetilde{a}_j$, $\widetilde{a}^\ast_j$ are the parameters corresponding
the dual basis. In essence, the equations (\ref{CT}) are defined
in a three-dimensional complex space $\C^3$. In turn, the space $\C^3$
is isometric to a 6-dimensional bivector space $\R^6$ (a parameter space
of the Lorentz group \cite{Kag26,Pet69}). The bivector space $\R^6$ is
a tangent space of the group manifold $\fL_6$ of the Lorentz group, that is,
the manifold $\fL_6$ in the each its point is equivalent locally to the
space $\R^6$. Thus, for all $\fg\in\fL_6$ we have $T_{\fg}\fL_6\simeq\R^6$.
There exists a close relationship between the metrics of the
Minkowski spacetime $\R^{1,3}$ and the metrics of $\R^6$ defined by the
formulae (see \cite{Pet69})
\begin{equation}\label{Metric}
g_{ab}\longrightarrow g_{\alpha\beta\gamma\delta}\equiv
g_{\alpha\gamma}g_{\beta\delta}-g_{\alpha\delta}g_{\beta\gamma},
\end{equation}
where $g_{\alpha\beta}$ is a metric tensor of the spacetime $\R^{1,3}$, and
collective indices are skewsymmetric pairs
$\alpha\beta\rightarrow a$, $\gamma\delta\rightarrow b$. In more detail, if
\[\ar
g_{\alpha\beta}=\begin{pmatrix}
-1 & 0 & 0 & 0\\
0  & -1& 0 & 0\\
0  & 0 & -1& 0\\
0  & 0 & 0 & 1
\end{pmatrix},
\]
then in virtue of (\ref{Metric}) for the metric tensor of $\R^6$ we obtain
\begin{equation}\label{MetB}
g_{ab}=\ar\begin{pmatrix}
-1& 0 & 0 & 0 & 0 & 0\\
0 & -1& 0 & 0 & 0 & 0\\
0 & 0 & -1& 0 & 0 & 0\\
0 & 0 & 0 & 1 & 0 & 0\\
0 & 0 & 0 & 0 & 1 & 0\\
0 & 0 & 0 & 0 & 0 & 1
\end{pmatrix},
\end{equation}
where the order of collective indices in $\R^6$ is
$23\rightarrow 0$, $10\rightarrow 1$, $20\rightarrow 2$,
$30\rightarrow 3$, $31\rightarrow 4$, $12\rightarrow 5$.

Let us write an invariance condition for the system (\ref{CT}).
As it is shown in \cite{Kag26}, the Lorentz transformations can be represented
by linear transformations of the space $\R^6$. Let
$\fg: a^\prime=\fg^{-1} a$ be a transformation of the bivector space $\R^6$,
that is, $a^\prime=\sum^6_{b=1}g_{ba} a_b$, where
$a=(a_1,a_2,a_3,a^\ast_1,a^\ast_2,a^\ast_3)$ and
$g_{ba}$ is the metric tensor (\ref{MetB}). We can write the tensor
(\ref{MetB}) in the form
$g_{ab}=\ar\begin{pmatrix}
g^-_{ik} & \\
& g^+_{ik}
\end{pmatrix}$,
then $a^\prime=\sum^3_{k=1}g^-_{ki}a_k$,
$a^\ast{}^\prime=\sum^3_{k=1}g^+_{ki}a^\ast_k$.
Replacing $\psi$ via $T^{-1}_{l\dot{l}}(\fg)\psi^\prime$,
and differentiation on
$a_k$ ($a^\ast_k$) by differentiation on $a^\prime_k$
($a^\ast_k{}^\prime$) via the formulae
\[
\frac{\partial}{\partial a_k}=\sum g^-_{ik}
\frac{\partial}{\partial a^\prime_i},\quad
\frac{\partial}{\partial a^\ast_k}=\sum g^+_{ik}
\frac{\partial}{\partial a^\ast_i{}^\prime},
\]
we obtain
\begin{multline}
\sum^3_{i=1}\left[
g^-_{i1}\sL^{l\dot{l}}_1\frac{\partial(T^{-1}_{l\dot{l}}(\fg)\psi^\prime)}
{\partial a^\prime_i}+
g^-_{i2}\sL^{l\dot{l}}_2\frac{\partial(T^{-1}_{l\dot{l}}(\fg)\psi^\prime)}
{\partial a^\prime_i}+
g^-_{i3}\sL^{l\dot{l}}_3\frac{\partial(T^{-1}_{l\dot{l}}(\fg)\psi^\prime)}
{\partial a^\prime_i}-
\right.\\
\shoveright{\left.
-ig^-_{i1}\sL^{l\dot{l}}_1\frac{\partial(T^{-1}_{l\dot{l}}(\fg)\psi^\prime)}
{\partial\overset{\ast}{a}_i{}^\prime}-
ig^-_{i2}\sL^{l\dot{l}}_2\frac{\partial(T^{-1}_{l\dot{l}}(\fg)\psi^\prime)}
{\partial\overset{\ast}{a}_i{}^\prime}-
ig^-_{i3}\sL^{l\dot{l}}_3\frac{\partial(T^{-1}_{l\dot{l}}(\fg)\psi^\prime)}
{\partial\overset{\ast}{a}_i{}^\prime}\right]+
\dot{\kappa}^cT^{-1}_{l\dot{l}}(\fg)\psi^\prime=0,}\\
\shoveleft{
\sum^3_{i=1}\left[
g^+_{i1}\overset{\ast}{\sL}{}^{l\dot{l}}_1
\frac{\partial(\overset{\ast}{T}_{l\dot{l}}\!\!{}^{-1}(\fg)
\dot{\psi}^\prime)}{\partial\widetilde{a}^\prime_i}+
g^+_{i2}\overset{\ast}{\sL}{}^{l\dot{l}}_2
\frac{\partial(\overset{\ast}{T}_{l\dot{l}}\!\!{}^{-1}(\fg)
\dot{\psi}^\prime)}{\partial\widetilde{a}^\prime_i}+
g^+_{i3}\overset{\ast}{\sL}{}^{l\dot{l}}_3
\frac{\partial(\overset{\ast}{T}_{l\dot{l}}\!\!{}^{-1}(\fg)
\dot{\psi}^\prime)}{\partial\widetilde{a}^\prime_i}+\right.}\nonumber\\
\left.
+ig^+_{i1}\overset{\ast}{\sL}{}^{l\dot{l}}_1
\frac{\partial(\overset{\ast}{T}_{l\dot{l}}\!\!{}^{-1}(\fg)
\dot{\psi}^\prime)}{\partial\widetilde{a^\ast}_i{}^\prime}+
ig^+_{i2}\overset{\ast}{\sL}{}^{l\dot{l}}_2
\frac{\partial(\overset{\ast}{T}_{l\dot{l}}\!\!{}^{-1}(\fg)
\dot{\psi}^\prime)}{\partial\widetilde{a^\ast}_i{}^\prime}+
ig^+_{i3}\overset{\ast}{\sL}{}^{l\dot{l}}_3
\frac{\partial(\overset{\ast}{T}_{l\dot{l}}\!\!{}^{-1}(\fg)
\dot{\psi}^\prime)}{\partial\widetilde{a^\ast}_i{}^\prime}\right]+
\kappa^c\overset{\ast}{T}_{l\dot{l}}\!\!{}^{-1}(\fg)\psi^\prime=0.\nonumber
\end{multline}
Or,
\begin{multline}
\sum_i\left[
g^-_{i1}\sL^{l\dot{l}}_1T^{-1}_{l\dot{l}}(\fg)
\frac{\partial\psi^\prime}{\partial a^\prime_i}+
g^-_{i2}\sL^{l\dot{l}}_2T^{-1}_{l\dot{l}}(\fg)
\frac{\partial\psi^\prime}{\partial a^\prime_i}+
g^-_{i3}\sL^{l\dot{l}}_3T^{-1}_{l\dot{l}}(\fg)
\frac{\partial\psi^\prime}{\partial a^\prime_i}-
\right.\\
\shoveright{\left.
-ig^-_{i1}\sL^{l\dot{l}}_1T^{-1}_{l\dot{l}}(\fg)\frac{\partial\psi^\prime}
{\partial a^\ast_i{}^\prime}-
ig^-_{i2}\sL^{l\dot{l}}_2T^{-1}_{l\dot{l}}(\fg)\frac{\partial\psi^\prime}
{\partial a^\ast_i{}^\prime}-
ig^-_{i3}\sL^{l\dot{l}}_3T^{-1}_{l\dot{l}}(\fg)\frac{\partial\psi^\prime}
{\partial a^\ast_i{}^\prime}\right]+
\dot{\kappa}^cT^{-1}_{l\dot{l}}(\fg)\dot{\psi}^\prime=0,}\\
\shoveleft{
\sum_i\left[
g^+_{i1}\overset{\ast}{\sL}{}^{l\dot{l}}_1
\overset{\ast}{T}_{l\dot{l}}\!\!{}^{-1}(\fg)
\frac{\partial\dot{\psi}^\prime}{\partial\widetilde{a}^\prime_i}+
g^+_{i2}\overset{\ast}{\sL}{}^{l\dot{l}}_2
\overset{\ast}{T}_{l\dot{l}}\!\!{}^{-1}(\fg)
\frac{\partial\dot{\psi}^\prime}{\partial\widetilde{a}^\prime_i}+
g^+_{i3}\overset{\ast}{\sL}{}^{l\dot{l}}_3
\overset{\ast}{T}_{l\dot{l}}\!\!{}^{-1}(\fg)
\frac{\partial\dot{\psi}^\prime}{\partial\widetilde{a}^\prime_i}+\right.}\\
\left.
+ig^+_{i1}\overset{\ast}{\sL}{}^{l\dot{l}}_1
\overset{\ast}{T}_{l\dot{l}}\!\!{}^{-1}(\fg)
\frac{\partial\dot{\psi}^\prime}{\partial\widetilde{a^\ast}_i{}^\prime}+
ig^+_{i2}\overset{\ast}{\sL}{}^{l\dot{l}}_2
\overset{\ast}{T}_{l\dot{l}}\!\!{}^{-1}(\fg)
\frac{\partial\dot{\psi}^\prime}{\partial\widetilde{a^\ast}_i{}^\prime}+
ig^+_{i3}\overset{\ast}{\sL}{}^{l\dot{l}}_3
\overset{\ast}{T}_{l\dot{l}}\!\!{}^{-1}(\fg)
\frac{\partial\dot{\psi}^\prime}{\partial\widetilde{a^\ast}_i{}^\prime}\right]+
\kappa^c\overset{\ast}{T}_{l\dot{l}}\!\!{}^{-1}(\fg)\psi^\prime=0.\nonumber
\end{multline}
For coincidence of the latter system with (\ref{CT}) we must multiply
this system by $T_{l\dot{l}}(\fg)$
($\overset{\ast}{T}_{l\dot{l}}(\fg)$) from the left,
\begin{eqnarray}
\sum_i\sum_k g^-_{ik}T_{l\dot{l}}(\fg)\sL^{l\dot{l}}_kT^{-1}_{l\dot{l}}(\fg)
\frac{\partial\psi^\prime}{\partial a^\prime_i}
-i\sum_i\sum_k g^-_{ik}T_{l\dot{l}}(\fg)\sL^{l\dot{l}}_kT^{-1}_{l\dot{l}}(\fg)
\frac{\partial\psi^\prime}{\partial a^\ast_i{}^\prime}+
\dot{\kappa}^c\dot{\psi}^\prime&=&0,\nonumber\\
\sum_i\sum_k g^+_{ik}\overset{\ast}{T}_{l\dot{l}}(\fg)
\overset{\ast}{\sL}{}^{l\dot{l}}_k
\overset{\ast}{T}_{l\dot{l}}\!\!{}^{-1}(\fg)\frac{\partial\dot{\psi}^\prime}
{\partial\widetilde{a}^\prime_i}+
i\sum_i\sum_k g^+_{ik}\overset{\ast}{T}_{l\dot{l}}(\fg)
\overset{\ast}{\sL}{}^{l\dot{l}}_k
\overset{\ast}{T}_{l\dot{l}}\!\!{}^{-1}(\fg)\frac{\partial\dot{\psi}^\prime}
{\partial\widetilde{a^\ast}_i{}^\prime}+
\kappa^c\psi^\prime&=&0.\nonumber
\end{eqnarray}
The requirement of invariance means that for any transformation $\fg$
between the matrices $\sL^{l\dot{l}}_k$
($\overset{\ast}{\sL}{}^{l\dot{l}}_k$) the following relations hold:
\begin{eqnarray}
\sum_k g^-_{ik}T_{l\dot{l}}(\fg)\sL^{l\dot{l}}_k
T^{-1}_{l\dot{l}}(\fg)&=&\sL^{l\dot{l}}_i,\nonumber\\
\sum_k g^+_{ik}\overset{\ast}{T}_{l\dot{l}}(\fg)
\overset{\ast}{\sL}{}^{l\dot{l}}_k
\overset{\ast}{T}_{l\dot{l}}\!\!\!{}^{-1}(\fg)
&=&\overset{\ast}{\sL}{}^{l\dot{l}}_i,\label{IC}
\end{eqnarray}
where $\overset{\ast}{\sL}{}^{l\dot{l}}_i$
are the matrices of the equations in the dual
representation space, $\kappa^c$ is a complex number,
$\partial/\partial\widetilde{a}_i$ mean covariant derivatives in the dual
space.
\section{The structure of the matrices $\sL^{l\dot{l}}_i$}
First of all,
let us find commutation relations between the matrices
$\sL^{l\dot{l}}_i$, $\overset{\ast}{\sL}{}^{l\dot{l}}_i$
and infinitesimal operators (\ref{Op1})--(\ref{Op12})
defined in the helicity basis.
Let us represent transformations $T_{l\dot{l}}(\fg)$
($\overset{\ast}{T}_{l\dot{l}}(\fg)$) in the infinitesimal form,
\begin{eqnarray}
\sI+\sA^{l\dot{l}}_i\xi+\ldots, &&\quad\sI+\sB^{l\dot{l}}_i\xi+\ldots,
\nonumber\\
\sI+\widetilde{\sA}^{l\dot{l}}_i\xi+\ldots, &&\quad \sI+
\widetilde{\sB}^{l\dot{l}}_i\xi+\ldots.
\nonumber
\end{eqnarray}
It is easy to see that the bivector space $\R^6$ contains two
three-dimensional subspaces $\R^3_-$ and $\R^3_+$ with the metric tensors
$g^-_{ik}$ and $g^+_{ik}$, respectively. Let us consider a rotation
$\fg=e+a_1\xi+\ldots$ in the subspace $\R^3_-$. The matrix of this
rotation can be represented in the form
\[
\begin{pmatrix}
1 & 0 & 0\\
0 & 1 &-\xi\\
0 &\xi& 1
\end{pmatrix}.
\]
Substituting these transformations into invariance conditions (\ref{IC}),
we obtain with an accuracy of the terms of second order the following
three equalities:
\begin{eqnarray}
(\sI+\sA^{l\dot{l}}_1\xi)\Lambda^{l\dot{l}}_1(\sI-\sA^{l\dot{l}}_1\xi)&=&
\Lambda^{l\dot{l}}_1,\nonumber\\
(\sI+\sA^{l\dot{l}}_1\xi)(\Lambda^{l\dot{l}}_2-\xi\Lambda^{l\dot{l}}_3)
(\sI-\sA^{l\dot{l}}_1\xi)&=&\Lambda^{l\dot{l}}_2,\nonumber\\
(\sI+\sA^{l\dot{l}}_1\xi)(\xi\Lambda^{l\dot{l}}_2+\Lambda^{l\dot{l}}_3)
(\sI-\sA^{l\dot{l}}_1\xi)&=&\Lambda^{l\dot{l}}_3.\nonumber
\end{eqnarray}
Hence it follows that
\begin{eqnarray}
\sA^{l\dot{l}}_1\Lambda^{l\dot{l}}_1-\Lambda^{l\dot{l}}_1\sA^{l\dot{l}}_1&=&0,
\nonumber\\
\sA^{l\dot{l}}_1\Lambda^{l\dot{l}}_2-\Lambda^{l\dot{l}}_2\sA^{l\dot{l}}_1-
\Lambda^{l\dot{l}}_3&=&0,\nonumber\\
\sA^{l\dot{l}}_1\Lambda^{l\dot{l}}_3-\Lambda^{l\dot{l}}_3\sA^{l\dot{l}}_1+
\Lambda^{l\dot{l}}_2&=&0,\nonumber
\end{eqnarray}
or
\begin{eqnarray}
\ld\sA^{l\dot{l}}_1,\Lambda^{l\dot{l}}_1\rd&=&0,\nonumber\\
\ld\sA^{l\dot{l}}_1,\Lambda^{l\dot{l}}_2\rd&=&\Lambda^{l\dot{l}}_3,\nonumber\\
\ld\sA^{l\dot{l}}_1,\Lambda^{l\dot{l}}_3\rd&=&-\Lambda^{l\dot{l}}_2.\nonumber
\end{eqnarray}
Analogously, for a rotation $\fg=e+a_2\xi+\ldots$ with the matrix
\[
\begin{pmatrix}
1 & 0 & \xi\\
0 & 1 & 0\\
-\xi & 0 & 1
\end{pmatrix}
\]
we have
\begin{eqnarray}
(1+\sA^{l\dot{l}}_2\xi)(\Lambda^{l\dot{l}}_1+\xi\Lambda^{l\dot{l}}_3)
(1-\sA^{l\dot{l}}_2\xi)&=&\Lambda^{l\dot{l}}_1,\nonumber\\
(1+\sA^{l\dot{l}}_2\xi)\Lambda^{l\dot{l}}_2(1-\sA^{l\dot{l}}_2\xi)&=&
\Lambda^{l\dot{l}}_2,\nonumber\\
(1+\sA^{l\dot{l}}_2\xi)(\xi\Lambda^{l\dot{l}}_1+\Lambda^{l\dot{l}}_3)
(1-\sA^{l\dot{l}}_2\xi)&=&\Lambda^{l\dot{l}}_3.\nonumber
\end{eqnarray}
From the latter relations we see that
\begin{eqnarray}
\ld\sA^{l\dot{l}}_2,\Lambda^{l\dot{l}}_1\rd&=&-\Lambda^{l\dot{l}}_3,\nonumber\\
\ld\sA^{l\dot{l}}_2,\Lambda^{l\dot{l}}_2\rd&=&0,\nonumber\\
\ld\sA^{l\dot{l}}_2,\Lambda^{l\dot{l}}_3\rd&=&\Lambda^{l\dot{l}}_1.\nonumber
\end{eqnarray}
Further, taking into account all possible transformations (rotations)
in the subspaces $\R^3_-$ and $\R^3_+$, we obtain the following commutation
relations:
\begin{equation}\label{AL}
\begin{array}{rcl}
\ld\sA^{l\dot{l}}_1,\sL^{l\dot{l}}_1\rd &=& 0,\\
\ld\sA^{l\dot{l}}_2,\sL^{l\dot{l}}_1\rd &=&-\sL^{l\dot{l}}_3,\\
\ld\sA^{l\dot{l}}_3,\sL^{l\dot{l}}_1\rd &=& \sL^{l\dot{l}}_2,
\end{array}\;\;\;
\begin{array}{rcl}
\ld\sA^{l\dot{l}}_1,\sL^{l\dot{l}}_2\rd &=& \sL^{l\dot{l}}_3,\\
\ld\sA^{l\dot{l}}_2,\sL^{l\dot{l}}_2\rd &=& 0,\\
\ld\sA^{l\dot{l}}_3,\sL^{l\dot{l}}_2\rd &=&-\sL^{l\dot{l}}_1,
\end{array}\;\;\;
\begin{array}{rcl}
\ld\sA^{l\dot{l}}_1,\sL^{l\dot{l}}_3\rd &=&-\sL^{l\dot{l}}_2,\\
\ld\sA^{l\dot{l}}_2,\sL^{l\dot{l}}_3\rd &=& \sL^{l\dot{l}}_1,\\
\ld\sA^{l\dot{l}}_3,\sL^{l\dot{l}}_3\rd &=& 0.
\end{array}
\end{equation}
\begin{equation}\label{BL}
\begin{array}{rcl}
\ld\sB^{l\dot{l}}_1,\sL^{l\dot{l}}_1\rd &=& 0,\\
\ld\sB^{l\dot{l}}_2,\sL^{l\dot{l}}_1\rd &=&i\sL^{l\dot{l}}_3,\\
\ld\sB^{l\dot{l}}_3,\sL^{l\dot{l}}_1\rd &=& -i\sL^{l\dot{l}}_2,
\end{array}\;\;\;
\begin{array}{rcl}
\ld\sB^{l\dot{l}}_1,\sL^{l\dot{l}}_2\rd &=& -i\sL^{l\dot{l}}_3,\\
\ld\sB^{l\dot{l}}_2,\sL^{l\dot{l}}_2\rd &=& 0,\\
\ld\sB^{l\dot{l}}_3,\sL^{l\dot{l}}_2\rd &=&i\sL^{l\dot{l}}_1,
\end{array}\;\;\;
\begin{array}{rcl}
\ld\sB^{l\dot{l}}_1,\sL^{l\dot{l}}_3\rd &=&i\sL^{l\dot{l}}_2,\\
\ld\sB^{l\dot{l}}_2,\sL^{l\dot{l}}_3\rd &=& -i\sL^{l\dot{l}}_1,\\
\ld\sB^{l\dot{l}}_3,\sL^{l\dot{l}}_3\rd &=& 0.
\end{array}
\end{equation}
\begin{equation}\label{DAL'}
{\renewcommand{\arraystretch}{1.2}
\begin{array}{rcl}
\ld\widetilde{\sA}^{l\dot{l}}_1,\sL^{l\dot{l}}_1\rd &=& 0,\\
\ld\widetilde{\sA}^{l\dot{l}}_2,\sL^{l\dot{l}}_1\rd &=& -\sL^{l\dot{l}}_3,\\
\ld\widetilde{\sA}^{l\dot{l}}_3,\sL^{l\dot{l}}_1\rd &=&\sL^{l\dot{l}}_2,
\end{array}\;\;\;
\begin{array}{rcl}
\ld\widetilde{\sA}^{l\dot{l}}_1,\sL^{l\dot{l}}_2\rd &=&\sL^{l\dot{l}}_3,\\
\ld\widetilde{\sA}^{l\dot{l}}_2,\sL^{l\dot{l}}_2\rd &=& 0,\\
\ld\widetilde{\sA}^{l\dot{l}}_3,\sL^{l\dot{l}}_2\rd &=& -\sL^{l\dot{l}}_1,
\end{array}\;\;\;
\begin{array}{rcl}
\ld\widetilde{\sA}^{l\dot{l}}_1,\sL^{l\dot{l}}_3\rd &=& -\sL^{l\dot{l}}_2,\\
\ld\widetilde{\sA}^{l\dot{l}}_2,\sL^{l\dot{l}}_3\rd &=&\sL^{l\dot{l}}_1,\\
\ld\widetilde{\sA}^{l\dot{l}}_3,\sL^{l\dot{l}}_3\rd &=& 0.
\end{array}}
\end{equation}
\begin{equation}\label{DBL'}
{\renewcommand{\arraystretch}{1.2}
\begin{array}{rcl}
\ld\widetilde{\sB}^{l\dot{l}}_1,\sL^{l\dot{l}}_1\rd &=& 0,\\
\ld\widetilde{\sB}^{l\dot{l}}_2,\sL^{l\dot{l}}_1\rd &=&-i\sL^{l\dot{l}}_3,\\
\ld\widetilde{\sB}^{l\dot{l}}_3,\sL^{l\dot{l}}_1\rd &=&i\sL^{l\dot{l}}_2,
\end{array}\;\;\;
\begin{array}{rcl}
\ld\widetilde{\sB}^{l\dot{l}}_1,\sL^{l\dot{l}}_2\rd &=&i\sL^{l\dot{l}}_3,\\
\ld\widetilde{\sB}^{l\dot{l}}_2,\sL^{l\dot{l}}_2\rd &=& 0,\\
\ld\widetilde{\sB}^{l\dot{l}}_3,\sL^{l\dot{l}}_2\rd &=&-i\sL^{l\dot{l}}_1,
\end{array}\;\;\;
\begin{array}{rcl}
\ld\widetilde{\sB}^{l\dot{l}}_1,\sL^{l\dot{l}}_3\rd &=&-i\sL^{l\dot{l}}_2,\\
\ld\widetilde{\sB}^{l\dot{l}}_2,\sL^{l\dot{l}}_3\rd &=&i\sL^{l\dot{l}}_1,\\
\ld\widetilde{\sB}^{l\dot{l}}_3,\sL^{l\dot{l}}_3\rd &=& 0.
\end{array}}
\end{equation}
\begin{equation}\label{DAL}
{\renewcommand{\arraystretch}{1.6}
\begin{array}{rcl}
\ld\widetilde{\sA}^{l\dot{l}}_1,\overset{\ast}{\sL}\!{}^{l\dot{l}}_1\rd &=& 0,\\
\ld\widetilde{\sA}^{l\dot{l}}_2,
\overset{\ast}{\sL}{}^{l\dot{l}}_1\rd &=& -\overset{\ast}{\sL}{}^{l\dot{l}}_3,\\
\ld\widetilde{\sA}^{l\dot{l}}_3,
\overset{\ast}{\sL}{}^{l\dot{l}}_1\rd &=&\overset{\ast}{\sL}{}^{l\dot{l}}_2,
\end{array}\;\;\;
\begin{array}{rcl}
\ld\widetilde{\sA}^{l\dot{l}}_1,
\overset{\ast}{\sL}{}^{l\dot{l}}_2\rd &=&\overset{\ast}{\sL}^{l\dot{l}}_3,\\
\ld\widetilde{\sA}^{l\dot{l}}_2,\overset{\ast}{\sL}{}^{l\dot{l}}_2\rd &=& 0,\\
\ld\widetilde{\sA}^{l\dot{l}}_3,
\overset{\ast}{\sL}{}^{l\dot{l}}_2\rd &=& -\overset{\ast}{\sL}{}^{l\dot{l}}_1,
\end{array}\;\;\;
\begin{array}{rcl}
\ld\widetilde{\sA}^{l\dot{l}}_1,
\overset{\ast}{\sL}{}^{l\dot{l}}_3\rd &=& -\overset{\ast}{\sL}{}^{l\dot{l}}_2,\\
\ld\widetilde{\sA}^{l\dot{l}}_2,
\overset{\ast}{\sL}{}^{l\dot{l}}_3\rd &=&\overset{\ast}{\sL}{}^{l\dot{l}}_1,\\
\ld\widetilde{\sA}^{l\dot{l}}_3,\overset{\ast}{\sL}{}^{l\dot{l}}_3\rd &=& 0.
\end{array}}
\end{equation}
\begin{equation}\label{DBL}
{\renewcommand{\arraystretch}{1.2}
\begin{array}{rcl}
\ld\widetilde{\sB}^{l\dot{l}}_1,\overset{\ast}{\sL}{}^{l\dot{l}}_1\rd &=& 0,\\
\ld\widetilde{\sB}^{l\dot{l}}_2,
\overset{\ast}{\sL}{}^{l\dot{l}}_1\rd &=&-i\overset{\ast}{\sL}{}^{l\dot{l}}_3,\\
\ld\widetilde{\sB}^{l\dot{l}}_3,
\overset{\ast}{\sL}{}^{l\dot{l}}_1\rd &=&i\overset{\ast}{\sL}{}^{l\dot{l}}_2,
\end{array}\;\;\;
\begin{array}{rcl}
\ld\widetilde{\sB}^{l\dot{l}}_1,
\overset{\ast}{\sL}{}^{l\dot{l}}_2\rd &=&i\overset{\ast}{\sL}{}^{l\dot{l}}_3,\\
\ld\widetilde{\sB}^{l\dot{l}}_2,\overset{\ast}{\sL}{}^{l\dot{l}}_2\rd &=& 0,\\
\ld\widetilde{\sB}^{l\dot{l}}_3,
\overset{\ast}{\sL}{}^{l\dot{l}}_2\rd &=&-i\overset{\ast}{\sL}{}^{l\dot{l}}_1,
\end{array}\;\;\;
\begin{array}{rcl}
\ld\widetilde{\sB}^{l\dot{l}}_1,
\overset{\ast}{\sL}{}^{l\dot{l}}_3\rd &=&-i\overset{\ast}{\sL}{}^{l\dot{l}}_2,\\
\ld\widetilde{\sB}^{l\dot{l}}_2,
\overset{\ast}{\sL}{}^{l\dot{l}}_3\rd &=&i\overset{\ast}{\sL}{}^{l\dot{l}}_1,\\
\ld\widetilde{\sB}^{l\dot{l}}_3,\overset{\ast}{\sL}{}^{l\dot{l}}_3\rd &=& 0.
\end{array}}
\end{equation}
\begin{equation}\label{AL'}
\begin{array}{rcl}
\ld\sA^{l\dot{l}}_1,\overset{\ast}{\sL}{}^{l\dot{l}}_1\rd &=& 0,\\
\ld\sA^{l\dot{l}}_2,\overset{\ast}{\sL}{}^{l\dot{l}}_1\rd
&=&-\overset{\ast}{\sL}{}^{l\dot{l}}_3,\\
\ld\sA^{l\dot{l}}_3,\overset{\ast}{\sL}{}^{l\dot{l}}_1\rd
&=& \overset{\ast}{\sL}{}^{l\dot{l}}_2,
\end{array}\;\;\;
\begin{array}{rcl}
\ld\sA^{l\dot{l}}_1,\overset{\ast}{\sL}{}^{l\dot{l}}_2\rd
&=& \overset{\ast}{\sL}{}^{l\dot{l}}_3,\\
\ld\sA^{l\dot{l}}_2,\overset{\ast}{\sL}{}^{l\dot{l}}_2\rd &=& 0,\\
\ld\sA^{l\dot{l}}_3,\overset{\ast}{\sL}{}^{l\dot{l}}_2\rd
&=&-\overset{\ast}{\sL}{}^{l\dot{l}}_1,
\end{array}\;\;\;
\begin{array}{rcl}
\ld\sA^{l\dot{l}}_1,\overset{\ast}{\sL}{}^{l\dot{l}}_3\rd
&=&-\overset{\ast}{\sL}{}^{l\dot{l}}_2,\\
\ld\sA^{l\dot{l}}_2,\overset{\ast}{\sL}{}^{l\dot{l}}_3\rd
&=& \overset{\ast}{\sL}{}^{l\dot{l}}_1,\\
\ld\sA^{l\dot{l}}_3,\overset{\ast}{\sL}{}^{l\dot{l}}_3\rd &=& 0.
\end{array}
\end{equation}
\begin{equation}\label{BL'}
\begin{array}{rcl}
\ld\sB^{l\dot{l}}_1,\overset{\ast}{\sL}{}^{l\dot{l}}_1\rd &=& 0,\\
\ld\sB^{l\dot{l}}_2,\overset{\ast}{\sL}{}^{l\dot{l}}_1\rd
&=&i\overset{\ast}{\sL}{}^{l\dot{l}}_3,\\
\ld\sB^{l\dot{l}}_3,\overset{\ast}{\sL}{}^{l\dot{l}}_1\rd
&=& -i\overset{\ast}{\sL}{}^{l\dot{l}}_2,
\end{array}\;\;\;
\begin{array}{rcl}
\ld\sB^{l\dot{l}}_1,\overset{\ast}{\sL}{}^{l\dot{l}}_2\rd
&=& -i\overset{\ast}{\sL}{}^{l\dot{l}}_3,\\
\ld\sB^{l\dot{l}}_2,\overset{\ast}{\sL}{}^{l\dot{l}}_2\rd &=& 0,\\
\ld\sB^{l\dot{l}}_3,\overset{\ast}{\sL}{}^{l\dot{l}}_2\rd
&=&i\overset{\ast}{\sL}{}^{l\dot{l}}_1,
\end{array}\;\;\;
\begin{array}{rcl}
\ld\sB^{l\dot{l}}_1,\overset{\ast}{\sL}{}^{l\dot{l}}_3\rd
&=&i\overset{\ast}{\sL}{}^{l\dot{l}}_2,\\
\ld\sB^{l\dot{l}}_2,\overset{\ast}{\sL}{}^{l\dot{l}}_3\rd
&=& -i\overset{\ast}{\sL}{}^{l\dot{l}}_1,\\
\ld\sB^{l\dot{l}}_3,\overset{\ast}{\sL}{}^{l\dot{l}}_3\rd &=& 0.
\end{array}
\end{equation}
\begin{sloppypar}
From the latter relations and definition (\ref{SL25}) it immediately follows
that commutation relations between $\sL^{l\dot{l}}_3$,
$\overset{\ast}{\sL}{}^{l\dot{l}}_3$ and
generators $\sY^{l\dot{l}}_\pm$, $\sY^{l\dot{l}}_3$,
$\sX^{l\dot{l}}_\pm$, $\sX^{l\dot{l}}_3$ are of the form
\end{sloppypar}
\begin{equation}\label{LX}
{\renewcommand{\arraystretch}{1.5}
\left.\begin{array}{l}
\ld\ld\sL^{l\dot{l}}_3,\sX^{l\dot{l}}_-\rd,\sX^{l\dot{l}}_+\rd =
2\sL^{l\dot{l}}_3,\\
\ld\ld\sL^{l\dot{l}}_3,\sY^{l\dot{l}}_-\rd,\sY^{l\dot{l}}_+\rd=
2\sL^{l\dot{l}}_3,\\
\ld\sL^{l\dot{l}}_3,\sX^{l\dot{l}}_3\rd =0,\\
\ld\sL^{l\dot{l}}_3,\sY^{l\dot{l}}_3\rd =0,
\end{array}\right.
}
{\renewcommand{\arraystretch}{1.5}
\left.\begin{array}{l}
\ld\ld\overset{\ast}{\sL}{}^{l\dot{l}}_3,\sY^{l\dot{l}}_-\rd,
\sY^{l\dot{l}}_+\rd =2\overset{\ast}{\sL}{}^{l\dot{l}}_3,\\
\ld\ld\overset{\ast}{\sL}{}^{l\dot{l}}_3,\sX^{l\dot{l}}_-\rd,
\sX^{l\dot{l}}_+\rd =2\overset{\ast}{\sL}{}^{l\dot{l}}_3,\\
\ld\overset{\ast}{\sL}{}^{l\dot{l}}_3,\sY^{l\dot{l}}_3\rd =0,\\
\ld\overset{\ast}{\sL}{}^{l\dot{l}}_3,\sX^{l\dot{l}}_3\rd =0.
\end{array}\right.
}
\end{equation}
Using the relations (\ref{LX}) we will find an explicit form of the matrices
$\sL^{l\dot{l}}_3$ and $\overset{\ast}{\sL}{}^{l\dot{l}}_3$,
and further we will find
$\sL^{l\dot{l}}_1$, $\sL^{l\dot{l}}_2$ and $\overset{\ast}{\sL}{}^{l\dot{l}}_1$,
$\overset{\ast}{\sL}{}^{l\dot{l}}_2$.

The wave function\index{function!wave}
$\boldsymbol{\psi}$ is transformed within
some representation $T_{l\dot{l}}(\fg)$ of the group $\fG_+$.
We assume that $T_{l\dot{l}}(\fg)$
is decomposed into irreducible representations.
The components of the
function $\boldsymbol{\psi}$ we will numerate by the indices
$l$, $\dot{l}$ and $m$, $\dot{m}$, where $l$ ($\dot{l}$)
is a weight of irreducible representation,
$m$ ($\dot{m}$)
is a number of the components in the
representation of the weight $l$ ($\dot{l}$).
In the case when a representation with one and the
same weight $l$ ($\dot{l}$) at the
decomposition of $\boldsymbol{\psi}$ occurs more than one time, then with
the aim to distinguish these representations we will add the index
$k$ ($\dot{k}$),
which indicates a number of the
representations of the weight $l$ ($\dot{l}$).
Denoting
$\zeta_{lm;\dot{l}\dot{m}}=\mid lm;\dot{l}\dot{m}\rangle$ and coming to the
helicity basis,\index{basis!helicity}
we obtain a following decomposition for the wave function:
\[
\boldsymbol{\psi}(a_1,a_2,a_3,a^\ast_1,a^\ast_2,a^\ast_3)=
\sum_{l,m,k,\dot{l},\dot{m},\dot{k}}
\psi^{k\dot{k}}_{lm;\dot{l}\dot{m}}(a_1,a_2,a_3,a^\ast_1,a^\ast_2,a^\ast_3)
\zeta^{k\dot{k}}_{lm;\dot{l}\dot{m}},
\]
where $a_1,a_2,a_3,a^\ast_1,a^\ast_2,a^\ast_3$
are the coordinates of the complex space
$\C^3\simeq\R^6$
(parameters of $SL(2,\C)$)\footnote{Recall that the wave function
$\boldsymbol{\psi}(a_j,a^\ast_j)$ is defined on the group manifold
$\fL_6$, that is, $\boldsymbol{\psi}$ is a function on the Lorentz group.}.
Analogously, for the dual representation we have
\[
\dot{\boldsymbol{\psi}}(\widetilde{a}_1,\widetilde{a}_2,\widetilde{a}_3,
\widetilde{a^\ast}_1,\widetilde{a^\ast}_2,\widetilde{a^\ast}_3)=
\sum_{l,m,k,\dot{l},\dot{m},\dot{k}}
\dot{\psi}^{k\dot{k}}_{lm;\dot{l}\dot{m}}
(\widetilde{a}_1,\widetilde{a}_2,\widetilde{a}_3,\widetilde{a^\ast}_1,
\widetilde{a^\ast}_2,\widetilde{a^\ast}_3)
\zeta^{k\dot{k}}_{lm;\dot{l}\dot{m}}.
\]
The transformation $\sL^{l\dot{l}}_3$ in the helicity basis has the form
\[
\sL^{l\dot{l}}_3\zeta^{k\dot{k}}_{lm;\dot{l}\dot{m}}=
\sum_{l^\prime,m^\prime,k^\prime,\dot{l}^\prime,\dot{m}^\prime,\dot{k}^\prime}
c^{k^\prime k;\dot{k}^\prime\dot{k}}_{l^\prime l,m^\prime m;\dot{l}^\prime,
\dot{l},\dot{m}^\prime\dot{m}}
\zeta^{k^\prime\dot{k}^\prime}_{l^\prime m^\prime;\dot{l}^\prime\dot{m}^\prime}.
\]
Using the commutators (\ref{LX}), we will find the numbers
$c^{k^\prime k;\dot{k}^\prime\dot{k}}_{l^\prime l,m^\prime m;\dot{l}^\prime,
\dot{l},\dot{m}^\prime\dot{m}}$.
First of all, recalling that
\begin{eqnarray}
\sX^{l\dot{l}}_-\zeta^{k\dot{k}}_{lm;\dot{l}\dot{m}}&=&\boldsymbol{\alpha}^l_m
\zeta^{k\dot{k}}_{l,m-1;\dot{l}\dot{m}},\nonumber\\
\sX^{l\dot{l}}_+\zeta^{k\dot{k}}_{lm;\dot{l}\dot{m}}&=&\boldsymbol{\alpha}^{l}_{m+1}
\zeta^{k\dot{k}}_{l,m+1;\dot{l}\dot{m}},\nonumber\\
\sX^{l\dot{l}}_3\zeta^{k\dot{k}}_{lm;\dot{l}\dot{m}}&=&
m\zeta^{k\dot{k}}_{lm;\dot{l}\dot{m}},\nonumber
\end{eqnarray}
where $(\boldsymbol{\alpha}^l_m)^2=(l+m)(l-m+1)$, we obtain
\begin{eqnarray}
\sL^{l\dot{l}}_3\sX^{l\dot{l}}_3
\zeta^{k\dot{k}}_{lm;\dot{l}\dot{m}}&=&m\sL^{l\dot{l}}_3
\zeta^{k\dot{k}}_{lm;\dot{l}\dot{m}}=
m\sum_{l^\prime,m^\prime,k^\prime,\dot{l}^\prime,\dot{m}^\prime,\dot{k}^\prime}
c^{k^\prime k;\dot{k}^\prime\dot{k}}_{l^\prime l,m^\prime m;\dot{l}^\prime,
\dot{l},\dot{m}^\prime\dot{m}}
\zeta^{k^\prime\dot{k}^\prime}_{l^\prime m^\prime;\dot{l}^\prime\dot{m}^\prime},
\nonumber\\
\sX^{l\dot{l}}_3\sL^{l\dot{l}}_3
\zeta^{k\dot{k}}_{lm;\dot{l}\dot{m}}&=&\sX^{l\dot{l}}_3
\sum_{l^\prime,m^\prime,k^\prime,\dot{l}^\prime,\dot{m}^\prime,\dot{k}^\prime}
c^{k^\prime k;\dot{k}^\prime\dot{k}}_{l^\prime l,m^\prime m;\dot{l}^\prime,
\dot{l},\dot{m}^\prime\dot{m}}
\zeta^{k^\prime\dot{k}^\prime}_{l^\prime m^\prime;\dot{l}^\prime\dot{m}^\prime}=
m^\prime\sum_{l^\prime,m^\prime,k^\prime,\dot{l}^\prime,\dot{m}^\prime,\dot{k}^\prime}
c^{k^\prime k;\dot{k}^\prime\dot{k}}_{l^\prime l,m^\prime m;\dot{l}^\prime,
\dot{l},\dot{m}^\prime\dot{m}}
\zeta^{k^\prime\dot{k}^\prime}_{l^\prime m^\prime;\dot{l}^\prime\dot{m}^\prime}.
\nonumber
\end{eqnarray}
From the second equation of (\ref{LX}) we have
$(\sL^{l\dot{l}}_3\sX^{l\dot{l}}_3-
\sX^{l\dot{l}}_3\sL^{l\dot{l}}_3)\zeta^{k\dot{k}}_{lm;\dot{l}\dot{m}}=0$.
Therefore,
\[
\sum_{l^\prime,m^\prime,k^\prime,\dot{l}^\prime,\dot{m}^\prime,\dot{k}^\prime}
(m-m^\prime)
c^{k^\prime k;\dot{k}^\prime\dot{k}}_{l^\prime l,m^\prime m;\dot{l}^\prime,
\dot{l},\dot{m}^\prime\dot{m}}
\zeta^{k^\prime\dot{k}^\prime}_{l^\prime m^\prime;\dot{l}^\prime
\dot{m}^\prime}=0.
\]
Hence it immediately follows that $m^\prime=m$. By this reason we can
denote the coefficients
$c^{k^\prime k;\dot{k}^\prime\dot{k}}_{l^\prime l, m^\prime m;\dot{l}^\prime
\dot{l},\dot{m}^\prime\dot{m}}$ via
$c^{k^\prime k;\dot{k}^\prime\dot{k}}_{l^\prime l, m;
\dot{l}^\prime\dot{l},\dot{m}^\prime\dot{m}}$. Analogously, from the
relations
\begin{eqnarray}
\sY^{l\dot{l}}_-\zeta^{k\dot{k}}_{lm;\dot{l}\dot{m}}
&=&\balpha^{\dot{l}}_{\dot{m}}\zeta^{k\dot{k}}_{lm;\dot{l},\dot{m}-1},
\nonumber\\
\sY^{l\dot{l}}_+\zeta^{k\dot{k}}_{lm;\dot{l}\dot{m}}
&=&\balpha^{\dot{l}}_{\dot{m}+1}\zeta^{k\dot{k}}_{lm;\dot{l},\dot{m}+1},
\nonumber\\
\sY^{l\dot{l}}_3\zeta^{k\dot{k}}_{lm;\dot{l}\dot{m}}&=&
\dot{m}\zeta^{k\dot{k}}_{lm;\dot{l}\dot{m}}\nonumber
\end{eqnarray}
and equation $(\Lambda^{l\dot{l}}_3\sY^{l\dot{l}}_3-\sY^{l\dot{l}}_3
\Lambda^{l\dot{l}}_3)\zeta^{k\dot{k}}_{lm;\dot{l}\dot{m}}=0$ we see that
the coefficients $c^{k^\prime k;\dot{k}^\prime\dot{k}}_{l^\prime l,m;
\dot{l}^\prime\dot{l},\dot{m}^\prime\dot{m}}$ can be replaced by
$c^{k^\prime k;\dot{k}^\prime\dot{k}}_{l^\prime l,m;\dot{l}^\prime\dot{l},
\dot{m}}$.

Let us use now the first equation of the system (\ref{LX}):
\begin{eqnarray}
\sL^{l\dot{l}}_3\sX^{l\dot{l}}_-
\zeta^{k\dot{k}}_{lm;\dot{l}\dot{m}}&=&\sL^{l\dot{l}}_3\boldsymbol{\alpha}^l_m
\zeta^{k\dot{k}}_{l,m-1;\dot{l}\dot{m}}=
\boldsymbol{\alpha}^l_m\sum_{l^\prime,k^\prime,\dot{l}^\prime,\dot{k}^\prime}
c^{k^\prime k;\dot{k}^\prime\dot{k}}_{l^\prime l,m-1;\dot{l}^\prime\dot{l}
\dot{m}}
\zeta^{k^\prime\dot{k}^\prime}_{l^\prime,m-1;\dot{l}^\prime\dot{m}},
\nonumber\\
\sX^{l\dot{l}}_-\sL^{l\dot{l}}_3
\zeta^{k\dot{k}}_{lm;\dot{l}\dot{m}}&=&\sX^{l\dot{l}}_-
\sum_{l^\prime,k^\prime,\dot{l}^\prime,\dot{k}^\prime}
c^{k^\prime k;\dot{k}^\prime\dot{k}}_{l^\prime l,m;\dot{l}^\prime\dot{l},\dot{m}}
\zeta^{k^\prime\dot{k}^\prime}_{l^\prime m;\dot{l}^\prime\dot{m}}=
\sum_{l^\prime,k^\prime,\dot{l}^\prime,\dot{k}^\prime}
\boldsymbol{\alpha}^{l^\prime}_m
c^{k^\prime k;\dot{k}^\prime\dot{k}}_{l^\prime l,m;\dot{l}^\prime\dot{l},\dot{m}}
\zeta^{k^\prime\dot{k}^\prime}_{l^\prime,m-1;\dot{l}^\prime\dot{m}},
\nonumber\\
\ld\sL^{l\dot{l}}_3,\sX^{l\dot{l}}_-\rd
\zeta^{k\dot{k}}_{lm;\dot{l}\dot{m}}&=&
\sum_{l^\prime,k^\prime,\dot{l}^\prime,\dot{k}^\prime}
\ld\boldsymbol{\alpha}^l_m
c^{k^\prime k;\dot{k}^\prime\dot{k}}_{l^\prime l,m-1;\dot{l}^\prime\dot{l},\dot{m}}-
\boldsymbol{\alpha}^{l^\prime}_m
c^{k^\prime k;\dot{k}^\prime\dot{k}}_{l^\prime l,m;\dot{l}^\prime\dot{l},
\dot{m}}\rd
\zeta^{k^\prime\dot{k}^\prime}_{l^\prime,m-1;\dot{l}^\prime\dot{m}}.\nonumber
\end{eqnarray}
Further,
\begin{eqnarray}
\ld\sL^{l\dot{l}}_3,\sX^{l\dot{l}}_-\rd\sX^{l\dot{l}}_+
\zeta^{k\dot{k}}_{lm;\dot{l}\dot{m}}&=&
\ld\sL^{l\dot{l}}_3,\sX^{l\dot{l}}_-\rd
\boldsymbol{\alpha}^l_{m+1}\zeta^{k\dot{k}}_{l,m+1;\dot{l}\dot{m}}=
\nonumber\\
&=&\boldsymbol{\alpha}^l_{m+1}
\sum_{l^\prime,k^\prime,\dot{l}^\prime,\dot{k}^\prime}\ld
\boldsymbol{\alpha}^l_{m+1}
c^{k^\prime k;\dot{k}^\prime\dot{k}}_{l^\prime l,m;\dot{l}^\prime\dot{l},\dot{m}}-
\boldsymbol{\alpha}^{l^\prime}_{m+1}
c^{k^\prime k;\dot{k}^\prime\dot{k}}_{l^\prime l,m+1;\dot{l}^\prime\dot{l},
\dot{m}}\rd
\zeta^{k^\prime\dot{k}^\prime}_{l^\prime,m;\dot{l}^\prime\dot{m}},\nonumber\\
\sX^{l\dot{l}}_+\ld\sL^{l\dot{l}}_3,\sX^{l\dot{l}}_-\rd
\zeta^{k\dot{k}}_{lm;\dot{l}\dot{m}}&=&
\sX^{l\dot{l}}_+
\sum_{l^\prime,k^\prime,\dot{l}^\prime,\dot{k}^\prime}
\ld\boldsymbol{\alpha}^l_m
c^{k^\prime k;\dot{k}^\prime\dot{k}}_{l^\prime l,m-1;\dot{l}^\prime\dot{l},
\dot{m}}-\boldsymbol{\alpha}^{l^\prime}_m
c^{k^\prime k;\dot{k}^\prime\dot{k}}_{l^\prime l,m;\dot{l}^\prime\dot{l},
\dot{m}}\rd
\zeta^{k^\prime\dot{k}^\prime}_{l^\prime,m-1;\dot{l}^\prime\dot{m}}=\nonumber\\
&=&\sum_{l^\prime,k^\prime,\dot{l}^\prime,\dot{k}^\prime}
\boldsymbol{\alpha}^{l^\prime}_m
\ld\boldsymbol{\alpha}^l_m
c^{k^\prime k;\dot{k}^\prime\dot{k}}_{l^\prime l,m-1;\dot{l}^\prime\dot{l},
\dot{m}}-
\boldsymbol{\alpha}^{l^\prime}_m
c^{k^\prime k;\dot{k}^\prime\dot{k}}_{l^\prime l,m;\dot{l}^\prime\dot{l},
\dot{m}}\rd
\zeta^{k^\prime\dot{k}^\prime}_{l^\prime,m;\dot{l}^\prime\dot{m}},\nonumber
\end{eqnarray}
\begin{multline}
\ld\ld\sL^{l\dot{l}}_3,\sX^{l\dot{l}}_-\rd,\sX^{l\dot{l}}_+\rd
\zeta^{k\dot{k}}_{lm;\dot{l}\dot{m}}=\\
=\sum_{l^\prime,k^\prime,\dot{l}^\prime,\dot{k}^\prime}
\left\{
\ld(\boldsymbol{\alpha}^l_{m+1})^2+
(\boldsymbol{\alpha}^{l^\prime}_m)^2\rd
c^{k^\prime k;\dot{k}^\prime\dot{k}}_{l^\prime l,m;\dot{l}^\prime\dot{l},
\dot{m}}-
\boldsymbol{\alpha}^{l^\prime}_m\boldsymbol{\alpha}^l_m
c^{k^\prime k;\dot{k}^\prime\dot{k}}_{l^\prime l,m-1;\dot{l}^\prime
\dot{l},\dot{m}}-\right.\\
-\left.\boldsymbol{\alpha}^{l^\prime}_{m+1}\boldsymbol{\alpha}^l_{m+1}
c^{k^\prime k;\dot{k}^\prime\dot{k}}_{l^\prime l,m+1;\dot{l}^\prime\dot{l},
\dot{m}}
\right\}
\zeta^{k^\prime\dot{k}^\prime}_{l^\prime m;\dot{l}^\prime\dot{m}}.
\nonumber
\end{multline}
Thus, the first commutator from (\ref{LX}) gives a system of equations
with respect to the coefficients
$c^{k^\prime k;\dot{k}^\prime\dot{k}}_{l^\prime l,m;\dot{l}^\prime\dot{l},
\dot{m}}$:
\[
2c^{k^\prime k;\dot{k}^\prime\dot{k}}_{l^\prime l,m;\dot{l}^\prime\dot{l},
\dot{m}}=
\ld(\boldsymbol{\alpha}^l_{m+1})^2+
(\boldsymbol{\alpha}^{l^\prime}_m)^2\rd
c^{k^\prime k;\dot{k}^\prime\dot{k}}_{l^\prime l,m;\dot{l}^\prime\dot{l},
\dot{m}}-
\boldsymbol{\alpha}^{l^\prime}_m\boldsymbol{\alpha}^l_m
c^{k^\prime k;\dot{k}^\prime\dot{k}}_{l^\prime l,m-1;\dot{l}^\prime
\dot{l},\dot{m}}
-\boldsymbol{\alpha}^{l^\prime}_{m+1}\boldsymbol{\alpha}^l_{m+1}
c^{k^\prime k;\dot{k}^\prime\dot{k}}_{l^\prime l,m+1;\dot{l}^\prime\dot{l},
\dot{m}}.
\]
Or, substituting instead $\boldsymbol{\alpha}^l_m$ their values, we obtain
\begin{multline}
2c^{k^\prime k;\dot{k}^\prime\dot{k}}_{l^\prime l, m;\dot{l}^\prime\dot{l},
\dot{m}}
=\ld(l+m+1)(l-m)+(l^\prime+m)(l^\prime-m+1)\rd
c^{k^\prime k;\dot{k}^\prime\dot{k}}_{l^\prime l,m;\dot{l}^\prime\dot{l},
\dot{m}}-\\
-\sqrt{(l^\prime+m)(l^\prime-m+1)(l+m)(l-m+1)}
c^{k^\prime k;\dot{k}^\prime\dot{k}}_{l^\prime l,m-1;\dot{l}^\prime\dot{l},
\dot{m}}-\\
-\sqrt{(l^\prime+m+1)(l^\prime-m)(l+m+1)(l-m)}
c^{k^\prime k;\dot{k}^\prime\dot{k}}_{l^\prime l,m+1;\dot{l}^\prime\dot{l},
\dot{m}}.\label{Sys}
\end{multline}
This system can be solved at the fixed indices
$l^\prime$, $l$, $\dot{l}^\prime$, $\dot{l}$, $k^\prime$,
$k$, $\dot{k}^\prime$, $\dot{k}$.
Let us fix some indices
$l^\prime$, $l$, $\dot{l}^\prime$, $\dot{l}$, $k^\prime$,
$k$, $\dot{k}^\prime$, $\dot{k}$
and denote
$c^{k^\prime k;\dot{k}^\prime\dot{k}}_{l^\prime l,m;\dot{l}^\prime\dot{l},
\dot{m}}$
via $c_m$. Then we obtain a system
of homogeneous equations for $c_m$, where $-\min(l^\prime,l)\leq m
\leq \min(l^\prime,l)$. We solve these equations using the Gauss method.
When $m$ has the value $m_0=\min(l^\prime,l)$, we obtain an equation
containing two unknown variables $c_{m_0}$ and $c_{m_0-1}$, from which
$c_{m_0-1}$ is defined via $c_{m_0}$. Further, when $m$ has a value
$m_0-1$, we obtain an equation with $c_{m_0-2}$, $c_{m_0-1}$, $c_{m_0}$,
from which we can define $c_{m_0-2}$ via $c_{m_0}$ again. In doing so, we
see that the coefficients
$c^{k^\prime k;\dot{k}^\prime\dot{k}}_{l^\prime l,m;\dot{l}^\prime\dot{l},
\dot{m}}$ different from
zero when $|l^\prime-l|\leq 1$, that is, at $l^\prime=l$,
$l^\prime=l-1$ and $l^\prime=l+1$. For other values of $l^\prime$
the coefficients
$c^{k^\prime k;\dot{k}^\prime\dot{k}}_{l^\prime l,m;\dot{l}^\prime\dot{l},
\dot{m}}$ are equal to zero.
First, we take $l^\prime=l$
($\dot{l}^\prime$, $\dot{l}$, $k^\prime$, $k$, $\dot{k}^\prime$,
$\dot{k}$ are arbitrary), then
the equations (\ref{Sys}) are rewritten as follows
\begin{multline}
\ld 2-(l+m+1)(l-m)-(l-m)(l-m+1)\rd
c^{k^\prime k;\dot{k}^\prime\dot{k}}_{ll,m;\dot{l}^\prime\dot{l},\dot{m}}+\\
+(l+m)(l-m+1)
c^{k^\prime k;\dot{k}^\prime\dot{k}}_{ll,m-1;\dot{l}^\prime\dot{l},\dot{m}}+
(l+m+1)(l-m)
c^{k^\prime k;\dot{k}^\prime\dot{k}}_{ll,m+1;\dot{l}^\prime\dot{l},\dot{m}}=0.
\nonumber
\end{multline}
Supposing $m=l$, we find that
$(1-l)c^{k^\prime k;\dot{k}^\prime\dot{k}}_{ll,l;\dot{l}^\prime\dot{l},\dot{m}}+
lc^{k^\prime k;\dot{k}^\prime\dot{k}}_{ll,l-1;\dot{l}^\prime\dot{l},\dot{m}}=0$.
Whence
$c^{k^\prime k;\dot{k}^\prime\dot{k}}_{ll,l;\dot{l}^\prime\dot{l},\dot{m}}=
c^{k^\prime k;\dot{k}^\prime\dot{k}}_{ll;\dot{l}^\prime\dot{l},\dot{m}}\cdot l$;
$c^{k^\prime k;\dot{k}^\prime\dot{k}}_{ll,l-1;\dot{l}^\prime\dot{l},\dot{m}}
=c^{k^\prime k;\dot{k}^\prime\dot{k}}_{ll;\dot{l}^\prime\dot{l},\dot{m}}(l-1)$,
where the constant
$c^{k^\prime k;\dot{k}^\prime\dot{k}}_{ll;\dot{l}^\prime\dot{l},\dot{m}}$
does not depend on $m$. Supposing $m=l-1$, we find analogously that
$c^{k^\prime k;\dot{k}^\prime\dot{k}}_{ll,l-2;\dot{l}^\prime\dot{l},\dot{m}}=
c^{k^\prime k;\dot{k}^\prime\dot{k}}_{ll;\dot{l}^\prime\dot{l},\dot{m}}(l-2)$.
It is easy to verify that for any $m$ there is an equality
\[
c^{k^\prime k;\dot{k}^\prime\dot{k}}_{ll,m;\dot{l}^\prime\dot{l},\dot{m}}=
c^{k^\prime k;\dot{k}^\prime\dot{k}}_{ll;\dot{l}^\prime\dot{l},\dot{m}}\cdot m.
\]
Let us suppose now $l^\prime=l-1$, then the equations (\ref{Sys}) take
the form
\begin{multline}
\ld 2-(l+m+1)(l-m)-(l+m-1)(l-m)\rd
c^{k^\prime k;\dot{k}^\prime\dot{k}}_{l-1,l,m;\dot{l}^\prime\dot{l},\dot{m}}+\\
+\sqrt{(l+m-1)(l-m)(l+m)(l-m+1)}
c^{k^\prime k;\dot{k}^\prime\dot{k}}_{l-1,l,m-1;\dot{l}^\prime\dot{l},\dot{m}}+\\
+\sqrt{(l+m)(l-m-1)(l+m+1)(l-m)}
c^{k^\prime k;\dot{k}^\prime\dot{k}}_{l-1,l,m+1;\dot{l}^\prime\dot{l},\dot{m}}
=0.\nonumber
\end{multline}
Making in these equations the substitutions
\begin{eqnarray}
c^{k^\prime k;\dot{k}^\prime\dot{k}}_{l-1,l,m;\dot{l}^\prime\dot{l},\dot{m}}&=&
\widetilde{c}^{k^\prime k;\dot{k}^\prime\dot{k}}_{l-1,l,m;\dot{l}^\prime
\dot{l},\dot{m}}\sqrt{(l+m)(l-m)},\nonumber\\
c^{k^\prime k;\dot{k}^\prime\dot{k}}_{l-1,l,m-1;\dot{l}^\prime\dot{l},\dot{m}}
&=&\widetilde{c}^{k^\prime k;\dot{k}^\prime\dot{k}}_{l-1,l,m-1;\dot{l}^\prime
\dot{l},\dot{m}}\sqrt{(l+m-1)(l-m+1)},\nonumber\\
c^{k^\prime k;\dot{k}^\prime\dot{k}}_{l-1,l,m+1;\dot{l}^\prime\dot{l},\dot{m}}
&=&\widetilde{c}^{k^\prime k;\dot{k}^\prime\dot{k}}_{l-1,l,m+1;\dot{l}^\prime
\dot{l},\dot{m}}\sqrt{(l+m+1)(l-m-1)},\nonumber
\end{eqnarray}
we obtain
\begin{multline}
\ld 2-(l+m+1)(l-m)-(l+m-1)(l-m)\rd
\widetilde{c}^{k^\prime k;\dot{k}^\prime\dot{k}}_{l-1,l,m;\dot{l}^\prime
\dot{l},\dot{m}}\sqrt{(l+m)(l-m)}+\\
+(l+m-1)(l-m+1)
\widetilde{c}^{k^\prime k;\dot{k}^\prime\dot{k}}_{l-1,l,m-1;\dot{l}^\prime
\dot{l},\dot{m}}\sqrt{(l-m)(l+m)}+\\
+(l-m-1)(l+m+1)
\widetilde{c}^{k^\prime k;\dot{k}^\prime\dot{k}}_{l-1,l,m+1;\dot{l}^\prime
\dot{l},\dot{m}}\sqrt{(l+m)(l-m)}=0.
\nonumber
\end{multline}
Whence
\begin{multline}
2\ld 1-l^2+m^2\rd\widetilde{c}^{k^\prime k;\dot{k}^\prime\dot{k}}_{l-1,l,m;
\dot{l}^\prime\dot{l},\dot{m}}+
\ld l^2-(m-1)^2\rd\widetilde{c}^{k^\prime k;\dot{k}^\prime\dot{k}}_{l-1,l,m-1;
\dot{l}^\prime\dot{l},\dot{m}}+\\
+\ld l^2-(m+1)^2\rd\widetilde{c}^{k^\prime k;\dot{k}^\prime\dot{k}}_{l-1,l,m+1;
\dot{l}^\prime\dot{l},\dot{m}}=0.\nonumber
\end{multline}
It is easy to verify that this system can be solved at
$\widetilde{c}^{k^\prime k;\dot{k}^\prime\dot{k}}_{l-1,l,m;\dot{l}^\prime
\dot{l},\dot{m}}$ (this coefficient does not depend on
$m$). For that reason we can suppose
$\widetilde{c}^{k^\prime k;\dot{k}^\prime\dot{k}}_{l-1,l,m;\dot{l}^\prime
\dot{l},\dot{m}}=c^{k^\prime k;\dot{k}^\prime\dot{k}}_{l-1,l;\dot{l}^\prime
\dot{l},\dot{m}}$.
Coming back to the old variables, we find that
\[
c^{k^\prime k;\dot{k}^\prime\dot{k}}_{l-1,l,m;\dot{l}^\prime\dot{l},\dot{m}}=
c^{k^\prime k;\dot{k}^\prime\dot{k}}_{l-1,l;\dot{l}^\prime\dot{l},\dot{m}}
\sqrt{l^2-m^2}.
\]

Finally, let us suppose $l^\prime=l+1$. In this case the system
(\ref{Sys}) takes the form
\begin{multline}
\ld 2-(l+m+1)(l-m)-(l+m+1)(l-m+2)\rd
c^{k^\prime k;\dot{k}^\prime\dot{k}}_{l+1,l,m;\dot{l}^\prime\dot{l},\dot{m}}+\\
+\sqrt{(l+m+1)(l-m+2)(l+m)(l-m+1)}
c^{k^\prime k;\dot{k}^\prime\dot{k}}_{l+1,l,m-1;\dot{l}^\prime\dot{l},\dot{m}}+\\
+\sqrt{(l+m+2)(l-m+1)(l+m+1)(l-m)}
c^{k^\prime k;\dot{k}^\prime\dot{k}}_{l+1,l,m+1;\dot{l}^\prime\dot{l},\dot{m}}
=0.\nonumber
\end{multline}
Making the substitutions
\begin{eqnarray}
c^{k^\prime k;\dot{k}^\prime\dot{k}}_{l+1,l,m;\dot{l}^\prime\dot{l},\dot{m}}&=&
\sqrt{(l+m)(l-m+1)}
\widetilde{c}^{k^\prime k;\dot{k}^\prime\dot{k}}_{l+1,l,m;\dot{l}^\prime\dot{l},
\dot{m}},\nonumber\\
c^{k^\prime k;\dot{k}^\prime\dot{k}}_{l+1,l,m-1;\dot{l}^\prime\dot{l},\dot{m}}
&=&
\sqrt{(l+m)(l-m+2)}\widetilde{c}^{k^\prime k;\dot{k}^\prime\dot{k}}_{l+1,l,m-1;
\dot{l}^\prime\dot{l},\dot{m}},\nonumber\\
c^{k^\prime k;\dot{k}^\prime\dot{k}}_{l+1,l,m+1;\dot{l}^\prime\dot{l},\dot{m}}
&=&
\sqrt{(l+m+2)(l-m)}
\widetilde{c}^{k^\prime k;\dot{k}^\prime\dot{k}}_{l+1,l,m+1;\dot{l}^\prime
\dot{l},\dot{m}},\nonumber
\end{eqnarray}
we obtain
\begin{multline}
\ld 2-(l+m+1)(l-m)-(l+m+1)(l-m+2)\rd\sqrt{(l+m+1)(l-m+1)}
\widetilde{c}^{k^\prime k;\dot{k}^\prime\dot{k}}_{l+1,l,m;\dot{l}^\prime
\dot{l},\dot{m}}+\\
+(l+m)(l-m+2)\sqrt{(l+m+1)(l-m+1)}
\widetilde{c}^{k^\prime k;\dot{k}^\prime\dot{k}}_{l+1,l,m-1;\dot{l}^\prime
\dot{l},\dot{m}}+\\
+(l+m+2)(l-m)\sqrt{(l+m+1)(l-m+1)}
\widetilde{c}^{k^\prime k;\dot{k}^\prime\dot{k}}_{l+1,l,m-1;\dot{l}^\prime
\dot{l},\dot{m}}=0.
\nonumber
\end{multline}
Whence
\begin{multline}
2\ld m^2-l^2-2l\rd\widetilde{c}^{k^\prime k;\dot{k}^\prime\dot{k}}_{l+1,l,m;
\dot{l}^\prime\dot{l},\dot{m}}+
\ld l^2-m^2+2l+2m\rd\widetilde{c}^{k^\prime k;\dot{k}^\prime\dot{k}}_{l+1,l,
m-1;\dot{l}^\prime\dot{l},\dot{m}}+\\
+\ld l^2-m^2+2l-2m\rd\widetilde{c}^{k^\prime k;\dot{k}^\prime\dot{k}}_{l+1,l,
m+1;\dot{l}^\prime\dot{l},\dot{m}}=0.\nonumber
\end{multline}
A solution $c^{k^\prime k;\dot{k}^\prime\dot{k}}_{l+1,l;\dot{l}^\prime\dot{l},
\dot{m}}$ of the latter equation does not depend
on $m$ also.

Thus, the action of the commutator $\ld\ld\Lambda^{l\dot{l}}_3,
\sX^{l\dot{l}}_-\rd,\sX^{l\dot{l}}_+\rd$ on the basis vectors
$\zeta^{k\dot{k}}_{lm;\dot{l}\dot{m}}$ gives us the following solutions:
\begin{eqnarray}
c^{k^\prime k;\dot{k}^\prime\dot{k}}_{l-1,l,m;\dot{l}^\prime\dot{l},\dot{m}}
&=&c^{k^\prime k;\dot{k}^\prime\dot{k}}_{l-1,l;\dot{l}^\prime\dot{l},\dot{m}}
\sqrt{l^2-m^2},\nonumber\\
c^{k^\prime k;\dot{k}^\prime\dot{k}}_{ll,m;\dot{l}^\prime\dot{l},\dot{m}}&=&
c^{k^\prime k;\dot{k}^\prime\dot{k}}_{ll;\dot{l}^\prime\dot{l},\dot{m}}m,
\nonumber\\
c^{k^\prime k;\dot{k}^\prime\dot{k}}_{l+1,l,m;\dot{l}^\prime\dot{l},\dot{m}}
&=&c^{k^\prime k;\dot{k}^\prime\dot{k}}_{l+1,l;\dot{l}^\prime\dot{l},\dot{m}}
\sqrt{(l+1)^2-m^2}.\nonumber
\end{eqnarray}
With the aim to find the final form for
non-zero elements of $\Lambda^{l\dot{l}}_3$
we must apply the commutator $\ld\ld\Lambda^{l\dot{l}}_3,\sY^{l\dot{l}}_-\rd,
\sY^{l\dot{l}}_+\rd$. In the result we have
\begin{multline}
\ld\ld\sL^{l\dot{l}}_3,\sY^{l\dot{l}}_-\rd,\sY^{l\dot{l}}_+\rd
\zeta^{k\dot{k}}_{lm;\dot{l}\dot{m}}=\\
=\sum_{l^\prime,k^\prime,\dot{l}^\prime,\dot{k}^\prime}
\left\{
\ld(\boldsymbol{\alpha}^{\dot{l}}_{\dot{m}+1})^2+
(\boldsymbol{\alpha}^{\dot{l}^\prime}_{\dot{m}})^2\rd
c^{k^\prime k;\dot{k}^\prime\dot{k}}_{l^\prime l,m;\dot{l}^\prime\dot{l},
\dot{m}}-
\boldsymbol{\alpha}^{\dot{l}^\prime}_{\dot{m}}
\boldsymbol{\alpha}^{\dot{l}}_{\dot{m}}
c^{k^\prime k;\dot{k}^\prime\dot{k}}_{l^\prime l,m;\dot{l}^\prime
\dot{l},\dot{m}-1}-\right.\\
-\left.\boldsymbol{\alpha}^{\dot{l}^\prime}_{\dot{m}+1}
\boldsymbol{\alpha}^{\dot{l}}_{\dot{m}+1}
c^{k^\prime k;\dot{k}^\prime\dot{k}}_{l^\prime l,m;\dot{l}^\prime\dot{l},
\dot{m}+1}
\right\}
\zeta^{k^\prime\dot{k}^\prime}_{l^\prime m;\dot{l}^\prime\dot{m}}.
\nonumber
\end{multline}
Hence it follows
\[
2c^{k^\prime k;\dot{k}^\prime\dot{k}}_{l^\prime l,m;\dot{l}^\prime\dot{l},
\dot{m}}=
\ld(\boldsymbol{\alpha}^{\dot{l}}_{\dot{m}+1})^2+
(\boldsymbol{\alpha}^{\dot{l}^\prime}_{\dot{m}})^2\rd
c^{k^\prime k;\dot{k}^\prime\dot{k}}_{l^\prime l,m;\dot{l}^\prime\dot{l},
\dot{m}}-
\boldsymbol{\alpha}^{\dot{l}^\prime}_{\dot{m}}
\boldsymbol{\alpha}^{\dot{l}}_{\dot{m}}
c^{k^\prime k;\dot{k}^\prime\dot{k}}_{l^\prime l,m;\dot{l}^\prime
\dot{l},\dot{m}-1}
-\boldsymbol{\alpha}^{\dot{l}^\prime}_{\dot{m}+1}
\boldsymbol{\alpha}^{\dot{l}}_{\dot{m}+1}
c^{k^\prime k;\dot{k}^\prime\dot{k}}_{l^\prime l,m;\dot{l}^\prime\dot{l},
\dot{m}+1}.
\]
Solutions of the latter system are derived by means of the analogous
calculations presented in the previous case of the commutator
$\ld\ld\Lambda^{l\dot{l}}_3,\sX^{l\dot{l}}_-\rd,\sX^{l\dot{l}}_+\rd$.
They have the form
\begin{eqnarray}
c^{k^\prime k;\dot{k}^\prime\dot{k}}_{l^\prime,l,m;\dot{l}-1,
\dot{l},\dot{m}}
&=&c^{k^\prime k;\dot{k}^\prime\dot{k}}_{l^\prime l,m;\dot{l}-1,\dot{l}}
\sqrt{\dot{l}^2-\dot{m}^2},\nonumber\\
c^{k^\prime k;\dot{k}^\prime\dot{k}}_{l^\prime l,m;\dot{l},\dot{l},\dot{m}}&=&
c^{k^\prime k;\dot{k}^\prime\dot{k}}_{l^\prime l,m;\dot{l}\dot{l}}\dot{m},
\nonumber\\
c^{k^\prime k;\dot{k}^\prime\dot{k}}_{l^\prime l,m;\dot{l}+1,\dot{l},\dot{m}}
&=&c^{k^\prime k;\dot{k}^\prime\dot{k}}_{l^\prime l,m;\dot{l}+1,\dot{l}}
\sqrt{(\dot{l}+1)^2-\dot{m}^2}.\nonumber
\end{eqnarray}
Thus, matrix elements of $\sL^{l\dot{l}}_3$ are
\begin{equation}\label{L3}
{\renewcommand{\arraystretch}{1.5}
\sL^{l\dot{l}}_3:\quad\left\{\begin{array}{ccc}
c^{k^\prime k;\dot{k}^\prime\dot{k}}_{l-1,l,m;\dot{l}-1,\dot{l},\dot{m}}
&=&c^{k^\prime k;\dot{k}^\prime\dot{k}}_{l-1,l;\dot{l}-1,\dot{l}}
\sqrt{(l^2-m^2)(\dot{l}^2-\dot{m}^2)},\\
c^{k^\prime k;\dot{k}^\prime\dot{k}}_{ll,m;\dot{l}\dot{l},\dot{m}}&=&
c^{k^\prime k;\dot{k}^\prime\dot{k}}_{ll;\dot{l}\dot{l}}m\dot{m},\\
c^{k^\prime k;\dot{k}^\prime\dot{k}}_{l+1,l,m;\dot{l}+1,\dot{l},\dot{m}}
&=&c^{k^\prime k;\dot{k}^\prime\dot{k}}_{l+1,l;\dot{l}+1,\dot{l}}
\sqrt{((l+1)^2-m^2)((\dot{l}+1)^2-\dot{m}^2)}.
\end{array}\right.}
\end{equation}
All other elements of the matrix $\sL^{l\dot{l}}_3$ are equal to zero.

Let us define now
elements of the matrices $\sL^{l\dot{l}}_1$ and
$\sL^{l\dot{l}}_2$. For the transformation
$\sL^{l\dot{l}}_1$ in the helicity basis we have
\[
\sL^{l\dot{l}}_1\zeta^{k\dot{k}}_{lm;\dot{l}\dot{m}}=
\sum_{l^\prime,m^\prime,\dot{l}^\prime,\dot{m}^\prime,k^\prime,\dot{k}^\prime}
a^{k^\prime k;\dot{k}^\prime\dot{k}}_{l^\prime l,m^\prime m;\dot{l}^\prime
\dot{l},\dot{m}^\prime\dot{m}}
\zeta^{k^\prime\dot{k}^\prime}_{l^\prime m^\prime;\dot{l}^\prime\dot{m}^\prime}.
\]
We will find the numbers
$a^{k^\prime k;\dot{k}^\prime\dot{k}}_{l^\prime l,m^\prime m;\dot{l}^\prime
\dot{l},\dot{m}^\prime\dot{m}}$
using the relations $\sL^{l\dot{l}}_1=\ld\sA^{l\dot{l}}_2,\sL^{l\dot{l}}_3\rd$
(or $\sL^{l\dot{l}}_1=i\ld\sB^{l\dot{l}}_2,\sL^{l\dot{l}}_3\rd$)
and (\ref{Op2}) (or (\ref{Op5})). Indeed,
\begin{multline}
\sL^{l\dot{l}}_1\zeta^{k\dot{k}}_{lm;\dot{l}\dot{m}}=
\sA^{l\dot{l}}_2\sL^{l\dot{l}}_3\zeta^{k\dot{k}}_{lm;\dot{l}\dot{m}}-
\sL^{l\dot{l}}_3\sA^{l\dot{l}}_2\zeta^{k\dot{k}}_{lm;\dot{l}\dot{m}}=\\
=\sA^{l\dot{l}}_2
\sum_{l^\prime,m^\prime,\dot{l}^\prime,\dot{m}^\prime k^\prime,\dot{k}^\prime}
c^{k^\prime k;\dot{k}^\prime\dot{k}}_{l^\prime l,m^\prime m;\dot{l}^\prime
\dot{l},\dot{m}^\prime\dot{m}}
\zeta^{k^\prime \dot{k}^\prime}_{l^\prime m^\prime;\dot{l}^\prime
\dot{m}^\prime}-
\frac{1}{2}\sL^{l\dot{l}}_3\left(\boldsymbol{\alpha}^l_m
\zeta^{k\dot{k}}_{l,m-1;\dot{l},\dot{m}}-
\boldsymbol{\alpha}^l_{m+1}\zeta^{k\dot{k}}_{l,m+1;\dot{l}\dot{m}}\right.-\\
\left.-\balpha^{\dot{l}}_{\dot{m}}\zeta^{k\dot{k}}_{l,m;\dot{l},\dot{m}-1}+
+\balpha^{\dot{l}}_{\dot{m}+1}\zeta^{k\dot{k}}_{l,m;\dot{l},\dot{m}+1}\right)=
\frac{1}{2}
\sum_{l^\prime,m^\prime,\dot{l}^\prime,\dot{m}^\prime, k^\prime,\dot{k}^\prime}
c^{k^\prime k;\dot{k}^\prime\dot{k}}_{l^\prime l,m^\prime m;\dot{l}^\prime
\dot{l},\dot{m}^\prime\dot{m}}
\left(\boldsymbol{\alpha}^{l^\prime}_{m^\prime}
\zeta^{k^\prime}_{l^\prime,m^\prime-1;\dot{l}^\prime\dot{m}^\prime}-\right.\\
-\left.\boldsymbol{\alpha}^{l^\prime}_{m^\prime+1}
\zeta^{k^\prime}_{l^\prime,m^\prime+1;\dot{l}^\prime\dot{m}^\prime}
-\balpha^{\dot{l}^\prime}_{\dot{m}^\prime}
\zeta^{k^\prime\dot{k}^\prime}_{l^\prime,m^\prime;\dot{l}^\prime,\dot{m}^\prime
-1}+\balpha^{\dot{l}^\prime}_{\dot{m}^\prime+1}
\zeta^{k^\prime\dot{k}^\prime}_{l^\prime,m^\prime;\dot{l}^\prime,\dot{m}^\prime
+1}\right)-\\
-\frac{1}{2}\boldsymbol{\alpha}^l_m
\sum_{l^\prime,m^\prime,\dot{l}^\prime,\dot{m}^\prime,k^\prime\dot{k}^\prime}
c^{k^\prime k;\dot{k}^\prime\dot{k}}_{l^\prime l,m^\prime,m-1;\dot{l}^\prime
dot{l},\dot{m}^\prime\dot{m}}
\zeta^{k^\prime\dot{k}^\prime}_{l^\prime m^\prime;\dot{l}^\prime\dot{m}^\prime}+
\frac{1}{2}\boldsymbol{\alpha}^l_{m+1}
\sum_{l^\prime,m^\prime,\dot{l}^\prime,\dot{m}^\prime,k^\prime,\dot{k}^\prime}
c^{k^\prime k;\dot{k}^\prime\dot{k}}_{l^\prime l,m^\prime, m+1;\dot{l}^\prime
\dot{l},\dot{m}^\prime\dot{m}}
\zeta^{k^\prime\dot{k}^\prime}_{l^\prime m^\prime;\dot{l}^\prime\dot{m}^\prime}
+\\
+\frac{1}{2}\balpha^{\dot{l}}_{\dot{m}}
\sum_{l^\prime,m^\prime,\dot{l}^\prime,\dot{m}^\prime,k^\prime,\dot{k}^\prime}
c^{k^\prime k;\dot{k}^\prime\dot{k}}_{l^\prime l,m^\prime, m;\dot{l}^\prime
\dot{l},\dot{m}^\prime\dot{m}-1}
\zeta^{k^\prime\dot{k}^\prime}_{l^\prime m^\prime;\dot{l}^\prime\dot{m}^\prime}-
\frac{1}{2}\balpha^{\dot{l}}_{\dot{m}+1}
\sum_{l^\prime,m^\prime,\dot{l}^\prime,\dot{m}^\prime,k^\prime,\dot{k}^\prime}
c^{k^\prime k;\dot{k}^\prime\dot{k}}_{l^\prime l,m^\prime, m;\dot{l}^\prime
\dot{l},\dot{m}^\prime\dot{m}+1}
\zeta^{k^\prime\dot{k}^\prime}_{l^\prime m^\prime;\dot{l}^\prime\dot{m}^\prime}.
\nonumber
\end{multline}
Dividing the first sum on the four and changing the
summation index in the each sums obtained, we come to the following
expression:
\begin{multline}
\sL^{l\dot{l}}_1\zeta^{k\dot{k}}_{lm;\dot{l}\dot{m}}=\frac{1}{2}
\sum_{l^\prime,m^\prime,\dot{l}^\prime,\dot{m}^\prime,k^\prime,\dot{k}^\prime}
\left(\boldsymbol{\alpha}^{l^\prime}_{m^\prime+1}
c^{k^\prime k;\dot{k}^\prime\dot{k}}_{l^\prime l,m^\prime+1,m;\dot{l}^\prime
\dot{l},\dot{m}^\prime\dot{m}}-
\boldsymbol{\alpha}^{l^\prime}_{m^\prime}
c^{k^\prime k;\dot{k}^\prime\dot{k}}_{l^\prime l,m^\prime-1,m;\dot{l}^\prime
\dot{l},\dot{m}^\prime\dot{m}}-\right.\\
-\balpha^{\dot{l}^\prime}_{\dot{m}^\prime+1}
c^{k^\prime k;\dot{k}^\prime\dot{k}}_{l^\prime l,m^\prime m;\dot{l}^\prime
\dot{l},\dot{m}^\prime+1,\dot{m}}
+\balpha^{\dot{l}^\prime}_{\dot{m}^\prime}
c^{k^\prime k;\dot{k}^\prime\dot{k}}_{l^\prime l,m^\prime m;\dot{l}^\prime
\dot{l},\dot{m}^\prime-1,\dot{m}}
-\boldsymbol{\alpha}^l_m
c^{k^\prime k;\dot{k}^\prime\dot{k}}_{l^\prime l,m^\prime, m-1;\dot{l}^\prime
\dot{l},\dot{m}^\prime\dot{m}}+
\boldsymbol{\alpha}^l_{m+1}
c^{k^\prime k;\dot{k}^\prime\dot{k}}_{l^\prime l,m^\prime, m+1;\dot{l}^\prime
\dot{l},\dot{m}^\prime\dot{m}}+\\
+\left.\balpha^{\dot{l}}_{\dot{m}}
c^{k^\prime k;\dot{k}^\prime\dot{k}}_{l^\prime l,m^\prime m;\dot{l}^\prime
\dot{l},\dot{m}^\prime,\dot{m}-1}-
\balpha^{\dot{l}}_{\dot{m}+1}
c^{k^\prime k;\dot{k}^\prime\dot{k}}_{l^\prime l,m^\prime m;\dot{l}^\prime
\dot{l},\dot{m}^\prime,\dot{m}+1}\right)
\zeta^{k^\prime\dot{k}^\prime}_{l^\prime m^\prime;\dot{l}^\prime\dot{m}^\prime}.
\nonumber
\end{multline}
Therefore, elements of the matrix $\sL^{l\dot{l}}_1$ have the form
\begin{multline}
a^{k^\prime k;\dot{k}^\prime\dot{k}}_{l^\prime l,m^\prime m;\dot{l}^\prime
\dot{l},\dot{m}^\prime\dot{m}}=\frac{1}{2}
\left(\boldsymbol{\alpha}^{l^\prime}_{m^\prime+1}
c^{k^\prime k;\dot{k}^\prime\dot{k}}_{l^\prime l,m^\prime+1,m;\dot{l}^\prime
\dot{l},\dot{m}^\prime\dot{m}}-
\boldsymbol{\alpha}^{l^\prime}_{m^\prime}
c^{k^\prime k;\dot{k}^\prime\dot{k}}_{l^\prime l,m^\prime-1,m;\dot{l}^\prime
\dot{l},\dot{m}^\prime\dot{m}}
-\balpha^{\dot{l}^\prime}_{\dot{m}^\prime+1}
c^{k^\prime k;\dot{k}^\prime\dot{k}}_{l^\prime l,m^\prime m;\dot{l}^\prime
\dot{l},\dot{m}^\prime+1,\dot{m}}+\right.\\
+\balpha^{\dot{l}^\prime}_{\dot{m}^\prime}
c^{k^\prime k;\dot{k}^\prime\dot{k}}_{l^\prime l,m^\prime m;\dot{l}^\prime
\dot{l},\dot{m}^\prime-1,\dot{m}}
-\boldsymbol{\alpha}^l_m
c^{k^\prime k;\dot{k}^\prime\dot{k}}_{l^\prime l,m^\prime, m-1;\dot{l}^\prime
\dot{l},\dot{m}^\prime\dot{m}}+
\boldsymbol{\alpha}^l_{m+1}
c^{k^\prime k;\dot{k}^\prime\dot{k}}_{l^\prime l,m^\prime, m+1;\dot{l}^\prime
\dot{l},\dot{m}^\prime\dot{m}}+\\
+\left.\balpha^{\dot{l}}_{\dot{m}}
c^{k^\prime k;\dot{k}^\prime\dot{k}}_{l^\prime l,m^\prime m;\dot{l}^\prime
\dot{l},\dot{m}^\prime,\dot{m}-1}-
\balpha^{\dot{l}}_{\dot{m}+1}
c^{k^\prime k;\dot{k}^\prime\dot{k}}_{l^\prime l,m^\prime m;\dot{l}^\prime
\dot{l},\dot{m}^\prime,\dot{m}+1}\right)
\zeta^{k^\prime\dot{k}^\prime}_{l^\prime m^\prime;\dot{l}^\prime\dot{m}^\prime}.
\label{AM}
\end{multline}\begin{sloppypar}\noindent
Since $c^{k^\prime k;\dot{k}^\prime\dot{k}}_{l^\prime l,m^\prime m;
\dot{l}^\prime\dot{l},\dot{m}^\prime\dot{m}}\neq 0$ only at
$m^\prime=m$, $\dot{m}^\prime=\dot{m}$ and
$l^\prime=l-1,\,l,\,l+1$, $\dot{l}^\prime=\dot{l}-1$,
$\dot{l}$, $\dot{l}+1$, then at the fixed indices
$m$, $\dot{m}$, $l$, $\dot{l}$ $k^\prime$, $k$, $\dot{k}^\prime$, $\dot{k}$
we have twelve numbers
$a^{k^\prime k;\dot{k}^\prime\dot{k}}_{l^\prime l,m^\prime m;\dot{l}^\prime
\dot{l},\dot{m}^\prime\dot{m}}$
which are different from zero.
Substituting $\boldsymbol{\alpha}^l_m=\sqrt{(l+m)(l-m+1)}$,
$\boldsymbol{\alpha}^{\dot{l}}_{\dot{m}}=\sqrt{(\dot{l}+\dot{m})(\dot{l}-
\dot{m}+1)}$
into (\ref{AM})
and using $c^{k^\prime k;\dot{k}^\prime\dot{k}}_{l^\prime lm;\dot{l}^\prime
\dot{l},\dot{m}}$ from (\ref{L3}), we find that
\end{sloppypar}
\begin{eqnarray}
a^{k^\prime k;\dot{k}^\prime\dot{k}}_{l-1,l,m-1,m;\dot{l}-1,\dot{l},\dot{m}
\dot{m}}&=&
-\frac{1}{2}c^{k^\prime k;\dot{k}^\prime\dot{k}}_{l-1,l;\dot{l}-1,\dot{l}}
\sqrt{(l+m)(l+m-1)(\dot{l}^2-\dot{m}^2)},\nonumber\\
a^{k^\prime k;\dot{k}^\prime\dot{k}}_{ll,m-1,m;\dot{l}\dot{l},\dot{m}\dot{m}}
&=&
\frac{1}{2}c^{k^\prime k;\dot{k}^\prime\dot{k}}_{ll;\dot{l}\dot{l}}
\dot{m}\sqrt{(l+m)(l-m+1)},\nonumber\\
a^{k^\prime k;\dot{k}^\prime\dot{k}}_{l+1,l,m-1,m;\dot{l}+1,\dot{l},\dot{m}
\dot{m}}&=&
\frac{1}{2}c^{k^\prime k;\dot{k}^\prime\dot{k}}_{l+1,l;\dot{l}+1,\dot{l}}
\sqrt{(l-m+1)(l-m+2)((\dot{l}+1)^2-\dot{m}^2)},\nonumber
\end{eqnarray}
\begin{eqnarray}
a^{k^\prime k;\dot{k}^\prime\dot{k}}_{l-1,l,m+1,m;\dot{l}-1,\dot{l},\dot{m}
\dot{m}}&=&
\frac{1}{2}c^{k^\prime k;\dot{k}^\prime\dot{k}}_{l-1,l;\dot{l}-1,\dot{l}}
\sqrt{(l-m)(l-m-1)(\dot{l}^2-\dot{m}^2)},\nonumber\\
a^{k^\prime k;\dot{k}^\prime\dot{k}}_{ll,m+1,m;\dot{l}\dot{l},\dot{m}
\dot{m}}&=&
\frac{1}{2}c^{k^\prime k;\dot{k}^\prime\dot{k}}_{ll;\dot{l}\dot{l}}
\dot{m}\sqrt{(l+m+1)(l-m)},\nonumber\\
a^{k^\prime k;\dot{k}^\prime\dot{k}}_{l+1,l,m+1,m;\dot{l}+1,\dot{l},\dot{m}
\dot{m}}&=&
-\frac{1}{2}c^{k^\prime k;\dot{k}^\prime\dot{k}}_{l+1,l;\dot{l}+1,\dot{l}}
\sqrt{(l+m+1)(l+m+2)((\dot{l}+1)^2-\dot{m}^2)},\nonumber
\end{eqnarray}
\begin{eqnarray}
a^{k^\prime k;\dot{k}^\prime\dot{k}}_{l-1,l,m,m;\dot{l}-1,\dot{l},\dot{m}-1,
\dot{m}}&=&
\frac{1}{2}c^{k^\prime k;\dot{k}^\prime\dot{k}}_{l-1,l;\dot{l}-1,\dot{l}}
\sqrt{(l^2-m^2)(\dot{l}+\dot{m})(\dot{l}+\dot{m}-1)},\nonumber\\
a^{k^\prime k;\dot{k}^\prime\dot{k}}_{ll,m,m;\dot{l}\dot{l},\dot{m}-1,\dot{m}}
&=&
-\frac{1}{2}c^{k^\prime k;\dot{k}^\prime\dot{k}}_{ll;\dot{l}\dot{l}}
m\sqrt{(\dot{l}+\dot{m})(\dot{l}-\dot{m}+1)},\nonumber\\
a^{k^\prime k;\dot{k}^\prime\dot{k}}_{l+1,l,m,m;\dot{l}+1,\dot{l},\dot{m}-1,
\dot{m}}&=&
-\frac{1}{2}c^{k^\prime k;\dot{k}^\prime\dot{k}}_{l+1,l;\dot{l}+1,\dot{l}}
\sqrt{((l+1)^2-m^2)(\dot{l}-\dot{m}+1)(\dot{l}-\dot{m}+2)},\nonumber
\end{eqnarray}
\begin{eqnarray}
a^{k^\prime k;\dot{k}^\prime\dot{k}}_{l-1,l,m,m;\dot{l}-1,\dot{l},\dot{m}+1,
\dot{m}}&=&
-\frac{1}{2}c^{k^\prime k;\dot{k}^\prime\dot{k}}_{l-1,l;\dot{l}-1,\dot{l}}
\sqrt{(l^2-m^2)(\dot{l}-\dot{m})(\dot{l}-\dot{m}-1)},\nonumber\\
a^{k^\prime k;\dot{k}^\prime\dot{k}}_{ll,m,m;\dot{l}\dot{l},\dot{m}+1,
\dot{m}}&=&
-\frac{1}{2}c^{k^\prime k;\dot{k}^\prime\dot{k}}_{ll;\dot{l}\dot{l}}
m\sqrt{(\dot{l}+\dot{m}+1)(\dot{l}-\dot{m})},\nonumber\\
a^{k^\prime k;\dot{k}^\prime\dot{k}}_{l+1,l,m,m;\dot{l}+1,\dot{l},\dot{m}+1,
\dot{m}}&=&
\frac{1}{2}c^{k^\prime k;\dot{k}^\prime\dot{k}}_{l+1,l;\dot{l}+1,\dot{l}}
\sqrt{((l+1)^2-m^2)(\dot{l}+\dot{m}+1)(\dot{l}+\dot{m}+2)}.\label{L1}
\end{eqnarray}

Let us define elements of the matrix $\sL^{l\dot{l}}_2$. From
$\sL^{l\dot{l}}_2=-\ld\sA^{l\dot{l}}_1,\sL^{l\dot{l}}_3\rd$
(or $\sL^{l\dot{l}}_2=-i\ld\sB^{l\dot{l}}_1,\sL^{l\dot{l}}_3\rd$) and
\[
\sL^{l\dot{l}}_2\zeta^{k\dot{k}}_{lm;\dot{l}\dot{m}}=
\sum_{l^\prime,m^\prime,\dot{l}^\prime,\dot{m}^\prime,k^\prime,\dot{k}^\prime}
b^{k^\prime k;\dot{k}^\prime\dot{k}}_{l^\prime l,m^\prime m;\dot{l}^\prime
\dot{l},\dot{m}^\prime\dot{m}}
\zeta^{k^\prime\dot{k}^\prime}_{l^\prime m^\prime;\dot{l}\dot{m}}
\]
and also the relations
(\ref{Op1}) (or (\ref{Op4})), (\ref{L3}) it follows that
\begin{eqnarray}
b^{k^\prime k;\dot{k}^\prime\dot{k}}_{l-1,l,m-1,m;\dot{l}-1,\dot{l},\dot{m}
\dot{m}}&=&
-\frac{i}{2}c^{k^\prime k;\dot{k}^\prime\dot{k}}_{l-1,l;\dot{l}-1,\dot{l}}
\sqrt{(l+m)(l+m-1)(\dot{l}^2-\dot{m}^2)},\nonumber\\
b^{k^\prime k;\dot{k}^\prime\dot{k}}_{ll,m-1,m;\dot{l}\dot{l},\dot{m}\dot{m}}
&=&
\frac{i}{2}c^{k^\prime k;\dot{k}^\prime\dot{k}}_{ll;\dot{l}\dot{l}}
\dot{m}\sqrt{(l+m)(l-m+1)},\nonumber\\
b^{k^\prime k;\dot{k}^\prime\dot{k}}_{l+1,l,m-1,m;\dot{l}+1,\dot{l},\dot{m}
\dot{m}}&=&
\frac{i}{2}c^{k^\prime k;\dot{k}^\prime\dot{k}}_{l+1,l;\dot{l}+1,\dot{l}}
\sqrt{(l-m+1)(l-m+2)((\dot{l}+1)^2-\dot{m}^2)},\nonumber
\end{eqnarray}
\begin{eqnarray}
b^{k^\prime k;\dot{k}^\prime\dot{k}}_{l-1,l,m+1,m;\dot{l}-1,\dot{l},\dot{m}
\dot{m}}&=&
-\frac{i}{2}c^{k^\prime k;\dot{k}^\prime\dot{k}}_{l-1,l;\dot{l}-1,\dot{l}}
\sqrt{(l-m)(l-m-1)(\dot{l}^2-\dot{m}^2)},\nonumber\\
b^{k^\prime k;\dot{k}^\prime\dot{k}}_{ll,m+1,m;\dot{l}\dot{l},\dot{m}
\dot{m}}&=&
-\frac{i}{2}c^{k^\prime k;\dot{k}^\prime\dot{k}}_{ll;\dot{l}\dot{l}}
\dot{m}\sqrt{(l+m+1)(l-m)},\nonumber\\
b^{k^\prime k;\dot{k}^\prime\dot{k}}_{l+1,l,m+1,m;\dot{l}+1,\dot{l},\dot{m}
\dot{m}}&=&
\frac{i}{2}c^{k^\prime k;\dot{k}^\prime\dot{k}}_{l+1,l;\dot{l}+1,\dot{l}}
\sqrt{(l+m+1)(l+m+2)((\dot{l}+1)^2-\dot{m}^2)},\nonumber
\end{eqnarray}
\begin{eqnarray}
b^{k^\prime k;\dot{k}^\prime\dot{k}}_{l-1,l,m,m;\dot{l}-1,\dot{l},\dot{m}-1,
\dot{m}}&=&
\frac{i}{2}c^{k^\prime k;\dot{k}^\prime\dot{k}}_{l-1,l;\dot{l}-1,\dot{l}}
\sqrt{(l^2-m^2)(\dot{l}+\dot{m})(\dot{l}+\dot{m}-1)},\nonumber\\
b^{k^\prime k;\dot{k}^\prime\dot{k}}_{ll,m,m;\dot{l}\dot{l},\dot{m}-1,\dot{m}}
&=&
-\frac{i}{2}c^{k^\prime k;\dot{k}^\prime\dot{k}}_{ll;\dot{l}\dot{l}}
m\sqrt{(\dot{l}+\dot{m})(\dot{l}-\dot{m}+1)},\nonumber\\
b^{k^\prime k;\dot{k}^\prime\dot{k}}_{l+1,l,m,m;\dot{l}+1,\dot{l},\dot{m}-1,
\dot{m}}&=&
-\frac{i}{2}c^{k^\prime k;\dot{k}^\prime\dot{k}}_{l+1,l;\dot{l}+1,\dot{l}}
\sqrt{((l+1)^2-m^2)(\dot{l}-\dot{m}+1)(\dot{l}-\dot{m}+2)},\nonumber
\end{eqnarray}
\begin{eqnarray}
b^{k^\prime k;\dot{k}^\prime\dot{k}}_{l-1,l,m,m;\dot{l}-1,\dot{l},\dot{m}+1,
\dot{m}}&=&
\frac{i}{2}c^{k^\prime k;\dot{k}^\prime\dot{k}}_{l-1,l;\dot{l}-1,\dot{l}}
\sqrt{(l^2-m^2)(\dot{l}-\dot{m})(\dot{l}-\dot{m}-1)},\nonumber\\
b^{k^\prime k;\dot{k}^\prime\dot{k}}_{ll,m,m;\dot{l}\dot{l},\dot{m}+1,
\dot{m}}&=&
\frac{i}{2}c^{k^\prime k;\dot{k}^\prime\dot{k}}_{ll;\dot{l}\dot{l}}
m\sqrt{(\dot{l}+\dot{m}+1)(\dot{l}-\dot{m})},\nonumber\\
b^{k^\prime k;\dot{k}^\prime\dot{k}}_{l+1,l,m,m;\dot{l}+1,\dot{l},\dot{m}+1,
\dot{m}}&=&
-\frac{i}{2}c^{k^\prime k;\dot{k}^\prime\dot{k}}_{l+1,l;\dot{l}+1,\dot{l}}
\sqrt{((l+1)^2-m^2)(\dot{l}+\dot{m}+1)(\dot{l}+\dot{m}+2)}.\label{L2}
\end{eqnarray}

Coming to the dual representations,\index{representation!dual}
we find elements of the matrices
$\overset{\ast}{\sL}{}^{l\dot{l}}_1$,
$\overset{\ast}{\sL}{}^{l\dot{l}}_2$ and
$\overset{\ast}{\sL}{}^{l\dot{l}}_3$.
The dual transformations $\overset{\ast}{\sL}{}^{l\dot{l}}_i$ in the
helicity basis are
\begin{eqnarray}
\overset{\ast}{\sL}{}^{l\dot{l}}_1\zeta^{k\dot{k}}_{lm;\dot{l}\dot{m}}&=&
\sum_{l^\prime,m^\prime,\dot{l}^\prime,\dot{m}^\prime,k^\prime,\dot{k}^\prime}
\overset{\ast}{a}{}^{k^\prime k;\dot{k}^\prime\dot{k}}_{l^\prime l,m^\prime m;
\dot{l}^\prime\dot{l},\dot{m}^\prime\dot{m}}
\zeta^{k^\prime\dot{k}^\prime}_{lm;\dot{l}^\prime\dot{m}^\prime},\nonumber\\
\overset{\ast}{\sL}{}^{l\dot{l}}_2\zeta^{k\dot{k}}_{lm;\dot{l}\dot{m}}&=&
\sum_{l^\prime,m^\prime,\dot{l}^\prime,\dot{m}^\prime,k^\prime,\dot{k}^\prime}
\overset{\ast}{b}{}^{k^\prime k;\dot{k}^\prime\dot{k}}_{l^\prime l,m^\prime m;
\dot{l}^\prime\dot{l},\dot{m}^\prime\dot{m}}
\zeta^{k^\prime\dot{k}^\prime}_{lm;\dot{l}^\prime\dot{m}^\prime},\nonumber\\
\overset{\ast}{\sL}{}^{l\dot{l}}_3\zeta^{k\dot{k}}_{lm;\dot{l}\dot{m}}&=&
\sum_{l^\prime,m^\prime,\dot{l}^\prime,\dot{m}^\prime,k^\prime,\dot{k}^\prime}
\overset{\ast}{c}{}^{k^\prime k;\dot{k}^\prime\dot{k}}_{l^\prime l,m^\prime m;
\dot{l}^\prime\dot{l},\dot{m}^\prime\dot{m}}
\zeta^{k^\prime\dot{k}^\prime}_{lm;\dot{l}^\prime\dot{m}^\prime}.\nonumber
\end{eqnarray}
\begin{sloppypar}\noindent
Calculating the commutators $\ld\overset{\ast}{\sL}{}^{l\dot{l}}_3,
\sY^{l\dot{l}}_3\rd$,
$\ld\ld\overset{\ast}{\sL}{}^{l\dot{l}}_3,\sY^{l\dot{l}}_-\rd,
\sY^{l\dot{l}}_+\rd$ with respect to the vectors
$\zeta^{k\dot{k}}_{lm;\dot{l}\dot{m}}$, we find elements of
the matrix $\overset{\ast}{\sL}{}^{l\dot{l}}_3$:
\begin{equation}\label{L3'}
{\renewcommand{\arraystretch}{1.5}
\overset{\ast}{\sL}{}^{l\dot{l}}_3:\quad\left\{\begin{array}{ccc}
\overset{\ast}{c}{}^{k^\prime k;\dot{k}^\prime\dot{k}}_{l-1,l,m;
\dot{l}-1,\dot{l},\dot{m}}
&=&\overset{\ast}{c}{}^{k^\prime k;\dot{k}^\prime\dot{k}}_{l-1,l;\dot{l}-1,\dot{l}}
\sqrt{(l^2-m^2)(\dot{l}^2-\dot{m}^2)},\\
\overset{\ast}{c}{}^{k^\prime k;\dot{k}^\prime\dot{k}}_{ll,m;
\dot{l}\dot{l},\dot{m}}&=&
\overset{\ast}{c}{}^{k^\prime k;\dot{k}^\prime\dot{k}}_{ll;
\dot{l}\dot{l}}m\dot{m},\\
\overset{\ast}{c}{}^{k^\prime k;\dot{k}^\prime\dot{k}}_{l+1,l,m;
\dot{l}+1,\dot{l},\dot{m}}
&=&\overset{\ast}{c}{}^{k^\prime k;\dot{k}^\prime\dot{k}}_{l+1,l;
\dot{l}+1,\dot{l}}
\sqrt{((l+1)^2-m^2)((\dot{l}+1)^2-\dot{m}^2)}.
\end{array}\right.}
\end{equation}
Using the relations
$\overset{\ast}{\sL}{}^{l\dot{l}}_1=\ld\widetilde{\sA}^{l\dot{l}}_2,
\overset{\ast}{\sL}{}^{l\dot{l}}_3\rd$ (or
$\overset{\ast}{\sL}{}^{l\dot{l}}_1=-i\ld\widetilde{\sB}^{l\dot{l}}_2,
\overset{\ast}{\sL}{}^{l\dot{l}}_3\rd$) and (\ref{Op8})
(or (\ref{Op11})), we find elements
$\overset{\ast}{a}{}^{k^\prime k;\dot{k}^\prime
\dot{k}}_{l^\prime l,m^\prime m;\dot{l}^\prime\dot{l},\dot{m}^\prime\dot{m}}$
of the matrix $\overset{\ast}{\sL}{}^{l\dot{l}}_1$:
\begin{eqnarray}
\overset{\ast}{a}{}^{k^\prime k;\dot{k}^\prime\dot{k}}_{l-1,l,m-1,m;
\dot{l}-1,\dot{l},\dot{m}\dot{m}}&=&
\frac{1}{2}\overset{\ast}{c}{}^{k^\prime k;\dot{k}^\prime\dot{k}}_{l-1,l;
\dot{l}-1,\dot{l}}\sqrt{(l+m)(l+m-1)(\dot{l}^2-\dot{m}^2)},\nonumber\\
\overset{\ast}{a}{}^{k^\prime k;\dot{k}^\prime\dot{k}}_{ll,m-1,m;
\dot{l}\dot{l},\dot{m}\dot{m}}&=&
-\frac{1}{2}\overset{\ast}{c}{}^{k^\prime k;\dot{k}^\prime\dot{k}}_{ll;
\dot{l}\dot{l}}\dot{m}\sqrt{(l+m)(l-m+1)},\nonumber\\
\overset{\ast}{a}{}^{k^\prime k;\dot{k}^\prime\dot{k}}_{l+1,l,m-1,m;
\dot{l}+1,\dot{l},\dot{m}\dot{m}}&=&
-\frac{1}{2}\overset{\ast}{c}{}^{k^\prime k;\dot{k}^\prime\dot{k}}_{l+1,l;
\dot{l}+1,\dot{l}}\sqrt{(l-m+1)(l-m+2)((\dot{l}+1)^2-\dot{m}^2)},\nonumber
\end{eqnarray}
\begin{eqnarray}
\overset{\ast}{a}{}^{k^\prime k;\dot{k}^\prime\dot{k}}_{l-1,l,m+1,m;
\dot{l}-1,\dot{l},\dot{m}\dot{m}}&=&
-\frac{1}{2}\overset{\ast}{c}{}^{k^\prime k;\dot{k}^\prime\dot{k}}_{l-1,l;
\dot{l}-1,\dot{l}}\sqrt{(l-m)(l-m-1)(\dot{l}^2-\dot{m}^2)},\nonumber\\
\overset{\ast}{a}{}^{k^\prime k;\dot{k}^\prime\dot{k}}_{ll,m+1,m;
\dot{l}\dot{l},\dot{m}\dot{m}}&=&
-\frac{1}{2}\overset{\ast}{c}{}^{k^\prime k;\dot{k}^\prime\dot{k}}_{ll;
\dot{l}\dot{l}}\dot{m}\sqrt{(l+m+1)(l-m)},\nonumber\\
\overset{\ast}{a}{}^{k^\prime k;\dot{k}^\prime\dot{k}}_{l+1,l,m+1,m;
\dot{l}+1,\dot{l},\dot{m}\dot{m}}&=&
\frac{1}{2}\overset{\ast}{c}{}^{k^\prime k;\dot{k}^\prime\dot{k}}_{l+1,l;
\dot{l}+1,\dot{l}}\sqrt{(l+m+1)(l+m+2)((\dot{l}+1)^2-\dot{m}^2)},\nonumber
\end{eqnarray}
\begin{eqnarray}
\overset{\ast}{a}{}^{k^\prime k;\dot{k}^\prime\dot{k}}_{l-1,l,m,m;
\dot{l}-1,\dot{l},\dot{m}-1,\dot{m}}&=&
-\frac{1}{2}\overset{\ast}{c}{}^{k^\prime k;\dot{k}^\prime\dot{k}}_{l-1,l;
\dot{l}-1,\dot{l}}\sqrt{(l^2-m^2)(\dot{l}+\dot{m})(\dot{l}+\dot{m}-1)},\nonumber\\
\overset{\ast}{a}{}^{k^\prime k;\dot{k}^\prime\dot{k}}_{ll,m,m;
\dot{l}\dot{l},\dot{m}-1,\dot{m}}&=&
\frac{1}{2}\overset{\ast}{c}{}^{k^\prime k;\dot{k}^\prime\dot{k}}_{ll;
\dot{l}\dot{l}}m\sqrt{(\dot{l}+\dot{m})(\dot{l}-\dot{m}+1)},\nonumber\\
\overset{\ast}{a}{}^{k^\prime k;\dot{k}^\prime\dot{k}}_{l+1,l,m,m;\dot{l}+1,
\dot{l},\dot{m}-1,\dot{m}}&=&
\frac{1}{2}\overset{\ast}{c}{}^{k^\prime k;\dot{k}^\prime\dot{k}}_{l+1,l;
\dot{l}+1,\dot{l}}
\sqrt{((l+1)^2-m^2)(\dot{l}-\dot{m}+1)(\dot{l}-\dot{m}+2)},\nonumber
\end{eqnarray}
\begin{eqnarray}
\overset{\ast}{a}{}^{k^\prime k;\dot{k}^\prime\dot{k}}_{l-1,l,m,m;\dot{l}-1,
\dot{l},\dot{m}+1,\dot{m}}&=&
\frac{1}{2}\overset{\ast}{c}{}^{k^\prime k;\dot{k}^\prime\dot{k}}_{l-1,l;
\dot{l}-1,\dot{l}}
\sqrt{(l^2-m^2)(\dot{l}-\dot{m})(\dot{l}-\dot{m}-1)},\nonumber\\
\overset{\ast}{a}{}^{k^\prime k;\dot{k}^\prime\dot{k}}_{ll,m,m;
\dot{l}\dot{l},\dot{m}+1,\dot{m}}&=&
\frac{1}{2}\overset{\ast}{c}{}^{k^\prime k;\dot{k}^\prime\dot{k}}_{ll;
\dot{l}\dot{l}}m\sqrt{(\dot{l}+\dot{m}+1)(\dot{l}-\dot{m})},\nonumber\\
\overset{\ast}{a}{}^{k^\prime k;\dot{k}^\prime\dot{k}}_{l+1,l,m,m;
\dot{l}+1,\dot{l},\dot{m}+1,\dot{m}}&=&
-\frac{1}{2}\overset{\ast}{c}{}^{k^\prime k;\dot{k}^\prime\dot{k}}_{l+1,l;
\dot{l}+1,\dot{l}}
\sqrt{((l+1)^2-m^2)(\dot{l}+\dot{m}+1)(\dot{l}+\dot{m}+2)}.\label{L1'}
\end{eqnarray}

Further, from the relations
$\overset{\ast}{\sL}{}^{l\dot{l}}_2=-\ld\widetilde{\sA}^{l\dot{l}}_1,
\overset{\ast}{\sL}{}^{l\dot{l}}_3\rd$ (or
$\overset{\ast}{\sL}^{l\dot{l}}_2=i\ld\widetilde{\sB}^{l\dot{l}}_1,
\overset{\ast}{\sL}{}^{l\dot{l}}_3\rd$) we obtain elements
$\overset{\ast}{b}{}^{k^\prime k;\dot{k}^\prime
\dot{k}}_{l^prime l,m^\prime m;\dot{l}^\prime\dot{l},\dot{m}^\prime\dot{m}}$ of
$\overset{\ast}{\sL}{}^{l\dot{l}}_2$.
All calculations are analogous to the calculations presented
for the case of $\sL^{l\dot{l}}_i$. In the result we have \end{sloppypar}
\begin{eqnarray}
\overset{\ast}{b}{}^{k^\prime k;\dot{k}^\prime\dot{k}}_{l-1,l,m-1,m;
\dot{l}-1,\dot{l},\dot{m}\dot{m}}&=&
\frac{i}{2}\overset{\ast}{c}{}^{k^\prime k;\dot{k}^\prime\dot{k}}_{l-1,l;
\dot{l}-1,\dot{l}}\sqrt{(l+m)(l+m-1)(\dot{l}^2-\dot{m}^2)},\nonumber\\
\overset{\ast}{b}{}^{k^\prime k;\dot{k}^\prime\dot{k}}_{ll,m-1,m;
\dot{l}\dot{l},\dot{m}\dot{m}}&=&
-\frac{i}{2}\overset{\ast}{c}{}^{k^\prime k;\dot{k}^\prime\dot{k}}_{ll;
\dot{l}\dot{l}}\dot{m}\sqrt{(l+m)(l-m+1)},\nonumber\\
\overset{\ast}{b}{}^{k^\prime k;\dot{k}^\prime\dot{k}}_{l+1,l,m-1,m;
\dot{l}+1,\dot{l},\dot{m}\dot{m}}&=&
-\frac{i}{2}\overset{\ast}{c}{}^{k^\prime k;\dot{k}^\prime\dot{k}}_{l+1,l;
\dot{l}+1,\dot{l}}\sqrt{(l-m+1)(l-m+2)((\dot{l}+1)^2-\dot{m}^2)},\nonumber
\end{eqnarray}
\begin{eqnarray}
\overset{\ast}{b}{}^{k^\prime k;\dot{k}^\prime\dot{k}}_{l-1,l,m+1,m;
\dot{l}-1,\dot{l},\dot{m}\dot{m}}&=&
\frac{i}{2}\overset{\ast}{c}{}^{k^\prime k;\dot{k}^\prime\dot{k}}_{l-1,l;
\dot{l}-1,\dot{l}}\sqrt{(l-m)(l-m-1)(\dot{l}^2-\dot{m}^2)},\nonumber\\
\overset{\ast}{b}{}^{k^\prime k;\dot{k}^\prime\dot{k}}_{ll,m+1,m;
\dot{l}\dot{l},\dot{m}\dot{m}}&=&
\frac{i}{2}\overset{\ast}{c}{}^{k^\prime k;\dot{k}^\prime\dot{k}}_{ll;
\dot{l}\dot{l}}\dot{m}\sqrt{(l+m+1)(l-m)},\nonumber\\
\overset{\ast}{b}{}^{k^\prime k;\dot{k}^\prime\dot{k}}_{l+1,l,m+1,m;
\dot{l}+1,\dot{l},\dot{m}\dot{m}}&=&
-\frac{i}{2}\overset{\ast}{c}{}^{k^\prime k;\dot{k}^\prime\dot{k}}_{l+1,l;
\dot{l}+1,\dot{l}}\sqrt{(l+m+1)(l+m+2)((\dot{l}+1)^2-\dot{m}^2)},\nonumber
\end{eqnarray}
\begin{eqnarray}
\overset{\ast}{b}{}^{k^\prime k;\dot{k}^\prime\dot{k}}_{l-1,l,m,m;
\dot{l}-1,\dot{l},\dot{m}-1,\dot{m}}&=&
-\frac{i}{2}\overset{\ast}{c}{}^{k^\prime k;\dot{k}^\prime\dot{k}}_{l-1,l;
\dot{l}-1,\dot{l}}
\sqrt{(l^2-m^2)(\dot{l}+\dot{m})(\dot{l}+\dot{m}-1)},\nonumber\\
\overset{\ast}{b}{}^{k^\prime k;\dot{k}^\prime\dot{k}}_{ll,m,m;
\dot{l}\dot{l},\dot{m}-1,\dot{m}}&=&
\frac{i}{2}\overset{\ast}{c}{}^{k^\prime k;\dot{k}^\prime\dot{k}}_{ll;
\dot{l}\dot{l}}m\sqrt{(\dot{l}+\dot{m})(\dot{l}-\dot{m}+1)},\nonumber\\
\overset{\ast}{b}{}^{k^\prime k;\dot{k}^\prime\dot{k}}_{l+1,l,m,m;
\dot{l}+1,\dot{l},\dot{m}-1,\dot{m}}&=&
\frac{i}{2}\overset{\ast}{c}{}^{k^\prime k;\dot{k}^\prime\dot{k}}_{l+1,l;
\dot{l}+1,\dot{l}}
\sqrt{((l+1)^2-m^2)(\dot{l}-\dot{m}+1)(\dot{l}-\dot{m}+2)},\nonumber
\end{eqnarray}
\begin{eqnarray}
\overset{\ast}{b}{}^{k^\prime k;\dot{k}^\prime\dot{k}}_{l-1,l,m,m;
\dot{l}-1,\dot{l},\dot{m}+1,\dot{m}}&=&
-\frac{i}{2}\overset{\ast}{c}{}^{k^\prime k;\dot{k}^\prime\dot{k}}_{l-1,l;
\dot{l}-1,\dot{l}}
\sqrt{(l^2-m^2)(\dot{l}-\dot{m})(\dot{l}-\dot{m}-1)},\nonumber\\
\overset{\ast}{b}{}^{k^\prime k;\dot{k}^\prime\dot{k}}_{ll,m,m;\dot{l}
\dot{l},\dot{m}+1,\dot{m}}&=&
-\frac{i}{2}\overset{\ast}{c}{}^{k^\prime k;\dot{k}^\prime\dot{k}}_{ll;
\dot{l}\dot{l}}m\sqrt{(\dot{l}+\dot{m}+1)(\dot{l}-\dot{m})},\nonumber\\
\overset{\ast}{b}{}^{k^\prime k;\dot{k}^\prime\dot{k}}_{l+1,l,m,m;
\dot{l}+1,\dot{l},\dot{m}+1,\dot{m}}&=&
\frac{i}{2}\overset{\ast}{c}{}^{k^\prime k;\dot{k}^\prime\dot{k}}_{l+1,l;
\dot{l}+1,\dot{l}}
\sqrt{((l+1)^2-m^2)(\dot{l}+\dot{m}+1)(\dot{l}+\dot{m}+2)}.\label{L2'}
\end{eqnarray}

In general, the matrix $\sL^{l\dot{l}}_3$ must be a reducible
representation of the proper Lorentz group $\fG_+$, and can always be
written in the form
\begin{equation}\label{Decomp}
\sL^{l\dot{l}}_3=\begin{bmatrix}
\sL^{l_1\dot{l}_1}_3 & & & &\\
& \sL^{l_2\dot{l}_2}_3 & &\text{\huge 0} &\\
& & \sL^{l_3\dot{l}_3}_3 & &\\
&\text{\huge 0} & & \ddots &\\
& & & & \sL^{l_n\dot{l}_n}_3
\end{bmatrix},
\end{equation}
where $\sL^{l_i\dot{l}_i}_3$ is a spin block
(the matrix $\overset{\ast}{\sL}{}^{l\dot{l}}_3$ has the same
decompositions). It is obvious that the matrices $\sL^{l\dot{l}}_1$,
$\sL^{l\dot{l}}_2$ and $\overset{\ast}{\sL}{}^{l\dot{l}}_1$,
$\overset{\ast}{\sL}{}^{l\dot{l}}_2$ admit also the decompositions of the
type (\ref{Decomp}) by definition.
If the spin block $\sL^{l_i\dot{l}_i}_3$ has non--null
roots, then the particle possesses the spin $s_i=|l_i-\dot{l}_i|$.
The spin block
$\sL^s_3$ in (\ref{Decomp}) consists of the elements
$c^s_{\boldsymbol{\tau}\boldsymbol{\tau}^\prime}$, where
$\boldsymbol{\tau}_{l_1,\dot{l}_2}$ and
$\boldsymbol{\tau}_{l^\prime_1,\dot{l}^\prime_2}$
are interlocking irreducible
representations\index{representation!interlocking} of the Lorentz group,
that is, such representations, for which
$l^\prime_1=l_1\pm\frac{1}{2}$,
$\dot{l}^\prime_2=\dot{l}_2\pm\frac{1}{2}$.
At this point, the block $\sL^s_3$ contains only the elements
$c^s_{\boldsymbol{\tau}\boldsymbol{\tau}^\prime}$ corresponding to such
interlocking representations $\boldsymbol{\tau}_{l_1,\dot{l}_2}$,
$\boldsymbol{\tau}_{l^\prime_1,\dot{l}^\prime_2}$ which satisfy the conditions
\[
|l_1-\dot{l}_2|\leq s\leq l_1+\dot{l}_2,\quad
|l^\prime_1-\dot{l}^\prime_2|\leq s \leq l^\prime_1+\dot{l}^\prime_2.
\]
The interlocking irreducible representations
of the Lorentz group also called as Bhabha--Gel'fand--Yaglom chains
\cite{Bha45,GY48}.

Corresponding to the decomposition (\ref{Decomp}), the wave function
also decomposes into a direct sum of component wave functions which we
write
\[
\boldsymbol{\psi}=\psi_{l_1m_1;\dot{l}_1\dot{m}_1}+
\psi_{l_2m_2;\dot{l}_2\dot{m}_2}+\psi_{l_3m_3;\dot{l}_3\dot{m}_3}+\ldots.
\]
According to a de Broglie theory of fusion \cite{Bro43},
interlocking representations give rise to
indecomposable RWE.\index{equation!wave!indecomposable}
Otherwise, we have decomposable
equations. \index{equation!wave!decomposable}
As is known, the indecomposable
RWE correspond to composite particles.
A relation between indecomposable RWE and composite particles will be
studied in a separate work.
\section{Separation of variables in RWE}
\subsection{Boundary value problem}
Following to the classical methods of mathematical physics \cite{CH31},
it is quite natural to set up {\it a boundary value problem for the
relativistic wave equations (relativistically invariant system)}.
It is well known that all the physically meaningful requirements,
which follow from the experience, are contained in the boundary value
problem.

Let us construct in $\C^3$ the two--dimensional complex sphere $\dS^2$
from the
quantities $z_k=x_k+iy_k$, $\overset{\ast}{z}_k=x_k-iy_k$
as follows
\begin{equation}\label{CS}
\bz^2=z^2_1+z^2_2+z^2_3=\bx^2-\by^2+2i\bx\by=r^2
\end{equation}
and its complex conjugate (dual) sphere $\dot{\dS}^2$,
\begin{equation}\label{DS}
\overset{\ast}{\bz}{}^2=\overset{\ast}{z}_1{}^2+\overset{\ast}{z}_2{}^2+
\overset{\ast}{z}_3{}^2=\bx^2-\by^2-2i\bx\by=\overset{\ast}{r}{}^2.
\end{equation}
For more details about the two-dimensional complex sphere see
\cite{Hus70,HS70,SH70}.
It is well-known that both quantities $\bx^2-\by^2$, $\bx\by$ are
invariant with respect to the Lorentz transformations, since a surface of
the complex sphere is invariant
(Casimir operators\index{operator!Casimir} of the Lorentz group are
constructed from such quantities, see also (\ref{KO})).
Moreover, since the real and imaginary parts of the complex two-sphere
transform like the electric and magnetic fields, respectively,
the invariance of $\bz^2\sim(\bE+i\bB)^2$ under proper Lorentz
transformations is evident. At this point, the quantities
$\bx^2-\by^2$, $\bx\by$ are similar to the well known electromagnetic
invariants $E^2-B^2$, $\bE\bB$. This intriguing relationship between
the Laplace-Beltrami operators (\ref{KO}), Casimir operators of the
Lorentz group and electromagnetic invariants
$E^2-B^2\sim\bx^2-\by^2$, $\bE\bB\sim\bx\by$ leads naturally to a
Riemann-Silberstein representation of the electromagnetic field
(see, for example, \cite{Web01,Sil07,Bir96}).
In other words, the two-dimensional sphere,
considered as a homogeneous space of the Poincar\'{e} group, is the most
suitable arena for the subsequent investigations in quantum electrodynamics.

We will set up a boundary value problem for the two-dimensional complex
sphere $\dS^2$ (this problem can be considered as a relativistic
generalization of the classical Dirichlet problem for the sphere $S^2$).\\
{\it Let $T$ be an unbounded region in $\C^3\simeq\R^6$
and let $\Sigma$ be a surface of the complex two-sphere (correspondingly,
$\dot{\Sigma}$, for the dual two-sphere), then it needs to find
a function $\boldsymbol{\psi}(\fg)=(\psi_{m\dot{m}}(\fg),
\dot{\psi}_{m\dot{m}}(\fg))^T$
satisfying the following conditions:\\
1) $\boldsymbol{\psi}(\fg)$ is a solution of the system
\begin{eqnarray}
\sum^3_{j=1}\Lambda^{l\dot{l}}_j\frac{\partial\psi}{\partial a_j}-
i\sum^3_{j=1}\Lambda^{l\dot{l}}_j\frac{\partial\psi}{\partial a^\ast_j}+
\dot{\kappa}^c\dot{\psi}&=&0,\nonumber\\
\sum^3_{j=1}\overset{\ast}{\Lambda}{}^{l\dot{l}}_j\frac{\partial\dot{\psi}}
{\partial\widetilde{a}_j}+i\sum^3_{j=1}
\overset{\ast}{\Lambda}{}^{l\dot{l}}_j
\frac{\partial\dot{\psi}}{\partial\widetilde{a}^\ast_j}+
\kappa^c\psi&=&0,\nonumber
\end{eqnarray}
in the all region $T$;\\
2) $\boldsymbol{\psi}(\fg)$ is a continuous function (everywhere in $T$),
including the surfaces $\Sigma$ and $\dot{\Sigma}$;\\
3) $\left.\phantom{\frac{x}{x}}\psi_{m\dot{m}}(\fg)\right|_\Sigma=
F_{m\dot{m}}(\fg)$,
$\left.\phantom{\frac{x}{x}}\dot{\psi}_{m\dot{m}}(\fg)\right|_{\dot{\Sigma}}=
\dot{F}_{m\dot{m}}(\fg)$, where $F_{m\dot{m}}(\fg)$ and
$\dot{F}_{m\dot{m}}(\fg)$ are square integrable
functions defined on the surfaces $\Sigma$ and $\dot{\Sigma}$, respectively.}

In particular, boundary conditions can be represented by constants,
\[
\left.\phantom{\frac{x}{x}}\psi(\fg)\right|_\Sigma=\text{const}=F_0,\quad
\left.\phantom{\frac{x}{x}}\dot{\psi}(\fg)\right|_{\dot{\Sigma}}=\text{const}=
\dot{F}_0.
\]
It is obvious that an explicit form of the boundary conditions follows
from the experience. For example, they can describe a distribution of
energy in the experiment.

With the aim to solve the boundary value problem we come to the complex
Euler angles (\ref{CEA}) and represent the function
$\boldsymbol{\psi}(r,\theta^c,\varphi^c)=(\psi_{m\dot{m}}(r,\theta^c,\varphi^c),
\dot{\psi}_{m\dot{m}}(r^\ast,\dot{\theta}^c,\dot{\varphi}^c))^T$ in the
form of following series:
\begin{eqnarray}
\psi_{m\dot{m}}(r,\theta^c,\varphi^c)&=&\sum^\infty_{l,\dot{l}=0}
\sum_{k,\dot{k}}
\boldsymbol{f}_{lmk;\dot{l}\dot{m}\dot{k}}(r)\sum^l_{n=-l}
\sum^{\dot{l}}_{\dot{n}=-\dot{l}}\alpha^{m\dot{m}}_{ln;\dot{l}\dot{n}}
\fM^{l\dot{l}}_{mn;\dot{m}\dot{n}}(\varphi,\epsilon,\theta,\tau,0,0),
\label{Fourier1}\\
\dot{\psi}_{m\dot{m}}(r^\ast,\dot{\theta}^c,\dot{\varphi}^c)&=&
\sum^\infty_{l,\dot{l}=0}\sum_{k,\dot{k}}
\overset{\ast}{\boldsymbol{f}}_{lmk;\dot{l}\dot{m}\dot{k}}(r^\ast)
\sum^l_{n=-l}
\sum^{\dot{l}}_{\dot{n}=-\dot{l}}
\alpha^{m\dot{m}}_{ln;\dot{l}\dot{n}}
\overset{\ast}{\fM}{}^{l\dot{l}}_{mn;\dot{m}\dot{n}}(\varphi,
\epsilon,\theta,\tau,0,0),
\label{Fourier2}
\end{eqnarray}
where
\begin{eqnarray}
\alpha^{m\dot{m}}_{ln;\dot{l}\dot{n}}&=&
\frac{(-1)^n(2l+1)(2\dot{l}+1)}{32\pi^4}
\int\limits_{\dS^2}
F_{m\dot{m}}(\theta^c,\varphi^c)
\fM^{l\dot{l}}_{mn;\dot{m}\dot{n}}(\varphi,\epsilon,\theta,\tau,0,0)
\sin\theta^c\sin\dot{\theta}^cd\theta d\varphi d\tau d\epsilon,\nonumber\\
\alpha^{m\dot{m}}_{ln;\dot{l}\dot{n}}&=
&\frac{(-1)^{\dot{n}}(2l+1)(2\dot{l}+1)}{32\pi^4}
\int\limits_{\dS^2}
\dot{F}_{m\dot{m}}(\dot{\theta}^c,\dot{\varphi}^c)
\overset{\ast}{\fM}{}^{l\dot{l}}_{mn;\dot{m}\dot{n}}
(\varphi,\epsilon,\theta,\tau,0,0)
\sin\theta^c\sin\dot{\theta}^cd\theta d\varphi d\tau d\epsilon,\nonumber
\end{eqnarray}
\begin{sloppypar}\noindent
The indices $k$ and $\dot{k}$ numerate equivalent representations.
$\fM^{l\dot{l}}_{mn;\dot{m}\dot{n}}(\varphi,\epsilon,\theta,\tau,0,0)$
($\overset{\ast}{\fM}{}^{l\dot{l}}_{mn;\dot{m}\dot{n}}
(\varphi,\epsilon,\theta,\tau,0,0)$) are
hyperspherical functions defined on
the surface $\Sigma$ ($\dot{\Sigma}$) of the
two-dimensional complex sphere of the radius $r$ ($r^\ast$),
$\boldsymbol{f}_{lmk;\dot{l}\dot{m}\dot{k}}(r)$ and
$\overset{\ast}{\boldsymbol{f}}_{lmk;\dot{l}\dot{m}\dot{k}}(r^\ast)$ are radial
functions. It is easy to see that we come here to the harmonic analysis
on the complex two-sphere, since the series (\ref{Fourier1}) and
(\ref{Fourier2}) have the structure of the Fourier series on $\dS^2$.
\end{sloppypar}
Let us introduce now hyperspherical coordinates on the surfaces of the
complex and dual spheres,
\begin{equation}\label{HC}
\begin{array}{ccl}
z_1&=&r\sin\theta^c\cos\varphi^c,\\
z_2&=&r\sin\theta^c\sin\varphi^c,\\
z_3&=&r\cos\theta^c,
\end{array}\quad
\begin{array}{ccl}
z^\ast_1&=&r^\ast\sin\dot{\theta}^c\cos\dot{\varphi}^c,\\
z^\ast_2&=&r^\ast\sin\dot{\theta}^c\sin\dot{\varphi}^c,\\
z^\ast_3&=&r^\ast\cos\dot{\theta}^c,
\end{array}
\end{equation}
where $\theta^c$, $\varphi^c$ are the complex Euler angles.
Let us show that solutions of the equations
(\ref{CT}) can be found in therms of expansions
in generalized hyperspherical functions considered in the
previous section.

With this end in view let us transform the system (\ref{CT}) as
follows. First of all, let us define the derivatives
$\dfrac{\partial}{\partial a_i}$, $\dfrac{\partial}{\partial a^\ast_i}$
on the surface of the two-dimensional complex sphere (\ref{CS})
and write them in the
hyperspherical coordinates (\ref{HC}) as
\begin{eqnarray}
\frac{\partial}{\partial a_1}&=&-\frac{\sin\varphi^c}{r\sin\theta^c}
\frac{\partial}{\partial\varphi}+\frac{\cos\varphi^c\cos\theta^c}{r}
\frac{\partial}{\partial\theta}+\cos\varphi^c\sin\theta^c\frac{\partial}
{\partial r},\label{CD1}\\
\frac{\partial}{\partial a_2}&=&\frac{\cos\varphi^c}{r\sin\theta^c}
\frac{\partial}{\partial\varphi}+\frac{\sin\varphi^c\cos\theta^c}{r}
\frac{\partial}{\partial\theta}+\sin\varphi^c\sin\theta^c\frac{\partial}
{\partial r},\label{CD2}\\
\frac{\partial}{\partial a_3}&=&-\frac{\sin\theta^c}{r}\frac{\partial}
{\partial\theta}+\cos\theta^c\frac{\partial}{\partial r}.\label{CD3}
\end{eqnarray}
\begin{eqnarray}
\frac{\partial}{\partial a^\ast_1}&=&i\frac{\partial}{\partial a_1}=
-\frac{\sin\dot{\varphi}^c}{r\sin\dot{\theta}^c}\frac{\partial}
{\partial\epsilon}+
\frac{\cos\dot{\varphi}^c\sin\dot{\theta}^c}{r}\frac{\partial}{\partial\tau}+
i\cos\dot{\varphi}^c\sin\dot{\theta}^c\frac{\partial}{\partial r},\label{CD4}\\
\frac{\partial}{\partial a^\ast_2}&=&i\frac{\partial}{\partial a_2}=
\frac{\cos\dot{\varphi}^c}{r\sin\dot{\theta}^c}
\frac{\partial}{\partial\epsilon}+
\frac{\sin\dot{\varphi}^c\cos\dot{\theta}^c}{r}\frac{\partial}{\partial\tau}+
i\sin\dot{\varphi}^c\sin\dot{\theta}^c\frac{\partial}{\partial r},\label{CD5}\\
\frac{\partial}{\partial a^\ast_3}&=&i\frac{\partial}{\partial a_3}=
-\frac{\sin\dot{\theta}^c}{r}\frac{\partial}{\partial\tau}+
i\cos\dot{\theta}^c\frac{\partial}{\partial r}.\label{CD6}
\end{eqnarray}
Analogously, on the surface of the dual sphere (\ref{DS}) we have
\begin{eqnarray}
\frac{\partial}{\partial\widetilde{a}_1}&=&
-\frac{\sin\varphi^c}{r^\ast\sin\theta^c}\frac{\partial}
{\partial\varphi}+\frac{\cos\varphi^c\cos\theta^c}{r^\ast}
\frac{\partial}{\partial\theta}+\cos\varphi^c\sin\theta^c
\frac{\partial}{\partial r^\ast},\label{CDD1}\\
\frac{\partial}{\partial\widetilde{a}_2}&=&
\frac{\cos\varphi^c}{r^\ast\sin\theta^c}\frac{\partial}
{\partial\varphi}+\frac{\sin\varphi^c\cos\theta^c}{r^\ast}
\frac{\partial}{\partial\theta}+\sin\varphi^c\sin\theta^c
\frac{\partial}{\partial r^\ast},\label{CDD2}\\
\frac{\partial}{\partial\widetilde{a}_3}&=&
-\frac{\sin\theta^c}{r^\ast}\frac{\partial}{\partial\theta}+
\cos\theta^c\frac{\partial}{\partial r^\ast}.\label{CDD3}
\end{eqnarray}
\begin{eqnarray}
\frac{\partial}{\partial\widetilde{a}^\ast_1}&=&
-i\frac{\partial}{\partial\widetilde{a}_1}=
\frac{\sin\dot{\varphi}^c}{r^\ast\sin\dot{\theta}^c}\frac{\partial}
{\partial\epsilon}-\frac{\cos\dot{\varphi}^c\cos\dot{\theta}^c}{r^\ast}
\frac{\partial}{\partial\tau}-i\cos\dot{\varphi}^c\sin\dot{\theta}^c
\frac{\partial}{\partial r^\ast},\label{CDD4}\\
\frac{\partial}{\partial\widetilde{a}^\ast_2}&=&
-i\frac{\partial}{\partial\widetilde{a}_2}=
-\frac{\cos\dot{\varphi}^c}{r^\ast\sin\dot{\theta}^c}\frac{\partial}
{\partial\epsilon}-\frac{\sin\dot{\varphi}^c\cos\dot{\theta}^c}{r^\ast}
\frac{\partial}{\partial\tau}-i\sin\dot{\varphi}^c\sin\dot{\theta}^c
\frac{\partial}{\partial r^\ast},\label{CDD5}\\
\frac{\partial}{\partial\widetilde{a}^\ast_3}&=&
-i\frac{\partial}{\partial\widetilde{a}_3}=
\frac{\sin\dot{\theta}^c}{r^\ast}\frac{\partial}{\partial\tau}
-i\cos\dot{\theta}^c\frac{\partial}{\partial r^\ast}.\label{CDD6}
\end{eqnarray}

Coming back to the equations (\ref{CT}), we see that the matrices
$\sL^{l\dot{l}}_j$
and $\overset{\ast}{\sL}{}^{l\dot{l}}_j$ inherit their tensor structures
from the infinitesimal operators (\ref{TenStr}),
\begin{eqnarray}
\sL^{l\dot{l}}_j&=&\sL^l_j\otimes\boldsymbol{1}_{2\dot{l}+1}-
\boldsymbol{1}_{2l+1}\otimes\sL^{\dot{l}}_j,\nonumber\\
\overset{\ast}{\sL}{}^{l\dot{l}}_j&=&
\overset{\ast}{\sL}{}^{\dot{l}}_j\otimes\boldsymbol{1}_{2l+1}-
\boldsymbol{1}_{2\dot{l}+1}\otimes\overset{\ast}{\sL}{}^l_j.\nonumber
\end{eqnarray}
Taking into account the latter expressions, we rewrite the system
(\ref{CT}) as follows
\begin{eqnarray}
\sum^3_{j=1}\left(\sL^l_j\otimes\boldsymbol{1}_{2\dot{l}+1}-
\boldsymbol{1}_{2l+1}\otimes\sL^{\dot{l}}_j\right)
\frac{\partial\boldsymbol{\psi}}{\partial a_j}+
i\sum^3_{j=1}\left(\sL^l_j\otimes\boldsymbol{1}_{2\dot{l}+1}-
\boldsymbol{1}_{2l+1}\otimes\sL^{\dot{l}}_j\right)
\frac{\partial\boldsymbol{\psi}}{\partial a^\ast_j}+
\kappa^c\boldsymbol{\psi}&=&0,\nonumber\\
\sum^3_{j=1}\left(\overset{\ast}{\sL}{}^{\dot{l}}_j\otimes
\boldsymbol{1}_{2l+1}-\boldsymbol{1}_{2\dot{l}+1}\otimes
\overset{\ast}{\sL}{}^l_j\right)\frac{\dot{\boldsymbol{\psi}}}
{\partial\widetilde{a}_j}-
i\sum^3_{j=1}\left(\overset{\ast}{\sL}{}^{\dot{l}}_j\otimes
\boldsymbol{1}_{2l+1}-\boldsymbol{1}_{2\dot{l}+1}\otimes
\overset{\ast}{\sL}{}^l_j\right)\frac{\dot{\boldsymbol{\psi}}}
{\partial\widetilde{a}^\ast_j}+\dot{\kappa}^c
\dot{\boldsymbol{\psi}}&=&0.\nonumber
\end{eqnarray}
Substituting the functions
$\boldsymbol{\psi}=T^{-1}_{l\dot{l}}(\fg)\boldsymbol{\psi}^\prime$
($\dot{\boldsymbol{\psi}}=\overset{\ast}{T}{}^{-1}_{l\dot{l}}(\fg)
\dot{\boldsymbol{\psi}}^\prime$) and the derivatives (\ref{CD1})--(\ref{CD6}),
(\ref{CDD1})--(\ref{CDD6}) into this system, and
multiply by $T_{l\dot{l}}(\fg)=T_{l\dot{l}}(\varphi^c,\theta^c,0)$
($\overset{\ast}{T}_{l\dot{l}}(\fg)=
\overset{\ast}{T}_{l\dot{l}}(\dot{\varphi}^c,\dot{\theta}^c,0)$) from the left,
we obtain
\begin{multline}
T_{l\dot{l}}(\fg)\left(\sL^l_1\otimes\boldsymbol{1}_{2\dot{1}+1}-
\boldsymbol{1}_{2l+1}\otimes\sL^{\dot{l}}_1\right)
\left[-\frac{\sin\varphi^c}{r\sin\theta^c}
\frac{\partial(T^{-1}_{l\dot{l}}(\fg)\boldsymbol{\psi}^\prime)}
{\partial\varphi}+\right.\\
\left.+\frac{\cos\varphi^c\cos\theta^c}{r}
\frac{\partial(T^{-1}_{l\dot{l}}(\fg)\boldsymbol{\psi}^\prime)}
{\partial\theta}+
\cos\varphi^c\sin\theta^c
\frac{\partial(T^{-1}_{l\dot{l}}(\fg)\boldsymbol{\psi}^\prime)}
{\partial r}\right]+\\
+T_{l\dot{l}}(\fg)\left(\sL^l_2\otimes\boldsymbol{1}_{2\dot{l}+1}-
\boldsymbol{1}_{2l+1}\otimes\sL^{\dot{l}}_2\right)
\left[\frac{\cos\varphi^c}{r\sin\theta^c}\frac{\partial(T^{-1}_{l\dot{l}}(\fg)
\boldsymbol{\psi}^\prime)}{\partial\varphi}+\right.\\
\left.+\frac{\sin\varphi^c\cos\theta^c}{r}\frac{\partial(T^{-1}_{l\dot{l}}(\fg)
\boldsymbol{\psi}^\prime)}{\partial\theta}+\sin\varphi^c\sin\theta^c
\frac{\partial(T^{-1}_{l\dot{l}}(\fg)\boldsymbol{\psi}^\prime)}
{\partial r}\right]+
\nonumber
\end{multline}
\begin{multline}
T_{l\dot{l}}(\fg)\sL^{l\dot{l}}_3
\left[-\frac{\sin\theta^c}{r}
\frac{\partial(T^{-1}_{l\dot{l}}(\fg)\boldsymbol{\psi}^\prime)}
{\partial\theta}+
\cos\theta^c
\frac{\partial(T^{-1}_{l\dot{l}}(\fg)\boldsymbol{\psi}^\prime)}
{\partial r}\right]+\\
+iT_{l\dot{l}}(\fg)\left(\sL^l_1\otimes\boldsymbol{1}_{2\dot{l}+1}-
\boldsymbol{1}_{2l+1}\otimes\sL^{\dot{l}}_1\right)
\left[-\frac{\sin\dot{\varphi}^c}{r\sin\dot{\theta}^c}
\frac{\partial(T^{-1}_{l\dot{l}}(\fg)
\boldsymbol{\psi}^\prime)}{\partial\epsilon}+\right.\\
\left.+\frac{\cos\dot{\varphi}^c\sin\dot{\theta}^c}{r}
\frac{\partial(T^{-1}_{l\dot{l}}(\fg)
\boldsymbol{\psi}^\prime)}{\partial\tau}+i\cos\dot{\varphi}^c\sin\dot{\theta}^c
\frac{\partial(T^{-1}_{l\dot{l}}(\fg)\boldsymbol{\psi}^\prime)}
{\partial r}\right]+
\nonumber
\end{multline}
\begin{multline}
+iT_{l\dot{l}}(\fg)\left(\sL^l_2\otimes\boldsymbol{1}_{2\dot{1}+1}-
\boldsymbol{1}_{2l+1}\otimes\sL^{\dot{l}}_2\right)
\left[-\frac{\cos\dot{\varphi}^c}{r\sin\dot{\theta}^c}
\frac{\partial(T^{-1}_{l\dot{l}}(\fg)\boldsymbol{\psi}^\prime)}
{\partial\epsilon}+\right.\\
\left.+\frac{\sin\dot{\varphi}^c\cos\dot{\theta}^c}{r}
\frac{\partial(T^{-1}_{l\dot{l}}(\fg)\boldsymbol{\psi}^\prime)}
{\partial\theta}+
i\sin\dot{\varphi}^c\sin\dot{\theta}^c
\frac{\partial(T^{-1}_{l\dot{l}}(\fg)\boldsymbol{\psi}^\prime)}
{\partial r}\right]+\\
+iT_{l\dot{l}}(\fg)\sL^{l\dot{l}}_3
\left[-\frac{\sin\dot{\varphi}^c}{r}\frac{\partial(T^{-1}_{l\dot{l}}(\fg)
\boldsymbol{\psi}^\prime)}{\partial\tau}+
i\cos\dot{\theta}^c\frac{\partial(T^{-1}_{l\dot{l}}(\fg)
\boldsymbol{\psi}^\prime)}{\partial r}\right]
+\kappa^c\boldsymbol{\psi}=0,
\nonumber
\end{multline}
\begin{multline}
\overset{\ast}{T}_{l\dot{l}}(\fg)\left(\overset{\ast}{\sL}{}^{\dot{l}}_1
\otimes\boldsymbol{1}_{2l+1}-\boldsymbol{1}_{2\dot{l}+1}\otimes
\overset{\ast}{\sL}{}^l_1\right)
\left[-\frac{\sin\varphi^c}{r^\ast\sin\theta^c}
\frac{(\overset{\ast}{T}{}^{-1}_{l\dot{l}}(\fg)\dot{\boldsymbol{\psi}}^\prime)}
{\partial\varphi}+\right.\\
\left.+\frac{\cos\varphi^c\cos\theta^c}{r^\ast}
\frac{(\overset{\ast}{T}{}^{-1}_{l\dot{l}}(\fg)\dot{\boldsymbol{\psi}}^\prime)}
{\partial\theta}+\cos\varphi^c\sin\theta^c
\frac{(\overset{\ast}{T}{}^{-1}_{l\dot{l}}(\fg)\dot{\boldsymbol{\psi}}^\prime)}
{\partial r^\ast}\right]+\\
\overset{\ast}{T}_{l\dot{l}}(\fg)\left(\overset{\ast}{\sL}{}^{\dot{l}}_2
\otimes\boldsymbol{1}_{2l+1}-\boldsymbol{1}_{2\dot{l}+1}\otimes
\overset{\ast}{\sL}{}^l_2\right)
\left[\frac{\cos\varphi^c}{r^\ast\sin\theta^c}
\frac{(\overset{\ast}{T}{}^{-1}_{l\dot{l}}(\fg)\dot{\boldsymbol{\psi}}^\prime)}
{\partial\varphi}+\right.\\
\left.+\frac{\sin\varphi^c\cos\theta^c}{r^\ast}
\frac{(\overset{\ast}{T}{}^{-1}_{l\dot{l}}(\fg)\dot{\boldsymbol{\psi}}^\prime)}
{\partial\theta}+\sin\varphi^c\sin\theta^c
\frac{(\overset{\ast}{T}{}^{-1}_{l\dot{l}}(\fg)\dot{\boldsymbol{\psi}}^\prime)}
{\partial r^\ast}\right]+
\nonumber
\end{multline}
\begin{multline}
+\overset{\ast}{T}_{l\dot{l}}(\fg)\overset{\ast}{\sL}{}^{l\dot{l}}_3
\left[-\frac{\sin\theta^c}{r^\ast}
\frac{(\overset{\ast}{T}{}^{-1}_{l\dot{l}}(\fg)\dot{\boldsymbol{\psi}}^\prime)}
{\partial\theta}+
\cos\theta^c
\frac{(\overset{\ast}{T}{}^{-1}_{l\dot{l}}(\fg)\dot{\boldsymbol{\psi}}^\prime)}
{\partial r^\ast}\right]-\\
-i\overset{\ast}{T}_{l\dot{l}}(\fg)\left(\overset{\ast}{\sL}{}^{\dot{l}}_1
\otimes\boldsymbol{1}_{2l+1}-\boldsymbol{1}_{2\dot{l}+1}\otimes
\overset{\ast}{\sL}{}^l_1\right)
\left[\frac{\sin\dot{\varphi}^c}{r^\ast\sin\dot{\theta}^c}
\frac{(\overset{\ast}{T}{}^{-1}_{l\dot{l}}(\fg)\dot{\boldsymbol{\psi}}^\prime)}
{\partial\epsilon}-\right.\\
\left.-\frac{\cos\dot{\varphi}^c\cos\dot{\theta}^c}{r^\ast}
\frac{(\overset{\ast}{T}{}^{-1}_{l\dot{l}}(\fg)\dot{\boldsymbol{\psi}}^\prime)}
{\partial\tau}-i\cos\dot{\varphi}^c\sin\dot{\theta}^c
\frac{(\overset{\ast}{T}{}^{-1}_{l\dot{l}}(\fg)\dot{\boldsymbol{\psi}}^\prime)}
{\partial r^\ast}\right]-
\nonumber
\end{multline}
\begin{multline}
-i\overset{\ast}{T}_{l\dot{l}}(\fg)\left(\overset{\ast}{\sL}{}^{\dot{l}}_2
\otimes\boldsymbol{1}_{2l+1}-\boldsymbol{1}_{2\dot{l}+1}\otimes
\overset{\ast}{\sL}{}^l_2\right)
\left[-\frac{\cos\dot{\varphi}^c}{r^\ast\sin\dot{\theta}^c}
\frac{(\overset{\ast}{T}{}^{-1}_{l\dot{l}}(\fg)\dot{\boldsymbol{\psi}}^\prime)}
{\partial\epsilon}-\right.\\
\left.-\frac{\sin\dot{\varphi}^c\cos\dot{\theta}^c}{r^\ast}
\frac{(\overset{\ast}{T}{}^{-1}_{l\dot{l}}(\fg)\dot{\boldsymbol{\psi}}^\prime)}
{\partial\tau}-i\sin\dot{\varphi}^c\sin\dot{\theta}^c
\frac{(\overset{\ast}{T}{}^{-1}_{l\dot{l}}(\fg)\dot{\boldsymbol{\psi}}^\prime)}
{\partial r^\ast}\right]-\\
-i\overset{\ast}{T}_{l\dot{l}}(\fg)\overset{\ast}{\sL}{}^{l\dot{l}}_3
\left[\frac{\sin\dot{\theta}^c}{r^\ast}
\frac{(\overset{\ast}{T}{}^{-1}_{l\dot{l}}(\fg)\dot{\boldsymbol{\psi}}^\prime)}
{\partial\tau}-
i\cos\dot{\theta}^c
\frac{(\overset{\ast}{T}{}^{-1}_{l\dot{l}}(\fg)\dot{\boldsymbol{\psi}}^\prime)}
{\partial r^\ast}
\right]
+\dot{\kappa}^c\dot{\boldsymbol{\psi}}=0.
\nonumber
\end{multline}
In virtue of the invariance conditions (\ref{IC}) we have
\begin{gather}
T_{l\dot{l}}(\fg)\left[-\sL^{l\dot{l}}_1\sin\varphi^c+
\sL^{l\dot{l}}_2\cos\varphi^c\right]
T^{-1}_{l\dot{l}}(\fg)=\sL^{l\dot{l}}_1,\nonumber\\
T_{l\dot{l}}(\fg)\left[\sL^{l\dot{l}}_1\cos\varphi^c\cos\theta^c+
\sL^{l\dot{l}}_2\sin\varphi^c
\cos\theta^c-\sL^{l\dot{l}}_3\sin\theta^c\right]
T^{-1}_{l\dot{l}}(\fg)=\sL_2,\nonumber\\
T_{l\dot{l}}(\fg)\left[2\sL^{l\dot{l}}_1\cos\varphi^c\sin\theta^c+
\sL^{l\dot{l}}_2\sin\varphi^c\sin\theta^c+
\sL^{l\dot{l}}_3\cos\theta^c\right]T^{-1}_{l\dot{l}}(\fg)=
\sL^{l\dot{l}}_3,\nonumber\\
T_{l\dot{l}}(\fg)\left[-\sL^{l\dot{l}}_1\sin\dot{\varphi}^c+
\sL^{l\dot{l}}_2\cos\dot{\varphi}^c\right]T^{-1}_{l\dot{l}}(\fg)=
\sL^{l\dot{l}}_1,\nonumber\\
T_{l\dot{l}}(\fg)\left[\sL^{l\dot{l}}_1\cos\dot{\varphi}^c\cos\dot{\theta}^c+
\sL^{l\dot{l}}_2\sin\dot{\varphi}^c
\cos\dot{\theta}^c-\sL^{l\dot{l}}_3\sin\dot{\theta}^c\right]
T^{-1}_{l\dot{l}}(\fg)=\sL^{l\dot{l}}_2,\nonumber\\
T_{l\dot{l}}(\fg)\left[\sL^{l\dot{l}}_1\cos\dot{\varphi}^c\sin\dot{\theta}^c+
\sL^{l\dot{l}}_2\sin\dot{\varphi}^c\sin\dot{\theta}^c+
\sL^{l\dot{l}}_3\cos\dot{\theta}^c\right]
T^{-1}_{l\dot{l}}(\fg)=\sL^{l\dot{l}}_3,\nonumber
\end{gather}
\begin{gather}
\overset{\ast}{T}_{l\dot{l}}(\fg)\left[
-\overset{\ast}{\sL}{}^{l\dot{l}}_1\sin\varphi^c+
\overset{\ast}{\sL}{}^{l\dot{l}}_2\cos\varphi^c\right]
\overset{\ast}{T}{}^{-1}_{l\dot{l}}(\fg)=
\overset{\ast}{\sL}{}^{l\dot{l}}_1,\nonumber\\
\overset{\ast}{T}_{l\dot{l}}(\fg)\left[
\overset{\ast}{\sL}{}^{l\dot{l}}_1\cos\varphi^c
\cos\theta^c+
\overset{\ast}{\sL}{}^{l\dot{l}}_2\sin\varphi^c\cos\theta^c-
\overset{\ast}{\sL}{}^{l\dot{l}}_3\sin\theta^c\right]
\overset{\ast}{T}{}^{-1}_{l\dot{l}}(\fg)=
\overset{\ast}{\sL}{}^{l\dot{l}}_2,\nonumber\\
\overset{\ast}{T}_{l\dot{l}}(\fg)\left[
\overset{\ast}{\sL}{}^{l\dot{l}}_1\cos\varphi^c
\sin\theta^c+
\overset{\ast}{\sL}{}^{l\dot{l}}_2\sin\varphi^c\sin\theta^c+
\overset{\ast}{\sL}{}^{l\dot{l}}_3\cos\theta^c\right]
\overset{\ast}{T}{}^{-1}_{l\dot{l}}(\fg)=
\overset{\ast}{\sL}{}^{l\dot{l}}_3,\nonumber\\
\overset{\ast}{T}_{l\dot{l}}(\fg)\left[
\overset{\ast}{\sL}{}^{l\dot{l}}_1\sin\dot{\varphi}^c-
\overset{\ast}{\sL}{}^{l\dot{l}}_2\cos\dot{\varphi}^c\right]
\overset{\ast}{T}{}^{-1}_{l\dot{l}}(\fg)=
\overset{\ast}{\sL}{}^{l\dot{l}}_1,\nonumber\\
\overset{\ast}{T}_{l\dot{l}}(\fg)\left[
-\overset{\ast}{\sL}{}^{l\dot{l}}_1\cos\dot{\varphi}^c
\cos\dot{\theta}^c-
\overset{\ast}{\sL}{}^{l\dot{l}}_2\sin\dot{\varphi}^c\cos\dot{\theta}^c+
\overset{\ast}{\sL}{}^{l\dot{l}}_3\sin\dot{\theta}^c\right]
\overset{\ast}{T}{}^{-1}_{l\dot{l}}(\fg)=
\overset{\ast}{\sL}{}^{l\dot{l}}_2,\nonumber\\
\overset{\ast}{T}_{l\dot{l}}(\fg)\left[
\overset{\ast}{\sL}{}^{l\dot{l}}_1\cos\dot{\varphi}^c
\sin\dot{\theta}^c+
\overset{\ast}{\sL}{}^{l\dot{l}}_2\sin\dot{\varphi}^c\sin\dot{\theta}^c+
\overset{\ast}{\sL}{}^{l\dot{l}}_3\cos\dot{\theta}^c\right]
\overset{\ast}{T}{}^{-1}_{l\dot{l}}(\fg)=
\overset{\ast}{\sL}{}^{l\dot{l}}_3.\nonumber
\end{gather}

Taking into account the latter relations we can write the system
(\ref{CT}) as follows
\begin{multline}
\frac{1}{r\sin\theta^c}\sL^{l}_1\otimes\boldsymbol{1}_{2\dot{l}+1}
T_{l\dot{l}}(\fg)
\frac{\partial(T^{-1}_{l\dot{l}}(\fg)
\boldsymbol{\psi}^\prime)}{\partial\varphi}-
\frac{1}{r\sin\dot{\theta}^c}\boldsymbol{1}_{2l+1}\otimes\sL^{\dot{l}}_1
T_{l\dot{l}}(\fg)\frac{\partial(T^{-1}_{l\dot{l}}(\fg)
\boldsymbol{\psi}^\prime)}{\partial\varphi}+\\
+\frac{i}{r\sin\theta^c}\sL^{l}_1\otimes\boldsymbol{1}_{2\dot{l}+1}
T_{l\dot{l}}(\fg)
\frac{\partial(T^{-1}_{l\dot{l}}(\fg)
\boldsymbol{\psi}^\prime)}{\partial\epsilon}+
\frac{i}{r\sin\dot{\theta}^c}\boldsymbol{1}_{2l+1}\otimes\sL^{\dot{l}}_1
T_{l\dot{l}}(\fg)\frac{\partial(T^{-1}_{l\dot{l}}(\fg)
\boldsymbol{\psi}^\prime)}{\partial\epsilon}-\\
-\frac{1}{r}\sL^{l}_2\otimes\boldsymbol{1}_{2\dot{l}+1}
T_{l\dot{l}}(\fg)\frac{\partial(T^{-1}_{l\dot{l}}(\fg)
\boldsymbol{\psi}^\prime)}{\partial\theta}+
\frac{1}{r}\boldsymbol{1}_{2l+1}\otimes
\sL^{\dot{l}}_2T_{l\dot{l}}(\fg)
\frac{\partial(T^{-1}_{l\dot{l}}(\fg)\boldsymbol{\psi}^\prime)}{\partial\theta}-
\nonumber
\end{multline}
\begin{multline}
-\frac{i}{r}\sL^{l}_2\otimes\boldsymbol{1}_{2\dot{l}+1}
T_{l\dot{l}}(\fg)\frac{\partial(T^{-1}_{l\dot{l}}(\fg)
\boldsymbol{\psi}^\prime)}{\partial\tau}-
\frac{i}{r}\boldsymbol{1}_{2l+1}\otimes
\sL^{\dot{l}}_2T_{l\dot{l}}(\fg)
\frac{\partial(T^{-1}_{l\dot{l}}(\fg)\boldsymbol{\psi}^\prime)}{\partial\tau}-
\\
-\sL^{l\dot{l}}_3T_{l\dot{l}}(\fg)
\frac{\partial(T^{-1}_{l\dot{l}}(\fg)\boldsymbol{\psi}^\prime)}{\partial r}+
i\sL^{l\dot{l}}_3T_{l\dot{l}}(\fg)
\frac{\partial(T^{-1}_{l\dot{l}}(\fg)\boldsymbol{\psi}^\prime)}{\partial r}+
\kappa^c\boldsymbol{\psi}^\prime=0,
\nonumber
\end{multline}
\begin{multline}
\frac{1}{r^\ast\sin\theta^c}\overset{\ast}{\sL}{}^{\dot{l}}_1
\otimes\boldsymbol{1}_{2l+1}
\overset{\ast}{T}_{l\dot{l}}(\fg)
\frac{\partial(\overset{\ast}{T}{}^{-1}_{l\dot{l}}(\fg)
\dot{\boldsymbol{\psi}}^\prime)}
{\partial\varphi}-
\frac{1}{r^\ast\sin\dot{\theta}^c}\boldsymbol{1}_{2\dot{l}+1}\otimes
\overset{\ast}{\sL}{}^{l}_1
\overset{\ast}{T}_{l\dot{l}}(\fg)
\frac{\partial(\overset{\ast}{T}{}^{-1}_{l\dot{l}}(\fg)
\dot{\boldsymbol{\psi}}^\prime)}
{\partial\varphi}+
\\
+\frac{i}{r^\ast\sin\theta^c}\overset{\ast}{\sL}{}^{\dot{l}}_1
\otimes\boldsymbol{1}_{2l+1}
\overset{\ast}{T}_{l\dot{l}}(\fg)
\frac{\partial(\overset{\ast}{T}{}^{-1}_{l\dot{l}}(\fg)
\dot{\boldsymbol{\psi}}^\prime)}
{\partial\epsilon}+
\frac{i}{r^\ast\sin\dot{\theta}^c}\boldsymbol{1}_{2\dot{l}+1}\otimes
\overset{\ast}{\sL}{}^{l}_1
\overset{\ast}{T}_{l\dot{l}}(\fg)
\frac{\partial(\overset{\ast}{T}{}^{-1}_{l\dot{l}}(\fg)
\dot{\boldsymbol{\psi}}^\prime)}
{\partial\epsilon}-
\\
-\frac{1}{r^\ast}\overset{\ast}{\sL}^{\dot{l}}_2
\otimes\boldsymbol{1}_{2l+1}
\overset{\ast}{T}_{l\dot{l}}(\fg)
\frac{\partial(\overset{\ast}{T}{}^{-1}_{l\dot{l}}(\fg)
\dot{\boldsymbol{\psi}}^\prime)}
{\partial\theta}+\frac{1}{r^\ast}
\boldsymbol{1}_{2\dot{l}+1}\otimes
\overset{\ast}{\sL}{}^{l}_2\overset{\ast}{T}_{l\dot{l}}(\fg)
\frac{\partial(\overset{\ast}{T}{}^{-1}_{l\dot{l}}(\fg)
\dot{\boldsymbol{\psi}}^\prime)}
{\partial\theta}-
\nonumber
\end{multline}
\begin{multline}
-\frac{i}{r^\ast}\overset{\ast}{\sL}^{\dot{l}}_2
\otimes\boldsymbol{1}_{2l+1}
\overset{\ast}{T}_{l\dot{l}}(\fg)
\frac{\partial(\overset{\ast}{T}{}^{-1}_{l\dot{l}}(\fg)
\dot{\boldsymbol{\psi}}^\prime)}
{\partial\tau}-\frac{i}{r^\ast}
\boldsymbol{1}_{2\dot{l}+1}\otimes
\overset{\ast}{\sL}{}^{l}_2\overset{\ast}{T}_{l\dot{l}}(\fg)
\frac{\partial(\overset{\ast}{T}{}^{-1}_{l\dot{l}}(\fg)
\dot{\boldsymbol{\psi}}^\prime)}
{\partial\tau}+\\
+\overset{\ast}{\sL}{}^{l\dot{l}}_3
\overset{\ast}{T}_{l\dot{l}}(\fg)
\frac{\partial(\overset{\ast}{T}{}^{-1}_{l\dot{l}}(\fg)
\dot{\boldsymbol{\psi}}^\prime)}
{\partial r^\ast}
-i\overset{\ast}{\sL}{}^{l\dot{l}}_3
\overset{\ast}{T}_{l\dot{l}}(\fg)
\frac{\partial(\overset{\ast}{T}{}^{-1}_{l\dot{l}}(\fg)
\dot{\boldsymbol{\psi}}^\prime)}
{\partial r^\ast}
+\dot{\kappa}^c\dot{\boldsymbol{\psi}}^\prime=0.
\label{Complex3}
\end{multline}
The matrices $T^{-1}_{l\dot{l}}(\fg)$,
$\overset{\ast}{T}{}^{-1}_{l\dot{l}}(\fg)$ depend
on $\varphi$, $\epsilon$, $\theta$, $\tau$. Therefore, we must
differentiate in $T^{-1}_{l\dot{l}}(\fg)\boldsymbol{\psi}^\prime$
($\overset{\ast}{T}{}^{-1}_{l\dot{l}}(\fg)\dot{\boldsymbol{\psi}}^\prime$)
the both factors. After
differentiation we come to the following system:
\begin{multline}
\frac{1}{r\sin\theta^c}\sL^{l}_1\otimes\boldsymbol{1}_{2\dot{l}+1}
\frac{\partial\boldsymbol{\psi}^\prime}
{\partial\varphi}-
\frac{1}{r\sin\dot{\theta}^c}\boldsymbol{1}_{2l+1}\otimes
\sL^{\dot{l}}_1
\frac{\partial\boldsymbol{\psi}^\prime}{\partial\varphi}+
\frac{i}{r\sin\theta^c}\sL^{l}_1\otimes\boldsymbol{1}_{2\dot{l}+1}
\frac{\partial\boldsymbol{\psi}^\prime}
{\partial\epsilon}-\\
-\frac{i}{r\sin\dot{\theta}^c}\boldsymbol{1}_{2l+1}\otimes
\sL^{\dot{l}}_1
\frac{\partial\boldsymbol{\psi}^\prime}{\partial\epsilon}-
\frac{1}{r}\sL^{l}_2\otimes\boldsymbol{1}_{2\dot{l}+1}
\frac{\partial\boldsymbol{\psi}^\prime}
{\partial\theta}
+\frac{1}{r}\boldsymbol{1}_{2l+1}\otimes
\sL^{\dot{l}}_2\frac{\partial\boldsymbol{\psi}^\prime}
{\partial\theta}-\\
-\frac{i}{r}\sL^{l}_2\otimes\boldsymbol{1}_{2\dot{l}+1}
\frac{\partial\boldsymbol{\psi}^\prime}
{\partial\tau}
-\frac{i}{r}\boldsymbol{1}_{2l+1}\otimes
\sL^{\dot{l}}_2\frac{\partial\boldsymbol{\psi}^\prime}
{\partial\tau}
+\sL^{l\dot{l}}_3
\frac{\partial\boldsymbol{\psi}^\prime}{\partial r}
+i\sL^{l\dot{l}}_3
\frac{\partial\boldsymbol{\psi}^\prime}{\partial r}+
\nonumber
\end{multline}
\begin{multline}
+\left[
\frac{1}{r\sin\theta^c}\sL^{l}_1\otimes\boldsymbol{1}_{2\dot{l}+1}
T_{l\dot{l}}(\fg)
\frac{\partial T^{-1}_{l\dot{l}}(\fg)}{\partial\varphi}-
\frac{1}{r\sin\dot{\theta}^c}\boldsymbol{1}_{2l+1}\otimes
\sL^{\dot{l}}_1T_{l\dot{l}}(\fg)
\frac{\partial T^{-1}_{l\dot{l}}(\fg)}{\partial\varphi}+\right.\\
+\frac{i}{r\sin\theta^c}\sL^{l}_1\otimes\boldsymbol{1}_{2\dot{l}+1}
T_{l\dot{l}}(\fg)
\frac{\partial T^{-1}_{l\dot{l}}(\fg)}{\partial\epsilon}-
\frac{i}{r\sin\dot{\theta}^c}\boldsymbol{1}_{2l+1}\otimes
\sL^{\dot{l}}_1T_{l\dot{l}}(\fg)
\frac{\partial T^{-1}_{l\dot{l}}(\fg)}{\partial\epsilon}-\\
-\frac{1}{r}\sL^{l}_2\otimes\boldsymbol{1}_{2\dot{l}+1}
T_{l\dot{l}}(\fg)
\frac{\partial T^{-1}_{l\dot{l}}(\fg)}{\partial\theta}
+\frac{1}{r}\boldsymbol{1}_{2l+1}\otimes
\sL^{\dot{l}}_2T_{l\dot{l}}(\fg)
\frac{\partial T^{-1}_{l\dot{l}}(\fg)}{\partial\theta}-\\
\left.-\frac{i}{r}\sL^{l}_2\otimes\boldsymbol{1}_{2\dot{l}+1}
T_{l\dot{l}}(\fg)
\frac{\partial T^{-1}_{l\dot{l}}(\fg)}{\partial\tau}
-\frac{i}{r}\boldsymbol{1}_{2l+1}\otimes
\sL^{\dot{l}}_2T_{l\dot{l}}(\fg)
\frac{\partial T^{-1}_{l\dot{l}}(\fg)}{\partial\tau}
+\kappa^cI\right]
\boldsymbol{\psi}^\prime=0,
\nonumber
\end{multline}
\begin{multline}
\frac{1}{r^\ast\sin\theta^c}\overset{\ast}{\sL}{}^{\dot{l}}_1
\otimes\boldsymbol{1}_{2l+1}
\frac{\partial\dot{\boldsymbol{\psi}}^\prime}{\partial\varphi}-
\frac{1}{r^\ast\sin\dot{\theta}^c}\boldsymbol{1}_{2\dot{l}+1}\otimes
\overset{\ast}{\sL}{}^{l}_1
\frac{\partial\dot{\boldsymbol{\psi}}^\prime}{\partial\varphi}+
\frac{i}{r^\ast\sin\theta^c}\overset{\ast}{\sL}{}^{\dot{l}}_1
\otimes\boldsymbol{1}_{2l+1}
\frac{\partial\dot{\boldsymbol{\psi}}^\prime}{\partial\epsilon}+\\
+\frac{i}{r^\ast\sin\dot{\theta}^c}\boldsymbol{1}_{2\dot{l}+1}\otimes
\overset{\ast}{\sL}{}^{l}_1
\frac{\partial\dot{\boldsymbol{\psi}}^\prime}{\partial\epsilon}-
\frac{1}{r^\ast}\overset{\ast}{\sL}{}^{\dot{l}}_2
\otimes\boldsymbol{1}_{2l+1}
\frac{\partial\dot{\boldsymbol{\psi}}^\prime}{\partial\theta}+
\frac{1}{r^\ast}\boldsymbol{1}_{2\dot{l}+1}\otimes
\overset{\ast}{\sL}{}^{l}_2
\frac{\partial\dot{\boldsymbol{\psi}}^\prime}{\partial\theta}-\\
-\frac{i}{r^\ast}\overset{\ast}{\sL}{}^{\dot{l}}_2
\otimes\boldsymbol{1}_{2l+1}
\frac{\partial\dot{\boldsymbol{\psi}}^\prime}{\partial\tau}
-\frac{i}{r^\ast}\boldsymbol{1}_{2\dot{l}+1}\otimes
\overset{\ast}{\sL}{}^{l}_2
\frac{\partial\dot{\boldsymbol{\psi}}^\prime}{\partial\tau}+
\overset{\ast}{\sL}{}^{l\dot{l}}_3
\frac{\partial\dot{\boldsymbol{\psi}}^\prime}{\partial r^\ast}-
i\overset{\ast}{\sL}{}^{l\dot{l}}_3
\frac{\partial\dot{\boldsymbol{\psi}}^\prime}{\partial r^\ast}+
\nonumber
\end{multline}
\begin{multline}
+\left[
\frac{1}{r^\ast\sin\theta^c}
\overset{\ast}{\sL}{}^{\dot{l}}_1\otimes\boldsymbol{1}_{2l+1}
\overset{\ast}{T}_{l\dot{l}}(\fg)
\frac{\partial\overset{\ast}{T}{}^{-1}_{l\dot{l}}(\fg)}{\partial\varphi}
-
\frac{1}{r^\ast\sin\dot{\theta}^c}\boldsymbol{1}_{2\dot{l}+1}\otimes
\overset{\ast}{\sL}{}^{l}_1
\overset{\ast}{T}_{l\dot{l}}(\fg)
\frac{\partial
\overset{\ast}{T}{}^{-1}_{l\dot{l}}(\fg)}{\partial\varphi}+\right.\\
+\frac{i}{r^\ast\sin\theta^c}
\overset{\ast}{\sL}{}^{\dot{l}}_1\otimes\boldsymbol{1}_{2l+1}
\overset{\ast}{T}_{l\dot{l}}(\fg)
\frac{\partial\overset{\ast}{T}{}^{-1}_{l\dot{l}}(\fg)}{\partial\epsilon}
+
\frac{i}{r^\ast\sin\dot{\theta}^c}\boldsymbol{1}_{2\dot{l}+1}\otimes
\overset{\ast}{\sL}{}^{l}_1
\overset{\ast}{T}_{l\dot{l}}(\fg)
\frac{\partial\overset{\ast}{T}{}^{-1}_{l\dot{l}}(\fg)}{\partial\epsilon}-\\
-\frac{1}{r^\ast}\overset{\ast}{\sL}{}^{\dot{l}}_2\otimes
\boldsymbol{1}_{2l+1}
\overset{\ast}{T}_{l\dot{l}}(\fg)
\frac{\partial\overset{\ast}{T}{}^{-1}_{l\dot{l}}(\fg)}{\partial\theta}+
\frac{1}{r^\ast}\boldsymbol{1}_{2\dot{l}+1}\otimes
\overset{\ast}{\sL}{}^{l}_2
\overset{\ast}{T}_{l\dot{l}}(\fg)
\frac{\partial\overset{\ast}{T}{}^{-1}_{l\dot{l}}(\fg)}{\partial\theta}\\
\left.-\frac{i}{r^\ast}\overset{\ast}{\sL}{}^{\dot{l}}_2\otimes
\boldsymbol{1}_{2l+1}
\overset{\ast}{T}_{l\dot{l}}(\fg)
\frac{\partial\overset{\ast}{T}{}^{-1}_{l\dot{l}}(\fg)}{\partial\tau}-
\frac{i}{r^\ast}\boldsymbol{1}_{2\dot{l}+1}\otimes
\overset{\ast}{\sL}{}^{l}_2
\overset{\ast}{T}_{l\dot{l}}(\fg)
\frac{\partial\overset{\ast}{T}{}^{-1}_{l\dot{l}}(\fg)}{\partial\tau}+
\dot{\kappa}^cI\right]\dot{\boldsymbol{\psi}}^\prime=0.
\label{Complex4}
\end{multline}
Let us show that the products
$T_{l\dot{l}}(\fg)\frac{\partial T^{-1}_{l\dot{l}}(\fg)}{\partial\varphi}$,
$\ldots$,
$\overset{\ast}{T}_{l\dot{l}}(\fg)
\frac{\partial\overset{\ast}{T}{}^{-1}_{l\dot{l}}(\fg)}
{\partial\tau}$ are expressed via linear combinations of
the infinitesimal operators. For example, let us consider the simplest
tensor representation $\boldsymbol{\tau}_{\frac{1}{2},\frac{1}{2}}=
\boldsymbol{\tau}_{\frac{1}{2},0}\otimes\boldsymbol{\tau}_{0,\frac{1}{2}}$.
The representation $\boldsymbol{\tau}_{\frac{1}{2},\frac{1}{2}}$ is
realized in the four-dimensional symmetric space $\Sym(1,1)$. The matrix
of $\boldsymbol{\tau}_{\frac{1}{2},\frac{1}{2}}$ in the space $\Sym(1,1)$
has the following form:
\[
T_{\frac{1}{2}\frac{1}{2}}(\fg)=\ar\begin{pmatrix}
e^\epsilon\cos\frac{\theta^c}{2}\cos\frac{\dot{\theta}^c}{2} &
-ie^{i\varphi}\cos\frac{\theta^c}{2}\sin\frac{\dot{\theta}^c}{2} &
ie^{-i\varphi}\sin\frac{\theta^c}{2}\cos\frac{\dot{\theta}^c}{2} &
e^{-\epsilon}\sin\frac{\theta^c}{2}\sin\frac{\dot{\theta}^c}{2} \\
-ie^\epsilon\cos\frac{\theta^c}{2}\sin\frac{\dot{\theta}^c}{2} &
e^{i\varphi}\cos\frac{\theta^c}{2}\cos\frac{\dot{\theta}^c}{2} &
e^{-i\varphi}\sin\frac{\theta^c}{2}\sin\frac{\dot{\theta}^c}{2} &
ie^{-\epsilon}\sin\frac{\theta^c}{2}\cos\frac{\dot{\theta}^c}{2} \\
ie^\epsilon\sin\frac{\theta^c}{2}\cos\frac{\dot{\theta}^c}{2} &
e^{i\varphi}\sin\frac{\theta^c}{2}\sin\frac{\dot{\theta}^c}{2} &
e^{-i\varphi}\cos\frac{\theta^c}{2}\cos\frac{\dot{\theta}^c}{2} &
-ie^{-\epsilon}\cos\frac{\theta^c}{2}\sin\frac{\dot{\theta}^c}{2} \\
e^\epsilon\sin\frac{\theta^c}{2}\sin\frac{\dot{\theta}^c}{2} &
ie^{i\varphi}\sin\frac{\theta^c}{2}\cos\frac{\dot{\theta}^c}{2} &
-ie^{-i\varphi}\cos\frac{\theta^c}{2}\sin\frac{\dot{\theta}^c}{2} &
e^{-\epsilon}\cos\frac{\theta^c}{2}\cos\frac{\dot{\theta}^c}{2}
\end{pmatrix}.
\]
\begin{sloppypar}\noindent
This matrix is obtained from (\ref{Tenrep1}) via replacing all the functions
$\fZ^{l\dot{l}}_{mn;\dot{m}\dot{n}}(\theta,\tau)$ by
$\fM^{l\dot{l}}_{mn;\dot{m}\dot{n}}(\varphi,\epsilon,\theta,\tau,0,0)=
e^{-m(\epsilon+i\varphi)-\dot{m}(\epsilon-i\varphi)}
\fZ^{l\dot{l}}_{mn;\dot{m}\dot{n}}(\theta,\tau)$.
An inverse matrix for $T_{\frac{1}{2}\frac{1}{2}}(\fg)$ is
\end{sloppypar}
\[
T^{-1}_{\frac{1}{2}\frac{1}{2}}(\fg)=\ar\begin{pmatrix}
e^{-\epsilon}\cos\frac{\theta^c}{2}\cos\frac{\dot{\theta}^c}{2} &
ie^{-\epsilon}\cos\frac{\theta^c}{2}\sin\frac{\dot{\theta}^c}{2} &
-ie^{-\epsilon}\sin\frac{\theta^c}{2}\cos\frac{\dot{\theta}^c}{2} &
e^{-\epsilon}\sin\frac{\theta^c}{2}\sin\frac{\dot{\theta}^c}{2} \\
ie^{-i\varphi}\cos\frac{\theta^c}{2}\sin\frac{\dot{\theta}^c}{2} &
e^{-i\varphi}\cos\frac{\theta^c}{2}\cos\frac{\dot{\theta}^c}{2} &
e^{-i\varphi}\sin\frac{\theta^c}{2}\sin\frac{\dot{\theta}^c}{2} &
-ie^{-i\varphi}\sin\frac{\theta^c}{2}\cos\frac{\dot{\theta}^c}{2} \\
-ie^{i\varphi}\sin\frac{\theta^c}{2}\cos\frac{\dot{\theta}^c}{2} &
e^{i\varphi}\sin\frac{\theta^c}{2}\sin\frac{\dot{\theta}^c}{2} &
e^{i\varphi}\cos\frac{\theta^c}{2}\cos\frac{\dot{\theta}^c}{2} &
ie^{i\varphi}\cos\frac{\theta^c}{2}\sin\frac{\dot{\theta}^c}{2} \\
e^\epsilon\sin\frac{\theta^c}{2}\sin\frac{\dot{\theta}^c}{2} &
-ie^{\epsilon}\sin\frac{\theta^c}{2}\cos\frac{\dot{\theta}^c}{2} &
ie^{\epsilon}\cos\frac{\theta^c}{2}\sin\frac{\dot{\theta}^c}{2} &
e^{\epsilon}\cos\frac{\theta^c}{2}\cos\frac{\dot{\theta}^c}{2}
\end{pmatrix}.
\]
Infinitesimal operators of the representations
$\boldsymbol{\tau}_{\frac{1}{2},0}$ and $\boldsymbol{\tau}_{0,\frac{1}{2}}$
are
\[
\sA^{\frac{1}{2}}_1=-\frac{i}{2}\begin{bmatrix}
0 & 1\\
1 & 0
\end{bmatrix},\quad
\sA^{\frac{1}{2}}_2=\frac{1}{2}\begin{bmatrix}
0 & 1\\
-1 & 0
\end{bmatrix},\quad
\sA^{\frac{1}{2}}_3=\frac{1}{2}\begin{bmatrix}
i & 0\\
0 &-i
\end{bmatrix},
\]
\[
\sB^{\frac{1}{2}}_1=-\frac{1}{2}\begin{bmatrix}
0 & 1\\
1 & 0
\end{bmatrix},\quad
\sB^{\frac{1}{2}}_2=\frac{1}{2}\begin{bmatrix}
0 & -i\\
i & 0
\end{bmatrix},\quad
\sB^{\frac{1}{2}}_3=\frac{1}{2}\begin{bmatrix}
1 & 0\\
0 &-1
\end{bmatrix},
\]
\[
\widetilde{\sA}^{\frac{1}{2}}_1=-\frac{i}{2}\begin{bmatrix}
0 & 1\\
1 & 0
\end{bmatrix},\quad
\widetilde{\sA}^{\frac{1}{2}}_2=\frac{1}{2}\begin{bmatrix}
0 & 1\\
-1 & 0
\end{bmatrix},\quad
\widetilde{\sA}^{\frac{1}{2}}_3=\frac{1}{2}\begin{bmatrix}
i & 0\\
0 &-i
\end{bmatrix},
\]
\begin{equation}\label{Infhalf}
\widetilde{\sB}^{\frac{1}{2}}_1=\frac{1}{2}\begin{bmatrix}
0 & 1\\
1 & 0
\end{bmatrix},\quad
\widetilde{\sB}^{\frac{1}{2}}_2=\frac{1}{2}\begin{bmatrix}
0 & i\\
-i & 0
\end{bmatrix},\quad
\widetilde{\sB}^{\frac{1}{2}}_3=\frac{1}{2}\begin{bmatrix}
-1 & 0\\
0 & 1
\end{bmatrix}.
\end{equation}
Taking into account the latter relations, we find
\begin{multline}
T_{\frac{1}{2}\frac{1}{2}}(\fg)
\frac{\partial T^{-1}_{\frac{1}{2}\frac{1}{2}}(\fg)}{\partial\varphi}=\\
\frac{1}{2}\ar\begin{pmatrix}
-i\cos\theta^c+i\cos\dot{\theta}^c & -\sin\dot{\theta}^c &
-\sin\theta^c & 0\\
\sin\dot{\theta}^c & -i\cos\theta^c-i\cos\dot{\theta}^c &
0 & -\sin\theta^c\\
\sin\theta^c & 0 & i\cos\theta^c+i\cos\dot{\theta}^c & -\sin\dot{\theta}^c \\
0 & \sin\theta^c & \sin\dot{\theta}^c & i\cos\theta^c-i\cos\dot{\theta}^c
\end{pmatrix}=\\
=-(\sA_3\otimes\boldsymbol{1}_2\cos\theta^c-
\boldsymbol{1}_2\otimes\widetilde{\sA}_3\cos\dot{\theta}^c)-
(\sA_2\otimes\boldsymbol{1}_2\sin\theta^c+
\boldsymbol{1}_2\otimes\widetilde{\sA}_2\sin\dot{\theta}^c),\label{Irel1}
\end{multline}
\begin{multline}
T_{\frac{1}{2}\frac{1}{2}}(\fg)
\frac{\partial T^{-1}_{\frac{1}{2}\frac{1}{2}}(\fg)}{\partial\epsilon}=\\
\frac{1}{2}\ar\begin{pmatrix}
-\cos\theta^c-\cos\dot{\theta}^c & -i\sin\dot{\theta}^c &
i\sin\theta^c & 0\\
i\sin\dot{\theta}^c & -\cos\theta^c+\cos\dot{\theta}^c &
0 & i\sin\theta^c\\
-i\sin\theta^c & 0 & \cos\theta^c-\cos\dot{\theta}^c & -i\sin\dot{\theta}^c \\
0 & -i\sin\theta^c & i\sin\dot{\theta}^c & \cos\theta^c+\cos\dot{\theta}^c
\end{pmatrix}=\\
=-(\sB_3\otimes\boldsymbol{1}_2\cos\theta^c-
\boldsymbol{1}_2\otimes\widetilde{\sB}_3\cos\dot{\theta}^c)-
(\sB_2\otimes\boldsymbol{1}_2\sin\theta^c+
\boldsymbol{1}_2\otimes\widetilde{\sB}_2\sin\dot{\theta}^c),\label{Irel2}
\end{multline}
\begin{equation}\label{Irel3}
T_{\frac{1}{2}\frac{1}{2}}(\fg)
\frac{\partial T^{-1}_{\frac{1}{2}\frac{1}{2}}(\fg)}{\partial\theta}=
\frac{1}{2}\ar\begin{pmatrix}
0 & i & -i & 0\\
i & 0 & 0 & -i\\
-i & 0 & 0 & i\\
0 & -i & i & 0
\end{pmatrix}=
\sA_1\otimes\boldsymbol{1}_2-\boldsymbol{1}_2\otimes\widetilde{\sA}_1,
\end{equation}
\begin{equation}\label{Irel4}
T_{\frac{1}{2}\frac{1}{2}}(\fg)
\frac{\partial T^{-1}_{\frac{1}{2}\frac{1}{2}}(\fg)}{\partial\tau}=
-\frac{1}{2}\ar\begin{pmatrix}
0 & 1 & 1 & 0\\
1 & 0 & 0 & 1\\
1 & 0 & 0 & 1\\
0 & 1 & 1 & 0
\end{pmatrix}=
\sB_1\otimes\boldsymbol{1}_2-\boldsymbol{1}_2\otimes\widetilde{\sB}_1.
\end{equation}
Further, a matrix of the conjugate representation
$\overset{\ast}{\boldsymbol{\tau}}_{\frac{1}{2},\frac{1}{2}}$ has the form
\[
\overset{\ast}{T}_{\frac{1}{2}\frac{1}{2}}(\fg)=\ar\begin{pmatrix}
e^\epsilon\cos\frac{\dot{\theta}^c}{2}\cos\frac{\theta^c}{2} &
ie^{-i\varphi}\cos\frac{\dot{\theta}^c}{2}\sin\frac{\theta^c}{2} &
-ie^{i\varphi}\sin\frac{\dot{\theta}^c}{2}\cos\frac{\theta^c}{2} &
e^{-\epsilon}\sin\frac{\dot{\theta}^c}{2}\sin\frac{\theta^c}{2} \\
ie^\epsilon\cos\frac{\dot{\theta}^c}{2}\sin\frac{\theta^c}{2} &
e^{-i\varphi}\cos\frac{\dot{\theta}^c}{2}\cos\frac{\theta^c}{2} &
e^{i\varphi}\sin\frac{\dot{\theta}^c}{2}\sin\frac{\theta^c}{2} &
-ie^{-\epsilon}\sin\frac{\dot{\theta}^c}{2}\cos\frac{\theta^c}{2} \\
-ie^\epsilon\sin\frac{\dot{\theta}^c}{2}\cos\frac{\theta^c}{2} &
e^{-i\varphi}\sin\frac{\dot{\theta}^c}{2}\sin\frac{\theta^c}{2} &
e^{i\varphi}\cos\frac{\dot{\theta}^c}{2}\cos\frac{\theta^c}{2} &
ie^{-\epsilon}\cos\frac{\dot{\theta}^c}{2}\sin\frac{\theta^c}{2} \\
e^\epsilon\sin\frac{\dot{\theta}^c}{2}\sin\frac{\theta^c}{2} &
-ie^{-i\varphi}\sin\frac{\dot{\theta}^c}{2}\cos\frac{\theta^c}{2} &
ie^{i\varphi}\cos\frac{\dot{\theta}^c}{2}\sin\frac{\theta^c}{2} &
e^{-\epsilon}\cos\frac{\dot{\theta}^c}{2}\cos\frac{\theta^c}{2}
\end{pmatrix},
\]
and its inverse matrix is
\[
\overset{\ast}{T}{}^{-1}_{\frac{1}{2}\frac{1}{2}}(\fg)=\ar\begin{pmatrix}
e^{-\epsilon}\cos\frac{\dot{\theta}^c}{2}\cos\frac{\theta^c}{2} &
-ie^{-\epsilon}\cos\frac{\dot{\theta}^c}{2}\sin\frac{\theta^c}{2} &
ie^{-\epsilon}\sin\frac{\dot{\theta}^c}{2}\cos\frac{\theta^c}{2} &
e^{-\epsilon}\sin\frac{\dot{\theta}^c}{2}\sin\frac{\theta^c}{2} \\
-ie^{i\varphi}\cos\frac{\dot{\theta}^c}{2}\sin\frac{\theta^c}{2} &
e^{i\varphi}\cos\frac{\dot{\theta}^c}{2}\cos\frac{\theta^c}{2} &
e^{i\varphi}\sin\frac{\dot{\theta}^c}{2}\sin\frac{\theta^c}{2} &
ie^{i\varphi}\sin\frac{\dot{\theta}^c}{2}\cos\frac{\theta^c}{2} \\
ie^{-i\varphi}\sin\frac{\dot{\theta}^c}{2}\cos\frac{\theta^c}{2} &
e^{-i\varphi}\sin\frac{\dot{\theta}^c}{2}\sin\frac{\theta^c}{2} &
e^{-i\varphi}\cos\frac{\dot{\theta}^c}{2}\cos\frac{\theta^c}{2} &
-ie^{-i\varphi}\cos\frac{\dot{\theta}^c}{2}\sin\frac{\theta^c}{2} \\
e^\epsilon\sin\frac{\dot{\theta}^c}{2}\sin\frac{\theta^c}{2} &
ie^{\epsilon}\sin\frac{\dot{\theta}^c}{2}\cos\frac{\theta^c}{2} &
-ie^{\epsilon}\cos\frac{\dot{\theta}^c}{2}\sin\frac{\theta^c}{2} &
e^{\epsilon}\cos\frac{\dot{\theta}^c}{2}\cos\frac{\theta^c}{2}
\end{pmatrix}.
\]
In this case we have
\begin{multline}
\overset{\ast}{T}_{\frac{1}{2}\frac{1}{2}}(\fg)
\frac{\partial\overset{\ast}{T}{}^{-1}_{\frac{1}{2}\frac{1}{2}}(\fg)}
{\partial\varphi}=\\
\frac{1}{2}\ar\begin{pmatrix}
-i\cos\theta^c+i\cos\dot{\theta}^c & -\sin\theta^c &
-\sin\dot{\theta}^c & 0\\
\sin\theta^c & i\cos\theta^c+i\cos\dot{\theta}^c &
0 & -\sin\dot{\theta}^c\\
\sin\dot{\theta}^c & 0 & -i\cos\theta^c-i\cos\dot{\theta}^c & -\sin\theta^c \\
0 & \sin\dot{\theta}^c & \sin\theta^c & i\cos\theta^c-i\cos\dot{\theta}^c
\end{pmatrix}=\\
=(\widetilde{\sA}_3\otimes\boldsymbol{1}_2\cos\dot{\theta}^c-
\boldsymbol{1}_2\otimes\sA_3\cos\theta^c)-
(\widetilde{\sA}_2\otimes\boldsymbol{1}_2\sin\dot{\theta}^c+
\boldsymbol{1}_2\otimes\sA_2\sin\theta^c),\label{Irel5}
\end{multline}
\begin{multline}
\overset{\ast}{T}_{\frac{1}{2}\frac{1}{2}}(\fg)
\frac{\partial\overset{\ast}{T}{}^{-1}_{\frac{1}{2}\frac{1}{2}}(\fg)}
{\partial\epsilon}=\\
\frac{1}{2}\ar\begin{pmatrix}
-\cos\theta^c-\cos\dot{\theta}^c & i\sin\theta^c &
-i\sin\dot{\theta}^c & 0\\
-i\sin\theta^c & \cos\theta^c-\cos\dot{\theta}^c &
0 & -i\sin\dot{\theta}^c\\
i\sin\dot{\theta}^c & 0 & -\cos\theta^c+\cos\dot{\theta}^c & i\sin\theta^c \\
0 & i\sin\dot{\theta}^c & -i\sin\theta^c & \cos\theta^c+\cos\dot{\theta}^c
\end{pmatrix}=\\
=(\widetilde{\sB}_3\otimes\boldsymbol{1}_2\cos\dot{\theta}^c-
\boldsymbol{1}_2\otimes\sB_3\cos\theta^c)-
(\widetilde{\sB}_2\otimes\boldsymbol{1}_2\sin\dot{\theta}^c+
\boldsymbol{1}_2\otimes\sB_2\sin\theta^c),\label{Irel6}
\end{multline}
\begin{equation}\label{Irel7}
\overset{\ast}{T}_{\frac{1}{2}\frac{1}{2}}(\fg)
\frac{\partial\overset{\ast}{T}{}^{-1}_{\frac{1}{2}\frac{1}{2}}(\fg)}
{\partial\theta}=
\frac{1}{2}\ar\begin{pmatrix}
0 & -i & i & 0\\
-i & 0 & 0 & i\\
i & 0 & 0 & -i\\
0 & i & -i & 0
\end{pmatrix}=
-(\widetilde{\sA}_1\otimes\boldsymbol{1}_2-\boldsymbol{1}_2\otimes\sA_1),
\end{equation}
\begin{equation}\label{Irel8}
\overset{\ast}{T}_{\frac{1}{2}\frac{1}{2}}(\fg)
\frac{\partial\overset{\ast}{T}{}^{-1}_{\frac{1}{2}\frac{1}{2}}(\fg)}
{\partial\tau}=
-\frac{1}{2}\ar\begin{pmatrix}
0 & 1 & 1 & 0\\
1 & 0 & 0 & 1\\
1 & 0 & 0 & 1\\
0 & 1 & 1 & 0
\end{pmatrix}=
-(\widetilde{\sB}_1\otimes\boldsymbol{1}_2-\boldsymbol{1}_2\otimes\sB_1).
\end{equation}
In the following example we consider the first nontrivial tensor
representation $\boldsymbol{\tau}_{1,\frac{1}{2}}=
\boldsymbol{\tau}_{1,0}\otimes\boldsymbol{\tau}_{0,\frac{1}{2}}$.
The representation $\boldsymbol{\tau}_{1,\frac{1}{2}}$ is realized
in the six-dimensional symmetric space $\Sym(2,1)$. The matrix of
$\boldsymbol{\tau}_{1,\frac{1}{2}}$ in the space $\Sym(2,1)$ has the
following form:
\begin{multline}
T_{1\frac{1}{2}}(\fg)=\left(\ar\begin{array}{ccc}
\scr e^{\frac{3\epsilon+i\varphi}{2}}
\cos^2\frac{\theta^c}{2}\cos\frac{\dot{\theta}^c}{2} &
\scr -ie^{\frac{\epsilon+3i\varphi}{2}}
\cos^2\frac{\theta^c}{2}\sin\frac{\dot{\theta}^c}{2} &
\scr\frac{ie^{\frac{\epsilon-i\varphi}{2}}}{\sqrt{2}}
\sin\theta^c\cos\frac{\dot{\theta}^c}{2} \\
\scr -ie^{\frac{3\epsilon+i\varphi}{2}}
\cos^2\frac{\theta^c}{2}\sin\frac{\dot{\theta}^c}{2} &
\scr e^{\frac{\epsilon+3i\varphi}{2}}
\cos^2\frac{\theta^c}{2}\cos\frac{\dot{\theta}^c}{2} &
\scr\frac{e^{\frac{\epsilon-i\varphi}{2}}}{\sqrt{2}}
\sin\theta^c\sin\frac{\dot{\theta}^c}{2} \\
\scr\frac{ie^{\frac{3\epsilon+i\varphi}{2}}}{\sqrt{2}}
\sin\theta^c\cos\frac{\dot{\theta}^c}{2} &
\scr\frac{e^{\frac{\epsilon+3i\varphi}{2}}}{\sqrt{2}}
\sin\theta^c\sin\frac{\dot{\theta}^c}{2} &
\scr e^{\frac{\epsilon-i\varphi}{2}}
\cos\theta^c\cos\frac{\dot{\theta}^c}{2} \\
\scr\frac{e^{\frac{3\epsilon+i\varphi}{2}}}{\sqrt{2}}
\sin\theta^c\sin\frac{\dot{\theta}^c}{2} &
\scr\frac{ie^{\frac{\epsilon+3i\varphi}{2}}}{\sqrt{2}}
\sin\theta^c\cos\frac{\dot{\theta}^c}{2} &
\scr -ie^{\frac{\epsilon-i\varphi}{2}}
\cos\theta^c\sin\frac{\dot{\theta}^c}{2} \\
\scr -e^{\frac{3\epsilon+i\varphi}{2}}
\sin^2\frac{\theta^c}{2}\cos\frac{\dot{\theta}^c}{2} &
\scr ie^{\frac{\epsilon+3i\varphi}{2}}
\sin^2\frac{\theta^c}{2}\sin\frac{\dot{\theta}^c}{2} &
\scr\frac{ie^{\frac{\epsilon-i\varphi}{2}}}{\sqrt{2}}
\sin\theta^c\cos\frac{\dot{\theta}^c}{2} \\
\scr ie^{\frac{3\epsilon+i\varphi}{2}}
\sin^2\frac{\theta^c}{2}\sin\frac{\dot{\theta}^c}{2} &
\scr -e^{\frac{\epsilon+3i\varphi}{2}}
\sin^2\frac{\theta^c}{2}\cos\frac{\dot{\theta}^c}{2} &
\scr\frac{e^{\frac{\epsilon-i\varphi}{2}}}{\sqrt{2}}
\sin\theta^c\sin\frac{\dot{\theta}^c}{2}
\end{array}\right.\\
\left.\ar\begin{array}{ccc}
\scr\frac{e^{\frac{-\epsilon+i\varphi}{2}}}{\sqrt{2}}
\sin\theta^c\sin\frac{\dot{\theta}^c}{2} &
\scr -e^{-\frac{\epsilon+3i\varphi}{2}}
\sin^2\frac{\theta^c}{2}\cos\frac{\dot{\theta}^c}{2} &
\scr ie^{-\frac{3\epsilon+i\varphi}{2}}
\sin^2\frac{\theta^c}{2}\sin\frac{\dot{\theta}^c}{2} \\
\scr\frac{ie^{\frac{-\epsilon+i\varphi}{2}}}{\sqrt{2}}
\sin\theta^c\cos\frac{\dot{\theta}^c}{2} &
\scr ie^{-\frac{\epsilon+3i\varphi}{2}}
\sin^2\frac{\theta^c}{2}\sin\frac{\dot{\theta}^c}{2} &
\scr -e^{-\frac{3\epsilon+i\varphi}{2}}
\sin^2\frac{\theta^c}{2}\cos\frac{\dot{\theta}^c}{2} \\
\scr -ie^{\frac{-\epsilon+i\varphi}{2}}
\cos\theta^c\sin\frac{\dot{\theta}^c}{2} &
\scr\frac{ie^{-\frac{\epsilon+3i\varphi}{2}}}{\sqrt{2}}
\sin\theta^c\cos\frac{\dot{\theta}^c}{2} &
\scr\frac{e^{-\frac{3\epsilon+i\varphi}{2}}}{\sqrt{2}}
\sin\theta^c\sin\frac{\dot{\theta}^c}{2} \\
\scr e^{\frac{-\epsilon+i\varphi}{2}}
\cos\theta^c\cos\frac{\dot{\theta}^c}{2} &
\scr\frac{e^{-\frac{\epsilon+3i\varphi}{2}}}{\sqrt{2}}
\sin\theta^c\sin\frac{\dot{\theta}^c}{2} &
\scr\frac{ie^{-\frac{3\epsilon+i\varphi}{2}}}{\sqrt{2}}
\sin\theta^c\cos\frac{\dot{\theta}^c}{2} \\
\scr\frac{e^{\frac{-\epsilon+i\varphi}{2}}}{\sqrt{2}}
\sin\theta^c\sin\frac{\dot{\theta}^c}{2} &
\scr e^{-\frac{\epsilon+3i\varphi}{2}}
\cos^2\frac{\theta^c}{2}\cos\frac{\dot{\theta}^c}{2} &
\scr -ie^{-\frac{3\epsilon+i\varphi}{2}}
\cos^2\frac{\theta^c}{2}\sin\frac{\dot{\theta}^c}{2} \\
\scr\frac{ie^{\frac{-\epsilon+i\varphi}{2}}}{\sqrt{2}}
\sin\theta^c\cos\frac{\dot{\theta}^c}{2} &
\scr -ie^{-\frac{\epsilon+3i\varphi}{2}}
\cos^2\frac{\theta^c}{2}\sin\frac{\dot{\theta}^c}{2} &
\scr e^{-\frac{3\epsilon+i\varphi}{2}}
\cos^2\frac{\theta^c}{2}\cos\frac{\dot{\theta}^c}{2}
\end{array}\right).\nonumber
\end{multline}
In turn, this matrix is obtained from (\ref{Tenrep2}) via replacing all
the functions $\fZ^{l\dot{l}}_{mn;\dot{m}\dot{n}}(\theta,\tau)$ by
$\fM^{l\dot{l}}_{mn;\dot{m}\dot{n}}(\varphi,\epsilon,\theta,\tau,0,0)$.
An inverse matrix for $T_{1\frac{1}{2}}(\fg)$ is
\begin{multline}
T^{-1}_{1\frac{1}{2}}(\fg)=\left(\ar\begin{array}{ccc}
\scr e^{-\frac{3\epsilon+i\varphi}{2}}
\cos^2\frac{\theta^c}{2}\cos\frac{\dot{\theta}^c}{2} &
\scr ie^{-\frac{3\epsilon+i\varphi}{2}}
\cos^2\frac{\theta^c}{2}\sin\frac{\dot{\theta}^c}{2} &
\scr-\frac{ie^{-\frac{3\epsilon+i\varphi}{2}}}{\sqrt{2}}
\sin\theta^c\cos\frac{\dot{\theta}^c}{2} \\
\scr ie^{-\frac{\epsilon+3i\varphi}{2}}
\cos^2\frac{\theta^c}{2}\sin\frac{\dot{\theta}^c}{2} &
\scr e^{-\frac{\epsilon+3i\varphi}{2}}
\cos^2\frac{\theta^c}{2}\cos\frac{\dot{\theta}^c}{2} &
\scr\frac{e^{-\frac{\epsilon+3i\varphi}{2}}}{\sqrt{2}}
\sin\theta^c\sin\frac{\dot{\theta}^c}{2} \\
\scr-\frac{ie^{\frac{-\epsilon+i\varphi}{2}}}{\sqrt{2}}
\sin\theta^c\cos\frac{\dot{\theta}^c}{2} &
\scr\frac{e^{\frac{-\epsilon+i\varphi}{2}}}{\sqrt{2}}
\sin\theta^c\sin\frac{\dot{\theta}^c}{2} &
\scr e^{\frac{-\epsilon+i\varphi}{2}}
\cos\theta^c\cos\frac{\dot{\theta}^c}{2} \\
\scr\frac{e^{\frac{\epsilon-i\varphi}{2}}}{\sqrt{2}}
\sin\theta^c\sin\frac{\dot{\theta}^c}{2} &
\scr-\frac{ie^{\frac{\epsilon-i\varphi}{2}}}{\sqrt{2}}
\sin\theta^c\cos\frac{\dot{\theta}^c}{2} &
\scr ie^{\frac{\epsilon-i\varphi}{2}}
\cos\theta^c\sin\frac{\dot{\theta}^c}{2} \\
\scr -e^{\frac{\epsilon+3i\varphi}{2}}
\sin^2\frac{\theta^c}{2}\cos\frac{\dot{\theta}^c}{2} &
\scr -ie^{\frac{\epsilon+3i\varphi}{2}}
\sin^2\frac{\theta^c}{2}\sin\frac{\dot{\theta}^c}{2} &
\scr-\frac{ie^{\frac{\epsilon+3i\varphi}{2}}}{\sqrt{2}}
\sin\theta^c\cos\frac{\dot{\theta}^c}{2} \\
\scr -ie^{\frac{3\epsilon+i\varphi}{2}}
\sin^2\frac{\theta^c}{2}\sin\frac{\dot{\theta}^c}{2} &
\scr -e^{\frac{3\epsilon+i\varphi}{2}}
\sin^2\frac{\theta^c}{2}\cos\frac{\dot{\theta}^c}{2} &
\scr\frac{e^{\frac{3\epsilon+i\varphi}{2}}}{\sqrt{2}}
\sin\theta^c\sin\frac{\dot{\theta}^c}{2}
\end{array}\right.\\
\left.\ar\begin{array}{ccc}
\scr\frac{e^{-\frac{3\epsilon+i\varphi}{2}}}{\sqrt{2}}
\sin\theta^c\sin\frac{\dot{\theta}^c}{2} &
\scr -e^{-\frac{3\epsilon+i\varphi}{2}}
\sin^2\frac{\theta^c}{2}\cos\frac{\dot{\theta}^c}{2} &
\scr -ie^{-\frac{3\epsilon+i\varphi}{2}}
\sin^2\frac{\theta^c}{2}\sin\frac{\dot{\theta}^c}{2} \\
\scr-\frac{ie^{-\frac{\epsilon+3i\varphi}{2}}}{\sqrt{2}}
\sin\theta^c\cos\frac{\dot{\theta}^c}{2} &
\scr -ie^{-\frac{\epsilon+3i\varphi}{2}}
\sin^2\frac{\theta^c}{2}\sin\frac{\dot{\theta}^c}{2} &
\scr -e^{-\frac{\epsilon+3i\varphi}{2}}
\sin^2\frac{\theta^c}{2}\cos\frac{\dot{\theta}^c}{2} \\
\scr ie^{\frac{-\epsilon+i\varphi}{2}}
\cos\theta^c\sin\frac{\dot{\theta}^c}{2} &
\scr-\frac{ie^{\frac{-\epsilon+i\varphi}{2}}}{\sqrt{2}}
\sin\theta^c\cos\frac{\dot{\theta}^c}{2} &
\scr\frac{e^{\frac{-\epsilon+i\varphi}{2}}}{\sqrt{2}}
\sin\theta^c\sin\frac{\dot{\theta}^c}{2} \\
\scr e^{\frac{\epsilon-i\varphi}{2}}
\cos\theta^c\cos\frac{\dot{\theta}^c}{2} &
\scr\frac{e^{\frac{\epsilon-i\varphi}{2}}}{\sqrt{2}}
\sin\theta^c\sin\frac{\dot{\theta}^c}{2} &
\scr-\frac{ie^{\frac{\epsilon-i\varphi}{2}}}{\sqrt{2}}
\sin\theta^c\cos\frac{\dot{\theta}^c}{2} \\
\scr\frac{e^{\frac{\epsilon+3i\varphi}{2}}}{\sqrt{2}}
\sin\theta^c\sin\frac{\dot{\theta}^c}{2} &
\scr e^{\frac{\epsilon+3i\varphi}{2}}
\cos^2\frac{\theta^c}{2}\cos\frac{\dot{\theta}^c}{2} &
\scr ie^{\frac{\epsilon+3i\varphi}{2}}
\cos^2\frac{\theta^c}{2}\sin\frac{\dot{\theta}^c}{2} \\
\scr-\frac{ie^{\frac{3\epsilon+i\varphi}{2}}}{\sqrt{2}}
\sin\theta^c\cos\frac{\dot{\theta}^c}{2} &
\scr ie^{\frac{3\epsilon+i\varphi}{2}}
\cos^2\frac{\theta^c}{2}\sin\frac{\dot{\theta}^c}{2} &
\scr e^{\frac{3\epsilon+i\varphi}{2}}
\cos^2\frac{\theta^c}{2}\cos\frac{\dot{\theta}^c}{2}
\end{array}\right).\nonumber
\end{multline}
In turn, infinitesimal operators of the representation
$\boldsymbol{\tau}_{1,0}$ are
\[
\sA^1_1=-\frac{i}{\sqrt{2}}\begin{bmatrix}
0 & 1 & 0\\
1 & 0 & 1\\
0 & 1 & 0
\end{bmatrix},\quad
\sA^1_2=\frac{1}{\sqrt{2}}\begin{bmatrix}
0 & 1 & 0\\
-1 & 0 & 1\\
0 &-1 & 0
\end{bmatrix},\quad
\sA^1_3=\begin{bmatrix}
i & 0 & 0\\
0 & 0 & 0\\
0 & 0 &-i
\end{bmatrix},
\]
\[
\sB^1_1=-\frac{1}{\sqrt{2}}\begin{bmatrix}
0 & 1 & 0\\
1 & 0 & 1\\
0 & 1 & 0
\end{bmatrix},\quad
\sB^1_2=\frac{1}{\sqrt{2}}\begin{bmatrix}
0 &-i & 0\\
i & 0 &-i\\
0 & i & 0
\end{bmatrix},\quad
\sB^1_3=\begin{bmatrix}
1 & 0 & 0\\
0 & 0 & 0\\
0 & 0 &-1
\end{bmatrix},
\]
\[
\widetilde{\sA}^1_1=-\frac{i}{\sqrt{2}}\begin{bmatrix}
0 & 1 & 0\\
1 & 0 & 1\\
0 & 1 & 0
\end{bmatrix},\quad
\widetilde{\sA}^1_2=\frac{1}{\sqrt{2}}\begin{bmatrix}
0 & 1 & 0\\
-1 & 0 & 1\\
0 & -1 & 0
\end{bmatrix},\quad
\widetilde{\sA}^1_3=\begin{bmatrix}
i & 0 & 0\\
0 & 0 & 0\\
0 & 0 & -i
\end{bmatrix},
\]
\[
\widetilde{\sB}^1_1=\frac{1}{\sqrt{2}}\begin{bmatrix}
0 & 1 & 0\\
1 & 0 & 1\\
0 & 1 & 0
\end{bmatrix},\quad
\widetilde{\sB}^1_2=\frac{1}{\sqrt{2}}\begin{bmatrix}
0 & i & 0\\
-i & 0 & i\\
0 & -i & 0
\end{bmatrix},\quad
\widetilde{\sB}^1_3=\begin{bmatrix}
-1 & 0 & 0\\
0 & 0 & 0\\
0 & 0 & 1
\end{bmatrix}.
\]
Taking into account the latter expressions and the operators
(\ref{Infhalf}), we obtain
\begin{gather}
T_{1\frac{1}{2}}(\fg)
\frac{\partial T^{-1}_{1\frac{1}{2}}(\fg)}{\partial\varphi}=\nonumber\\
\ar\begin{pmatrix}
\scr -i\cos\theta^c+\frac{i}{2}\cos\dot{\theta}^c &
\scr -\frac{1}{2}\sin\dot{\theta}^c &
\scr -\frac{1}{\sqrt{2}}\sin\theta^c & \scr 0 & \scr 0 & \scr 0 \\
\scr\frac{1}{2}\sin\dot{\theta}^c &
\scr -i\cos\theta^c-\frac{i}{2}\cos\dot{\theta}^c & \scr 0 &
\scr -\frac{1}{\sqrt{2}}\sin\theta^c & \scr 0 & \scr 0 \\
\scr \frac{1}{\sqrt{2}}\sin\theta^c & \scr 0 &
\scr \frac{i}{2}\cos\dot{\theta}^c &
\scr -\frac{1}{2}\sin\dot{\theta}^c &
\scr -\frac{1}{\sqrt{2}}\sin\theta^c & \scr 0 \\
\scr 0 & \scr \frac{1}{\sqrt{2}}\sin\theta^c &
\scr \frac{1}{2}\sin\dot{\theta}^c &
\scr -\frac{i}{2}\cos\dot{\theta}^c & \scr 0 &
\scr -\frac{1}{\sqrt{2}}\sin\theta^c \\
\scr 0 & \scr 0 & \scr \frac{1}{\sqrt{2}}\sin\theta^c & \scr 0 &
\scr i\cos\theta^c+\frac{i}{2}\cos\dot{\theta}^c &
\scr -\frac{1}{2}\sin\dot{\theta}^c \\
\scr 0 & \scr 0 & \scr 0 &
\scr \frac{1}{\sqrt{2}}\sin\theta^c &
\scr \frac{1}{2}\sin\dot{\theta}^c &
\scr i\cos\theta^c-\frac{i}{2}\cos\dot{\theta}^c
\end{pmatrix}=\nonumber\\
=-(\sA^1_3\otimes\boldsymbol{1}_2\cos\theta^c-
\boldsymbol{1}_3\otimes\widetilde{\sA}^{\frac{1}{2}}_3\cos\dot{\theta}^c)-
(\sA^1_2\otimes\boldsymbol{1}_2\sin\theta^c+
\boldsymbol{1}_3\otimes\widetilde{\sA}^{\frac{1}{2}}_2\sin\dot{\theta}^c),
\label{Irel9}
\end{gather}
\begin{gather}
T_{1\frac{1}{2}}(\fg)
\frac{\partial T^{-1}_{1\frac{1}{2}}(\fg)}{\partial\epsilon}=\nonumber\\
\ar\begin{pmatrix}
\scr -\cos\theta^c-\frac{1}{2}\cos\dot{\theta}^c &
\scr -\frac{i}{2}\sin\dot{\theta}^c &
\scr -\frac{i}{\sqrt{2}}\sin\theta^c & \scr 0 & \scr 0 & \scr 0 \\
\scr\frac{i}{2}\sin\dot{\theta}^c &
\scr -\cos\theta^c+\frac{1}{2}\cos\dot{\theta}^c & \scr 0 &
\scr \frac{i}{\sqrt{2}}\sin\theta^c & \scr 0 & \scr 0 \\
\scr -\frac{i}{\sqrt{2}}\sin\theta^c & \scr 0 &
\scr -\frac{1}{2}\cos\dot{\theta}^c &
\scr -\frac{i}{2}\sin\dot{\theta}^c &
\scr \frac{i}{\sqrt{2}}\sin\theta^c & \scr 0 \\
\scr 0 & \scr -\frac{i}{\sqrt{2}}\sin\theta^c &
\scr \frac{i}{2}\sin\dot{\theta}^c &
\scr \frac{1}{2}\cos\dot{\theta}^c & \scr 0 &
\scr \frac{i}{\sqrt{2}}\sin\theta^c \\
\scr 0 & \scr 0 & \scr -\frac{i}{\sqrt{2}}\sin\theta^c & \scr 0 &
\scr \cos\theta^c-\frac{1}{2}\cos\dot{\theta}^c &
\scr -\frac{i}{2}\sin\dot{\theta}^c \\
\scr 0 & \scr 0 & \scr 0 &
\scr -\frac{i}{\sqrt{2}}\sin\theta^c &
\scr \frac{i}{2}\sin\dot{\theta}^c &
\scr \cos\theta^c+\frac{1}{2}\cos\dot{\theta}^c
\end{pmatrix}=\nonumber\\
=-(\sB^1_3\otimes\boldsymbol{1}_2\cos\theta^c-
\boldsymbol{1}_3\otimes\widetilde{\sB}^{\frac{1}{2}}_3\cos\dot{\theta}^c)-
(\sB^1_2\otimes\boldsymbol{1}_2\sin\theta^c+
\boldsymbol{1}_3\otimes\widetilde{\sB}^{\frac{1}{2}}_2\sin\dot{\theta}^c),
\label{Irel10}
\end{gather}
\begin{equation}\label{Irel11}
T_{1\frac{1}{2}}(\fg)
\frac{\partial T^{-1}_{1\frac{1}{2}}(\fg)}{\partial\theta}=
\ar\begin{pmatrix}
\scr 0 & \scr \frac{i}{2} & \scr -\frac{i}{\sqrt{2}} & \scr 0 &
\scr 0 & \scr 0\\
\scr \frac{i}{2} & \scr 0 & \scr 0 & \scr -\frac{i}{\sqrt{2}} &
\scr 0 & \scr 0\\
\scr -\frac{i}{\sqrt{2}} & \scr 0 & \scr 0 & \scr \frac{i}{2} &
\scr -\frac{i}{\sqrt{2}} & \scr 0\\
\scr 0 & \scr -\frac{i}{\sqrt{2}} & \scr \frac{i}{2} & \scr 0 & \scr 0 &
\scr -\frac{i}{\sqrt{2}}\\
\scr 0 & \scr 0 & \scr -\frac{i}{\sqrt{2}} & \scr 0 & \scr 0 &
\scr \frac{i}{2}\\
\scr 0 & \scr 0 & \scr 0 & \scr -\frac{i}{\sqrt{2}} &
\scr \frac{i}{2} & \scr 0
\end{pmatrix}=
\sA^1_1\otimes\boldsymbol{1}_2-
\boldsymbol{1}_3\otimes\widetilde{\sA}^{\frac{1}{2}}_1,
\end{equation}
\begin{equation}\label{Irel12}
T_{1\frac{1}{2}}(\fg)
\frac{\partial T^{-1}_{1\frac{1}{2}}(\fg)}{\partial\tau}=
\ar\begin{pmatrix}
\scr 0 & \scr -\frac{1}{2} & \scr -\frac{1}{\sqrt{2}} & \scr 0 &
\scr 0 & \scr 0\\
\scr -\frac{1}{2} & \scr 0 & \scr 0 & \scr -\frac{1}{\sqrt{2}} &
\scr 0 & \scr 0\\
\scr -\frac{1}{\sqrt{2}} & \scr 0 & \scr 0 & \scr -\frac{1}{2} &
\scr -\frac{1}{\sqrt{2}} & \scr 0\\
\scr 0 & \scr -\frac{1}{\sqrt{2}} & \scr -\frac{1}{2} & \scr 0 & \scr 0 &
\scr -\frac{1}{\sqrt{2}}\\
\scr 0 & \scr 0 & \scr -\frac{1}{\sqrt{2}} & \scr 0 & \scr 0 &
\scr -\frac{1}{2}\\
\scr 0 & \scr 0 & \scr 0 & \scr -\frac{1}{\sqrt{2}} &
\scr -\frac{1}{2} & \scr 0
\end{pmatrix}
=\sB^1_1\otimes\boldsymbol{1}_2-
\boldsymbol{1}_3\otimes\widetilde{\sB}^{\frac{1}{2}}_1.
\end{equation}
Further, the conjugate representation
$\overset{\ast}{\boldsymbol{\tau}}_{1,\frac{1}{2}}$ acts in the space
$\Sym(1,2)$ and we have the following relations:
\begin{gather}
\overset{\ast}{T}_{1\frac{1}{2}}(\fg)
\frac{\partial
\overset{\ast}{T}{}^{-1}_{1\frac{1}{2}}(\fg)}{\partial\varphi}=\nonumber\\
\ar\begin{pmatrix}
\scr i\cos\dot{\theta}^c-\frac{i}{2}\cos\theta^c &
\scr -\frac{1}{2}\sin\theta^c &
\scr -\frac{1}{\sqrt{2}}\sin\dot{\theta}^c & \scr 0 & \scr 0 & \scr 0 \\
\scr\frac{1}{2}\sin\theta^c &
\scr i\cos\dot{\theta}^c+\frac{i}{2}\cos\theta^c & \scr 0 &
\scr -\frac{1}{\sqrt{2}}\sin\dot{\theta}^c & \scr 0 & \scr 0 \\
\scr \frac{1}{\sqrt{2}}\sin\dot{\theta}^c & \scr 0 &
\scr -\frac{i}{2}\cos\theta^c &
\scr -\frac{1}{2}\sin\theta^c &
\scr -\frac{1}{\sqrt{2}}\sin\dot{\theta}^c & \scr 0 \\
\scr 0 & \scr \frac{1}{\sqrt{2}}\sin\dot{\theta}^c &
\scr \frac{1}{2}\sin\theta^c &
\scr \frac{i}{2}\cos\theta^c & \scr 0 &
\scr -\frac{1}{\sqrt{2}}\sin\dot{\theta}^c \\
\scr 0 & \scr 0 & \scr \frac{1}{\sqrt{2}}\sin\dot{\theta}^c & \scr 0 &
\scr -i\cos\dot{\theta}^c-\frac{i}{2}\cos\theta^c &
\scr -\frac{1}{2}\sin\theta^c \\
\scr 0 & \scr 0 & \scr 0 &
\scr \frac{1}{\sqrt{2}}\sin\dot{\theta}^c &
\scr \frac{1}{2}\sin\theta^c &
\scr -i\cos\dot{\theta}^c+\frac{i}{2}\cos\theta^c
\end{pmatrix}=\nonumber\\
=(\widetilde{\sA}^1_3\otimes\boldsymbol{1}_2\cos\dot{\theta}^c-
\boldsymbol{1}_3\otimes\sA^{\frac{1}{2}}_3\cos\theta^c)-
(\widetilde{\sA}^1_2\otimes\boldsymbol{1}_2\sin\dot{\theta}^c+
\boldsymbol{1}_3\otimes\sA^{\frac{1}{2}}_2\sin\theta^c),
\label{Irel13}
\end{gather}
\begin{gather}
\overset{\ast}{T}_{1\frac{1}{2}}(\fg)
\frac{\partial
\overset{\ast}{T}{}^{-1}_{1\frac{1}{2}}(\fg)}{\partial\epsilon}=\nonumber\\
\ar\begin{pmatrix}
\scr -\cos\dot{\theta}^c-\frac{1}{2}\cos\theta^c &
\scr \frac{i}{2}\sin\theta^c &
\scr -\frac{i}{\sqrt{2}}\sin\dot{\theta}^c & \scr 0 & \scr 0 & \scr 0 \\
\scr-\frac{i}{2}\sin\theta^c &
\scr -\cos\dot{\theta}^c+\frac{1}{2}\cos\theta^c & \scr 0 &
\scr -\frac{i}{\sqrt{2}}\sin\dot{\theta}^c & \scr 0 & \scr 0 \\
\scr \frac{i}{\sqrt{2}}\sin\dot{\theta}^c & \scr 0 &
\scr -\frac{1}{2}\cos\theta^c &
\scr \frac{i}{2}\sin\theta^c &
\scr -\frac{i}{\sqrt{2}}\sin\dot{\theta}^c & \scr 0 \\
\scr 0 & \scr \frac{i}{\sqrt{2}}\sin\dot{\theta}^c &
\scr -\frac{i}{2}\sin\theta^c &
\scr \frac{1}{2}\cos\theta^c & \scr 0 &
\scr -\frac{i}{\sqrt{2}}\sin\dot{\theta}^c \\
\scr 0 & \scr 0 & \scr \frac{i}{\sqrt{2}}\sin\dot{\theta}^c & \scr 0 &
\scr \cos\dot{\theta}^c-\frac{1}{2}\cos\theta^c &
\scr \frac{i}{2}\sin\theta^c \\
\scr 0 & \scr 0 & \scr 0 &
\scr \frac{i}{\sqrt{2}}\sin\dot{\theta}^c &
\scr -\frac{i}{2}\sin\theta^c &
\scr \cos\dot{\theta}^c+\frac{1}{2}\cos\theta^c
\end{pmatrix}=\nonumber\\
=(\widetilde{\sB}^1_3\otimes\boldsymbol{1}_2\cos\dot{\theta}^c-
\boldsymbol{1}_3\otimes\sB^{\frac{1}{2}}_3\cos\theta^c)-
(\widetilde{\sB}^1_2\otimes\boldsymbol{1}_2\sin\dot{\theta}^c+
\boldsymbol{1}_3\otimes\sB^{\frac{1}{2}}_2\sin\theta^c),
\label{Irel14}
\end{gather}
\begin{equation}\label{Irel15}
\overset{\ast}{T}_{1\frac{1}{2}}(\fg)
\frac{\partial
\overset{\ast}{T}{}^{-1}_{1\frac{1}{2}}(\fg)}{\partial\theta}=
\ar\begin{pmatrix}
\scr 0 & \scr -\frac{i}{2} & \scr \frac{i}{\sqrt{2}} & \scr 0 &
\scr 0 & \scr 0\\
\scr -\frac{i}{2} & \scr 0 & \scr 0 & \scr \frac{i}{\sqrt{2}} &
\scr 0 & \scr 0\\
\scr \frac{i}{\sqrt{2}} & \scr 0 & \scr 0 & \scr -\frac{i}{2} &
\scr \frac{i}{\sqrt{2}} & \scr 0\\
\scr 0 & \scr \frac{i}{\sqrt{2}} & \scr -\frac{i}{2} & \scr 0 & \scr 0 &
\scr \frac{i}{\sqrt{2}}\\
\scr 0 & \scr 0 & \scr \frac{i}{\sqrt{2}} & \scr 0 & \scr 0 &
\scr -\frac{i}{2}\\
\scr 0 & \scr 0 & \scr 0 & \scr \frac{i}{\sqrt{2}} &
\scr -\frac{i}{2} & \scr 0
\end{pmatrix}
=-(\widetilde{\sA}^1_1\otimes\boldsymbol{1}_2-
\boldsymbol{1}_3\otimes\sA^{\frac{1}{2}}_1),
\end{equation}
\begin{equation}\label{Irel16}
\overset{\ast}{T}_{1\frac{1}{2}}(\fg)
\frac{\partial
\overset{\ast}{T}{}^{-1}_{1\frac{1}{2}}(\fg)}{\partial\tau}=
\ar\begin{pmatrix}
\scr 0 & \scr -\frac{1}{2} & \scr -\frac{1}{\sqrt{2}} & \scr 0 &
\scr 0 & \scr 0\\
\scr -\frac{1}{2} & \scr 0 & \scr 0 & \scr -\frac{1}{\sqrt{2}} &
\scr 0 & \scr 0\\
\scr -\frac{1}{\sqrt{2}} & \scr 0 & \scr 0 & \scr -\frac{1}{2} &
\scr -\frac{1}{\sqrt{2}} & \scr 0\\
\scr 0 & \scr -\frac{1}{\sqrt{2}} & \scr -\frac{1}{2} & \scr 0 & \scr 0 &
\scr -\frac{1}{\sqrt{2}}\\
\scr 0 & \scr 0 & \scr -\frac{1}{\sqrt{2}} & \scr 0 & \scr 0 &
\scr -\frac{1}{2}\\
\scr 0 & \scr 0 & \scr 0 & \scr -\frac{1}{\sqrt{2}} &
\scr -\frac{1}{2} & \scr 0
\end{pmatrix}
=-(\widetilde{\sB}^1_1\otimes\boldsymbol{1}_2-
\boldsymbol{1}_3\otimes\sB^{\frac{1}{2}}_1.
\end{equation}
It is easy to verify that relations of the type (\ref{Irel1})--(\ref{Irel16})
take place for any representation $T_{l\dot{l}}(\fg)$ of the group $\fG_+$.
Therefore,
\begin{multline}
T_{l\dot{l}}(\fg)\frac{\partial T^{-1}_{l\dot{l}}(\fg)}{\partial\varphi}=
-(\sA^l_3\otimes\boldsymbol{1}_{2\dot{l}+1}\cos\theta^c-
\boldsymbol{1}_{2l+1}\otimes\sA^{\dot{l}}_3\cos\dot{\theta}^c)-\\
-(\sA^l_2\otimes\boldsymbol{1}_{2\dot{l}+1}\sin\theta^c+
\boldsymbol{1}_{2l+1}\otimes\sA^{\dot{l}}_2\sin\dot{\theta}^c),
\nonumber
\end{multline}
\begin{multline}
T_{l\dot{l}}(\fg)\frac{\partial T^{-1}_{l\dot{l}}(\fg)}{\partial\epsilon}=
-(\sB^l_3\otimes\boldsymbol{1}_{2\dot{l}+1}\cos\theta^c-
\boldsymbol{1}_{2l+1}\otimes\sB^{\dot{l}}_3\cos\dot{\theta}^c)-\\
-(\sB^l_2\otimes\boldsymbol{1}_{2\dot{l}+1}\sin\theta^c+
\boldsymbol{1}_{2l+1}\otimes\sB^{\dot{l}}_2\sin\dot{\theta}^c),
\nonumber
\end{multline}
\[
T_{l\dot{l}}(\fg)\frac{\partial T^{-1}_{l\dot{l}}(\fg)}{\partial\theta}=
\sA^l_1\otimes\boldsymbol{1}_{2\dot{l}+1}-
\boldsymbol{1}_{2l+1}\otimes\sA^{\dot{l}}_1,
\]
\[
T_{l\dot{l}}(\fg)\frac{\partial T^{-1}_{l\dot{l}}(\fg)}{\partial\tau}=
\sB^l_1\otimes\boldsymbol{1}_{2\dot{l}+1}-
\boldsymbol{1}_{2l+1}\otimes\sB^{\dot{l}}_1,
\]
\begin{multline}
\overset{\ast}{T}_{l\dot{l}}(\fg)
\frac{\partial\overset{\ast}{T}{}^{-1}_{l\dot{l}}(\fg)}{\partial\varphi}=
(\sA^{\dot{l}}_3\otimes\boldsymbol{1}_{2l+1}\cos\dot{\theta}^c-
\boldsymbol{1}_{2\dot{l}+1}\otimes\sA^l_3\cos\theta^c)-\\
-(\sA^{\dot{l}}_2\otimes\boldsymbol{1}_{2l+1}\sin\dot{\theta}^c+
\boldsymbol{1}_{2\dot{l}+1}\otimes\sA^l_2\sin\theta^c),
\nonumber
\end{multline}
\begin{multline}
\overset{\ast}{T}_{l\dot{l}}(\fg)
\frac{\partial\overset{\ast}{T}{}^{-1}_{l\dot{l}}(\fg)}{\partial\epsilon}=
(\sB^{\dot{l}}_3\otimes\boldsymbol{1}_{2l+1}\cos\dot{\theta}^c-
\boldsymbol{1}_{2\dot{l}+1}\otimes\sB^l_3\cos\theta^c)-\\
-(\sB^{\dot{l}}_2\otimes\boldsymbol{1}_{2l+1}\sin\dot{\theta}^c+
\boldsymbol{1}_{2\dot{l}+1}\otimes\sB^l_2\sin\theta^c),
\nonumber
\end{multline}
\[
\overset{\ast}{T}_{l\dot{l}}(\fg)
\frac{\partial\overset{\ast}{T}{}^{-1}_{l\dot{l}}(\fg)}{\partial\theta}=
-(\sA^{\dot{l}}_1\otimes\boldsymbol{1}_{2l+1}-
\boldsymbol{1}_{2\dot{l}+1}\otimes\sA^l_1),
\]
\[
\overset{\ast}{T}_{l\dot{l}}(\fg)
\frac{\partial\overset{\ast}{T}{}^{-1}_{l\dot{l}}(\fg)}{\partial\tau}=
-(\sB^{\dot{l}}_1\otimes\boldsymbol{1}_{2l+1}-
\boldsymbol{1}_{2\dot{l}+1}\otimes\sB^l_1).
\]
Substituting these relations into the system (\ref{Complex4}), we obtain
\begin{multline}
\frac{1}{r\sin\theta^c}\sL^l_1\otimes\boldsymbol{1}_{2\dot{l}+1}
\frac{\partial\boldsymbol{\psi}^\prime}
{\partial\varphi}-
\frac{1}{r\sin\dot{\theta}^c}\boldsymbol{1}_{2l+1}\otimes\sL^{\dot{l}}_1
\frac{\partial\boldsymbol{\psi}^\prime}{\partial\varphi}+
\frac{i}{r\sin\theta^c}\sL^l_1\otimes\boldsymbol{1}_{2\dot{l}+1}
\frac{\partial\boldsymbol{\psi}^\prime}
{\partial\epsilon}+\\
+\frac{i}{r\sin\dot{\theta}^c}\boldsymbol{1}_{2l+1}\otimes\sL^{\dot{l}}_1
\frac{\partial\boldsymbol{\psi}^\prime}
{\partial\epsilon}-
\frac{1}{r}\sL^l_2\otimes\boldsymbol{1}_{2\dot{l}+1}
\frac{\partial\boldsymbol{\psi}^\prime}{\partial\theta}+
\frac{1}{r}\boldsymbol{1}_{2l+1}\otimes\sL^{\dot{l}}_2
\frac{\partial\boldsymbol{\psi}^\prime}{\partial\theta}-\\
-\frac{i}{r}\sL^l_2\otimes\boldsymbol{1}_{2\dot{l}+1}
\frac{\partial\boldsymbol{\psi}^\prime}{\partial\tau}
-\frac{i}{r}\boldsymbol{1}_{2l+1}\otimes\sL^{\dot{l}}_2
\frac{\partial\boldsymbol{\psi}^\prime}{\partial\tau}
+(1+i)\sL^{l\dot{l}}_3\frac{\partial\boldsymbol{\psi}^\prime}{\partial r}+\\
+\frac{1}{r}\left[\sL^{l\dot{l}}_1\sA^{l\dot{l}}_2-
\sL^{l\dot{l}}_2\sA^{l\dot{l}}_1+\sL^{l\dot{l}}_1\sB^{l\dot{l}}_2-
\sL^{l\dot{l}}_2\sB^{l\dot{l}}_1-
2\ctg\theta^c\sL^l_1\otimes\boldsymbol{1}_{2\dot{l}+1}\sA^l_3\otimes
\boldsymbol{1}_{2\dot{l}+1}-\right.
\\
\left.-2\ctg\dot{\theta}^c\boldsymbol{1}_{2l+1}\otimes
\sL^{\dot{l}}_1\boldsymbol{1}_{2l+1}\otimes\sA^{\dot{l}}_3
\right]
\boldsymbol{\psi}^\prime+
\kappa^c\boldsymbol{\psi}=0,
\nonumber
\end{multline}
\begin{multline}
\frac{1}{r^\ast\sin\theta^c}\overset{\ast}{\sL}{}^{\dot{l}}_1
\otimes\boldsymbol{1}_{2l+1}
\frac{\partial\dot{\boldsymbol{\psi}}^\prime}
{\partial\varphi}-
\frac{1}{r^\ast\sin\dot{\theta}^c}
\boldsymbol{1}_{2\dot{l}+1}\otimes\overset{\ast}{\sL}{}^l_1
\frac{\partial\dot{\boldsymbol{\psi}}^\prime}
{\partial\varphi}+
\frac{i}{r^\ast\sin\theta^c}\overset{\ast}{\sL}{}^{\dot{l}}_1
\otimes\boldsymbol{1}_{2l+1}
\frac{\partial\dot{\boldsymbol{\psi}}^\prime}{\partial\epsilon}+\\
+\frac{i}{r^\ast\sin\dot{\theta}^c}
\boldsymbol{1}_{2\dot{l}+1}\otimes\overset{\ast}{\sL}{}^l_1
\frac{\partial\dot{\boldsymbol{\psi}}^\prime}{\partial\epsilon}-
\frac{1}{r^\ast}\overset{\ast}{\sL}{}^{\dot{l}}_2
\otimes\boldsymbol{1}_{2l+1}
\frac{\partial\dot{\boldsymbol{\psi}}^\prime}{\partial\theta}+
\frac{1}{r^\ast}
\boldsymbol{1}_{2\dot{l}+1}\otimes\overset{\ast}{\sL}{}^l_2
\frac{\partial\dot{\boldsymbol{\psi}}^\prime}{\partial\theta}-\\
-\frac{i}{r^\ast}\overset{\ast}{\sL}{}^{\dot{l}}_2
\otimes\boldsymbol{1}_{2l+1}
\frac{\partial\dot{\boldsymbol{\psi}}^\prime}{\partial\tau}-
\frac{i}{r^\ast}
\boldsymbol{1}_{2\dot{l}+1}\otimes\overset{\ast}{\sL}{}^l_2
\frac{\partial\dot{\boldsymbol{\psi}}^\prime}{\partial\tau}+
(1-i)\overset{\ast}{\sL}{}^{l\dot{l}}_3
\frac{\partial\dot{\boldsymbol{\psi}}^\prime}{\partial r^\ast}+\\
+\frac{1}{r^\ast}\left[\overset{\ast}{\sL}{}^{l\dot{l}}_1
\widetilde{\sA}^{l\dot{l}}_2-
\overset{\ast}{\sL}{}^{l\dot{l}}_2\widetilde{\sA}^{l\dot{l}}_1+
\overset{\ast}{\sL}{}^{l\dot{l}}_1\widetilde{\sB}^{l\dot{l}}_2-
\overset{\ast}{\sL}{}^{l\dot{l}}_2\widetilde{\sB}^{l\dot{l}}_1-
2\ctg\theta^c\overset{\ast}{\sL}{}^{\dot{l}}_1\otimes
\boldsymbol{1}_{2l+1}\sA^{\dot{l}}_3\otimes
\boldsymbol{1}_{2l+1}-
\right.
\\
\left.-
2\ctg\dot{\theta}^c\boldsymbol{1}_{2\dot{l}+1}\otimes
\overset{\ast}{\sL}{}^l_1\boldsymbol{1}_{2\dot{l}+1}\otimes\sA^l_3
\right]
\dot{\boldsymbol{\psi}}^\prime+
\dot{\kappa}^c\dot{\boldsymbol{\psi}}=0.
\label{Complex5}
\end{multline}
Now we can separate the variables in the
relativistically invariant system. Namely, we represent the each component
$\psi^{k\dot{k}}_{lm;\dot{l}\dot{m}}$
of the wave function $\boldsymbol{\psi}$ in the form of an expansion
in the generalized hyperspherical
functions
$\fM^{l\dot{l}}_{mn;\dot{m}\dot{n}}$.\index{function!hypersperical!generalized}
This procedure gives rise
to separation of variables, that is, it reduces the relativistically
invariant system\index{relativistically invariant system}
to the system of ordinary differential equations.
Preliminarily, we will calculate elements of the matrices
$D=\sL^{l\dot{l}}_1\sA^{l\dot{l}}_2-\sL^{l\dot{l}}_2\sA^{l\dot{l}}_1$,
$E=\sL^{l\dot{l}}_1\sB^{l\dot{l}}_2-\sL^{l\dot{l}}_2\sB^{l\dot{l}}_1$,
$\overset{\ast}{D}=\overset{\ast}{\sL}{}^{l\dot{l}}_1
\widetilde{\sA}^{l\dot{l}}_2-
\overset{\ast}{\sL}{}^{l\dot{l}}_2\widetilde{\sA}^{l\dot{l}}_1$,
$\overset{\ast}{E}=\overset{\ast}{\sL}{}^{l\dot{l}}_1
\widetilde{\sB}^{l\dot{l}}_2-
\overset{\ast}{\sL}{}^{l\dot{l}}_2\widetilde{\sB}^{l\dot{l}}_1$. First of all,
let us find elements of the matrix
$D=\sL^{l\dot{l}}_1\sA^{l\dot{l}}_2-
\sL^{l\dot{l}}_2\sA^{l\dot{l}}_1$. Using the relations (\ref{AL}), we can write
$D=2\sL^{l\dot{l}}_3+\sA^{l\dot{l}}_2\sL^{l\dot{l}}_1-\sA^{l\dot{l}}_1
\sL^{l\dot{l}}_2$. As usual,
the action of the transformation $D$ in the
helicity basis has the following form:
\[
D\zeta^{k\dot{k}}_{lm;\dot{l}\dot{m}}=
\sum_{l^\prime,m^\prime,\dot{l}^\prime,\dot{m}^\prime,k^\prime,\dot{k}^\prime}
d^{k^\prime k;\dot{k}^\prime\dot{k}}_{l^\prime l,m^\prime m;\dot{l}^\prime
\dot{l},\dot{m}^\prime\dot{m}}
\zeta^{k^\prime\dot{k}^\prime}_{l^\prime,m^\prime;\dot{l}^\prime\dot{m}^\prime}.
\]
Taking into account (\ref{L3}), (\ref{L1}), (\ref{L2}) and
(\ref{Op1})--(\ref{Op2}), we obtain
\begin{multline}
D\zeta^{k\dot{k}}_{lm;\dot{l}\dot{m}}=(
2\sL^{l\dot{l}}_3+\sA^{l\dot{l}}_2\sL^{l\dot{l}}_1-\sA^{l\dot{l}}_1
\sL^{l\dot{l}}_2)
\zeta^{k\dot{k}}_{lm;\dot{l}\dot{m}}
=2\sum_{l^\prime,m^\prime,\dot{l}^\prime,\dot{m}^\prime,k^\prime,\dot{k}^\prime}
c^{k^\prime k;\dot{k}^\prime\dot{k}}_{l^\prime l,m^\prime m;
\dot{l}^\prime\dot{l},\dot{m}^\prime\dot{m}}
\zeta^{k^\prime\dot{k}^\prime}_{l^\prime m^\prime;\dot{l}^\prime
\dot{m}^\prime}+\\
+\sA^{l\dot{l}}_2
\sum_{l^\prime,m^\prime,\dot{l}^\prime,\dot{m}^\prime,k^\prime,\dot{k}^\prime}
a^{k^\prime k;\dot{k}^\prime\dot{k}}_{l^\prime l,m^\prime m;
\dot{l}^\prime\dot{l},\dot{m}^\prime\dot{m}}
\zeta^{k^\prime\dot{k}^\prime}_{l^\prime m^\prime;
\dot{l}^\prime\dot{m}^\prime}-
\sA^{l\dot{l}}_1
\sum_{l^\prime,m^\prime,\dot{l}^\prime,\dot{m}^\prime,k^\prime,\dot{k}^\prime}
b^{k^\prime k;\dot{k}^\prime\dot{k}}_{l^\prime l,m^\prime m;
\dot{l}^\prime\dot{l},\dot{m}^\prime\dot{m}}
\zeta^{k^\prime\dot{k}^\prime}_{l^\prime m^\prime;
\dot{l}^\prime\dot{m}^\prime}=\\
=2\sum_{l^\prime,m^\prime,\dot{l}^\prime,\dot{m}^\prime,k^\prime,\dot{k}^\prime}
c^{k^\prime k;\dot{k}^\prime\dot{k}}_{l^\prime l,m^\prime m;
\dot{l}^\prime\dot{l},\dot{m}^\prime\dot{m}}
\zeta^{k^\prime\dot{k}^\prime}_{l^\prime m^\prime;
\dot{l}^\prime\dot{m}^\prime}+
\frac{1}{2}
\sum_{l^\prime,m^\prime,\dot{l}^\prime,\dot{m}^\prime,k^\prime,\dot{k}^\prime}
a^{k^\prime k;\dot{k}^\prime\dot{k}}_{l^\prime l,m^\prime m;
\dot{l}^\prime\dot{l},\dot{m}^\prime\dot{m}}
\left(\boldsymbol{\alpha}^{l^\prime}_{m^\prime}
\zeta^{k^\prime\dot{k}^\prime}_{l^\prime,m^\prime-1;
\dot{l}^\prime\dot{m}^\prime}-\right.\\
-\boldsymbol{\alpha}^{l^\prime}_{m^\prime+1}
\zeta^{k^\prime\dot{k}^\prime}_{l^\prime,m^\prime+1;
\dot{l}^\prime\dot{m}^\prime}
\left.-\boldsymbol{\alpha}^{\dot{l}^\prime}_{\dot{m}^\prime}
\zeta^{k^\prime\dot{k}^\prime}_{l^\prime m^\prime;\dot{l}^\prime,
\dot{m}^\prime-1}+
\boldsymbol{\alpha}^{\dot{l}^\prime}_{\dot{m}^\prime+1}
\zeta^{k^\prime\dot{k}^\prime}_{l^\prime m^\prime;\dot{l}^\prime,
\dot{m}^\prime+1}\right)-\\
-\frac{i}{2}
\sum_{l^\prime,m^\prime,\dot{l}^\prime,\dot{m}^\prime,k^\prime,\dot{k}^\prime}
b^{k^\prime k;\dot{k}^\prime\dot{k}}_{l^\prime l,m^\prime m;
\dot{l}^\prime\dot{l},\dot{m}^\prime\dot{m}}
\left(-\boldsymbol{\alpha}^{l^\prime}_{m^\prime}
\zeta^{k^\prime\dot{k}^\prime}_{l^\prime,m^\prime-1;
\dot{l}^\prime\dot{m}^\prime}-
\boldsymbol{\alpha}^{l^\prime}_{m^\prime+1}
\zeta^{k^\prime\dot{k}^\prime}_{l^\prime,m^\prime+1;
\dot{l}^\prime\dot{m}^\prime}\right.+\\
\left.+\boldsymbol{\alpha}^{\dot{l}^\prime}_{\dot{m}^\prime}
\zeta^{k^\prime\dot{k}^\prime}_{l^\prime m^\prime;\dot{l}^\prime,
\dot{m}^\prime-1}+
\boldsymbol{\alpha}^{\dot{l}^\prime}_{\dot{m}^\prime+1}
\zeta^{k^\prime\dot{k}^\prime}_{l^\prime m^\prime;\dot{l}^\prime,
\dot{m}^\prime+1}\right).
\nonumber
\end{multline}
Dividing the each of the two latter sums on the four and changing the
summation index in the each eight sums obtained, we come to the following
expression:
\begin{multline}
D\zeta^{k\dot{k}}_{lm;\dot{l}\dot{m}}=
\sum_{l^\prime,m^\prime,\dot{l}^\prime,\dot{m}^\prime,k^\prime,\dot{k}^\prime}
\left(
2c^{k^\prime k;\dot{k}^\prime\dot{k}}_{l^\prime l,m^\prime m;
\dot{l}^\prime\dot{l},\dot{m}^\prime\dot{m}}+
\frac{1}{2}\boldsymbol{\alpha}^{l^\prime}_{m^\prime+1}
a^{k^\prime k;\dot{k}^\prime\dot{k}}_{l^\prime l,m^\prime+1,m;
\dot{l}^\prime\dot{l},\dot{m}^\prime\dot{m}}-\right.\\
\left.-\frac{1}{2}\boldsymbol{\alpha}^{l^\prime}_{m^\prime}
a^{k^\prime k;\dot{k}^\prime\dot{k}}_{l^\prime l,m^\prime-1,m;
\dot{l}^\prime\dot{l},\dot{m}^\prime\dot{m}}-
\frac{1}{2}\boldsymbol{\alpha}^{\dot{l}^\prime}_{\dot{m}^\prime+1}
a^{k^\prime k;\dot{k}^\prime\dot{k}}_{l^\prime l,m^\prime m;
\dot{m}^\prime+1,\dot{m}}+\right.\\
+\frac{1}{2}\boldsymbol{\alpha}^{\dot{l}^\prime}_{\dot{m}^\prime}
a^{k^\prime k;\dot{k}^\prime\dot{k}}_{l^\prime l,m^\prime m;
\dot{l}^\prime\dot{l},\dot{m}^\prime-1,\dot{m}}+
\frac{i}{2}\boldsymbol{\alpha}^{l^\prime}_{m^\prime+1}
b^{k^\prime k;\dot{k}^\prime\dot{k}}_{l^\prime l,m^\prime+1,m;
\dot{l}^\prime\dot{l},\dot{m}^\prime\dot{m}}+\\
+\frac{i}{2}\boldsymbol{\alpha}^{l^\prime}_{m^\prime}
b^{k^\prime k;\dot{k}^\prime\dot{k}}_{l^\prime l,m^\prime-1,m;
\dot{l}^\prime\dot{l},\dot{m}^\prime\dot{m}}-
\frac{i}{2}\boldsymbol{\alpha}^{\dot{l}^\prime}_{\dot{m}^\prime+1}
b^{k^\prime k;\dot{k}^\prime\dot{k}}_{l^\prime l,m^\prime m;
\dot{l}^\prime\dot{l},\dot{m}^\prime+1,\dot{m}}-\\
\left.-\frac{i}{2}\boldsymbol{\alpha}^{\dot{l}^\prime}_{\dot{m}^\prime}
b^{k^\prime k;\dot{k}^\prime\dot{k}}_{l^\prime l,m^\prime m;
\dot{l}^\prime\dot{l},\dot{m}^\prime-1,\dot{m}}\right)
\zeta^{k^\prime\dot{k}^\prime}_{l^\prime m^\prime;
\dot{l}^\prime\dot{m}^\prime}.\nonumber
\end{multline}
Therefore, a general element of the matrix $D$ has the form
\begin{multline}
d^{k^\prime k;\dot{k}^\prime\dot{k}}_{l^\prime l,m^\prime m;
\dot{l}^\prime\dot{l},\dot{m}^\prime\dot{m}}=
2c^{k^\prime k;\dot{k}^\prime\dot{k}}_{l^\prime l,m^\prime m;
\dot{l}^\prime\dot{l},\dot{m}^\prime\dot{m}}+
\frac{1}{2}\boldsymbol{\alpha}^{l^\prime}_{m^\prime+1}
a^{k^\prime k;\dot{k}^\prime\dot{k}}_{l^\prime l,m^\prime+1,m;
\dot{l}^\prime\dot{l},\dot{m}^\prime\dot{m}}-\\
-\frac{1}{2}\boldsymbol{\alpha}^{l^\prime}_{m^\prime}
a^{k^\prime k;\dot{k}^\prime\dot{k}}_{l^\prime l,m^\prime-1,m;
\dot{l}^\prime\dot{l},\dot{m}^\prime\dot{m}}-
\frac{1}{2}\boldsymbol{\alpha}^{\dot{l}^\prime}_{\dot{m}^\prime+1}
a^{k^\prime k;\dot{k}^\prime\dot{k}}_{l^\prime l,m^\prime m;
\dot{m}^\prime+1,\dot{m}}+\\
+\frac{1}{2}\boldsymbol{\alpha}^{\dot{l}^\prime}_{\dot{m}^\prime}
a^{k^\prime k;\dot{k}^\prime\dot{k}}_{l^\prime l,m^\prime m;
\dot{l}^\prime\dot{l},\dot{m}^\prime-1,\dot{m}}+
\frac{i}{2}\boldsymbol{\alpha}^{l^\prime}_{m^\prime+1}
b^{k^\prime k;\dot{k}^\prime\dot{k}}_{l^\prime l,m^\prime+1,m;
\dot{l}^\prime\dot{l},\dot{m}^\prime\dot{m}}+\\
+\frac{i}{2}\boldsymbol{\alpha}^{l^\prime}_{m^\prime}
b^{k^\prime k;\dot{k}^\prime\dot{k}}_{l^\prime l,m^\prime-1,m;
\dot{l}^\prime\dot{l},\dot{m}^\prime\dot{m}}-
\frac{i}{2}\boldsymbol{\alpha}^{\dot{l}^\prime}_{\dot{m}^\prime+1}
b^{k^\prime k;\dot{k}^\prime\dot{k}}_{l^\prime l,m^\prime m;
\dot{l}^\prime\dot{l},\dot{m}^\prime+1,\dot{m}}-
\frac{i}{2}\boldsymbol{\alpha}^{\dot{l}^\prime}_{\dot{m}^\prime}
b^{k^\prime k;\dot{k}^\prime\dot{k}}_{l^\prime l,m^\prime m;
\dot{l}^\prime\dot{l},\dot{m}^\prime-1,\dot{m}}
\nonumber
\end{multline}
Using the formulae (\ref{L3}), (\ref{L1}) and (\ref{L2}), we find that
\begin{equation}\label{VM}
D:\quad
{\renewcommand{\arraystretch}{1.3}
\left\{\begin{array}{ccc}
d^{k^\prime k;\dot{k}^\prime\dot{k}}_{l-1,l,mm;\dot{l}-1,\dot{l},
\dot{m}\dot{m}}&=&
c^{k^\prime k;\dot{k}^\prime\dot{k}}_{l-1,l;\dot{l}-1,\dot{l}}
(l+\dot{l})\sqrt{(l^2-m^2)(\dot{l}^2-\dot{m}^2)},\\
d^{k^\prime k;\dot{k}^\prime\dot{k}}_{l+1,l,mm;\dot{l}+1,\dot{l},
\dot{m}\dot{m}}&=&
-c^{k^\prime k;\dot{k}^\prime\dot{k}}_{l+1,l;\dot{l}+1,\dot{l}}
(l+\dot{l}+2)\sqrt{((l+1)^2-m^2)((\dot{l}+1)^2-\dot{m}^2)}.
\end{array}\right.}
\end{equation}
All other elements
$d^{k^\prime k;\dot{k}^\prime\dot{k}}_{l^\prime l,m^\prime m;
\dot{l}^\prime\dot{l},\dot{m}^\prime\dot{m}}$ are
equal to zero.

Analogously, using the relations (\ref{BL}), (\ref{DAL})--(\ref{DBL})
and the operators
(\ref{Op1})--(\ref{Op12}), we find that elements of the matrices
$E=-2i\sL^{l\dot{l}}_3+\sB^{l\dot{l}}_2\sL^{l\dot{l}}_1-
\sB^{l\dot{l}}_1\sL^{l\dot{l}}_2$,
$\overset{\ast}{D}=2\overset{\ast}{\sL}{}^{l\dot{l}}_3+
\widetilde{\sA}^{l\dot{l}}_2\overset{\ast}{\sL}{}^{l\dot{l}}_1-
\widetilde{\sA}^{l\dot{l}}_1\overset{\ast}{\sL}{}^{l\dot{l}}_2$,
$\overset{\ast}{E}=2i\overset{\ast}{\sL}{}^{l\dot{l}}_3+
\widetilde{\sB}^{l\dot{l}}_2\overset{\ast}{\sL}{}^{l\dot{l}}_1-
\widetilde{\sB}^{l\dot{l}}_1\overset{\ast}{\sL}{}^{l\dot{l}}_2$
are
\begin{equation}\label{UM}
E:\quad
{\renewcommand{\arraystretch}{1.3}
\left\{\begin{array}{ccc}
e^{k^\prime k;\dot{k}^\prime\dot{k}}_{l-1,l,mm;
\dot{l}-1,\dot{l},\dot{m}\dot{m}}&=&
ic^{k^\prime k;\dot{k}^\prime\dot{k}}_{l-1,l;
\dot{l}-1,\dot{l}}(\dot{l}-l-2)
\sqrt{(l^2-m^2)(\dot{l}^2-\dot{m}^2)},\\
e^{k^\prime k;\dot{k}^\prime\dot{k}}_{ll,mm;
\dot{l}\dot{l},\dot{m}\dot{m}}&=&
-2ic^{k^\prime k;\dot{k}^\prime\dot{k}}_{ll;\dot{l}\dot{l}}m\dot{m},\\
e^{k^\prime k;\dot{k}^\prime\dot{k}}_{l+1,l,mm;
\dot{l}+1,\dot{l},\dot{m}\dot{m}}&=&
ic^{k^\prime k;\dot{k}^\prime\dot{k}}_{l+1,l;
\dot{l}+1,\dot{l}}(l-\dot{l}-2)
\sqrt{((l+1)^2-m^2)((\dot{l}+1)^2-\dot{m}^2)}.
\end{array}\right.}
\end{equation}
\begin{equation}\label{DM}
\overset{\ast}{D}:\quad
{\renewcommand{\arraystretch}{1.3}
\left\{\begin{array}{ccc}
\overset{\ast}{d}{}^{k^\prime k;\dot{k}^\prime\dot{k}}_{l-1,l,mm;
\dot{l}-1,\dot{l},\dot{m}\dot{m}}&=&
\overset{\ast}{c}{}^{k^\prime k;\dot{k}^\prime\dot{k}}_{l-1,l;
\dot{l}-1,\dot{l}}(l+\dot{l})
\sqrt{(l^2-m^2)(\dot{l}^2-\dot{m}^2)},\\
\overset{\ast}{d}{}^{k^\prime k;\dot{k}^\prime\dot{k}}_{l+1,l,mm;
\dot{l}+1,\dot{l},\dot{m}\dot{m}}&=&
-\overset{\ast}{c}{}^{k^\prime k;\dot{k}^\prime\dot{k}}_{l+1,l;
\dot{l}+1,\dot{l}}(l+\dot{l}+2)
\sqrt{((l+1)^2-m^2)((\dot{l}+1)^2-\dot{m}^2)}.
\end{array}\right.}
\end{equation}
\begin{equation}\label{EM}
\overset{\ast}{E}:\quad
{\renewcommand{\arraystretch}{1.3}
\left\{\begin{array}{ccc}
\overset{\ast}{e}{}^{k^\prime k;\dot{k}^\prime\dot{k}}_{l-1,l,mm;
\dot{l}-1,\dot{l},\dot{m}\dot{m}}&=&
i\overset{\ast}{c}{}^{k^\prime k;\dot{k}^\prime\dot{k}}_{l-1,l;
\dot{l}-1,\dot{l}}(\dot{l}-l+2)
\sqrt{(l^2-m^2)(\dot{l}^2-\dot{m}^2)},\\
\overset{\ast}{e}{}^{k^\prime k;\dot{k}^\prime\dot{k}}_{ll,mm;
\dot{l}\dot{l},\dot{m}\dot{m}}&=&
2i\overset{\ast}{c}{}^{k^\prime k;\dot{k}^\prime\dot{k}}_{ll;
\dot{l}\dot{l}}m\dot{m},\\
\overset{\ast}{e}{}^{k^\prime k;\dot{k}^\prime\dot{k}}_{l+1,l,mm;
\dot{l}+1,\dot{l},\dot{m}\dot{m}}&=&
i\overset{\ast}{c}{}^{k^\prime k;\dot{k}^\prime\dot{k}}_{l+1,l;
\dot{l}+1,\dot{l}}(l-\dot{l}+2)
\sqrt{((l+1)^2-m^2)((\dot{l}+1)^2-\dot{m}^2)}.
\end{array}\right.}
\end{equation}

The system (\ref{Complex5}) in the components
$\psi^{k\dot{k}}_{lm;\dot{l}\dot{m}}$ can be written
as
\begin{multline}
\sum_{l^\prime,m^\prime,\dot{l}^\prime,\dot{m}^\prime,k^\prime,\dot{k}^\prime}
\left[
\frac{1}{r\sin\theta^c}
a^{kk^\prime}_{ll^\prime,mm^\prime}\otimes\boldsymbol{1}_{2\dot{l}+1}
\frac{\partial\psi^{k^\prime\dot{k}^\prime}_{l^\prime m^\prime;
\dot{l}^\prime\dot{m}^\prime}}{\partial\varphi}-
\frac{1}{r\sin\dot{\theta}^c}\boldsymbol{1}_{2l+1}\otimes
a^{\dot{k}\dot{k}^\prime}_{\dot{l}\dot{l}^\prime,\dot{m}\dot{m}^\prime}
\frac{\partial\psi^{k^\prime\dot{k}^\prime}_{l^\prime m^\prime;
\dot{l}^\prime\dot{m}^\prime}}{\partial\varphi}+\right.\\
+\frac{i}{r\sin\theta^c}
a^{kk^\prime}_{ll^\prime,mm^\prime}\otimes\boldsymbol{1}_{2\dot{l}+1}
\frac{\partial\psi^{k^\prime\dot{k}^\prime}_{l^\prime m^\prime;
\dot{l}^\prime\dot{m}^\prime}}{\partial\epsilon}+
\frac{i}{r\sin\dot{\theta}^c}\boldsymbol{1}_{2l+1}\otimes
a^{\dot{k}\dot{k}^\prime}_{\dot{l}\dot{l}^\prime,\dot{m}\dot{m}^\prime}
\frac{\partial\psi^{k^\prime\dot{k}^\prime}_{l^\prime m^\prime;
\dot{l}^\prime\dot{m}^\prime}}{\partial\epsilon}-\\
-\frac{1}{r}
b^{kk^\prime}_{ll^\prime,mm^\prime}\otimes\boldsymbol{1}_{2\dot{l}+1}
\frac{\partial\psi^{k^\prime\dot{k}^\prime}_{l^\prime m^\prime;
\dot{l}^\prime\dot{m}^\prime}}{\partial\theta}+
\frac{1}{r}\boldsymbol{1}_{2l+1}\otimes
b^{\dot{k}\dot{k}^\prime}_{\dot{l}\dot{l}^\prime,\dot{m}\dot{m}^\prime}
\frac{\partial\psi^{k^\prime\dot{k}^\prime}_{l^\prime m^\prime;
\dot{l}^\prime\dot{m}^\prime}}{\partial\theta}-
\nonumber
\end{multline}
\begin{multline}
-\frac{i}{r}b^{kk^\prime}_{ll^\prime,mm^\prime}\otimes
\boldsymbol{1}_{2\dot{l}+1}
\frac{\partial\psi^{k^\prime\dot{k}^\prime}_{l^\prime m^\prime;
\dot{l}^\prime\dot{m}^\prime}}{\partial\tau}-
\frac{i}{r}\boldsymbol{1}_{2l+1}\otimes
b^{\dot{k}\dot{k}^\prime}_{\dot{l}\dot{l}^\prime,\dot{m}\dot{m}^\prime}
\frac{\partial\psi^{k^\prime\dot{k}^\prime}_{l^\prime m^\prime;
\dot{l}^\prime\dot{m}^\prime}}{\partial\tau}+\\
\left.+(1+i)
c^{kk^\prime;\dot{k}\dot{k}^\prime}_{ll^\prime,mm^\prime;
\dot{l}\dot{l}^\prime,\dot{m}\dot{m}^\prime}
\frac{\partial\psi^{k^\prime\dot{k}^\prime}_{l^\prime m^\prime;
\dot{l}^\prime\dot{m}^\prime}}{\partial r}+
\frac{1}{r}
d^{kk^\prime;\dot{k}\dot{k}^\prime}_{ll^\prime,mm^\prime;
\dot{l}\dot{l}^\prime,\dot{m}\dot{m}^\prime}
\psi^{k^\prime\dot{k}^\prime}_{l^\prime m^\prime;
\dot{l}^\prime\dot{m}^\prime}+
\frac{1}{r}
e^{kk^\prime;\dot{k}\dot{k}^\prime}_{ll^\prime,mm^\prime;
\dot{l}\dot{l}^\prime,\dot{m}\dot{m}^\prime}
\psi^{k^\prime\dot{k}^\prime}_{l^\prime m^\prime;
\dot{l}^\prime\dot{m}^\prime}+\right.\\
\left.+\frac{2i}{r}\ctg\theta^c
a^{kk^\prime}_{ll^\prime,mm^\prime}\otimes\boldsymbol{1}_{2\dot{l}+1}
m^\prime
\psi^{k^\prime\dot{k}^\prime}_{l^\prime m^\prime;
\dot{l}^\prime\dot{m}^\prime}+
\frac{2i}{r}\ctg\dot{\theta}^c\boldsymbol{1}_{2l+1}\otimes
a^{\dot{k}\dot{k}^\prime}_{\dot{l}\dot{l}^\prime,\dot{m}\dot{m}^\prime}
\dot{m}^\prime
\psi^{k^\prime\dot{k}^\prime}_{l^\prime m^\prime;
\dot{l}^\prime\dot{m}^\prime}\right]+\\
+\kappa^c\psi^{k\dot{k}}_{lm;\dot{l}\dot{m}}=0,
\nonumber
\end{multline}
\begin{multline}
\sum_{l^\prime,m^\prime,\dot{l}^\prime,\dot{m}^\prime,k^\prime,\dot{k}^\prime}
\left[
\frac{1}{r^\ast\sin\theta^c}
\overset{\ast}{a}{}^{\dot{k}\dot{k}^\prime}_{\dot{l}\dot{l}^\prime,
\dot{m}\dot{m}^\prime}\otimes\boldsymbol{1}_{2l+1}
\frac{\partial\dot{\psi}^{k^\prime\dot{k}^\prime}_{l^\prime m^\prime;
\dot{l}^\prime\dot{m}^\prime}}
{\partial\varphi}-
\frac{1}{r^\ast\sin\dot{\theta}^c}\boldsymbol{1}_{2\dot{l}+1}\otimes
\overset{\ast}{a}{}^{kk^\prime}_{ll^\prime,
mm^\prime}
\frac{\partial\dot{\psi}^{k^\prime\dot{k}^\prime}_{l^\prime m^\prime;
\dot{l}^\prime\dot{m}^\prime}}
{\partial\varphi}+\right.\\
+\frac{i}{r^\ast\sin\theta^c}
\overset{\ast}{a}{}^{\dot{k}\dot{k}^\prime}_{\dot{l}\dot{l}^\prime,
\dot{m}\dot{m}^\prime}\otimes\boldsymbol{1}_{2l+1}
\frac{\partial\dot{\psi}^{k^\prime\dot{k}^\prime}_{l^\prime m^\prime;
\dot{l}^\prime\dot{m}^\prime}}
{\partial\epsilon}+
\frac{i}{r^\ast\sin\dot{\theta}^c}\boldsymbol{1}_{2\dot{l}+1}\otimes
\overset{\ast}{a}{}^{kk^\prime}_{ll^\prime,
mm^\prime}
\frac{\partial\dot{\psi}^{k^\prime\dot{k}^\prime}_{l^\prime m^\prime;
\dot{l}^\prime\dot{m}^\prime}}
{\partial\epsilon}-
\\
-\frac{1}{r^\ast}
\overset{\ast}{b}{}^{\dot{k}\dot{k}^\prime}_{\dot{l}\dot{l}^\prime,
\dot{m}\dot{m}^\prime}\otimes\boldsymbol{1}_{2l+1}
\frac{\partial\dot{\psi}^{k^\prime\dot{k}^\prime}_{l^\prime m^\prime;
\dot{l}^\prime\dot{m}^\prime}}
{\partial\theta}+
\frac{1}{r^\ast}\boldsymbol{1}_{2\dot{l}+1}\otimes
\overset{\ast}{b}{}^{kk^\prime}_{ll^\prime,
mm^\prime}
\frac{\partial\dot{\psi}^{k^\prime\dot{k}^\prime}_{l^\prime m^\prime;
\dot{l}^\prime\dot{m}^\prime}}
{\partial\theta}-
\nonumber
\end{multline}
\begin{multline}
-\frac{i}{r^\ast}
\overset{\ast}{b}{}^{\dot{k}\dot{k}^\prime}_{\dot{l}\dot{l}^\prime,
\dot{m}\dot{m}^\prime}\otimes\boldsymbol{1}_{2l+1}
\frac{\partial\dot{\psi}^{k^\prime\dot{k}^\prime}_{l^\prime m^\prime;
\dot{l}^\prime\dot{m}^\prime}}
{\partial\tau}-
\frac{i}{r^\ast}\boldsymbol{1}_{2\dot{l}+1}\otimes
\overset{\ast}{b}{}^{kk^\prime}_{ll^\prime,
mm^\prime}
\frac{\partial\dot{\psi}^{k^\prime\dot{k}^\prime}_{l^\prime m^\prime;
\dot{l}^\prime\dot{m}^\prime}}
{\partial\tau}+\\
+(1-i)
\overset{\ast}{c}{}^{kk^\prime;\dot{k}\dot{k}^\prime}_{ll^\prime,mm^\prime;
\dot{l}\dot{l}^\prime,\dot{m}\dot{m}^\prime}
\frac{\partial\dot{\psi}^{k^\prime\dot{k}^\prime}_{l^\prime m^\prime;
\dot{l}^\prime\dot{m}^\prime}}
{\partial r^\ast}
+\frac{1}{r^\ast}
\overset{\ast}{d}{}^{kk^\prime;\dot{k}\dot{k}^\prime}_{ll^\prime,mm^\prime;
\dot{l}\dot{l}^\prime,\dot{m}\dot{m}^\prime}
\dot{\psi}^{k^\prime\dot{k}^\prime}_{l^\prime m^\prime;
\dot{l}^\prime\dot{m}^\prime}+
\frac{1}{r^\ast}
\overset{\ast}{e}{}^{kk^\prime;\dot{k}\dot{k}^\prime}_{ll^\prime,mm^\prime;
\dot{l}\dot{l}^\prime,\dot{m}\dot{m}^\prime}
\dot{\psi}^{k^\prime\dot{k}^\prime}_{l^\prime m^\prime;
\dot{l}^\prime\dot{m}^\prime}+\\
\left.+\frac{2i}{r^\ast}\ctg\theta^c
\overset{\ast}{a}{}^{\dot{k}\dot{k}^\prime}_{\dot{l}\dot{l}^\prime,
\dot{m}\dot{m}^\prime}\otimes\boldsymbol{1}_{2l+1}\dot{m}^\prime
\dot{\psi}^{k^\prime\dot{k}^\prime}_{l^\prime m^\prime;\dot{l}^\prime\dot{m}^\prime}+
\frac{2i}{r^\ast}\ctg\dot{\theta}^c\boldsymbol{1}_{2\dot{l}+1}\otimes
\overset{\ast}{a}{}^{kk^\prime}_{ll^\prime,
mm^\prime}m^\prime
\dot{\psi}^{k^\prime\dot{k}^\prime}_{l^\prime m^\prime;
\dot{l}^\prime\dot{m}^\prime}\right]+\\
+\dot{\kappa}^c\dot{\psi}^{k\dot{k}}_{lm;\dot{l}\dot{m}}=0,
\label{Complex6}
\end{multline}\begin{sloppypar}\noindent
where the coefficients
$a^{k k^\prime;\dot{k}\dot{k}^\prime}_{l l^\prime,m m^\prime;
\dot{l}\dot{l}^\prime,\dot{m}\dot{m}^\prime}$, $\ldots$,
$\overset{\ast}{e}{}^{k k^\prime;\dot{k}\dot{k}^\prime}_{l l^\prime,
m m^\prime;\dot{l}\dot{l}^\prime,\dot{m}\dot{m}^\prime}$
are defined by the formulae (\ref{L1}), (\ref{L2}), (\ref{L3}), (\ref{VM}),
(\ref{L1'}), (\ref{L2'}), (\ref{L3'}), (\ref{UM}), respectively.
\end{sloppypar}

With the view to separate the variables in (\ref{Complex6}) let us assume that
\begin{eqnarray}
\psi^{k\dot{k}}_{lm;\dot{l}\dot{m}}&=&\boldsymbol{f}^{l_0\dot{l}_0}_{lmk;
\dot{l}\dot{m}\dot{k}}(r)
\fM^{l_0\dot{l}_0}_{mn;\dot{m}\dot{n}}
(\varphi,\epsilon,\theta,\tau,0,0),\nonumber\\
\dot{\psi}^{k\dot{k}}_{lm;\dot{l}\dot{m}}&=&
\overset{\ast}{\boldsymbol{f}}{}^{l_0\dot{l}_0}_{lmk;\dot{l}\dot{m}\dot{k}}
(r^\ast)
\overset{\ast}{\fM}{}^{l_0\dot{l}_0}_{mn;\dot{m}\dot{n}}
(\varphi,\epsilon,\theta,\tau,0,0),\label{F}
\end{eqnarray}
where $l_0\ge l$, $-l_0\le m,n\le l_0$ and $\dot{l}_0\ge\dot{l}$,
$-\dot{l}_0\le\dot{m},\dot{n}\le\dot{l}_0$.
Substituting the functions
(\ref{F}) into the system (\ref{Complex6}) and taking into account values
of the coefficients
$a^{k k^\prime;\dot{k}\dot{k}^\prime}_{l l^\prime,m m^\prime;
\dot{l}\dot{l}^\prime,\dot{m}\dot{m}^\prime}$, $\ldots$,
$\overset{\ast}{e}{}^{k k^\prime;\dot{k}\dot{k}^\prime}_{l l^\prime,
m m^\prime;\dot{l}\dot{l}^\prime,\dot{m}\dot{m}^\prime}$,
we collect together the terms with identical
radial functions.\index{function!radial}
In the result we obtain
\begin{multline}
\sum_{k^\prime,\dot{k}^\prime}
c^{kk^\prime;\dot{k}\dot{k}^\prime}_{l-1,l;\dot{l}-1,\dot{l}}
\left\{
\sqrt{(l^2-m^2)(\dot{l}^2-\dot{m}^2)}\left[
(1+i)\frac{\partial\boldsymbol{f}^{l_0\dot{l}_0}_{l-1,m,k^\prime;
\dot{l}-1,\dot{m}\dot{k}^\prime}}
{\partial r}+\right.\right.\\
+\left.\left.\frac{1}{r}(l-il+\dot{l}+i\dot{l}-2)
\boldsymbol{f}^{l_0}_{l-1,m,k^\prime}\right]
\fM^{l_0\dot{l}_0}_{mn;\dot{m}\dot{n}}(\varphi,\epsilon,\theta,\tau,0,0)
+\right.\\
+\frac{1}{2r}\sqrt{(l+m)(l+m-1)(\dot{l}^2-\dot{m}^2)}
\boldsymbol{f}^{l_0\dot{l}_0}_{l-1,m-1,k^\prime;\dot{l}-1,\dot{m}\dot{k}^\prime}
\left[-\frac{1}{\sin\theta^c}
\frac{\partial\fM^{l_0\dot{l}_0}_{m-1,n;\dot{m}\dot{n}}}{\partial\varphi}-
\right.\\
-\frac{i}{\sin\theta^c}
\frac{\partial\fM^{l_0\dot{l}_0}_{m-1,n;\dot{m}\dot{n}}}{\partial\epsilon}
\left.+i\frac{\partial\fM^{l_0\dot{l}_0}_{m-1,n;\dot{m}\dot{n}}}
{\partial\theta}-
\frac{\partial\fM^{l_0\dot{l}_0}_{m-1,n;\dot{m}\dot{n}}}{\partial\tau}-
\frac{2i(m-1)\cos\theta^c}{\sin\theta^c}
\fM^{l_0\dot{l}_0}_{m-1,n;\dot{m}\dot{n}}\right]+
\nonumber
\end{multline}
\begin{multline}
+\frac{1}{2r}\sqrt{(l-m)(l-m-1)(\dot{l}^2-\dot{m}^2)}
\boldsymbol{f}^{l_0\dot{l}_0}_{l-1,m+1,k^\prime;\dot{l}-1,\dot{m}\dot{k}^\prime}
\left[\frac{1}{\sin\theta^c}
\frac{\partial\fM^{l_0\dot{l}_0}_{m+1,n;\dot{m}\dot{n}}}{\partial\varphi}+
\right.\\
+\frac{i}{\sin\theta^c}
\frac{\partial\fM^{l_0\dot{l}_0}_{m+1,n;\dot{m}\dot{n}}}{\partial\epsilon}+
\left.
\left.+i\frac{\partial\fM^{l_0\dot{l}_0}_{m+1,n;\dot{m}\dot{n}}}{\partial\theta}-
\frac{\partial\fM^{l_0\dot{l}_0}_{m+1,n;\dot{m}\dot{n}}}{\partial\tau}+
\frac{2i(m+1)\cos\theta^c}{\sin\theta^c}
\fM^{l_0\dot{l}_0}_{m+1,n;\dot{m}\dot{n}}\right]\right.+
\\
+\frac{1}{2r}\sqrt{(l^2-m^2)(\dot{l}+\dot{m})(\dot{l}+\dot{m}-1)}
\boldsymbol{f}^{l_0\dot{l}_0}_{l-1,mk^\prime;\dot{l}-1,\dot{m}-1,\dot{k}^\prime}
\left[-\frac{1}{\sin\dot{\theta}^c}
\frac{\partial\fM^{l_0\dot{l}_0}_{mn;\dot{m}-1,\dot{n}}}{\partial\varphi}+
\right.
\nonumber
\end{multline}
\begin{multline}
\left.+\frac{i}{\sin\dot{\theta}^c}
\frac{\partial\fM^{l_0\dot{l}_0}_{mn;\dot{m}-1,\dot{n}}}{\partial\epsilon}
+i\frac{\partial\fM^{l_0\dot{l}_0}_{mn;\dot{m}-1,\dot{n}}}
{\partial\theta}+
\frac{\partial\fM^{l_0\dot{l}_0}_{mn;\dot{m}-1,\dot{n}}}{\partial\tau}+
\frac{2i(\dot{m}-1)\cos\dot{\theta}^c}{\sin\dot{\theta}^c}
\fM^{l_0\dot{l}_0}_{mn;\dot{m}-1,\dot{n}}\right]+
\\
+\frac{1}{2r}\sqrt{(l^2-m^2)(\dot{l}-\dot{m})(\dot{l}-\dot{m}-1)}
\boldsymbol{f}^{l_0\dot{l}_0}_{l-1,mk^\prime;\dot{l}-1,\dot{m}+1,\dot{k}^\prime}
\left[\frac{1}{\sin\dot{\theta}^c}
\frac{\partial\fM^{l_0\dot{l}_0}_{mn;\dot{m}+1,\dot{n}}}{\partial\varphi}-
\right.\\
-\frac{i}{\sin\dot{\theta}^c}
\frac{\partial\fM^{l_0\dot{l}_0}_{mn;\dot{m}+1,\dot{n}}}{\partial\epsilon}+
\left.
\left.+i\frac{\partial\fM^{l_0\dot{l}_0}_{mn;\dot{m}+1,\dot{n}}}{\partial\theta}+
\frac{\partial\fM^{l_0\dot{l}_0}_{mn;\dot{m}+1,\dot{n}}}{\partial\tau}-
\frac{2i(\dot{m}+1)\cos\dot{\theta}^c}{\sin\dot{\theta}^c}
\fM^{l_0\dot{l}_0}_{mn;\dot{m}+1,\dot{n}}\right]\right\}+
\nonumber
\end{multline}
\[
+\sum_{k^\prime,\dot{k}^\prime}
c^{kk^\prime;\dot{k}\dot{k}^\prime}_{ll;\dot{l}\dot{l}}\left\{\left[
(1+i)m\dot{m}
\frac{\partial\boldsymbol{f}^{l_0\dot{l}_0}_{lmk^\prime;
\dot{l}\dot{m}\dot{k}^\prime}}{\partial r}
-\frac{2i}{r}m\dot{m}
\boldsymbol{f}^{l_0\dot{l}_0}_{lmk^\prime;\dot{l}\dot{m}\dot{k}^\prime}\right]
\fM^{l_0\dot{l}_0}_{mn;\dot{m}\dot{n}}(\varphi,\epsilon,\theta,\tau,0,0)+\right.
\]
\begin{multline}
+\frac{1}{2r}\dot{m}\sqrt{(l+m)(l-m+1)}
\boldsymbol{f}^{l_0\dot{l}_0}_{l,m-1,k^\prime;\dot{l}\dot{m}\dot{k}^\prime}
\left[\frac{1}{\sin\theta^c}
\frac{\partial\fM^{l_0\dot{l}_0}_{m-1,n;\dot{m}\dot{n}}}{\partial\varphi}+
\frac{i}{\sin\theta^c}
\frac{\partial\fM^{l_0\dot{l}_0}_{m-1,n;\dot{m}\dot{n}}}{\partial\epsilon}
\right.-\\
\left.-i\frac{\partial\fM^{l_0\dot{l}_0}_{m-1,n;\dot{m}\dot{n}}}{\partial\theta}+
\frac{\partial\fM^{l_0\dot{l}_0}_{m-1,n;\dot{m}\dot{n}}}{\partial\tau}+
\frac{2i(m-1)\cos\theta^c}{\sin\theta^c}
\fM^{l_0\dot{l}_0}_{m-1,n;\dot{m}\dot{n}}\right]+
\\
+\frac{1}{2r}\dot{m}\sqrt{(l+m+1)(l-m)}
\boldsymbol{f}^{l_0\dot{l}_0}_{l,m+1,k^\prime;\dot{l}\dot{m}\dot{k}^\prime}
\left[\frac{1}{\sin\theta^c}
\frac{\partial\fM^{l_0\dot{l}_0}_{m+1,n;\dot{m}\dot{n}}}{\partial\varphi}+
\frac{i}{\sin\theta^c}
\frac{\partial\fM^{l_0\dot{l}_0}_{m+1,n;\dot{m}\dot{n}}}{\partial\epsilon}+
\right.
\nonumber
\end{multline}
\begin{multline}
\left.
\left.+i\frac{\partial\fM^{l_0\dot{l}_0}_{m+1,n;\dot{m}\dot{n}}}{\partial\theta}-
\frac{\partial\fM^{l_0\dot{l}_0}_{m+1,n;\dot{m}\dot{n}}}{\partial\tau}+
\frac{2i(m+1)\cos\theta^c}{\sin\theta^c}
\fM^{l_0\dot{l}_0}_{m+1,n;\dot{m}\dot{n}}\right]\right\}+
\\
+\frac{1}{2r}m\sqrt{(\dot{l}+\dot{m})(\dot{l}-\dot{m}+1)}
\boldsymbol{f}^{l_0\dot{l}_0}_{lmk^\prime;\dot{l},\dot{m}-1,\dot{k}^\prime}
\left[\frac{1}{\sin\dot{\theta}^c}
\frac{\partial\fM^{l_0\dot{l}_0}_{mn;\dot{m}-1,\dot{n}}}{\partial\varphi}-
\frac{i}{\sin\dot{\theta}^c}
\frac{\partial\fM^{l_0\dot{l}_0}_{mn;\dot{m}-1,\dot{n}}}{\partial\epsilon}
\right.-\\
\left.-i\frac{\partial\fM^{l_0\dot{l}_0}_{mn;\dot{m}-1,\dot{n}}}{\partial\theta}-
\frac{\partial\fM^{l_0\dot{l}_0}_{mn;\dot{m}-1,\dot{n}}}{\partial\tau}-
\frac{2i(\dot{m}-1)\cos\dot{\theta}^c}{\sin\dot{\theta}^c}
\fM^{l_0\dot{l}_0}_{mn;\dot{m}-1,\dot{n}}\right]+
\nonumber
\end{multline}
\begin{multline}
+\frac{1}{2r}m\sqrt{(\dot{l}+\dot{m}+1)(\dot{l}-\dot{m})}
\boldsymbol{f}^{l_0\dot{l}_0}_{lmk^\prime;\dot{l},\dot{m}+1,\dot{k}^\prime}
\left[\frac{1}{\sin\dot{\theta}^c}
\frac{\partial\fM^{l_0\dot{l}_0}_{mn;\dot{m}+1,\dot{n}}}{\partial\varphi}-
\frac{i}{\sin\dot{\theta}^c}
\frac{\partial\fM^{l_0\dot{l}_0}_{mn;\dot{m}+1,\dot{n}}}{\partial\epsilon}+
\right.\\
\left.
\left.+i\frac{\partial\fM^{l_0\dot{l}_0}_{mn;\dot{m}+1,\dot{n}}}{\partial\theta}+
\frac{\partial\fM^{l_0\dot{l}_0}_{mn;\dot{m}+1,\dot{n}}}{\partial\tau}-
\frac{2i(\dot{m}+1)\cos\dot{\theta}^c}{\sin\dot{\theta}^c}
\fM^{l_0\dot{l}_0}_{mn;\dot{m}+1,\dot{n}}\right]\right\}+
\nonumber
\end{multline}
\begin{multline}
+\sum_{k^\prime,\dot{k}^\prime}
c^{kk^\prime;\dot{k}\dot{k}^\prime}_{l+1,l;\dot{l}+1,\dot{l}}\left\{
\sqrt{((l+1)^2-m^2)((\dot{l}+1)^2-\dot{m}^2)}\left[
(1+i)\frac{\partial
\boldsymbol{f}^{l_0\dot{l}_0}_{l+1,m,k^\prime;\dot{l}+1,\dot{m}\dot{k}^\prime}}
{\partial r}+\right.\right.\\
+\left.\left.\frac{1}{r}(il-l-\dot{l}-i\dot{l}-2)
\boldsymbol{f}^{l_0\dot{l}_0}_{l+1,m,k^\prime;\dot{l}+1,\dot{m}\dot{k}^\prime}
\right]
\fM^{l_0\dot{l}_0}_{mn;\dot{m}\dot{n}}
(\varphi,\epsilon,\theta,\tau,0,0)+\right.\\
+\frac{1}{2r}\sqrt{(l-m+1)(l-m+2)((\dot{l}+1)^2-\dot{m}^2)}
\boldsymbol{f}^{l_0\dot{l}_0}_{l+1,m-1,k^\prime;\dot{l}-1,\dot{m}\dot{k}^\prime}
\left[\frac{1}{\sin\theta^c}
\frac{\partial\fM^{l_0\dot{l}_0}_{m-1,n;\dot{m}\dot{n}}}{\partial\varphi}+
\right.\nonumber
\end{multline}
\begin{multline}
\left.
+\frac{i}{\sin\theta^c}
\frac{\partial\fM^{l_0\dot{l}_0}_{m-1,n;\dot{m}\dot{n}}}{\partial\epsilon}
-i\frac{\partial\fM^{l_0\dot{l}_0}_{m-1,n;\dot{m}\dot{n}}}
{\partial\theta}+
\frac{\partial\fM^{l_0\dot{l}_0}_{m-1,n;\dot{m}\dot{n}}}{\partial\tau}+
\frac{2i(m-1)\cos\theta^c}{\sin\theta^c}
\fM^{l_0\dot{l}_0}_{m-1,n;\dot{m}\dot{n}}\right]+
\\
+\frac{1}{2r}\sqrt{(l+m+1)(l+m+2)((\dot{l}+1)^2-\dot{m}^2)}
\boldsymbol{f}^{l_0\dot{l}_0}_{l+1,m+1,k^\prime;\dot{l}+1,\dot{m}\dot{k}^\prime}
\left[-\frac{1}{\sin\theta^c}
\frac{\partial\fM^{l_0\dot{l}_0}_{m+1,n;\dot{m}\dot{n}}}{\partial\varphi}-
\right.\\
-\frac{i}{\sin\theta^c}
\frac{\partial\fM^{l_0\dot{l}_0}_{m+1,n;\dot{m}\dot{n}}}{\partial\epsilon}-
\left.
\left.-i\frac{\partial\fM^{l_0\dot{l}_0}_{m+1,n;\dot{m}\dot{n}}}
{\partial\theta}+
\frac{\partial\fM^{l_0\dot{l}_0}_{m+1,n;\dot{m}\dot{n}}}{\partial\tau}-
\frac{2i(m+1)\cos\theta^c}{\sin\theta^c}
\fM^{l_0\dot{l}_0}_{m+1,n;\dot{m}\dot{n}}\right]\right.+
\nonumber
\end{multline}
\begin{multline}
+\frac{1}{2r}\sqrt{((l+1)^2-m^2)(\dot{l}-\dot{m}+1)(\dot{l}-\dot{m}+2)}
\boldsymbol{f}^{l_0\dot{l}_0}_{l+1,mk^\prime;\dot{l}-1,\dot{m}-1,\dot{k}^\prime}
\left[\frac{1}{\sin\dot{\theta}^c}
\frac{\partial\fM^{l_0\dot{l}_0}_{mn;\dot{m}-1,\dot{n}}}{\partial\varphi}-
\right.\\
-\frac{i}{\sin\dot{\theta}^c}
\frac{\partial\fM^{l_0\dot{l}_0}_{mn;\dot{m}-1,\dot{n}}}{\partial\epsilon}
\left.-i\frac{\partial\fM^{l_0\dot{l}_0}_{mn;\dot{m}-1,\dot{n}}}
{\partial\theta}-
\frac{\partial\fM^{l_0\dot{l}_0}_{mn;\dot{m}-1,\dot{n}}}{\partial\tau}+
\frac{2i(\dot{m}-1)\cos\dot{\theta}^c}{\sin\dot{\theta}^c}
\fM^{l_0\dot{l}_0}_{mn;\dot{m}-1,\dot{n}}\right]+
\\
+\frac{1}{2r}\sqrt{((l+1)^2-m^2)(\dot{l}+\dot{m}+1)(\dot{l}+\dot{m}+2)}
\boldsymbol{f}^{l_0\dot{l}_0}_{l+1,mk^\prime;\dot{l}+1,\dot{m}+1,\dot{k}^\prime}
\left[-\frac{1}{\sin\dot{\theta}^c}
\frac{\partial\fM^{l_0\dot{l}_0}_{mn;\dot{m}+1,\dot{n}}}{\partial\varphi}+
\right.
\nonumber
\end{multline}
\begin{multline}
+\frac{i}{\sin\dot{\theta}^c}
\frac{\partial\fM^{l_0\dot{l}_0}_{mn;\dot{m}+1,\dot{n}}}{\partial\epsilon}-
\left.
\left.-i\frac{\partial\fM^{l_0\dot{l}_0}_{mn;\dot{m}+1,\dot{n}}}
{\partial\theta}-
\frac{\partial\fM^{l_0\dot{l}_0}_{mn;\dot{m}+1,\dot{n}}}{\partial\tau}+
\frac{2i(\dot{m}+1)\cos\dot{\theta}^c}{\sin\dot{\theta}^c}
\fM^{l_0\dot{l}_0}_{mn;\dot{m}+1,\dot{n}}\right]\right\}+
\\
+\kappa^c\boldsymbol{f}^{l_0\dot{l}_0}_{lm;\dot{l}\dot{m}}
\fM^{l_0\dot{l}_0}_{mn;\dot{m}\dot{n}}(\varphi,\epsilon,\theta,\tau,0,0)=0,
\nonumber
\end{multline}
\begin{multline}
\sum_{\dot{k}^\prime}\overset{\ast}{c}{}^{kk^\prime;\dot{k}
\dot{k}^\prime}_{l-1,l;\dot{l}-1,\dot{l}}
\left\{
\sqrt{(l^2-m^2)(\dot{l}^2-\dot{m}^2)}\left[(1-i)
\frac{\partial\overset{\ast}{\boldsymbol{f}}{}^{l_0\dot{l}_0}_{
l-1,mk^\prime;\dot{l}-1,\dot{m}\dot{k}^\prime}}
{\partial r^\ast}-\right.\right.\\
-\left.\left.\frac{1}{r^\ast}(l-il+\dot{l}+i\dot{l}+2)
\overset{\ast}{\boldsymbol{f}}{}^{l_0\dot{l}_0}_{
l-1,mk^\prime;\dot{l}-1,\dot{m}\dot{k}^\prime}\right]
\overset{\ast}{\fM}{}^{l_0\dot{l}_0}_{mn;\dot{m}\dot{n}}
(\varphi,\epsilon,\theta,\tau,0,0)+\right.\\
+\frac{1}{2r^\ast}\sqrt{(l^2-m^2)(\dot{l}+\dot{m})(\dot{l}+\dot{m}-1)}
\overset{\ast}{\boldsymbol{f}}{}^{l_0\dot{l}_0}_{
l-1,mk^\prime;\dot{l}-1,\dot{m}-1,\dot{k}^\prime}\left[
-\frac{1}{\sin\theta^c}
\frac{\partial\overset{\ast}{\fM}{}^{l_0\dot{l}_0}_{mn;\dot{m}-1,\dot{n}}}
{\partial\varphi}-\right.\\
-\frac{i}{\sin\theta^c}
\frac{\partial\overset{\ast}{\fM}{}^{l_0\dot{l}_0}_{mn;\dot{m}-1,\dot{n}}}
{\partial\epsilon}
\left.+i\frac{\partial\overset{\ast}{\fM}{}^{l_0\dot{l}_0}_{mn;\dot{m}-1,
\dot{n}}}{\partial\theta}-
\frac{\partial\overset{\ast}{\fM}{}^{l_0\dot{l}_0}_{mn;\dot{m}-1,\dot{n}}}
{\partial\tau}-
\frac{2i(\dot{m}-1)\cos\theta^c}{\sin\theta^c}
\overset{\ast}{\fM}{}^{l_0\dot{l}_0}_{mn;\dot{m}-1,\dot{n}}\right]+
\nonumber
\end{multline}
\begin{multline}
+\frac{1}{2r^\ast}\sqrt{(l^2-m^2)(\dot{l}-\dot{m})(\dot{l}-\dot{m}-1)}
\overset{\ast}{\boldsymbol{f}}{}^{l_0\dot{l}_0}_{l-1,mk^\prime;
\dot{l}-1,\dot{m}+1,\dot{k}^\prime}\left[
\frac{1}{\sin\theta^c}
\frac{\partial\overset{\ast}{\fM}{}^{l_0\dot{l}_0}_{mn;\dot{m}+1,\dot{n}}}
{\partial\varphi}+\right.\\
+\frac{i}{\sin\theta^c}
\frac{\partial\overset{\ast}{\fM}{}^{l_0\dot{l}_0}_{mn;\dot{m}+1,\dot{n}}}
{\partial\epsilon}
%
\left.
\left.+i\frac{\partial\overset{\ast}{\fM}{}^{l_0\dot{l}_0}_{mn;\dot{m}+1,
\dot{n}}}{\partial\theta}-
\frac{\partial\overset{\ast}{\fM}{}^{l_0\dot{l}_0}_{mn;\dot{m}+1,\dot{n}}}
{\partial\tau}+
\frac{2i(\dot{m}+1)\cos\theta^c}{\sin\theta^c}
\overset{\ast}{\fM}{}^{l_0\dot{l}_0}_{mn;\dot{m}+1,\dot{n}}\right]\right.+
\\
+\frac{1}{2r^\ast}\sqrt{(l+m)(l+m-1)(\dot{l}^2-\dot{m}^2)}
\overset{\ast}{\boldsymbol{f}}{}^{l_0\dot{l}_0}_{
l-1,m-1,k^\prime;\dot{l}-1,\dot{m}\dot{k}^\prime}\left[
-\frac{1}{\sin\dot{\theta}^c}
\frac{\partial\overset{\ast}{\fM}{}^{l_0\dot{l}_0}_{m-1,n;\dot{m}\dot{n}}}
{\partial\varphi}+\right.
\nonumber
\end{multline}
\begin{multline}
+\frac{i}{\sin\dot{\theta}^c}
\frac{\partial\overset{\ast}{\fM}{}^{l_0\dot{l}_0}_{m-1,n;\dot{m}\dot{n}}}
{\partial\epsilon}
\left.+i\frac{\partial\overset{\ast}{\fM}{}^{l_0\dot{l}_0}_{m-1,n;\dot{m}
\dot{n}}}{\partial\theta}+
\frac{\partial\overset{\ast}{\fM}{}^{l_0\dot{l}_0}_{m-1,n;\dot{m}\dot{n}}}
{\partial\tau}+
\frac{2i(m-1)\cos\dot{\theta}^c}{\sin\dot{\theta}^c}
\overset{\ast}{\fM}{}^{l_0\dot{l}_0}_{m-1,n;\dot{m}\dot{n}}\right]+
\\
+\frac{1}{2r^\ast}\sqrt{(l-m)(l-m-1)(\dot{l}^2-\dot{m}^2)}
\overset{\ast}{\boldsymbol{f}}{}^{l_0\dot{l}_0}_{l-1,m+1,k^\prime;
\dot{l}-1,\dot{m}\dot{k}^\prime}\left[
\frac{1}{\sin\dot{\theta}^c}
\frac{\partial\overset{\ast}{\fM}{}^{l_0\dot{l}_0}_{m+1,n;\dot{m}\dot{n}}}
{\partial\varphi}-\right.\\
-\frac{i}{\sin\dot{\theta}^c}
\frac{\partial\overset{\ast}{\fM}{}^{l_0\dot{l}_0}_{m+1,n;\dot{m}\dot{n}}}
{\partial\epsilon}
\left.
\left.+i\frac{\partial\overset{\ast}{\fM}{}^{l_0\dot{l}_0}_{m+1,n;\dot{m}
\dot{n}}}{\partial\theta}+
\frac{\partial\overset{\ast}{\fM}{}^{l_0\dot{l}_0}_{m+1,n;\dot{m}\dot{n}}}
{\partial\tau}-
\frac{2i(m+1)\cos\dot{\theta}^c}{\sin\dot{\theta}^c}
\overset{\ast}{\fM}{}^{l_0\dot{l}_0}_{m+1,n;\dot{m}\dot{n}}\right]\right\}+
\\
+\sum_{k^\prime,\dot{k}^\prime}
\overset{\ast}{c}^{kk^\prime;\dot{k}\dot{k}^\prime}_{ll;
\dot{l}\dot{l}}\left\{\left[
(1-i)m\dot{m}
\frac{\partial\overset{\ast}{\boldsymbol{f}}{}^{l_0\dot{l}_0}_{lmk^\prime;
\dot{l},\dot{m},\dot{k}^\prime}}
{\partial r^\ast}+\frac{2i}{r^\ast}m\dot{m}
\overset{\ast}{\boldsymbol{f}}{}^{l_0\dot{l}_0}_{lmk^\prime;
\dot{l},\dot{m},\dot{k}^\prime}\right]
\overset{\ast}{\fM}{}^{l_0\dot{l}_0}_{mn;\dot{m},\dot{n}}
(\varphi,\epsilon,\theta,\tau,0,0)+\right.
\nonumber
\end{multline}
\begin{multline}
+\frac{1}{2r^\ast}m\sqrt{(\dot{l}+\dot{m})(\dot{l}-\dot{m}+1)}
\overset{\ast}{\boldsymbol{f}}{}^{l_0\dot{l}_0}_{lmk^prime;
\dot{l},\dot{m}-1,\dot{k}^\prime}\left[
\frac{1}{\sin\theta^c}
\frac{\partial\overset{\ast}{\fM}{}^{l_0\dot{l}_0}_{mn;\dot{m}-1,\dot{n}}}
{\partial\varphi}+
\frac{i}{\sin\theta^c}
\frac{\partial\overset{\ast}{\fM}{}^{l_0\dot{l}_0}_{mn;\dot{m}-1,\dot{n}}}
{\partial\epsilon}-\right.\\
\left.-i\frac{\partial\overset{\ast}{\fM}{}^{l_0\dot{l}_0}_{mn;\dot{m}-1,
\dot{n}}}{\partial\theta}+
\frac{\partial\overset{\ast}{\fM}{}^{l_0\dot{l}_0}_{mn;\dot{m}-1,\dot{n}}}
{\partial\tau}+
\frac{2i(\dot{m}-1)\cos\theta^c}{\sin\theta^c}
\overset{\ast}{\fM}{}^{l_0\dot{l}_0}_{mn;\dot{m}-1,\dot{n}}\right]+
\\
+\frac{1}{2r^\ast}m\sqrt{(\dot{l}+\dot{m}+1)(\dot{l}-\dot{m})}
\overset{\ast}{\boldsymbol{f}}{}^{l_0\dot{l}_0}_{lmk^\prime;\dot{l},\dot{m}+1,
\dot{k}^\prime}\left[
\frac{1}{\sin\theta^c}
\frac{\partial\overset{\ast}{\fM}{}^{l_0\dot{l}_0}_{mn;\dot{m}+1,\dot{n}}}
{\partial\varphi}+
\frac{i}{\sin\theta^c}
\frac{\partial\overset{\ast}{\fM}{}^{l_0\dot{l}_0}_{mn;\dot{m}+1,\dot{n}}}
{\partial\epsilon}+\right.
\nonumber
\end{multline}
\begin{multline}
\left.
\left.+i\frac{\partial\overset{\ast}{\fM}{}^{l_0\dot{l}_0}_{mn;\dot{m}+1,
\dot{n}}}{\partial\theta}-
\frac{\partial\overset{\ast}{\fM}{}^{l_0\dot{l}_0}_{mn;\dot{m}+1,\dot{n}}}
{\partial\tau}+
\frac{2i(\dot{m}+1)\cos\theta^c}{\sin\theta^c}
\overset{\ast}{\fM}{}^{l_0\dot{l}_0}_{mn;\dot{m}+1,\dot{n}}\right]\right.+
\\
+\frac{1}{2r^\ast}\dot{m}\sqrt{(l+m)(l-m+1)}
\overset{\ast}{\boldsymbol{f}}{}^{l_0\dot{l}_0}_{l,m-1,k^prime;
\dot{l}\dot{m}\dot{k}^\prime}\left[
\frac{1}{\sin\dot{\theta}^c}
\frac{\partial\overset{\ast}{\fM}{}^{l_0\dot{l}_0}_{m-1,n;\dot{m}\dot{n}}}
{\partial\varphi}-
\frac{i}{\sin\dot{\theta}^c}
\frac{\partial\overset{\ast}{\fM}{}^{l_0\dot{l}_0}_{m-1,n;\dot{m}\dot{n}}}
{\partial\epsilon}-\right.\\
\left.-i\frac{\partial\overset{\ast}{\fM}{}^{l_0\dot{l}_0}_{m-1,n;\dot{m}
\dot{n}}}{\partial\theta}-
\frac{\partial\overset{\ast}{\fM}{}^{l_0\dot{l}_0}_{m-1,n;\dot{m}\dot{n}}}
{\partial\tau}-
\frac{2i(m-1)\cos\dot{\theta}^c}{\sin\dot{\theta}^c}
\overset{\ast}{\fM}{}^{l_0\dot{l}_0}_{m-1,n;\dot{m}\dot{n}}\right]+
\nonumber
\end{multline}
\begin{multline}
+\frac{1}{2r^\ast}\dot{m}\sqrt{(l+m+1)(l-m)}
\overset{\ast}{\boldsymbol{f}}{}^{l_0\dot{l}_0}_{l,m+1,k^\prime;\dot{l}\dot{m}
\dot{k}^\prime}\left[
\frac{1}{\sin\dot{\theta}^c}
\frac{\partial\overset{\ast}{\fM}{}^{l_0\dot{l}_0}_{m+1,n;\dot{m}\dot{n}}}
{\partial\varphi}-
\frac{i}{\sin\dot{\theta}^c}
\frac{\partial\overset{\ast}{\fM}{}^{l_0\dot{l}_0}_{m+1,n;\dot{m}\dot{n}}}
{\partial\epsilon}+\right.
\\
\left.
\left.+i\frac{\partial\overset{\ast}{\fM}{}^{l_0\dot{l}_0}_{m+1,n;\dot{m}
\dot{n}}}{\partial\theta}+
\frac{\partial\overset{\ast}{\fM}{}^{l_0\dot{l}_0}_{m+1,n;\dot{m}\dot{n}}}
{\partial\tau}-
\frac{2i(m+1)\cos\dot{\theta}^c}{\sin\dot{\theta}^c}
\overset{\ast}{\fM}{}^{l_0\dot{l}_0}_{m+1,n;\dot{m}\dot{n}}\right]\right\}+
\\
+\sum_{k^\prime,\dot{k}^\prime}
\overset{\ast}{c}^{kk^\prime;\dot{k}\dot{k}^\prime}_{l+1,l;
\dot{l}+1,\dot{l}}\left\{\sqrt{((l+1)^2-m^2)((\dot{l}+1)^2-\dot{m}^2)}\left[
(1-i)
\frac{\partial\overset{\ast}{\boldsymbol{f}}{}^{l_0\dot{l}_0}_{l+1,mk^\prime;
\dot{l}+1,\dot{m},\dot{k}^\prime}}
{\partial r^\ast}+\right.\right.
\\
+\left.\left.\frac{1}{r^\ast}(il-l-\dot{l}-i\dot{l}-2+i2)
\overset{\ast}{\boldsymbol{f}}{}^{l_0\dot{l}_0}_{l+1,mk^\prime;
\dot{l}+1,\dot{m},\dot{k}^\prime}\right]
\overset{\ast}{\fM}{}^{l_0\dot{l}_0}_{mn;\dot{m}\dot{n}}
(\varphi,\epsilon,\theta,\tau,0,0)+\right.\\
+\frac{1}{2r^\ast}\sqrt{((l+1)^2-m^2)(\dot{l}-\dot{m}+1)(\dot{l}-\dot{m}+2)}
\overset{\ast}{\boldsymbol{f}}{}^{l_0\dot{l}_0}_{l+1,mk^\prime;
\dot{l}+1,\dot{m}-1,\dot{k}^\prime}\left[
\frac{1}{\sin\theta^c}
\frac{\partial\overset{\ast}{\fM}{}^{l_0\dot{l}_0}_{mn;\dot{m}-1,\dot{n}}}
{\partial\varphi}+\right.
\nonumber
\end{multline}
\begin{multline}
+\frac{i}{\sin\theta^c}
\frac{\partial\overset{\ast}{\fM}{}^{l_0\dot{l}_0}_{mn;\dot{m}-1,\dot{n}}}
{\partial\epsilon}
\left.-i\frac{\partial\overset{\ast}{\fM}{}^{l_0\dot{l}_0}_{mn;
\dot{m}-1,\dot{n}}}{\partial\theta}+
\frac{\partial\overset{\ast}{\fM}{}^{l_0\dot{l}_0}_{mn;\dot{m}-1,\dot{n}}}
{\partial\tau}+
\frac{2i(\dot{m}-1)\cos\theta^c}{\sin\theta^c}
\overset{\ast}{\fM}{}^{l_0\dot{l}_0}_{mn;\dot{m}-1,\dot{n}}\right]+
\\
+\frac{1}{2r^\ast}\sqrt{((l+1)^2-m^2)(\dot{l}+\dot{m}+1)(\dot{l}+\dot{m}+2)}
\overset{\ast}{\boldsymbol{f}}{}^{l_0\dot{l}_0}_{l+1,mk^\prime;
\dot{l}+1,\dot{m}+1,\dot{k}^\prime}\left[
-\frac{1}{\sin\theta^c}
\frac{\partial\overset{\ast}{\fM}{}^{l_0\dot{l}_0}_{mn;\dot{m}+1,\dot{n}}}
{\partial\varphi}-\right.\\
-\frac{i}{\sin\theta^c}
\frac{\partial\overset{\ast}{\fM}{}^{l_0\dot{l}_0}_{mn;\dot{m}+1,\dot{n}}}
{\partial\epsilon}
\left.
\left.-i\frac{\partial\overset{\ast}{\fM}{}^{l_0\dot{l}_0}_{mn;
\dot{m}+1,\dot{n}}}{\partial\theta}+
\frac{\partial\overset{\ast}{\fM}{}^{l_0\dot{l}_0}_{mn;\dot{m}+1,\dot{n}}}
{\partial\tau}-
\frac{2i(\dot{m}+1)\cos\theta^c}{\sin\theta^c}
\overset{\ast}{\fM}{}^{l_0\dot{l}_0}_{mn;\dot{m}+1,\dot{n}}\right]\right.+
\nonumber
\end{multline}
\begin{multline}
+\frac{1}{2r^\ast}\sqrt{((l-m+1)(l-m+2)((\dot{l}+1)^2-\dot{m}^2)}
\overset{\ast}{\boldsymbol{f}}{}^{l_0\dot{l}_0}_{l+1,m-1,k^\prime;
\dot{l}+1,\dot{m}\dot{k}^\prime}\left[
\frac{1}{\sin\dot{\theta}^c}
\frac{\partial\overset{\ast}{\fM}{}^{l_0\dot{l}_0}_{m-1,n;\dot{m}\dot{n}}}
{\partial\varphi}-\right.\\
-\frac{i}{\sin\dot{\theta}^c}
\frac{\partial\overset{\ast}{\fM}{}^{l_0\dot{l}_0}_{m-1,n;\dot{m}\dot{n}}}
{\partial\epsilon}
\left.-i\frac{\partial\overset{\ast}{\fM}{}^{l_0\dot{l}_0}_{m-1,n;
\dot{m}\dot{n}}}{\partial\theta}-
\frac{\partial\overset{\ast}{\fM}{}^{l_0\dot{l}_0}_{m-1,n;\dot{m}\dot{n}}}
{\partial\tau}-
\frac{2i(m-1)\cos\dot{\theta}^c}{\sin\dot{\theta}^c}
\overset{\ast}{\fM}{}^{l_0\dot{l}_0}_{m-1,n;\dot{m}\dot{n}}\right]+
\\
+\frac{1}{2r^\ast}\sqrt{((l+m+1)(l+m+2)((\dot{l}+1)^2-\dot{m}^2)}
\overset{\ast}{\boldsymbol{f}}{}^{l_0\dot{l}_0}_{l+1,m+1,k^\prime;
\dot{l}+1,\dot{m}\dot{k}^\prime}\left[
-\frac{1}{\sin\dot{\theta}^c}
\frac{\partial\overset{\ast}{\fM}{}^{l_0\dot{l}_0}_{m+1,n;\dot{m}\dot{n}}}
{\partial\varphi}+\right.\\
+\frac{i}{\sin\dot{\theta}^c}
\frac{\partial\overset{\ast}{\fM}{}^{l_0\dot{l}_0}_{m+1,n;\dot{m}\dot{n}}}
{\partial\epsilon}
\left.
\left.-i\frac{\partial\overset{\ast}{\fM}{}^{l_0\dot{l}_0}_{m+1,n;
\dot{m}\dot{n}}}{\partial\theta}-
\frac{\partial\overset{\ast}{\fM}{}^{l_0\dot{l}_0}_{m+1,n;\dot{m}\dot{n}}}
{\partial\tau}+
\frac{2i(m+1)\cos\dot{\theta}^c}{\sin\dot{\theta}^c}
\overset{\ast}{\fM}{}^{l_0\dot{l}_0}_{m+1,n;\dot{m}\dot{n}}\right]\right\}+
\\
+\dot{\kappa}^c
\overset{\ast}{\boldsymbol{f}}{}^{l_0\dot{l}_0}_{lm;\dot{l}\dot{m}}
\overset{\ast}{\fM}{}^{l_0\dot{l}_0}_{mn;\dot{m}\dot{n}}
(\varphi,\epsilon,\theta,\tau,0,0)=0.
\label{Complex7}
\end{multline}
\begin{sloppypar}
\noindent
The each equation of the system obtained contains five generalized
hyperspherical functions\index{function!hyperspherical!generalized}
$\fM^{l_0\dot{l}_0}_{mn;\dot{m}\dot{n}}$,
$\fM^{l_0\dot{l}_0}_{m-1,n;\dot{m}\dot{n}}$,
$\fM^{l_0\dot{l}_0}_{m+1,n;\dot{m}\dot{n}}$,
$\fM^{l_0\dot{l}_0}_{mn;\dot{m}-1,\dot{n}}$,
$\fM^{l_0\dot{l}_0}_{mn;\dot{m}+1,\dot{n}}$
and their conjugate. Let us recall that
$\fM^{l_0\dot{l}_0}_{m\pm 1,n;\dot{m}\dot{n}}(\varphi,\epsilon,\theta,\tau,0,0)=
e^{-n(\epsilon+i\varphi)-\dot{n}(\epsilon-i\varphi)}
\fZ^{l_0\dot{l}_0}_{m\pm 1,n;\dot{m}\dot{n}}(\theta,\tau)$,
$\fM^{l_0\dot{l}_0}_{mn;\dot{m}\pm 1,\dot{n}}(\varphi,\epsilon,\theta,\tau,0,0)=
e^{-n(\epsilon+i\varphi)-\dot{n}(\epsilon-i\varphi)}
\fZ^{l_0\dot{l}_0}_{mn;\dot{m}\pm 1,\dot{n}}(\theta,\tau)$ and
$\overset{\ast}{\fM}{}^{l_0\dot{l}_0}_{m\pm 1,n;\dot{m}\dot{n}}
(\varphi,\epsilon,\theta,\tau,0,0)=
e^{-n(\epsilon-i\varphi)-\dot{n}(\epsilon-i\varphi)}
\overset{\ast}{\fZ}{}^{l_0\dot{l}_0}_{m\pm 1,n;\dot{m}\dot{n}}(\theta,\tau)$,
$\overset{\ast}{\fM}{}^{l_0\dot{l}_0}_{mn;\dot{m}\pm 1,\dot{n}}
(\varphi,\epsilon,\theta,\tau,0,0)=
e^{-n(\epsilon-i\varphi)-\dot{n}(\epsilon-i\varphi)}
\overset{\ast}{\fZ}{}^{l_0\dot{l}_0}_{mn;\dot{m}\pm 1,\dot{n}}(\theta,\tau)$.
Therefore,\end{sloppypar}
\[
\frac{\partial\fM^{l_0\dot{l}_0}_{m\pm 1,n;\dot{m}\dot{n}}}{\partial\varphi}=
(-in+i\dot{n})\fM^{l_0\dot{l}_0}_{m\pm 1,n;\dot{m}\dot{n}},\quad
\frac{\partial\fM^{l_0\dot{l}_0}_{m\pm 1,n;\dot{m}\dot{n}}}{\partial\epsilon}=
-(n+\dot{n})\fM^{l_0\dot{l}_0}_{m\pm 1,n;\dot{m}\dot{n}},
\]
\[
\frac{\partial\fM^{l_0\dot{l}_0}_{mn;\dot{m}\pm 1,\dot{n}}}{\partial\varphi}=
(-in+i\dot{n})\fM^{l_0\dot{l}_0}_{mn;\dot{m}\pm 1,\dot{n}},\quad
\frac{\partial\fM^{l_0\dot{l}_0}_{mn;\dot{m}\pm 1,\dot{n}}}{\partial\epsilon}=
-(n+\dot{n})\fM^{l_0\dot{l}_0}_{mn;\dot{m}\pm 1,\dot{n}},
\]
\[
\frac{\partial\overset{\ast}{\fM}{}^{l_0\dot{l}_0}_{m\pm 1,n;\dot{m}\dot{n}}}
{\partial\varphi}=
(in-i\dot{n})\overset{\ast}{\fM}{}^{l_0\dot{l}_0}_{m\pm 1,n;\dot{m}\dot{n}},
\quad
\frac{\partial\overset{\ast}{\fM}{}^{l_0\dot{l}_0}_{m\pm 1,n;\dot{m}\dot{n}}}
{\partial\epsilon}=
-(n+\dot{n})\overset{\ast}{\fM}{}^{l_0\dot{l}_0}_{m\pm 1,n;\dot{m}\dot{n}},
\]
\[
\frac{\partial\overset{\ast}{\fM}{}^{l_0\dot{l}_0}_{mn;\dot{m}\pm 1,\dot{n}}}
{\partial\varphi}=
(in-i\dot{n})\overset{\ast}{\fM}{}^{l_0\dot{l}_0}_{mn;\dot{m}\pm 1,\dot{n}},
\quad
\frac{\partial\overset{\ast}{\fM}{}^{l_0\dot{l}_0}_{mn;\dot{m}\pm 1,\dot{n}}}
{\partial\epsilon}=
-(n+\dot{n})\overset{\ast}{\fM}{}^{l_0\dot{l}_0}_{mn;\dot{m}\pm 1,\dot{n}}.
\]
We apply now the recurrence relations (\ref{RZ1})--(\ref{RZ4}),
(\ref{RZ5})--(\ref{RZ8}) to square brackets containing the
hyperspherical functions. For example,
in virtue of (\ref{RZ2}) the second bracket in (\ref{Complex7})
can be written as
\begin{multline}
e^{-n(\epsilon+i\varphi)-\dot{n}(\epsilon-i\varphi)}\left[
i\frac{\partial\fZ^{l_0\dot{l}_0}_{m-1,n;\dot{m}\dot{n}}}
{\partial\theta}-
\frac{\partial\fZ^{l_0\dot{l}_0}_{m-1,n;\dot{m}\dot{n}}}{\partial\tau}+\right.\\
\left.
+\frac{2i(n-(m-1)\cos\theta^c)}{\sin\theta^c}
\fZ^{l_0\dot{l}_0}_{m-1,n;\dot{m}\dot{n}}\right]=
2\sqrt{(l_0+m)(l_0-m+1)}\fM^{l_0\dot{l}_0}_{mn;\dot{m}\dot{n}}.\label{RR1}
\end{multline}
Further, in virtue of (\ref{RZ1}) for the third bracket
we have
\begin{multline}
e^{-n(\epsilon+i\varphi)-\dot{n}(\epsilon-i\varphi)}\left[
i\frac{\partial\fZ^{l_0\dot{l}_0}_{m+1,n;\dot{m}\dot{n}}}
{\partial\theta}-
\frac{\partial\fZ^{l_0\dot{l}_0}_{m+1,n;\dot{m}\dot{n}}}{\partial\tau}-\right.\\
\left.
-\frac{2i(n-(m+1)\cos\theta^c)}{\sin\theta^c}
\fZ^{l_0\dot{l}_0}_{m+1,n;\dot{m}\dot{n}}\right]=
2\sqrt{(l_0+m+1)(l_0-m)}\fM^{l_0\dot{l}_0}_{mn;\dot{m}\dot{n}}.\label{RR2}
\end{multline}
\begin{sloppypar}\noindent
and so on. In doing so,
we replace all the square brackets in the system (\ref{Complex7}) via
the relations of the type (\ref{RR1})--(\ref{RR2}) and
cancel all the equations by
$\fM^{l_0\dot{l}_0}_{mn;\dot{m}\dot{n}}$
($\overset{\ast}{\fM}{}^{l_0\dot{l}_0}_{mn;\dot{m}\dot{n}}$).
In the result we see that the relativistically invariant system is
reduced to a system of ordinary differential equations,\end{sloppypar}
\begin{multline}
\sum_{k^\prime,\dot{k}^\prime}c^{kk^\prime;\dot{k}\dot{k}^\prime}_{l-1,l;
\dot{l}-1,\dot{l}}\left[(1+i)
\sqrt{(l^2-m^2)(\dot{l}^2-\dot{m}^2)}
\frac{d\boldsymbol{f}^{l_0\dot{l}_0}_{l-1,mk^\prime;\dot{l}-1,
\dot{m}\dot{k}^\prime}}{d r}+\right.\\
+\frac{1}{r}(l-il+\dot{l}+i\dot{l}-2i)
\sqrt{(l^2-m^2)(\dot{l}^2-\dot{m}^2)}
\boldsymbol{f}^{l_0\dot{l}_0}_{l-1,mk^\prime;\dot{l}-1,\dot{m}\dot{k}^\prime}
+\\
+\frac{1}{r}\sqrt{(l+m)(l+m-1)(\dot{l}^2-\dot{m}^2)}
\sqrt{(l_0+m)(l_0-m+1)}
\boldsymbol{f}^{l_0\dot{l}_0}_{l-1,m-1,k^\prime;\dot{l}-1,\dot{m}\dot{k}^\prime}
+\\
\left.+\frac{1}{r}\sqrt{(l-m)(l-m-1)(\dot{l}^2-\dot{m}^2)}
\sqrt{(l_0+m+1)(l_0-m)}
\boldsymbol{f}^{l_0\dot{l}_0}_{l-1,m+1,k^\prime;\dot{l}-1,\dot{m}\dot{k}^\prime}
\right.+\\
+\frac{1}{r}\sqrt{(l^2-m^2)(\dot{l}+\dot{m})(\dot{l}+\dot{m}-1)}
\sqrt{(\dot{l}_0+\dot{m})(\dot{l}_0-\dot{m}+1)}
\boldsymbol{f}^{l_0\dot{l}_0}_{l-1,mk^\prime;\dot{l}-1,\dot{m}-1,\dot{k}^\prime}
+\\
\left.+\frac{1}{r}\sqrt{(l^2-m^2)(\dot{l}-\dot{m})(\dot{l}-\dot{m}-1)}
\sqrt{(\dot{l}_0+\dot{m}+1)(\dot{l}_0-\dot{m})}
\boldsymbol{f}^{l_0\dot{l}_0}_{l-1,mk^\prime;\dot{l}-1,\dot{m}+1,\dot{k}^\prime}
\right]+
\nonumber
\end{multline}
\begin{multline}
+\sum_{k^\prime,\dot{k}^\prime}c^{kk^\prime;\dot{k}\dot{k}^\prime}_{ll;
\dot{l}\dot{l}}\left[(1+i)
m\dot{m}\frac{d\boldsymbol{f}^{l_0\dot{l}_0}_{lmk^\prime;\dot{l}\dot{m}\dot{k}}}
{d r}-
\frac{2i}{r}m\dot{m}\boldsymbol{f}^{l_0\dot{l}_0}_{lmk^\prime;\dot{l}\dot{m}
\dot{k}^\prime}-\right.
\\
-\frac{1}{r}\dot{m}\sqrt{(l+m)(l-m+1)}
\sqrt{(l_0+m)(l_0-m+1)}
\boldsymbol{f}^{l_0\dot{l}_0}_{l,m-1,k^\prime;\dot{l}\dot{m}\dot{k}^\prime}+\\
\left.+\frac{1}{r}\dot{m}\sqrt{(l+m+1)(l-m)}
\sqrt{(l_0+m+1)(l_0-m)}
\boldsymbol{f}^{l_0\dot{l}_0}_{l,m+1,k^\prime;\dot{l}\dot{m}\dot{k}^\prime}
\right.+\\
-\frac{1}{r}m\sqrt{(\dot{l}+\dot{m})(\dot{l}-\dot{m}+1)}
\sqrt{(\dot{l}_0+\dot{m})(\dot{l}_0-\dot{m}+1)}
\boldsymbol{f}^{l_0\dot{l}_0}_{lmk^\prime;\dot{l},\dot{m}-1,\dot{k}^\prime}+\\
\left.+\frac{1}{r}m\sqrt{(\dot{l}+\dot{m}+1)}{(\dot{l}-\dot{m})}
\sqrt{(\dot{l}_0+\dot{m}+1)(\dot{l}_0-\dot{m})}
\boldsymbol{f}^{l_0\dot{l}_0}_{lmk^\prime;\dot{l},\dot{m}+1,\dot{k}^\prime}
\right]+
\nonumber
\end{multline}
\begin{multline}
+\sum_{k^\prime,\dot{k}^\prime}c^{kk^\prime;\dot{k}\dot{k}^\prime}_{l+1,l;
\dot{l}+1,\dot{l}}\left[(1+i)
\sqrt{((l+1)^2-m^2)((\dot{l}+1)^2-\dot{m}^2)}
\frac{d\boldsymbol{f}^{l_0\dot{l}_0}_{l+1,mk^\prime;\dot{l}+1,\dot{m}
\dot{k}^\prime}}{d r}+\right.\\
+\frac{1}{r}(il-l-\dot{l}-i\dot{l}-2)
\sqrt{((l+1)^2-m^2)((\dot{l}+1)^2-\dot{m}^2)}
\boldsymbol{f}^{l_0\dot{l}_0}_{l+1,mk^\prime;\dot{l}+1,\dot{m}\dot{k}^\prime}
-\\
-\frac{1}{r}\sqrt{(l-m+1)(l-m+2)((\dot{l}+1)^2-\dot{m}^2)}
\sqrt{(l_0+m)(l_0-m+1)}
\boldsymbol{f}^{l_0\dot{l}_0}_{l+1,m-1,k^\prime;\dot{l}-1,\dot{m}\dot{k}^\prime}
-\\
\left.-\frac{1}{r}\sqrt{(l+m+1)(l+m+2)((\dot{l}+1)^2-\dot{m}^2)}
\sqrt{(l_0+m+1)(l_0-m)}
\boldsymbol{f}^{l_0\dot{l}_0}_{l+1,m+1,k^\prime;\dot{l}+1,\dot{m}\dot{k}^\prime}
\right.-\\
-\frac{1}{r}\sqrt{((l+1)^2-m^2)(\dot{l}-\dot{m}+1)(\dot{l}-\dot{m}+2)}
\sqrt{(\dot{l}_0+\dot{m})(\dot{l}_0-\dot{m})}
\boldsymbol{f}^{l_0\dot{l}_0}_{l+1,mk^\prime;\dot{l}-1,\dot{m}-1,\dot{k}^\prime}
-\\
\left.-\frac{1}{r}\sqrt{((l+1)^2-m^2)(\dot{l}+\dot{m}+1)(\dot{l}+\dot{m}+2)}
\sqrt{(\dot{l}_0+\dot{m}+1)(\dot{l}_0-\dot{m})}
\boldsymbol{f}^{l_0\dot{l}_0}_{l+1,mk^\prime;\dot{l}+1,\dot{m}+1,\dot{k}^\prime}
\right]+
\\
+\kappa^c\boldsymbol{f}^{l_0}_{lmk}(r)=0,\nonumber
\end{multline}
\begin{multline}
\sum_{k^\prime,\dot{k}^\prime}
\overset{\ast}{c}{}^{kk^\prime;\dot{k}\dot{k}^\prime}_{l-1,l;
\dot{l}-1,\dot{l}}\left[(1-i)
\sqrt{(l^2-m^2)(\dot{l}^2-\dot{m}^2)}
\frac{d\overset{\ast}{\boldsymbol{f}}{}^{l_0\dot{l}_0}_{l-1,mk^\prime
\dot{l}-1,\dot{m}\dot{k}^\prime}}
{d r^\ast}+\right.\\
+\frac{1}{r^\ast}(l-il+\dot{l}+i\dot{l}+2i)
\sqrt{(l^2-m^2)(\dot{l}^2-\dot{m}^2)}
\overset{\ast}{\boldsymbol{f}}{}^{l_0\dot{l}_0}_{l-1,mk^\prime;
\dot{l}-1,\dot{m}\dot{k}^\prime}+\\
+\frac{1}{r^\ast}\sqrt{(l^2-m^2)(\dot{l}+\dot{m})(\dot{l}+\dot{m}-1)}
\sqrt{(\dot{l}_0+\dot{m})(\dot{l}_0-\dot{m}+1)}
\overset{\ast}{\boldsymbol{f}}{}^{l_0\dot{l}_0}_{l-1,mk^\prime;
\dot{l}-1,\dot{m}-1,\dot{k}^\prime}+\\
\left.
+\frac{1}{r^\ast}\sqrt{(l^2-m^2)(\dot{l}-\dot{m})(\dot{l}-\dot{m}-1)}
\sqrt{(\dot{l}_0+\dot{m}+1)(\dot{l}_0-\dot{m})}
\overset{\ast}{\boldsymbol{f}}{}^{l_0\dot{l}_0}_{l-1,mk^\prime;
\dot{l}-1,\dot{m}+1,\dot{k}^\prime}\right.+\\
+\frac{1}{r^\ast}\sqrt{(l+m)(l+m-1)(\dot{l}^2-\dot{m}^2)}
\sqrt{(l_0+m)(l_0-m+1)}
\overset{\ast}{\boldsymbol{f}}{}^{l_0\dot{l}_0}_{l-1,m-1,k^\prime;
\dot{l}-1,\dot{m}\dot{k}^\prime}+\\
\left.+\frac{1}{r^\ast}\sqrt{(l-m)(l-m-1)(\dot{l}^2-\dot{m}^2)}
\sqrt{(l_0+m+1)(l_0-m)}
\overset{\ast}{\boldsymbol{f}}{}^{l_0\dot{l}_0}_{l-1,m+1,k^\prime;
\dot{l}-1,\dot{m}\dot{k}^\prime}\right]+\\
\nonumber
\end{multline}
\begin{multline}
+\sum_{k^\prime,\dot{k}^\prime}
\overset{\ast}{c}{}^{kk^\prime;\dot{k}\dot{k}^\prime}_{ll;\dot{l}\dot{l}}\left[
(1-i)m\dot{m}
\frac{d\overset{\ast}{\boldsymbol{f}}{}^{l_0\dot{l}_0}_{lmk^\prime;
\dot{l}\dot{m}\dot{k}^\prime}}
{d r^\ast}
+\frac{2i}{r^\ast}m\dot{m}
\overset{\ast}{\boldsymbol{f}}{}^{l_0\dot{l}_0}_{lmk^\prime;
\dot{l}\dot{m}\dot{k}^\prime}-\right.
\\
-\frac{1}{r^\ast}m\sqrt{(\dot{l}+\dot{m})(\dot{l}-\dot{m}+1)}
\sqrt{(\dot{l}_0+\dot{m})(\dot{l}_0-\dot{m})}
\overset{\ast}{\boldsymbol{f}}{}^{l_0\dot{l}_0}_{lmk^\prime;
\dot{l},\dot{m}-1,\dot{k}^\prime}+\\
\left.
+\frac{1}{r^\ast}m\sqrt{(\dot{l}+\dot{m}+1)(\dot{l}-\dot{m})}
\sqrt{(\dot{l}_0+\dot{m}+1)(\dot{l}_0-\dot{m})}
\overset{\ast}{\boldsymbol{f}}{}^{l_0\dot{l}_0}_{lmk^\prime;
\dot{l},\dot{m}+1,\dot{k}^\prime}\right.-
\\
-\frac{1}{r^\ast}\dot{m}\sqrt{(l+m)(l-m+1)}
\sqrt{(l_0+m)(l_0-m+1)}
\overset{\ast}{\boldsymbol{f}}{}^{l_0\dot{l}_0}_{l,m-1,k^\prime;
\dot{l}\dot{m}\dot{k}^\prime}+\\
\left.
+\frac{1}{r^\ast}\dot{m}\sqrt{(l+m+1)(l-m)}
\sqrt{(l_0+m+1)(l_0-m)}
\overset{\ast}{\boldsymbol{f}}{}^{l_0\dot{l}_0}_{l,m+1,k^\prime;
\dot{l}\dot{m}\dot{k}^\prime}\right]+
\nonumber
\end{multline}
\begin{multline}
+\sum_{k^\prime,\dot{k}^\prime}
\overset{\ast}{c}{}^{kk^\prime;\dot{k}\dot{k}^\prime}_{l+1,l;
\dot{l}+1,\dot{l}}\left[(1-i)
\sqrt{((l+1)^2-m^2)((\dot{l}+1)^2-\dot{m}^2)}
\frac{d\overset{\ast}{\boldsymbol{f}}{}^{l_0\dot{l}_0}_{l+1,mk^\prime;
\dot{l}+1,\dot{m}\dot{k}^\prime}}
{d r^\ast}+\right.\\
+\frac{1}{r^\ast}(il-l-\dot{l}-i\dot{l}-2+2i)
\sqrt{((l+1)^2-m^2)((\dot{l}+1)^2-\dot{m}^2)}
\overset{\ast}{\boldsymbol{f}}{}^{l_0\dot{l}_0}_{l+1,mk^\prime;
\dot{l}+1,\dot{m}\dot{k}^\prime}+\\
+\frac{1}{r^\ast}\sqrt{((l+1)^2-m^2)(\dot{l}-\dot{m}+1)(\dot{l}-\dot{m}+2)}
\sqrt{(\dot{l}_0+\dot{m})(\dot{l}_0-\dot{m}+1)}
\overset{\ast}{\boldsymbol{f}}{}^{l_0\dot{l}_0}_{l+1,mk^\prime;
\dot{l}+1,\dot{m}-1,\dot{k}^\prime}-\\
\left.
-\frac{1}{r^\ast}\sqrt{((l+1)^2-m^2)(\dot{l}+\dot{m}+1)(\dot{l}+\dot{m}+2)}
\sqrt{(\dot{l}_0+\dot{m}+1)(\dot{l}_0-\dot{m})}
\overset{\ast}{\boldsymbol{f}}{}^{l_0\dot{l}_0}_{l+1,mk^\prime;
\dot{l}+1,\dot{m}+1,\dot{k}^\prime}\right.-
\\
-\frac{1}{r^\ast}\sqrt{(l-m+1)(l-m+2)((\dot{l}+1)^2-\dot{m}^2)}
\sqrt{(l_0+m)(l_0-m+1)}
\overset{\ast}{\boldsymbol{f}}{}^{l_0\dot{l}_0}_{l+1,m-1,k^\prime;
\dot{l}+1,\dot{m}\dot{k}^\prime}-\\
\left.
-\frac{1}{r^\ast}\sqrt{(l+m+1)(l+m+2)((\dot{l}+1)^2-\dot{m}^2)}
\sqrt{(l_0+m+1)(l_0-m)}
\overset{\ast}{\boldsymbol{f}}{}^{l_0\dot{l}_0}_{l+1,m+1,k^\prime;
\dot{l}+1,\dot{m}\dot{k}^\prime}\right]+
\\
+\dot{\kappa}^c
\overset{\ast}{\boldsymbol{f}}{}^{l_0\dot{l}_0}_{lm;
\dot{l}\dot{m}}=0.
\label{RFS}
\end{multline}
Substituting solutions of this system into the series (\ref{Fourier1}) and
(\ref{Fourier2}), we obtain a solution of the boundary value problem.

\end{document}